\documentclass[a4paper,12pt,twoside,openright]{book}

\usepackage{amsmath}
\usepackage{amsfonts}
\usepackage{amssymb}
\usepackage{graphicx}
\usepackage{psfrag}
\usepackage{hyphenat}

\def\nn{\nonumber}

\begin{document}
\frontmatter

\begin{titlepage}

\begin{centering}

\vspace{3cm}

{\Huge {\bf Spiky strings and the AdS/CFT correspondence}}

\vspace{3cm}

{\Large {\textsc{Manuel Losi}}}

\vspace{2cm}

{\large Dissertation submitted for the Degree of}\\
\vspace{1cm}
{\large Doctor of Philosophy}\\
\vspace{1cm}
{\large at the University of Cambridge}\\

\vspace{2cm}

{\large \emph{Department of Applied Mathematics and Theoretical Physics}}\\ 
\vspace{1cm}
{\large \emph{Gonville \& Caius College}}\\ 
\vspace{1cm}
{\large \emph{University of Cambridge, UK}}\\

\vspace{2cm}

{\large November 2010}

\end{centering}

\end{titlepage}

\newpage
$\phantom{a}$
\newpage

\section*{\begin{center} Acknowledgments \end{center}}

First of all, I am deeply grateful to my supervisor, Nicholas Dorey, for his advice and guidance throughout my PhD. His insight and constant support have proved invaluable in these years of research.

I would also like to thank the whole community of the Department of Applied Mathematics and Theoretical Physics for providing such a friendly,  active and stimulating working environment.

I am also grateful to Gonville and Caius College and to the Department of Applied Mathematics and Theoretical Physics for funding my conference trips.

This work was partly supported by the Isaac Newton Trust European Research Studentship (jointly funded by the Cambridge European Trust and the Isaac Newton Trust).

Last but not least, I would like to express my deepest gratitude to my parents and my sister, for constantly encouraging and supporting me, both morally and financially, throughout my studies. I would not have made it to this point without them.

\newpage
$\phantom{a}$
\newpage

\section*{\begin{center} Declaration \end{center}}

This dissertation is the result of my own work and includes nothing which is the outcome of work done in collaboration except where specifically indicated in the text. The research described in this dissertation was carried out in the Department of Applied Mathematics and Theoretical Physics, Cambridge University, between September 2006 and August 2010. Except where reference is made to the work of others, all the results are original and based on the following works of mine:

\begin{enumerate}
	\item \textbf{``Spiky Strings and Spin Chains''}
	
	      N. Dorey and M. Losi
	      
	      published in Int. J. Mod. Phys. A25:4641-4690, 2010, with the title ``Spiky Strings and Gauge Theory Partons'', ArXiv:0812.1704
	      
	\item \textbf{``Giant Holes''}
	
	      N. Dorey and M. Losi
	      
	      J. Phys. A43 (2010) 285402, ArXiv:1001.4750
	      
	\item \textbf{``Spiky Strings and Giant Holes''}
	
	      N. Dorey and M. Losi
	      
	      published in JHEP 12 (2010) 014, ArXiv:1008.5096
\end{enumerate}

These papers are referred to as \cite{Dorey:2008vp}, \cite{Dorey:2010iy} and \cite{Dorey:2010id} respectively in the bibliography. Chapter \ref{ch:FGA} is mostly based on \cite{Dorey:2010id}, while chapter \ref{sec:explicit_solutions} is mostly based on \cite{Dorey:2008vp} and \cite{Dorey:2010iy}. Section \ref{sec:glue_arcs} and chapter \ref{sec:pp-wave} are instead based on as yet unpublished material, which is also the result of work in collaboration with N. Dorey.

None of the original works contained in this dissertation has been submitted by me for any other degree, diploma or similar qualification.
\paragraph{}
\paragraph{}
\begin{flushright}
	Manuel Losi
	
	\emph{Cambridge, UK}
	
	\emph{29th November 2010}
\end{flushright}

\newpage

\section*{\begin{center} Abstract \end{center}}

In this dissertation, we explore some aspects of semiclassical type IIB string theory on $AdS_3 \times S^1$ and on pure $AdS_3$ in the limit of large angular momentum $S$.

We first focus on the integrability technique known as finite-gap formalism for strings in $AdS_3 \times S^1$, leading to the definition of a hyperelliptic Riemann surface, the spectral curve, which encodes, albeit in a rather implicit fashion, the semiclassical spectrum of a very large family of string solutions. Then, we show that, in the large angular momentum limit, the spectral curve separates into two distinct surfaces, allowing the derivation of an explicit expression for the spectrum, which is correspondingly characterised by two separate branches. The latter may be interpreted in terms of two kinds of spikes appearing on the strings: ``large'' spikes, yielding an infinite contribution to the energy and angular momentum of the string, and ``small'' spikes, representing finite excitations over the background of the ``large'' spikes.

According to the AdS/CFT correspondence, strings moving in $AdS_3 \times S^1$ should be dual to single trace operators in the $\mathfrak{sl}(2)$ sector of $\mathcal{N} = 4$ super Yang-Mills theory. The corresponding one-loop spectrum in perturbation theory may also be computed through integrability methods and, in the large conformal spin limit $S \to \infty$ (equivalent to the $AdS_3$ angular momentum in string theory) is also expressed in terms of a spectral curve and characterised in terms of the so-called holes. We show that, with the appropriate identifications and with the usual extrapolation from weak to strong 't Hooft coupling described by the cusp anomalous dimension, the large-$S$ spectra of gauge theory and of string theory coincide. Furthermore, we explain how ``small'' and ``large'' holes may be identified with ``small'' and ``large'' spikes.

Finally, we discuss several explicit spiky string solutions in $AdS_3$ which, at the leading semiclassical order, display the previously studied finite-gap spectrum. We compute the spectral curves of these strings in the large $S$ limit, finding that they correspond to specific regions of the moduli space of the finite-gap curves. We also explain how ``large'' spikes may be used in order to extract a discrete system of degrees of freedom from string theory, which can then be matched with the degrees of freedom of the dual gauge theory operators, and how ``small'' spikes are in fact very similar to the Giant Magnons living in $\mathbb{R} \times S^2$.

\tableofcontents

\mainmatter

\chapter{Introduction and literature review}

\section{The AdS/CFT correspondence}
\label{sec:AdSCFT}

String theory was originally devised as a model of strong interactions, describing the forces arising between quarks in terms of strings connecting them. After QCD proved to be much more successful at this task, interest in string theory declined until it was identified as a promising canditate for a unified field theory, providing a quantum model of all fundamental interactions, including gravity. While the progress made in this area is still insufficient to yield a conclusive answer, another aspect of string theory has been receiving increasing attention during the last decade: the gauge-string duality.

The first observation leading in this direction is due to 't Hooft \cite{'tHooft:1973jz}, who, in an attempt to solve QCD, and, more generally, any $U(N)$ Yang-Mills theory, proposed to consider the rank $N$ of the gauge group as a free parameter, together with the coupling $g_{YM}$. He then showed that, as $N \to \infty$, while the 't Hooft coupling $\lambda = g_{YM}^2 N$ is kept fixed, the Feynman diagrams appearing in the perturbative expansion of any $U(N)$ gauge theory are multiplied by a power of $N$ which is determined uniquely by the genus of the surface they span. It is then possible to rewrite the expansion as a sum over the genus: for instance, in the case of the free energy, we have
\begin{equation}
 \mathcal{F} = \sum_{g=0}^{\infty} N^{2-2g} \sum_{l=0}^{\infty} c_{g,l} \lambda^l
\label{eq:free_energy_U(N)_theory_tHooft_limit}
\end{equation}
This kind of expression is very similar to the sum over worldsheet topologies representing the string theory perturbative expansion.

Building on these considerations and on Witten's \cite{Witten:1995gx} and Klebanov's \cite{Klebanov:1997kc} results concerning stacks of coincident D-branes, Maldacena \cite{Maldacena:1997re} conjectured the now well-known AdS/CFT correspondence\footnote{For a recent review of the subject, see \cite{Serban:2010sr}, which we will loosely follow throughout this introductory chapter.}
. The statement, in its strongest form, is that $\mathcal{N} = 4$ super Yang-Mills theory on $\mathbb{R} \times S^3$ with gauge group $SU(N)$ is equivalent to type IIB superstring theory on $AdS_5 \times S^5$, provided we identify the parameters of the two models as follows:
 \begin{equation}
   4 \pi g_s = g^2_{YM} = \frac{\lambda}{N} \:, \qquad \frac{R^2}{\alpha'} = \sqrt{\lambda} \:,
\label{eq:identification_parameters_AdSCFT}
\end{equation}
where $g_s$ is the string coupling, $\alpha' = l_s^2$ is the square of the fundamental string length and $R$ is the common radius of $AdS_5$ and $S^5$. Both theories share the same global symmetry group $PSU(2,2|4)$, with bosonic subgroup $SO(2,4) \times SO(6)$. Super Yang-Mills theory is invariant under the conformal group $SO(4,2) \simeq SU(2,2)$, which is also the isometry group of $AdS_5$, and under rotations of its six scalar fields, represented by $SO(6) \simeq SU(4)$, which is the isometry group of $S^5$. The Cartan subalgebra of this group yields six conserved charges: three integer angular momenta $J_1, J_2, J_3$ from $SO(6)$ plus two half-integer angular momenta $S_1, S_2$ and one last charge from $SO(2,4)$, which is interpreted as either the energy $E$ in string theory or as the scaling dimension $\Delta$ in gauge theory. The equivalence of the models then implies the equivalence of the spectra, which is formulated in terms of the so-called ``dictionary'' of the correspondence: each gauge theory operator with a set of charges $(\Delta, S_1, S_2, J_1, J_2, J_3)$ is associated with a string state carrying the same charges $(E, S_1, S_2, J_1, J_2, J_3)$. In other words, the spectrum of scaling dimensions must coincide with the spectrum of string energies, for all values of $\lambda$ and $N$:
\begin{equation}
 \Delta (\lambda, N, S_1, S_2, J_1, J_2, J_3) = E (\lambda, N, S_1, S_2, J_1, J_2, J_3)
\label{eq:equivalence_of_spectra_AdSCFT}
\end{equation}
Due to the fact that gauge theory is defined on $\mathbb{R} \times S^3$, which is the boundary of $AdS_5$, this equivalence has been interpreted as a manifestation of the holographic principle \cite{'tHooft:1993gx}, according to which all the information related to a given volume of space can be represented on its boundary.

A very interesting feature of the correspondence is the fact that, due to the relationship between the parameters, Yang-Mills theory is in the perturbative regime for $\lambda \ll 1$, while in the case of string theory this happens for $\lambda \gg 1$. Hence, if it was possible to show that the conjecture is true, it could then be exploited in order to turn highly non-perturbative calculations in one of the models into simple perturbative computations in the other. Ideally, the next step would be to apply this reasoning to QCD, studying its strong-coupling behaviour by identifying its string dual and then considering its perturbative regime.

However, there is presently no proposed dual for QCD and, even in the AdS/CFT case, the two spectra are very hard to determine for generic $N$ and $\lambda$ and thus so far most efforts have been focusing on proving the weaker version of the correspondence, which states that the equivalence should hold in the 't Hooft limit, $N \to \infty$ with $\lambda$ fixed. As it is already apparent from the previous expression for the free energy, when the rank of the gauge group becomes large, the planar diagrams dominate over the higher-genus contributions, while the identification of the parameters of the two theories immediately tells us that the string coupling $g_s$ vanishes. Therefore, \emph{planar} $\mathcal{N} = 4$ super Yang-Mills theory on $\mathbb{R} \times S^3$ should be dual to \emph{free} type IIB string theory on $AdS_5 \times S^5$.
\paragraph{}

Even with these simplifications in place, the strong/weak nature of the correspondence makes it hard to test, since both spectra are only accessible in the perturbative regime and there is no region of parameter space in which this regime applies to both theories at the same time.

In order to circumvent this difficulty, the early tests of the correspondence focused on the so-called \emph{chiral primary operators}:
\begin{equation}
 \textrm{Tr } Z^{J_1} \qquad\qquad \textrm{Tr } Y^{J_2} \qquad\qquad \textrm{Tr } X^{J_3} \qquad\qquad
\label{eq:def_chiral_primaries}
\end{equation}
where $Z,Y,X$ are complex linear combinations of the scalar fields in super Yang-Mills theory, each of which carries one unit of R-charge $J_1$, $J_2$ and $J_3$ respectively. Such operators are BPS and thus their scaling dimensions are protected by supersymmetry and receive no quantum corrections, so that they are independent of the coupling $\lambda$. It is then possible to compare them directly to classical string energies in the $\lambda \gg 1$ limit, showing that their dual string states are point-particles rotating along a great circle of the $S^5$, with angular momentum $J_1, J_2$ and $J_3$ respectively.

Subsequently, Berenstein, Maldacena and Nastase \cite{Berenstein:2002jq} proposed to study operators with a very large R-charge $J = J_1$. The main idea was that, in the limit $J \to \infty$, the perturbative expansion of gauge theory should turn into a power series in $\lambda' = \lambda / J^2$ with corrections scaling as higher powers of $1/J$:
\begin{eqnarray}
 \Delta(\lambda) & = & \Delta_0 + \lambda \Delta_1 + \lambda^2 \Delta_2 + \ldots \nn \\
                 & = & \Delta_0 + \lambda' \left[ \Delta_1^{(0)} + O \left( \frac{1}{J} \right) \right]
                        + \lambda'^2 \left[ \Delta_2^{(0)} + O \left( \frac{1}{J} \right) \right] + \ldots \nn\\
                 &   & \qquad \textrm{as } J \to \infty
\label{eq:BMN_scaling_of_Delta}
\end{eqnarray}
this kind of behaviour is known as BMN scaling. Provided $\lambda'$ is sufficiently small, such an expression is still well-behaved even at strong coupling $\lambda \gg 1$ and can then be compared to perturbative string energies. In particular, these authors considered the semiclassical quantisation of a point-like string orbiting along the equator of $S^5$ with large angular momentum $J$, which is the dual of the chiral primary $\textrm{Tr} (Z^J)$ (the so-called BMN vacuum). In the large curvature limit $\lambda \gg 1$ string theory can be treated semiclassically and, for $J \to \infty$, i.e. when the particle approaches the speed of light, the target space metric reduces to the plane-wave background. The exact quantum spectrum of this system is described in terms of eight free bosons and eight free fermions and had already been computed in \cite{Metsaev:2001bj}. The result can then be Taylor expanded in $\lambda'$ and is $O(J^0)$ in the limit $J \to \infty$, and thus it can be compared to the super Yang-Mills spectrum at any loop order and at leading order in $J$.

The spectrum was then reproduced on the gauge side by evaluating a subset of the relevant Feynman diagrams (the others can be neglected for very large $J$) at one loop, exploiting the BMN scaling.

\section{The emergence of integrability}

The discovery of integrability on both sides of the correspondence initiated a period of rapid progress which allowed to go far beyond these early tests and towards the exact determination of the common spectrum of the two models. Bena, Polchinski and Roiban \cite{Bena:2003wd} proved that the Metsaev-Tseytlin sigma model \cite{Metsaev:1998it}, which describes type IIB string theory on $AdS_5 \times S^5$, is classically integrable by identifying an infinite tower of conserved charges. Meanwhile, Minahan and Zarembo \cite{Minahan:2002ve} showed that the 1-loop dilatation operator (which determines the 1-loop contribution to the scaling dimension) in the $\mathfrak{so}(6)$ sector, consisting of operators of the type $\textrm{Tr } (ZZZZXYZZXZ \ldots Z)$, is equivalent to the hamiltonian of an integrable $\mathfrak{so}(6)$ spin chain, where each elementary field inside the trace defines one site of the lattice and the orientation of the corresponding spin is determined by the type of field (i.e. $X$, $Y$ or $Z$).

The $\mathfrak{su}(2)$ subsector of this model, only containing operators with two different elementary fields inside the trace (e.g. $Z$ and $Y$), reduces to the Heisenberg $XXX_{1/2}$ ferromagnet, which can be diagonalised through the well-known technique of the Bethe ansatz \cite{Bethe:1931hc}. The method can be extended and applied to a much wider class of models (see \cite{Faddeev:1996iy} for a review), including the full $\mathfrak{so}(6)$ spin chain at 1-loop.

Integrability was later shown to hold also for the full $\mathfrak{psu}(2,2|4)$ dilatation operator at one loop \cite{Beisert:2003jj}. In \cite{Beisert:2003tq}, the 2-loop contribution in the $\mathfrak{so}(6)$ sector was computed and proved to be integrable, prompting the conjecture that the dilatation operator should be integrable at all loops, and an expression for the 3-loop term was proposed under this assumption; shortly afterwards that expression was confirmed by supersymmetry considerations \cite{Beisert:2003ys}. These higher-loop corrections to the dilatation operator, if further restricted to the $\mathfrak{su}(2)$ sector, can be reproduced by the Inozemtsev spin chain \cite{Serban:2004jf}, in terms of its higher conserved charges. The latter can then be diagonalised by Bethe ansatz in the asymptotic limit, i.e. in the limit in which the number of $Z$ fields inside the trace becomes infinitely large, while the number of impurities $Y$ remains finite. The diagonalisation method in this particular limit was called \emph{Asymptotic Bethe Ansatz} (ABA). 
\paragraph{}

These results allowed to carry out more detailed tests on the correspondence in the BMN limit, as they provided the perturbative expansion of the spectrum of Yang-Mills theory up to three loops in the $\mathfrak{su}(2)$ sector. This expansion has the expected BMN scaling and, in the $J \to \infty$ limit, yields a spectrum of the form \eqref{eq:BMN_scaling_of_Delta} up to $O(\lambda'^3)$, including corrections in $1/J$. As expected, this matched the semiclassical $O(J^0)$ plane-wave spectrum of the rotating point particle at all orders considered in the case of two $Y$ impurities appearing inside the trace. In order to go a step further, 1-loop corrections to the semiclassical approximation in string theory were calculated in the two- and three-impurity cases \cite{Callan:2003xr,Callan:2004ev}. The corresponding spectra could then be compared to the gauge theory spectrum obtained by ABA up to $O(1/J)$. A mismatch was found in the $O(1/J)$ coefficient at three loops, $\Delta_3^{(1)}$, in both cases.

This result was interpreted in terms of an order of limits problem due to the particular nature of the BMN limit. In fact, on the gauge theory side, we have to start with $\lambda \ll 1$ for the perturbative expansion to hold, then we take $J, \lambda \to \infty$ with $\lambda' = \textrm{const.}$, noticing that the expansion still makes sense in this limit. On the other hand, on the string side, we begin with $\lambda \gg 1$, so that we may work in the weak curvature regime, and then we simply take $J \to \infty$, with $\lambda' = \textrm{const.}$ as before. Clearly, the two procedures are different and thus there is no guarantee that the limits obtained on the two sides will be identical even if the spectra of the two theories are actually identical. In particular, the presence of terms that violate BMN scaling in the spectrum would lead to this kind of problem, and in fact it was later shown that such terms arise in the dilatation operator at four loops.
\paragraph{}

Nonetheless, the agreement of the low-order coefficients in the BMN limit helped strengthen the confidence in integrability techniques and encouraged attempts to extend such techniques to larger and larger families of objects on both sides of the correspondence.

As far as super Yang-Mills theory is concerned, a change of perspective was suggested by Staudacher \cite{Staudacher:2004tk} in analogy with the simple case of the coordinate Bethe ansatz which diagonalises the Heisenberg model corresponding to the 1-loop dilatation operator in the $\mathfrak{su}(2)$ sector. We define the spin chain configuration corresponding to the chiral primary $\textrm{Tr } (Z^L)$ as the vacuum state and then, when considering configurations in which one or more of the $Z$ fields is replaced by a $Y$ field, we identify each $Y$ impurity with an excitation of the spin chain, also called a ``magnon''. The eigenstates of the spin chain Hamiltonian, which yields the 1-loop contribution to the scaling dimension, are obtained as linear combinations of states with the same number of magnons. The coefficients of such linear combinations can then be interpreted as wavefunctions describing the positions of the magnons along the quantum spin chain.

In the asymptotic limit $J \to \infty$, the chain becomes infinitely long due to the presence of an infinite number of $Z$ fields inside the trace, and thus, as long as the number of impurities remains finite, it is possible to describe the eigenstates as configurations representing the scattering of magnons, which are considered quasi-particles in this picture. In fact, the Bethe ansatz procedure associates a conserved energy and a conserved momentum to each of these particles and then leads to the conclusion that they undergo \emph{factorised scattering} (i.e. the $n$-magnon S-matrix is just the product of $n(n-1)/2$ two-magnon S-matrices). This phenomenon is a well-known consequence of integrability.

The key point is that the Bethe ansatz equations, which determine the magnon momenta and, through these, the magnon scattering wavefunction and the magnon energies, are formulated in terms of the 2-magnon S-matrix. Staudacher's proposal was then to try to diagonalise the dilatation operator in the asymptotic limit (in other words, considering only operators having an infinite number of $Z$ fields inside the trace) by interpreting the corresponding spin chain configurations as magnon scattering processes, where the definition of magnon had of course to be extended in order to include all the other elementary fields as possible excitations.

This new approach presented one important advantage: the asymptotic limit eliminates the problem posed by length-changing interactions, which already appear at the 2-loop level and cause apparently insurmountable complications in the formulation of Bethe ans\"atze for chains of finite length. Furthermore, focusing on the S-matrix of the coordinate Bethe ansatz proved to be much more effective than the previously employed algebraic Bethe ansatz technique. In fact, the S-matrices associated with larger and larger sectors of super Yang-Mills theory were progressively determined out of symmetry considerations alone.
\paragraph{}

On the string side, remarkable progress was achieved with the formulation of the finite-gap construction \cite{Kazakov:2004qf}, which allowed to represent in terms of algebraic curves generic classical string solutions on $\mathbb{R} \times S^3$ (corresponding to the $\mathfrak{su}(2)$ sector in gauge theory) and, in particular, their spectra. The reduction to this subspace is only consistent at the semiclassical level, since quantum fluctuations of the string still involve the full background $AdS_5 \times S^5$, hence this integrability technique only provides us with the semiclassical spectrum of the string. This means that, once we take the $J \to \infty$ limit, we obtain the leading order contribution $O(J^0)$ at all orders in $\lambda'$, but we are unable to determine the $O(1/J)$ corrections in the BMN expansion. Semiclassical string theory on $\mathbb{R} \times S^3$ reduces to the semiclassical $SU(2)$ Principal Chiral Model and its Hamiltonian can be diagonalised by studying the corresponding algebraic curve (the spectral curve) or, equivalently, solving a set of integral equations.

In super Yang-Mills theory the spectrum of the $\mathfrak{su}(2)$ sector is determined by the $XXX_{1/2}$ Bethe ansatz at one loop and by the Inozemtsev asymptotic Bethe ansatz at two and three loops. In the asymptotic $J \to \infty$ limit, all these contributions can be expressed in terms of algebraic curves since the Bethe equations approach a continuum limit (also known as the thermodynamic limit) in which they also reduce to integral equations. If we then take $\lambda \gg 1$ with $\lambda' = \textrm{const.}$ as usual, we can compare the result with the above calculation from string theory. The two spectra were found to match at leading order in $J$ and up to $O(\lambda'^2)$.

This result allowed to confirm that the spectrum of the whole $\mathfrak{su}(2)$ sector in gauge theory is equal to the spectrum of the whole $\mathbb{R} \times S^3$ subspace in string theory, up to the orders considered, thus greatly extending the tests of the AdS/CFT conjecture in the BMN limit.

The finite-gap construction was later extended to strings in $AdS_3 \times S^1$ ($\mathfrak{sl}(2)$ sector) \cite{Kazakov:2004nh} and to the full target space $AdS_5 \times S^5$ (full $\mathfrak{psu}(2,2|4)$ algebra) \cite{Beisert:2005bm}.

\section{The all-loop asymptotic Bethe ansatz}

Following the successful comparison of the spectra through the finite-gap formalism, Beisert, Dippel and Staudacher \cite{Beisert:2004hm} proposed an asymptotic Bethe ansatz (known as the BDS ansatz), valid for the $\mathfrak{su}(2)$ sector at all orders in $\lambda$ and determined by assuming that integrability should hold at all loops, as previously conjectured, and that the spectrum should exhibit the BMN scaling at all loops. This ansatz reproduced the known weak coupling results, i.e. the $O(\lambda)$ term obtained from the $XXX_{1/2}$ Bethe ansatz and the $O(\lambda^2)$ and $O(\lambda^3)$ terms computed through the asymptotic Inozemtsev spin chain. It also yielded the BMN plane wave spectrum at strong coupling.

The BDS magnon dispersion relation has the interesting feature of being periodic in the magnon momentum $p$:
\begin{equation}
 E_{BDS} = \sqrt{1 + \frac{\lambda}{\pi^2} \sin^2 \frac{p}{2}} 
\label{eq:BDS_dispersion_relation}
\end{equation}
The periodicity appeared natural on the gauge side, as a manifest consequence of the discreteness of the spin chain (which is due to the discrete structure of the corresponding single-trace operator), but seemed rather puzzling on the string side.

Of course this property disappears in the usual BMN limit: we first need to Taylor-expand in $\lambda \ll 1$, reobtaining the perturbative contributions to the scaling dimension from one loop onwards, and then we must take $J, \lambda \to \infty$ with $\lambda' \ll 1$ fixed and expand again in powers of $1/J$. If we consider the 1-loop term, we see that, because of the quantisation conditions imposed by the Bethe ansatz, the magnon momentum $p = 2 \pi n/ (J+1)$ is inversely proportional to the length of the spin chain, and thus becomes infitesimally small in the asymptotic limit $J \to \infty$. Hence, when we perform the second expansion, we approximate $\sin (p/2) \simeq p/2$ and remove the periodicity.

However, since equation \eqref{eq:BDS_dispersion_relation} is valid at all values of the coupling $\lambda$, we can consider its limit as $\lambda \gg 1$:
\begin{equation}
 E_{BDS} \simeq \frac{\sqrt{\lambda}}{\pi} \left| \sin \frac{p}{2} \right|
\label{eq:giant_magnon_dispersion_relation}
\end{equation}
This new periodic dispersion relation should then correspond to some family of classical string configurations which was not touched by the previously discussed BMN analyses.

The mystery was solved in \cite{Hofman:2006xt}, where such objects were found to be classical arc-shaped strings with endpoints on the equator of $S^5$ and drooping towards one of the poles. These strings rigidly rotate around the north pole-south pole axis with a very large angular momentum and are infinitely long in worldsheet units. Their momentum parameter, appearing in the dispersion relation \eqref{eq:giant_magnon_dispersion_relation}, was identified with the angular separation between the two endpoints, thus explaining the periodicity in a natural way. Being the string duals of $\mathfrak{su}(2)$ magnons, these objects were named ``Giant Magnons''. They represent solitonic excitations above the BMN vacuum which yield a large but finite $O(\sqrt{\lambda})$ contribution to the vacuum energy ($E_0 = J$), given by $E_{BDS}$. In fact, they are continuosly connected to the small fluctuations of the rotating point-like string discussed earlier.

In \cite{Beisert:2004hm}, the BDS equations were also written as integral equations in the thermodynamic limit, in order to compare them with the equations determining the spectrum of the Principal Chiral Model in the BMN limit. A discrepancy was found at the 3-loop order $O(\lambda'^3)$. Again, this was interpreted as an order of limits problem and presented further evidence of the fact that BMN scaling should not hold at all loops due to the existence of higher order BMN-violating terms, which were associated with \emph{wrapping interactions}. These arise because of the fact that $n$-loop contributions to the dilatation operator are equivalent, in the spin chain picture, to conserved charges associated with operators which act on $(n+1)$ adjacent sites. It then follows that, when the length $L$ of the spin chain is lower than or equal to $(n-1)$, the interaction ``wraps'' around the spin chain touching all the sites at least once, invalidating the standard Bethe ansatz approach.

Wrapping terms do not contribute to the asymptotic Bethe ansatz in perturbative gauge theory, since the length $L$ of the spin chain, which equals the large R-charge $J$ plus the (small) number of impurities, is always assumed to be much larger than the loop order considered.

This is not the case at strong coupling, even in the asymptotic limit. If $\lambda$ is very large, the all-loop spectrum of gauge theory is required for a comparison with the string theory result, even in the BMN limit, and hence wrapping interactions will matter, since we consider $J$ to be large but still finite. Wrapping corrections for a chain of fixed length $L$ appear at the loop order $\lambda^{2L}$ and therefore they are exponentially suppressed with the size of the chain in perturbative super Yang-Mills theory. This type of behaviour can be easily seen to generate BMN-violating terms, but the exact structure of such terms is very hard to determine. These modifications of the spectrum due to wrapping are also known as \emph{finite-size effects}.
\paragraph{}

Meanwhile, the idea that the string spectrum could also be determined by a Bethe ansatz was being considered as well. In \cite{Arutyunov:2006yd} it was proved that the excitations of the string sigma model and the magnons of the super Yang-Mills spin chain shared the same symmetry and hence that the S-matrices of the two theories were identical, up to some technical details. This led to the conclusion that the Bethe equations determining the string theory spectrum should be the same as those associated with the gauge theory spectrum.

In \cite{Arutyunov:2004vx}, Arutyunov, Frolov and Staudacher (AFS) proposed a set of Bethe equations which should determine the quantum spectrum of string theory on $\mathbb{R} \times S^3$. In the thermodynamic limit, these equations reduce, by construction, to the integral equations previously found in \cite{Kazakov:2004qf}, correctly describing the semiclassical spectrum. They also yield the first quantum correction, $O(1/J)$, in the general case of $M$ excitations corresponding, on the gauge theory side, to $Y$-impurities inside the trace, correctly reproducing the results for $M=2,3$ obtained in \cite{Callan:2003xr,Callan:2004ev} through direct quantisation.

The new distinctive feature of these Bethe equations was an extra factor modifying the S-matrix describing the scattering of elementary excitations, which was crucial in providing the correct quantum correction to the spectrum at $O(1/J)$. The authors then proposed that such a factor should also be included in the gauge theory BDS ansatz in order to account for wrapping interactions and correctly reproduce the string spectrum at strong coupling $\lambda \gg 1$. They could not give a general expression for this factor but, using their results on the string side, they conjectured an approximate expression in the large $\lambda$ limit.
\paragraph{}

In \cite{Beisert:2005fw}, the all-loop asymptotic Bethe ansatz, extending the BDS ansatz to the whole $\mathfrak{psu}(2,2|4)$ algebra, was formulated on the grounds of symmetry considerations leading to a proposal for the complete magnon S-matrix, including the extra factor discussed in \cite{Arutyunov:2004vx}, which was called the \emph{dressing factor}. The latter was meant to allow to interpolate between the weak and strong 't Hooft coupling regimes, but its general expression was still unknown, apart from the fact that it was expected to reduce to $1 + O(\lambda^3)$ in the $\lambda \ll 1$ limit, since BMN scaling had already been found to hold up to the 2-loop level, and to match the AFS expression at large $\lambda$. The magnon dispersion relation obtained from these Bethe equations is the same as the BDS dispersion relation.

It was later proved in \cite{Beisert:2005tm} that supersymmetry constraints determine the S-matrix completely, except for the dressing factor, so that it has to match the above conjecture. It was also shown to satisfy the Yang-Baxter equation, which is a self-consistency requirement for factorised scattering, and hence for integrability.

This situation is similar to the case of integrable two-dimensional relativistic field theories, where the S-matrix of elementary excitations is fixed by the symmetries of the theory, together with unitarity, the Yang-Baxter equation and crossing symmetry. The latter is associated with the presence of anti-particles and relates scattering processes which differ from each other by charge conjugation of one of the particles involved.
\paragraph{}

In order to improve the knowledge of the dressing factor at strong coupling, the first one-loop corrections $O(1/J)$ to the AFS result in string theory were computed in \cite{Beisert:2005cw} in some particular cases. A conjecture for the complete one-loop result for all sectors was given by Hern\'andez and L\'opez in \cite{Hernandez:2006tk}. Evidence supporting this conjecture was given in \cite{Freyhult:2006vr} and also in \cite{Gromov:2007aq,Gromov:2007cd}, where a procedure for the semiclassical quantisation of the algebraic curves describing the spectrum of the string sigma-model was introduced. This procedure, while allowing to rigorously prove the validity of the one-loop conjecture, proved to be difficult to apply to higher orders.

The only other apparent way of determining the dressing factor was through crossing symmetry (as the S-matrix already satisfied the constraints of supersymmetry, unitarity and factorised scattering). In \cite{Janik:2006dc}, the crossing equation for the $AdS_5 \times S^5$ S-matrix was determined, overcoming difficulties related to the absence of conventional relativistic invariance (which is broken in the usual light-cone quantisation due to gauge fixing). It was then necessary to find a solution to this equation with the properties of reproducing the AFS and Hern\'andez-L\'opez results at strong coupling $\lambda \gg 1$ and of reducing to $1 + O(\lambda^3)$ at weak coupling.

Such a solution was proposed by Beisert, Hern\'andez and L\'opez (BHL) in \cite{Beisert:2006ib} in the form of a power series in $1/\sqrt{\lambda}$. Shortly afterwards, Beisert, Eden and Staudacher (BES) \cite{Beisert:2006ez} found its perturbative expansion in $\lambda \ll 1$ and then showed that it had the correct behaviour $1 + O(\lambda^3)$. At this stage, the result was still asymptotic and hence only valid in the limit $J \to \infty$, so that wrapping corrections in perturbative gauge theory could not be computed from it. 

In the same paper, the authors also derived a representation of the general solution in terms of scattering kernels, the so-called ``BES magic formula'', which was later rewritten as a double contour integral on the complex plane through an inverse Fourier transform in \cite{Dorey:2007xn}, where it was also shown that it presented the expected set of singularities.

\section{The all-loop asymptotic Bethe ansatz in the $\mathfrak{sl}(2)$ sector}

Before its final proof was obtained through the crossing equation, the Beisert-Staudacher all-loop asymptotic Bethe ansatz was also widely tested in the $\mathfrak{sl}(2)$ sector, both at strong and weak coupling. The operators in this family have the following structure:
\begin{equation}
 \textrm{Tr } ( D^{s_1} Z D^{s_2} Z \ldots D^{s_J} Z )
\label{eq:sl2_sector_generic_op}
\end{equation}
where $D$ stands for either $D_+$ or $D_-$, the light-cone components of the covariant derivative which carry conformal (or Lorentz) spin $+1$ and $-1$ respectively. Typically, operators with only one type of derivatives are considered, e.g. $D_+$, in which case we also have $S = \sum_{l = 1}^J s_j$. The twist $J$ of the operator equals the number of $Z$ fields inside the trace and coincides with the total R-charge.

This sector had already been investigated in previous works. First of all, the scaling dimension of the QCD analogue of the twist-two operator was shown to diverge like $\log S$ as $S \to \infty$ at one loop in the perturbative regime \cite{Gross:1974cs}. Some arguments were successively presented in \cite{Korchemsky:1988si,Korchemsky:1992xv}, indicating that the logarithmic scaling should persist at all loops. In \cite{Korchemsky:1988si}, it was also proved that the scaling dimension of the twist-two operator at large spin can be written as
\begin{equation}
 \Delta = S + 2 + 2 \Gamma(\lambda) \log S + \ldots
\label{eq:Delta_of_twist2_op_with_cuspanomdim}
\end{equation}
where $\Gamma(\lambda)$ is the so-called \emph{cusp anomalous dimension}, which appears in the divergences associated with Wilson loops with cusps \cite{Polyakov:1980ca}.

On the string side, the $\mathfrak{sl}(2)$ sector is represented by strings living in $AdS_3 \times S^1$. Gubser, Klebanov and Polyakov (GKP) \cite{Gubser:2002tv} considered a folded string rigidly rotating in $AdS_3$ and studied its classical spectrum in the large $S$ limit, obtaining
\begin{equation}
 E = S + 2 \Gamma(\lambda) \log S + \ldots
\label{eq:Delta_of_GKP_string}
\end{equation}
Since motion on the $S^1$ can be neglected in this limit ($S+J \simeq J$ for fixed finite $J$), this was identified as the string dual of the twist-two operator for large angular momentum\footnote{More precisely, the large $S$ spectrum of this string is indistinguishable from that of the proper string dual of the twist-two operator, which lives in $AdS_3 \times S^1$.}. The two results \eqref{eq:Delta_of_twist2_op_with_cuspanomdim} and \eqref{eq:Delta_of_GKP_string} then yielded the leading order term of the cusp anomalous dimension at weak and strong coupling:
\begin{equation}
 \Gamma(\lambda) = \left\{ \begin{array}{ll}
	\frac{\lambda}{4 \pi^2} + O(\lambda^2) & \textrm{for } \lambda \ll 1 \\
	\frac{\sqrt{\lambda}}{2 \pi} + O(\lambda^0) & \textrm{for } \lambda \gg 1
\end{array} \right. 
\label{eq:cuspanomdim_weak/strong_coupling}
\end{equation}

The generalisation of this analysis to operators of twist $J$ was extensively studied in \cite{Belitsky:2003ys,Belitsky:2004cz,Belitsky:2006en}, where it was found that the logarithmic term has a coefficient $K \Gamma(\lambda)$, with $K$ any integer between 2 and $J$. This result can be summarised as
\begin{equation}
 \Delta = S + J + f(\lambda) \log S + \ldots
\label{eq:def_univ_scaling_fn}
\end{equation}
where $f(\lambda)$ is called the \emph{universal scaling function} due to the fact that it is independent of the twist, at least when the latter is finite, both at weak and strong coupling. In \cite{Belitsky:2006en}, the logarithmic scaling was found to hold also in the $J \sim \log S$ regime, although in this case the scaling function depends on an additional parameter $j = L / \log S$; $f(\lambda,j)$ is also known as the \emph{generalised scaling function}.

The universal scaling function was calculated at weak coupling up to three loops $O(\lambda^3)$ analytically \cite{Kotikov:2004er}, exploiting a previous QCD calculation \cite{Moch:2004pa}, and up to four loops $O(\lambda^4)$ numerically \cite{Bern:2006ew,Cachazo:2007ad}. At strong coupling, the GKP result providing the leading classical order $O(\sqrt{\lambda})$ was extended to one loop $O(1)$ in \cite{Frolov:2002av} and to two loops $O(1/\sqrt{\lambda})$ in \cite{Roiban:2007jf,Roiban:2007dq} through string theory computations.
\paragraph{}

All of these results were reproduced by the asymptotic Bethe ansatz. An integral equation, known as the BES (Beisert, Eden, Staudacher) equation, for the determination of the spectrum of long operators starting from the asymptotic Bethe equations was proposed in \cite{Eden:2006rx}, and later refined in \cite{Beisert:2006ez} by including the dressing factor. This equation was based on scattering kernels like the ``BES magic formula'' for the dressing phase and, in \cite{Beisert:2006ez}, it was also used to obtain the first four orders of $f(\lambda)$ in the limit of small 't Hooft coupling $\lambda$. All the terms coincide with the results obtained from direct computation in \cite{Kotikov:2004er,Bern:2006ew,Cachazo:2007ad}. A four-loop analytic computation of the dilatation operator for the $\mathfrak{su}(2)$ sector \cite{Beisert:2007hz} was also found to be compatible with some features of the expansion obtained from the BES equation.

At strong coupling, expanding the BES equation in powers of $1/ \sqrt{\lambda}$ is a much harder task due to technical complications. The $O(\sqrt{\lambda})$ term was first computed numerically \cite{Benna:2006nd} and then analytically \cite{Kotikov:2006ts,Alday:2007qf,Kostov:2007kx,Beccaria:2007tk}. The $O(1)$ term was obtained in \cite{Casteill:2007ct} and \cite{Belitsky:2007kf}, while the $O(1/ \sqrt{\lambda})$ and a recursive procedure for the following terms were given in \cite{Basso:2007wd} and \cite{Kostov:2008ax}. Again, the results matched those obtained from direct string theory calculations.
\paragraph{}

The generalised scaling function was also studied in detail. Freyhult, Rej and Staudacher (FRS) \cite{Freyhult:2007pz}, employing the same methods used to obtain the BES equation, introduced an integral equation determining $f(\lambda,j)$ from the asymptotic Bethe ansatz. In \cite{Fioravanti:2008bh}, this equation was rederived and Taylor-expanded in the limit of small $j$ for all values of the 't Hooft coupling. It was also understood that, in the same limit, the FRS equation is equivalent to the equation yielding the energy of the $O(6)$ sigma model \cite{Basso:2008tx}, confirming the observations of \cite{Alday:2007mf}.

Away from this limit, the generalised scaling function was also studied in the $j \sim \sqrt{\lambda} \gg 1$ regime. The general structure of the expansion in $1/ \sqrt{\lambda}$ was predicted from the connection with the $O(6)$ sigma model \cite{Alday:2007mf}, while the coefficients were first computed from string theory in \cite{Roiban:2007ju}. These were later rederived directly from the asymptotic Bethe ansatz equations and a discrepancy in one of the two-loop coefficients with respect to the string theory calculation was found \cite{Gromov:2008en}. The ABA result was confirmed exploiting the relationship with the $O(6)$ sigma model \cite{Bajnok:2008it} and using the FRS equation \cite{Volin:2009uv}. The string theory calculation was later corrected and complete agreement with the Bethe ansatz result was established also at the two-loop order \cite{Giombi:2010fa}.

\section{Further developments}
\label{sec:further_developments}

The inversely Fourier-transformed version of the BHL/BES solution obtained in \cite{Dorey:2007xn} was proved to satisfy the crossing equation \cite{Arutyunov:2009kf}. Moreover, nearly at the same time, the crossing equation was solved in general in \cite{Volin:2009uv} and the double integral representation was identified as the solution with the minimum number of singularities in the physical strip. Thus, the all-loop asymptotic Bethe ansatz, together with the BHL/BES dressing factor, was finally proved to determine the \emph{planar} spectrum of gauge theory operators with large R-charge at all values of the 't Hooft coupling $\lambda$.
\paragraph{}

Extending this result to finite R-charge, i.e. to operators of finite length, away from the asymptotic limit, requires taking wrapping interactions into account. In the context of the AdS/CFT correspondence, finite size corrections were first studied in \cite{Ambjorn:2005wa}, in the framework of integrable field theories. Since then, two different approaches to this problem have been considered.

The first approach, introduced in \cite{Janik:2007wt}, exploits the L\"uscher method \cite{Luscher:1986pf}, which is applicable to generic field theories in two dimensions. This procedure was successfully employed in many cases, among the most recent of which are the wrapping corrections to the scaling dimension of the Konishi operator at four \cite{Bajnok:2008bm} and five \cite{Bajnok:2009vm} loops. The results coincide with the corresponding direct diagrammatic gauge theory calculations, which have been carried out up to four loops \cite{Fiamberti:2007rj,Velizhanin:2008jd}.

The second approach is based on the interpretation of the finite length of an integrable system as the inverse of the temperature in a copy of the same system with space and time interchanged, which is due to Zamolodchikov \cite{Zamolodchikov:1989cf,Zamolodchikov:1991et}. This procedure, originally proposed for the AdS/CFT correspondence in \cite{Ambjorn:2005wa}, was already known as the \emph{Thermodynamic Bethe Ansatz} (TBA) from the analysis of other integrable systems. In particular, Destri and DeVega \cite{Destri:1994bv} had already given an alternative formulation in terms of non-linear integral equations. In the case of AdS/CFT, the TBA was first applied to the so-called mirror model \cite{Arutyunov:2007tc}, which is equivalent to string theory on $AdS_5 \times S^5$, then to the $\mathfrak{su}(2)$ Principal Chiral Model \cite{Gromov:2008gj}, and finally to the full AdS/CFT system \cite{Gromov:2009tv}. This approach is conjectured to yield the full \emph{planar} spectrum of $\mathcal{N} = 4$ super Yang-Mills theory on $\mathbb{R} \times S^3$, for any operator length and at arbitrary values of the 't Hooft coupling $\lambda$.

In \cite{Gromov:2009tv}, it was checked that the TBA reduces to the asymptotic Bethe ansatz in the large length limit, and the same method was used to compute the wrapping correction to the scaling dimension of the Konishi operator at four loops in perturbative gauge theory. The calculation was extended to five loops in \cite{Arutyunov:2010gb,Balog:2010xa}. All the results are compatible with those obtained through the L\"uscher method.

In \cite{Bombardelli:2009ns,Gromov:2009bc}, by studying the TBA equations for the mirror model, an equivalent set of finite-difference equations, known as the Y-system, was obtained. The Y-system was then analysed numerically at strong coupling \cite{Gromov:2009zb} and the results were compared to the corresponding coefficients from the string theory expansion \cite{Roiban:2009aa}. A discrepancy was found in one of the higher order terms. The spectrum of strings moving in $AdS_3 \times S^1$ \cite{Gromov:2009tq} and in the full $AdS_5 \times S^5$ target space, as predicted from the algebraic curve formalism \cite{Gromov:2010vb}, was also correctly reproduced up to one loop through the Y-system.

\section{Topics covered in this thesis}

The research presented in this thesis focuses on the analysis of semiclassical strings in AdS space (more precisely in $AdS_3 \times S^1$ and pure $AdS_3$), in the limit of large angular momentum $S$, which will be carried out employing two methods, namely the finite-gap formalism and the direct analysis of explicit solutions. The former is of particular importance, since it is believed that finite-gap solutions are in fact generic, and therefore predictions obtained through this technique should describe the general behaviour of strings in AdS.

The results derived through both methods agree, leading to the conjecture that any string moving on this background should develop a certain number of spikes (or cusps) at large $S$, corresponding to solitonic objects on the worldsheet, and that the dynamics of these string solutions can be understood in terms of the dynamics of their spikes.

\begin{figure}
\centering
\includegraphics[width=100mm]{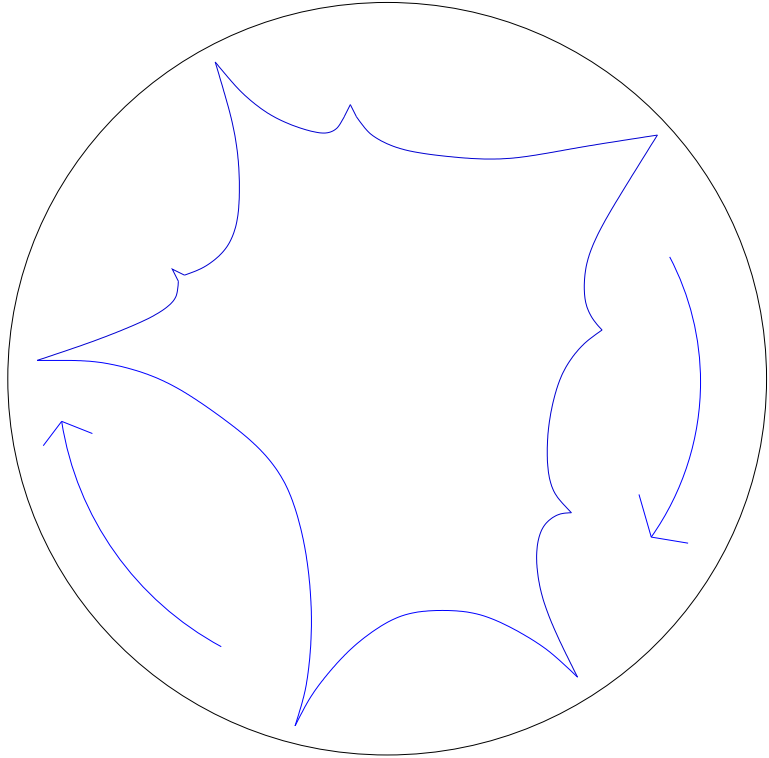}
\caption{A generic spiky string displaying both ``large'' and ``small'' spikes. The circle represents the boundary of $AdS_3$.}
\label{fig:generic_spiky_string}
\end{figure}

In particular, spikes branch out into two different sectors: ``large'' spikes, which become infinitely long, approaching the boundary of AdS in the large $S$ limit, and correspond to static solitons, and ``small'' spikes, which instead remain of finite length and are associated with solitons propagating along the background of the ``large'' spikes, as shown in Fig. \ref{fig:generic_spiky_string}. ``Large'' spikes are responsible for the leading diverging behaviour of the spectrum $E - S$, while ``small'' spikes yield a finite subleading contribution $E_{\rm sol}$. It is possible to associate a conserved momentum $P_{\rm sol}$ with each ``small'' spike and then to determine the dispersion relation $E_{\rm sol} (P_{\rm sol})$.

Both types of cusps are closely connected to the description of the operators in the dual gauge theory $\mathfrak{sl}(2)$ sector, in the large conformal spin $S$ limit. In particular, the one-loop spectrum of such operators is determined by the Hamiltonian of the quantum $SL(2,\mathbb{R})$ spin chain, which may be diagonalised by exploiting its parametrisation in terms of objects known as holes, which yield additive contributions to the total energy. As $S$ becomes infinite, holes display two types of behaviour: ``large'' holes have a diverging energy, whereas ``small'' holes maintain a finite energy.

``Large'' holes correspond to highly excited spins in the chain, which may be treated semiclassically. It is possible to show that a spin vector may be associated with each ``large'' spike of a given string solution and that such spin vectors are in a one-to-one correspondence with the highly excited spins in the gauge theory chain. Therefore, ``large'' holes correspond to ``large'' spikes. We have successfully tested this mapping on several explicit solutions, including a family of approximate solutions described in terms of a large number of parameters, which should be able to reproduce the large $S$ behaviour of a generic string in $AdS_3$.

``Small'' holes correspond to spins which are not highly excited and therefore are still quantised in terms of a set of Bethe equations. They may be interpreted as quasi-particles carrying energy $E_{\rm hole}$ and momentum $P_{\rm hole}$. Based on our results, we make three conjectures.

Firstly, due to the fact that the ``small'' hole dispersion relation $E_{\rm hole} (P_{\rm hole})$ matches the ``small'' spike dispersion relation in the large momentum limit, we propose that ``small'' spikes are the string theory duals of ``small'' holes, extrapolated at strong coupling.

Secondly, since the above agreement only holds up to the familiar discrepancy in the prefactor, which is $\lambda/4 \pi^2$ on the gauge side and $\sqrt{\lambda}/2 \pi$ on the string side, we conjecture that the proper prefactor for the dispersion relation of these objects at generic values of the coupling $\lambda$ is given by the cusp anomalous dimension $\Gamma (\lambda)$.

Lastly, we hypothesize that the ``small'' spikes are continuosly connected to the small quadratic transverse fluctuations of strings in $AdS_3$, or, in other words, that a ``small'' spike is simply the large-momentum and large-energy version of a quadratic fluctuation. All three conjectures have recently been proved true in \cite{Basso:2010in}, where the interpolating behaviour of these objects between weak and strong coupling was derived.

We will now analyse the above points in greater detail.
\paragraph{}

According to the AdS/CFT conjecture, strings living in $AdS_3 \times S^1$ are dual to gauge theory operators in the $\mathfrak{sl}(2)$ sector, which can be represented in terms of an $SL(2,\mathbb{R})$ spin chain. As explained earlier, this sector was studied, both on the gauge and on the string side, in order to determine the universal scaling function in the hope of reproducing it through the asymptotic Bethe ansatz with the inclusion of the dressing phase.

The one-loop spectrum of the corresponding scaling dimensions in gauge theory was given in \cite{Belitsky:2006en} in terms of algebraic curves, in a variety of large spin $S$ and large twist $J$ limits. In particular, the case $S \to \infty$ with a finite, fixed $J$ was not, strictly speaking, in the range of validity of the asymptotic Bethe ansatz. 

The interpretation of the operator \eqref{eq:sl2_sector_generic_op} in terms of the spin chain model associates each $Z$ field inside the trace with a site of the chain, and each derivative acting on that field with an excitation (or magnon) at the corresponding site, so that there are always $S$ magnons. The spin chain is diagonalised by the $SL(2, \mathbb{R})$ Bethe equations determining the rapidities of the magnons, which then yield the spectrum. Each magnon has a different rapidity, which must be chosen among a set of possible values. It turns out that there are always $J$ unassigned values, or in other words $J$ holes in the rapidity distribution, which, like magnons, may be treated as quasi-particles. The system can be equivalently described in terms of magnons or holes, and in both cases the total energy is given by the sum of the energy of the individual quasi-particles.

Due to the fact that there are always $S$ magnons and $J$ holes, the latter provide the most convenient parametrisation in the $S \to \infty$, $J = \textrm{const.}$ limit considered here. As mentioned above, at large $S$ holes may either yield a diverging $O (\log S)$ contribution to the total energy, in which case they are called ``large'' holes, or instead yield a finite $O(S^0)$ contribution, so that they are named ``small'' holes. The only constraint here is that there must always be at least two ``large'' holes.

``Large'' holes correspond to the spins of the quantum $SL(2, \mathbb{R})$ chain which become highly excited as $S \to \infty$. The leading order contribution to the spectrum can then be described in terms of a shorter semiclassical $SL(2, \mathbb{R})$ chain, which only contains these highly excited spins. The semiclassical analysis maps this system to an algebraic curve \cite{Belitsky:2006en}, which is in fact rather similar to the curves encountered in the string theory finite-gap construction. Semiclassical quantisation is then imposed directly on the moduli of this Riemann surface.

``Small'' holes correspond instead to spins at lower levels of excitation, which are therefore not visible at the leading semiclassical order and rather appear as small excitations, carrying conserved energy and momentum, which are quantised in terms of Bethe-type equations.

The resulting spectrum is given by
\begin{multline}
 \gamma_{\rm gauge} = \Delta - S - J = \frac{\lambda}{4 \pi^2} \left[ (J-M) \log S + H_K (l_1, \ldots, l_{J-M-1}) \phantom{\sum_{j=1}^M}
  \right. \\
 \left. + \sum_{j=1}^M E^{\rm hole}_j + O(1/S) \right] \:,
\label{eq:sl(2)_spectrum_gauge_1loop_BGK}
\end{multline}
where $M = 0, 1, \ldots, J-2$ is the number of ``small'' holes. We therefore see that the scaling dimension lies in a band of possible levels, where the specific level is determined by the number $J-M$ of ``large'' holes. The quantity $H_K$ represents the subleading contribution to the spectrum generated by the ``large'' holes, encodes the full dependence on their associated moduli $l_j$ and corresponds to the Hamiltonian governing the dynamics of these holes. The $l_i$ correspond to the filling fractions associated with the branch cuts on the Riemann surface and, according to the semiclassical quantisation conditions, they must be integer.
\paragraph{}

The leading $\log S$ part of this spectrum had already been reproduced in semiclassical string theory at $\lambda \gg 1$ through the finite-gap analysis \cite{Belitsky:2006en,Sakai:2006bp}, where $S \gg 1$ corresponds to angular momentum in $AdS_3$ and $J \sim \sqrt{\lambda}$ to angular momentum on $S^1$. In \cite{Dorey:2008zy}, this result was extended to include the subleading $O(S^0)$ term in the case of no ``small'' holes present (i.e. $M=0$):
\begin{equation}
 \gamma_{\rm string} = E - S - J = \frac{\sqrt{\lambda}}{2 \pi} \left[ K \log S + H_K (l_1, \ldots, l_{K-1}) + C_K + \ldots \right] \:.
\label{eq:sl(2)_spectrum_string_1loop_BGK}
\end{equation}
Specifically, the Hamiltonian $H_K$ is the same, while $C_K$ is a constant which is independent of the moduli. The spectrum of the string solutions is still described in terms of an algebraic curve, which, in the large $S$ limit, factorises into two separate Riemann surfaces: the first, $\tilde{\Sigma}_1$, has genus $K-2$, while the second, $\tilde{\Sigma}_2$, has genus $0$. All the filling fractions $l_j$, which must be integer due to the semiclassical quantisation conditions, are associated with the branch cuts of the first surface.

The main difference with respect to the gauge theory result is in the overall factor, which is $O(\lambda)$ on the gauge side and $O(\sqrt{\lambda})$ on the string side and corresponds to the leading contribution to the cusp anomalous dimension at weak and strong coupling, respectively. This suggests that configurations with the same moduli $l_j$ on both sides should be identified as dual states in the dictionary of the correspondence.

Another difference is that here the integer $K \geq 2$ is in fact unrelated to $J$, and hence a complete correspondence with the SYM spectrum requires us to choose $K = J$.

There is strong evidence \cite{Eden:2006rx,Beisert:2006ez} that this result could also have been obtained directly from the asymptotic Bethe ansatz, at least for the case $J=2$, due to the fact that the lowest dimension in the band is independent of $J$ and can therefore be evaluated in the limit $J \to \infty$, which then allows the use of the asymptotic Bethe equations. For fixed $J > 2$, none of the operators with $K > 2$ is covered by this property, but the fact that the spectrum \eqref{eq:sl(2)_spectrum_string_1loop_BGK} is also independent of $J$ may indicate that it still holds true, at least in the large $S$ limit.

A map between string theory and gauge theory degrees of freedom was also proposed in \cite{Dorey:2008zy} in order to shed further light on the agreement between the two spectra. The $SL(2, \mathbb{R})$ spin chain has one degree of freedom, in the form of a three-vector $\mathcal{L}_j \in \mathfrak{sl}(2,\mathbb{R})$, associated with each of its sites. Both the algebraic curve and the Hamiltonian $H_K$ of the chain are determined as functions of these parameters.

Such a discrete system can be extracted from a continuous string in the following way. The key hypothesis is that, as the angular momentum on AdS space becomes large, any string exhibiting the above spectrum should develop $K$ ``large'' spikes. As $S \to \infty$, the latter approach the boundary, while the worldsheet charge density for $E + S$ becomes $\delta$-function localised at the spikes:
\begin{equation}
 j^A_\tau (\tau, \sigma) \simeq \frac{8 \pi}{\sqrt{\lambda}} \sum_{k=1}^{K-1} L^A_k \delta (\sigma - \sigma_k) \:,
\label{eq:delta_fn_localisation_of_j}
\end{equation}
where $\sigma_k$ represents the worldsheet position of the $k$-th spike. The coefficients of these $\delta$-functions can then be decomposed onto the Lie algebra $\mathfrak{sl}(2,\mathbb{R})$ and the components $L^A_k$ should be identified with those of the three-vectors $\mathcal{L}_k$ of the spin chain:
\begin{equation}
 L^0_k \leftrightarrow \mathcal{L}^0_k  \qquad \qquad  L^\pm_k \leftrightarrow i \mathcal{L}^\pm_k
\label{eq:dof_identification}
\end{equation}
The main argument in support of this hypothesis is the fact that it accounts precisely for the matching of the two spectra \eqref{eq:sl(2)_spectrum_gauge_1loop_BGK} and \eqref{eq:sl(2)_spectrum_string_1loop_BGK}. In fact, it is possible to show that, with the above identification and under the assumption of $\delta$-function localisation, the algebraic curve obtained from the string theory finite-gap construction, more specifically $\tilde{\Sigma}_1$, matches the curve associated with the gauge theory spin chain, which also implies that the corresponding spectra are identical.

A more qualitative interpretation of this relationship between the degrees of freedom stems from the fact that the spikes, as they approach the boundary in the large $S$ limit, trace lightlike geodesics. These can be associated, in the gauge theory picture, with Wilson lines, which represent highly energetic gluons. The arcs of string joining the spikes are then identified with the chromomagnetic flux tubes connecting the gluons.
\paragraph{}

\begin{figure}%
\begin{center}
\includegraphics{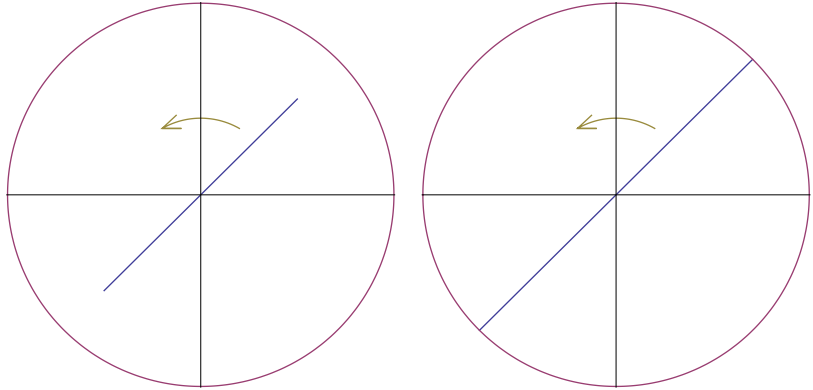}%
\end{center}
\caption{The GKP string is a folded straight line which rigidly rotates around its centre of symmetry, which coincides with the centre of $AdS_3$. Its length is controlled by a parameter and becomes infinite in the large $S$ limit (see the plot on the right-hand side). The arrows indicate the direction of rotation.}%
\label{fig:GKP_plot_intro}%
\end{figure}

\begin{figure}%
\begin{center}
\includegraphics{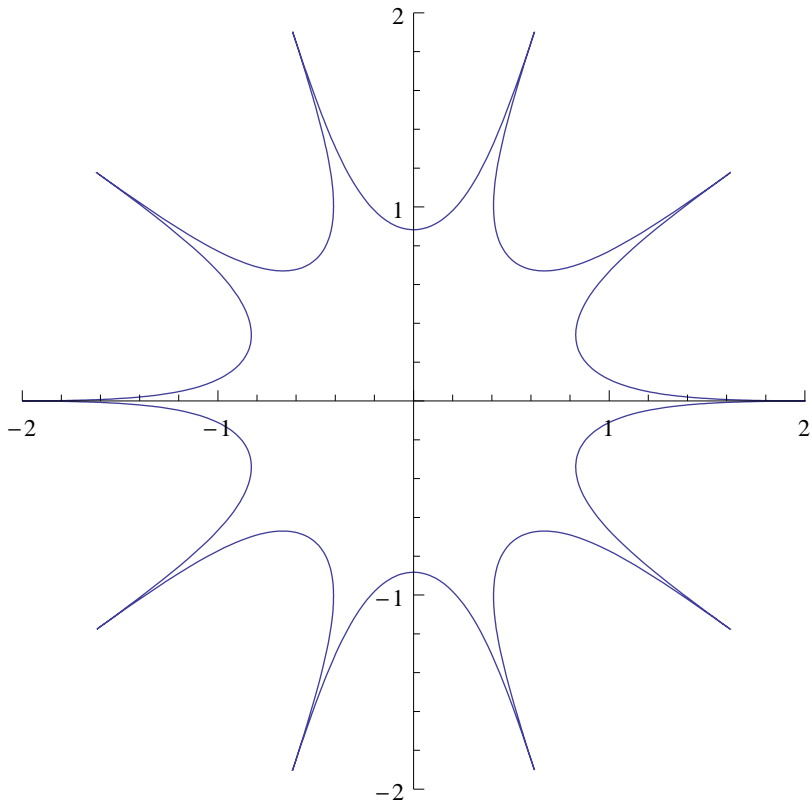}%
\end{center}
\caption{The Kruczenski spiky string consists of an arbitrary number of spikes which rigidly rotate around the centre of $AdS_3$. The angular separations between consecutive spikes are all identical. The spikes approach the boundary of $AdS_3$ in the large $S$ limit, exactly as in the GKP case.}%
\label{fig:K_plot_intro}%
\end{figure}

\begin{figure}%
\begin{center}
\includegraphics{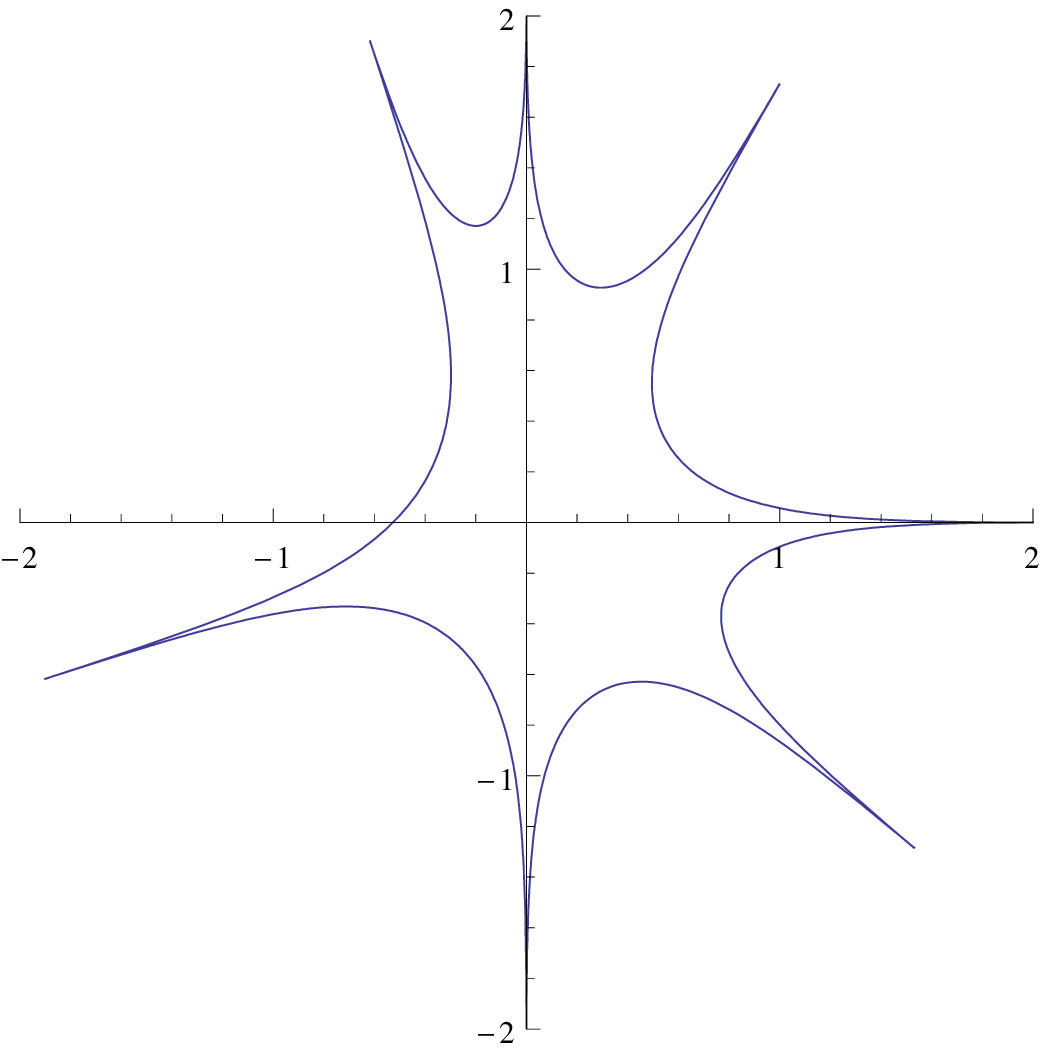}%
\end{center}
\caption{The general patched solution, with arbitary angular separations between consecutive spikes. This string is also rigidly rotating around the centre of $AdS_3$ and, as usual, the spikes approach the boundary for $S \to \infty$.}%
\label{fig:patch_K_plot_intro}%
\end{figure}

In \cite{Dorey:2008vp}, the above proposal was tested on two explicit spiky string solutions which had already appeared in the literature: the GKP folded rotating string \cite{Gubser:2002tv} (see Fig. \ref{fig:GKP_plot_intro}) and the symmetric spiky string found by Kruczenski \cite{Kruczenski:2004wg} (see Fig. \ref{fig:K_plot_intro}), which has spikes at the vertices of arbitrary regular polygons. Furthermore, a family of approximate solutions, which become exact in the $S \to \infty$ limit, was also constructed, describing spikes with arbitrary angular separations between each other (see Fig. \ref{fig:patch_K_plot_intro}).

For all these string solutions, the expected $\delta$-function localisation was observed and exploited in order to compute the algebraic curves and to show that they represent specific points in the moduli space of the gauge theory curve, thereby confirming its agreement with the finite-gap curve. Once the moduli on the gauge side were identified in terms of the string parameters, it was possible to calculate the corresponding spectra \eqref{eq:sl(2)_spectrum_gauge_1loop_BGK}. As predicted, these matched the spectra $E-S$ obtained from the explicit string solutions, directly from first principles calculations, up to the usual replacement of the prefactor.

These results support the map between the degrees of freedom in the two theories and the identification of ``large'' holes with ``large'' spikes.
\paragraph{}

We now move on to discuss the concept of ``small'' spikes. 

The GKP string, which so far has attracted our attention as an approximate dual of the twist-two operator in gauge theory, has another interesting feature: Alday and Maldacena \cite{Alday:2007mf} showed that it is a valid starting point for semiclassical quantisation and computed the full spectrum of small fluctuations. Since the string becomes infinitely long in the $S \to \infty$ limit, we can expect that it will also admit solitonic excitations carrying finite $O(\lambda)$ energy $E_{\rm sol}$, as opposed to the $O(1)$ energy of small fluctuations, and exhibiting factorised scattering.

This is suggested by the analogy with the BMN vacuum case discussed earlier. While on the string side this object is represented as a rotating point-particle, on the gauge side it corresponds to the operator $\textrm{Tr } (Z^J)$, which becomes infinitely long in the asymptotic $J \to \infty$ limit. Its excitations (magnons), corresponding to insertions of $Y$ fields inside the trace, undergo factorised scattering and their energies are given by the BDS dispersion relation \eqref{eq:BDS_dispersion_relation}. At strong coupling, the energy of a magnon becomes $O(\sqrt{\lambda})$ \eqref{eq:giant_magnon_dispersion_relation} and in fact corresponds to a solitonic object in string theory, the Giant Magnon identified by Hofman and Maldacena \cite{Hofman:2006xt}, which is continuously connected to the small fluctuations above the BMN point-particle.

According to Pohlmeyer reduction \cite{Pohlmeyer:1975nb,Mikhailov:2005qv,Mikhailov:2005zd}, the string equations of motion and Virasoro constraints on $\mathbb{R} \times S^2$ are equivalent to the sine-Gordon equation, which is known to be integrable. String solutions describing Giant Magnons correspond to multi-soliton configurations of the sine-Gordon field and the solitons can be seen to be in a one-to-one correspondence with the Giant Magnons, thus confirming the solitonic nature of these objects.

\begin{figure}%
\begin{center}
\includegraphics{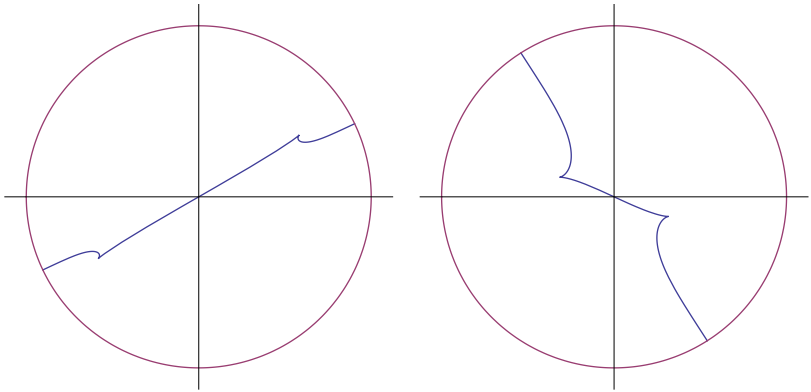}%
\end{center}
\caption{The two-soliton solution, describing two ``small'' spikes propagating along the infinite GKP string. As the string rotates around the centre of $AdS_3$, the two cusps approach each other, collide and then move asymptotically towards the endpoints on the boundary.}%
\label{fig:2s_plot_intro}%
\end{figure}

In \cite{Dorey:2010iy}, this picture was reconstructed in the case of the GKP vacuum. In fact, explicit string solutions representing the corresponding solitonic excitations had already been found in \cite{Jevicki:2007aa}. They describe ``small'' spikes propagating along the straight GKP string, extended up to the boundary in the radial direction, so that it has infinite energy and angular momentum (see Fig. \ref{fig:2s_plot_intro}). Strings in $AdS_3$ are Pohlmeyer-reduced to solutions of either the sinh-Gordon, the cosh-Gordon or the Liouville equation. The solutions described in \cite{Jevicki:2007aa} are associated with multi-soliton configurations of the sinh-Gordon field and, more specifically, the worldsheet position of each ``small'' spike coincides with the position of a soliton. This confirms that the cusps undergo factorised scattering.

As expected, each spike adds a finite $O(\sqrt{\lambda})$ contribution $E_{\rm sol} (v)$ to the quantity $E - S$ with respect to the vacuum value. Here $v$, with $|v| \leq 1$, represents the worldsheet velocity of the soliton associated with the cusp. Every spike also carries a conserved momentum $P_{\rm sol} (v)$, which is canonically conjugate to the worldsheet position of the spike and, together with the energy and upon elimination of the parameter $v$, yields the dispersion relation:
\begin{eqnarray}
 E_{\rm sol}(v) & = & \frac{\sqrt{\lambda}}{2\pi} \left[\frac{1}{2}\log\left(\frac{1+\sqrt{1-v^{2}}} {1- \sqrt{1-v^{2}}}\right)\,\,-\,\,
  \sqrt{1-v^{2}}\right] \nn \\
 P_{\rm sol}(v) & = & \frac{\sqrt{\lambda}}{2\pi}\left[\frac{\sqrt{1-v^{2}}}{v}\,\,-\,\, {\rm Tan}^{-1} \left( \frac{\sqrt{1-v^{2}}}{v}
  \right)\right] \:,
\label{eq:GH_dispersion_relation}
\end{eqnarray}
where the branch of the inverse tangent is chosen so that $-\pi/2\leq {\rm Tan}^{-1} x\leq +\pi/2$ for all real values of $x$. The discrete semiclassical spectrum can be obtained by imposing the momentum quantisation condition:
\begin{equation}
 P_{\rm sol}\cdot\, 2\log S \in 2\pi\,\,\mathbb{Z} \:.
\label{eq:GH_momentum_quantisation}
\end{equation}
The above dispersion relation is in general non-relativistic, due to the residual gauge fixing implemented in the construction of the string solutions, which breaks the Lorentz invariance of the sigma model in conformal gauge. The relativistic behaviour is however restored in the low momentum limit ($|v| \simeq 1$):
\begin{equation}
 E_{\rm sol} \simeq |P_{\rm sol}|\,\, +\,\,  O  \left(P^{\frac{5}{3}}_{\rm sol}\right) \:.
\label{eq:GH_disprel_low_P}
\end{equation}
Hence, at low momentum, the spikes become relativistic particles with a mass which is small compared to $\sqrt{\lambda}$. These properties are shared by the only transverse mode present in the spectrum of quadratic fluctuations of a string moving in $AdS_3$ \cite{Alday:2007mf}. Rather than conjecturing the appearance of a second transverse mode, it seems reasonable to hypothesize that the spikes are in fact continuously connected to the standard quadratic fluctuations, exactly as it happens in the case of the Giant Magnons. The appropriate mass term for the cusps, giving them the same mass as the transverse mode, should then be introduced by the leading quantum correction, $O(\lambda^0)$, to the semiclassical dispersion relation \eqref{eq:GH_dispersion_relation}.

This is the third conjecture concerning ``small'' spikes, which we mentioned earlier in this section, and which was confirmed in \cite{Basso:2010in}.




The large momentum $P_{\rm sol} \gg 1$ (i.e. $v \ll 1$) limit of the dispersion relation \eqref{eq:GH_dispersion_relation},
\begin{equation}
 E_{\rm sol} = \frac{\sqrt{\lambda}}{2\pi}\,\log\,|P_{\rm sol}|\,\, + \,\,  O \left(P^{0}_{\rm sol}\right)   \:,
\label{eq:GH_disprel_high_P}
\end{equation}
shows that ``small'' spikes become ``large'' when their worldsheet velocity vanishes. This also suggests that ``large'' spikes should be associated with static solitons. This idea is in fact confirmed by the results of \cite{Jevicki:2008mm}, which show that the spikes of the symmetric Kruczenski string coincide with static solitons. As we will see later, the GKP string may be obtained as a particular limit of the Kruczenski string, which also provides the fundamental building blocks for the general patched solution, and hence, this result applies to all the explicit string solutions with ``large'' spikes studied here.

Furthermore, \eqref{eq:GH_disprel_high_P} coincides with the dispersion relation for the ``small'' holes of the gauge theory spin chain in the limit of large momentum $P_{\rm hole} \gg 1$, in which they become ``large'', modulo the familiar replacement of the prefactor $\lambda/(4 \pi^2)$ with $\sqrt{\lambda}/(2 \pi)$, as in the usual interpolation of the cusp anomalous dimension from weak to strong coupling. This agreement also includes the semiclassical quantisation condition for the momenta, although this was only checked in the case of two ``large'' holes and $J-2$ ``small'' holes.

This result motivates the first and second conjectures on the ``small'' spikes, namely that they should represent the same objects as the holes of gauge theory, at strong coupling $\lambda \gg 1$ and at large momentum, and that, in the corresponding dispersion relation, the prefactor interpolating between the weak and strong coupling regimes should be given by the cusp anomalous dimension $\Gamma (\lambda)$. Both these conjectures were also confirmed in \cite{Basso:2010in}.

For this reason and due to the similarities with the Giant Magnons in $\mathbb{R} \times S^2$, the name ``Giant Holes'' was proposed for these solitonic excitations above the GKP vacuum.
\paragraph{}

Furthermore, in \cite{Dorey:2010id}, the finite-gap analysis of \cite{Dorey:2008zy} was extended to the general case $M \neq 0$ in which ``small'' spikes are present. The main difference with respect to the $M=0$ case lies in the fact that the surface $\tilde{\Sigma}_1$ loses branch cuts, so that its genus is now $K-M-2$, while $\tilde{\Sigma}_2$, despite remaining of genus $0$, acquires $M$ simple poles. The cuts on the first surface are still associated with integer filling fractions and, as in the previous case, control the leading order behaviour of the spectrum. The poles on the second surface instead generate subleading excitations in the spectrum, whose dispersion relation is in fact identical to the ``small'' spike dispersion relation \eqref{eq:GH_dispersion_relation}.

The resulting spectrum is given by:
\begin{multline}
 \gamma_{\rm string} = E - S - J = \frac{\sqrt{\lambda}}{2 \pi} \left[ (K-M) \log S + H_{(K-M)} (l_1, \ldots, l_{K-M-1}) \phantom{\sum_{j=1}^M}
  \right.\\
  \left. + \sum_{j=1}^M E_{\rm sol} (v_j) + C_{(K-M)} + \ldots \right] \:,
\label{eq:finite-gap_spectrum_complete}
\end{multline}
generalising \eqref{eq:sl(2)_spectrum_string_1loop_BGK} to include the presence of ``small'' spikes. All the previous results concerning $\tilde{\Sigma}_1$ still apply, and hence it still matches the gauge theory algebraic curve, leading to the previously mentioned identification between ``large'' spikes and ``large'' holes. Again we need to set $K = J$ in order to achieve this agreement.

An important aspect of this result is the fact that, since the dispersion relation \eqref{eq:GH_dispersion_relation} is obtained through the finite-gap construction, ``large'' and ``small'' spikes should appear in generic strings in AdS at large angular momentum. Furthermore, the above spectrum encompasses completely general solutions with arbitrary numbers of ``large'' and ``small'' spikes.

\paragraph{}

Finally, we will also present some as yet unpublished work (also in collaboration with N. Dorey) concerning some string solutions living in $AdS_3$-pp-wave space, which are obtained by applying an $AdS_3$ boost to the GKP vacuum and its excited states with one and two ``small'' spikes. These solutions appear as arcs of string with endpoints on the boundary of $AdS_3$ and drooping towards the interior and the excited states also display one or two ``small'' spikes propagating along the background arc. The dispersion relation of the excitations agrees with the Giant Hole dispersion relation \eqref{eq:GH_dispersion_relation}, up to a multiplicative factor in the expression for the energy, which depends on one of the additional parameters introduced by the pp-wave limit.

By virtue of the interpretation of such arcs of string as the chromomagnetic flux tubes connecting two highly energetic gluons located at the endpoints, such string solutions may be of interest, as they represent excited states of the two-gluon system.

\section{Outline}

The remainder of this thesis is structured as follows.

In chapter \ref{sec:gauge_theory_spin_chain} we review the main results concerning the gauge theory spin chain, which allows to diagonalise the one-loop dilatation operator in the $\mathfrak{sl}(2)$ sector. Starting from the quantum chain, we discuss the semiclassical limit of large conformal spin $S$, in which the spectrum is determined by an algebraic curve and described in terms of ``large'' and ``small'' holes. We also discuss the quantisation conditions for the semiclassical spin chain associated with the ``large'' holes and the quantisation conditions for the ``small'' holes.

In chapter \ref{sec:integrability_methods} we review two integrability methods: Pohlmeyer reduction, in the special case of the sinh-Gordon/cosh-Gordon/Liouville connection, which applies to string theory in pure $AdS_3$, and the finite-gap construction, which applies to string theory on $AdS_3 \times S^1$. In particular, the discussion of the second method will culminate with the formulation of the spectral problem, i.e. the definition of the spectral curve and the set of constraints placed upon it, which, if solved, would yield the semiclassical spectrum of a very large class of string solutions moving in this space.

In chapter \ref{ch:FGA} we discuss the large $AdS_3$ angular momentum limit of the finite-gap construction, showing that the spectral curve factorises into two separate Riemann surfaces and computing the semiclassical spectrum up to $O(S^0)$ and $O(\sqrt{\lambda})$. We argue that the moduli associated with each surface control a specific branch of the spectrum and that the latter coincides with the spectrum of the gauge theory spin chain, while one of the two Riemann surfaces matches the gauge theory spectral curve. Furthermore, we discuss an interpretation of this spectrum in terms of the appearance of two types of cusps on the corresponding string solutions, with ``large'' and ``small'' cusps being the respective duals of ``large'' and ``small'' holes in the spin chain. The material discussed in this chapter essentially amounts to the results of \cite{Dorey:2010id}.

In chapter \ref{sec:explicit_solutions}, based on \cite{Dorey:2008vp} and \cite{Dorey:2010iy}, we test the above interpretation on several explicit solutions displaying either ``large'' or ``small'' spikes. We study the former in order to apply the proposed procedure for extracting a discrete set of degrees of freedom, in the form of spin vectors, from the ``large'' spikes on the string side. Then, we use these spin vectors in order to reconstruct the first Riemann surface $\tilde{\Sigma}_1$ and, equivalently, the gauge theory spectral curve, thereby confirming the identification between ``large'' spikes and ``large'' holes. The second class of solutions is instead of interest since, apart from providing us with explicit examples of ``small'' spikes with the dispersion relation predicted from the finite-gap analysis, it allows us to argue that the ``small'' spikes are in fact the $AdS_3$ relatives of the Giant Magnons living in $\mathbb{R} \times S^2$. Furthermore, we develop a procedure which allows to construct a more general patched solution, albeit only at infinite angular momentum, containing both an arbitrary number of ``large'' spikes and (subject to limitations) an arbitrary number of ``small'' spikes.

In chapter \ref{sec:pp-wave}, we discuss some solutions living in the pp-wave region of $AdS_3$ space. We explain how they are obtained from certain explicit solutions considered in chapter \ref{sec:explicit_solutions}, namely those involving ``small'' spikes. We also compute the corresponding conserved charges, reobtaining the Giant Hole dispersion relation.

\chapter{The gauge theory integrable spin chain}
\label{sec:gauge_theory_spin_chain}

\section{Basic setup}

The action of the $\mathcal{N}=4$ super Yang-Mills theory in 4-dimensional spacetime is defined as
\begin{multline}
 S = \int d^4 x {\rm Tr} \: \left\{ \frac{1}{4} (F_{\mu\nu})^2 + \frac{1}{2} (D_\mu \phi_i)^2 - \frac{1}{4} g_{YM}^2 [\phi_i, \phi_j]
  [\phi^i, \phi^j] \right. \\
      \left. + \bar{\Psi}^a_{\dot{\alpha}} \sigma^{\dot{\alpha} \beta}_\mu D^\mu \Psi_{\beta a}
       - \frac{i}{2} g_{YM} \Psi_{\alpha a} \sigma^{ab}_i \epsilon^{\alpha \beta} [\phi^i, \Psi_{\beta b}]
       - \frac{i}{2} g_{YM} \bar{\Psi}^a_{\dot{\alpha}} \sigma^i_{ab} \epsilon^{\dot{\alpha} \dot{\beta}} [ \phi_i, \bar{\Psi}^b_{\dot{\beta}} ] \right\}
\label{eq:SYM_action}
\end{multline}
where we have the following ranges for the spacetime vector indices $\mu, \nu = 0, \ldots, 3$, for the internal vector indices $i, j = 1, \ldots, 6$, and finally for the spinor indices $\alpha, \beta, \dot{\alpha}, \dot{\beta} = 1,2$ and $a, b = 1, \ldots, 4$. The components of the vector field $A_\mu$ and of the spinors $\Psi_{\alpha,a}, \bar{\Psi}_{\dot{\alpha} a}$, together with the scalar fields $\phi_i$, are all in the adjoint representation of the gauge group $SU(N)$, while $\sigma^\mu$ and $\sigma^i$ are the chiral projections of the gamma matrices in four and six dimensions respectively. Finally, the covariant derivative $D_\mu$ and the field strength $F_{\mu\nu}$ are given by
\begin{equation}
 D_\mu = \partial_\mu - i g_{YM} [A_\mu, \:\:] \qquad \qquad F_{\mu\nu} = i g_{YM}^{-1} [D_\mu, D_\nu] 
  = \partial_\mu A_\nu - \partial_\nu A_\mu - i g_{YM} [A_\mu, A_\nu] \:.
\label{eq:def_SYM_covdev_fieldstr}
\end{equation}
The bosonic component of the global symmetry group $PSU(2,2|4)$ of the theory includes the conformal group $SO(2,4)$ and the group of rotations of the scalar fields $SO(6)$. The conformal invariance, which forces all the fields to be massless, survives at the quantum level, so that the coupling constant $g_{YM}$ is not renormalised and its $\beta$-function vanishes to all orders of perturbation theory. Super Yang-Mills theory is therefore called a conformal field theory (CFT).

For the purpose of defining the dictionary of the AdS/CFT correspondence in the planar limit, the objects we need to consider on the gauge side are the gauge-invariant single-trace local operators
\begin{equation}
 {\cal O}_I (x) = {\rm Tr} \: [ \chi_{i_1} (x) \chi_{i_2} (x) \ldots \chi_{i_n} (x) ],
\label{eq:generic_SYM_s-trace_op}
\end{equation}
which can be constructed out of any combination of the fundamental fields: $\chi_i \in \{ \phi_j, \Psi_a, \bar{\Psi}_b, F_{\mu\nu}, D_\mu \}$. The scaling dimensions of these objects can then be calculated by examining the corresponding two-point correlation functions:
\begin{equation}
 \left< {\cal O}_I (x) {\cal O}_I (y) \right> \sim \frac{1}{|x-y|^{2 \Delta^I}}.
\label{eq:scalingdim_from_correl_f-s}
\end{equation}
Since these operators are renormalised in the quantum theory, their dimensions receive quantum corrections, which, in the perturbative regime, can be organised in terms of a double expansion in the 't Hooft coupling $\lambda$ and in the genus of the corresponding Feynman diagrams. In the planar limit, this reduces to a power series in $\lambda$, due to the suppression of all the higher genus terms by inverse powers of the rank $N$ of the gauge group:
\begin{equation}
 \Delta^I (\lambda) = \Delta^I_0 + \gamma^I(\lambda) = \Delta^I_0 + \sum_{k=1}^\infty \Delta^I_k \lambda^k
\label{eq:scalingdim_perturbative_exp}
\end{equation}
The classical contribution $\Delta^I_0$ is given by the sum of the classical scaling dimensions of the fundamental fields appearing inside the trace:
\begin{equation}
 [A_\mu] = [\phi] = 1 \qquad \qquad [\Psi_a] = \frac{3}{2},
\label{eq:class_scalingdim_of_fund_fields}
\end{equation}
while the term $\gamma^I (\lambda)$ is usually referred to as the anomalous dimension.

A very useful tool in the computation of the scaling dimensions in super Yang-Mills theory is the dilatation operation ${\cal D}$, which is a Cartan generator of the $\mathfrak{psu}(2,2|4)$ symmetry algebra. It acts linearly on the operators,
\begin{equation}
 {\cal D} \circ {\cal O}_I (x) = \sum_J {\cal D}_{IJ} {\cal O}_J (x),
\label{eq:action_of_dilat-op_on_ops}
\end{equation}
its eigenvalues are the scaling dimensions and it also preserves all the other conserved charges $(J_i, S_j)$ associated with global symmetries.

For small values of the 't Hooft coupling, the dilatation operator can be computed perturbatively in the planar limit:
\begin{equation}
 {\cal D} = \sum_{n=0}^\infty {\cal D}_n \qquad \qquad {\cal D}_n = O( \lambda^n ).
\label{eq:dilat-op_perturb_exp}
\end{equation}
Due to the huge level of operator mixing, it is in general very hard to diagonalise the dilatation operator, even at one loop. However, the restrictions imposed by the preservation of the Lorentz spins $S_j$ and R-charges $J_i$ lead to the identification of closed sectors, i.e. families of operators which only mix with each other under the action of the dilatation operator, thus simplifying the problem.

The simplest class of operators we can study is given by the chiral primaries
\begin{equation}
 \textrm{Tr } Z^{J_1} \qquad\qquad \textrm{Tr } Y^{J_2} \qquad\qquad \textrm{Tr } X^{J_3}, \qquad\qquad
\label{eq:def_chiral_primaries_SYM}
\end{equation}
defined in terms of the complex scalar fields $Z = \phi_1 + i \phi_2$, $Y = \phi_3 + i \phi_4$, $X = \phi_5 + i \phi_6$, each of which carries one unit of R-charge $J_1$, $J_2$ and $J_3$ respectively. As mentioned in the introduction, these operators are BPS and hence their anomalous dimensions vanish, so that their scaling dimensions are simply equal to their R-charges. The BMN vacuum can be chosen to be any of the above operators, in the limit of very large R-charge, e.g. $\textrm{Tr } Z^{J_1}$, $J_1 \gg 1$.

If we allow two different types of complex scalars inside the trace, we obtain one of the three equivalent copies of the $\mathfrak{su}(2)$ sector, corresponding to objects of the following type, up to permutations of the elementary fields:
\begin{equation}
 \textrm{Tr } (Z^{J_1 } Y^{J_2}) \qquad \textrm{Tr } (Z^{J_1 } X^{J_3}) \qquad \textrm{Tr } (Y^{J_2 } X^{J_3})
\label{eq:def_su(2)_sectors_SYM}
\end{equation}
Operators representing elementary scalar excitations of a single type over the BMN vacuum, e.g. $\textrm{Tr } (ZZZYZYZZ \ldots Z)$ lie in the $\mathfrak{su}(2)$ sector.

\section{The $\mathfrak{sl}(2)$ integrable spin chain}
\label{sec:sl2(R)_spin_chain}

The $\mathfrak{sl}(2)$ sector consists of single-trace operators containing only one type of complex scalar field and only one of the light-cone components of the covariant derivative inside the trace. Among the various possible choices, we are interested in the following version of the $\mathfrak{sl}(2)$ sector:
\begin{equation}
 \textrm{Tr} \: ( D^{s_1}_+ Z D^{s_2}_+ Z \ldots D^{s_J}_+ Z ).
\label{eq:sl2_sector_specific_op}
\end{equation}
This operator carries a single Lorentz spin $S = \sum_{k=1}^J s_j$ (each derivative contributes one unit) and a single R-charge $J$, equal to the total number of $Z$ fields. The twist is defined as the classical scaling dimension minus the total spin, $\Delta_0 - S = J$.

The operators in this sector can be identified with configurations of the Heisenberg $\mathrm{XXX}_{- \frac{1}{2}}$ spin chain. Each $Z$ field inside the trace corresponds to a site of the chain, so that its total length equals $J$. Every site carries a representation of $SL(2, \mathbb{R})$ with quadratic Casimir equal to $-1/2$. Finally, each derivative acting on a $Z$ field is interpreted as a single excitation (also known as \emph{magnon}) of the spin at the corresponding site. In this picture, the ferromagnetic vacuum coincides with the BMN vacuum $\mathrm{Tr} \: Z^J$.

The 1-loop contribution to the anomalous dimension of operators in the $\mathfrak{sl}(2)$ sector is proportional to the $\mathrm{XXX}_{- \frac{1}{2}}$ spin chain Hamiltonian:
\begin{equation}
 \gamma (\lambda) = \frac{\lambda}{8 \pi^2} \mathcal{H}_{\mathrm{XXX}_{-1/2}} + O(\lambda^2)
\label{eq:gamma=H_XXX_-1/2}
\end{equation}
Hence, the diagonalisation of the dilatation operator in this subsector reduces to the problem of finding the energy spectrum of the spin chain, which, as a consequence of the integrability of the Heisenberg model, can be solved by the algebraic Bethe Ansatz technique\footnote{The  remainder of this chapter is based on the results of \cite{Belitsky:2003ys,Belitsky:2004cz,Belitsky:2006en} and references therein.}.

A quantum spin variable $\mathcal{L}^\pm_k, \mathcal{L}^0_k$, for $k = 1, \ldots, J$, satisfying appropriate commutation relations, is associated with each site of the chain. These variables are then used in order to introduce a Lax matrix $\mathbb{L}_k (u)$ at each site of the chain, given by
\begin{equation}
 \mathbb{L}_k (u) = 
  \begin{pmatrix}
	 u + i \mathcal{L}^0_k & i \mathcal{L}^+_k \\
	 i \mathcal{L}^-_k & u - i \mathcal{L}^0_k
	\end{pmatrix}  \in SU(1,1).
\label{eq:spin_chain_Lax_matrix}
\end{equation}
The monodromy matrix of the chain is then defined as the ordered product of all the Lax matrices,
\begin{equation}
 \Omega (u) = \mathbb{L}_1 (u) \mathbb{L}_2 (u) \ldots \mathbb{L}_J (u),
\label{eq:spin_chain_monodromy_matrix}
\end{equation}
and its trace yields a tower of conserved quantities $q_j$:
\begin{eqnarray}
 t_J (u) & = & \mathrm{Tr} \: [ \mathbb{L}_1 (u) \mathbb{L}_2 (u) \ldots \mathbb{L}_J (u) ] \nn\\
         & = & 2 u^J + q_2 u^{J-2} + \ldots + q_{J-1} u + q_J \:.
\label{eq:spin_chain_tr_Omega}
\end{eqnarray}
The conserved charges were shown to commute in \cite{Faddeev:1996iy}, leading to the conclusion that the quantum spin chain is integrable. From this point, the algebraic Bethe Ansatz equations can be derived by imposing the Baxter equation,
\begin{equation}
 \left( u + \frac{i}{2} \right)^J Q (u+i) + \left( u - \frac{i}{2} \right)^J Q (u-i) = Q (u) t_L (u),
\label{eq:Baxter_eq}
\end{equation}
where the Baxter Q-operator $Q(u)$ is a polynomial whose degree has to be equal to $S$, as a consequence of the large-$u$ behaviour of the equation:
\begin{equation}
 Q(u) = \prod_{k=1}^S (u - \lambda_k)
\label{eq:Baxter_Q}
\end{equation}
(we have omitted the overall normalisation factor). The variable $u$ is usually called the \emph{spectral parameter}. Substituting \eqref{eq:Baxter_Q} into \eqref{eq:Baxter_eq}, several poles appear on the left-hand side of the Baxter equation. However, since we know that $t_L (u)$ is a polynomial, we have to impose the cancellation of these apparent poles. This requirement forces the roots $\lambda_k$ to satisfy the same Bethe equations which can be derived directly from the algebraic Bethe Ansatz:
\begin{equation}
 \left( \frac{\lambda_k + \frac{i}{2}}{\lambda_k - \frac{i}{2}} \right)^J = \prod_{j=1, j \neq k}^S \frac{\lambda_k - \lambda_j - i}{\lambda_k - \lambda_j + i}.
\label{eq:ABAE_spin_chain}
\end{equation}
Each Bethe root $\lambda_k$ represents the rapidity of an individual magnon, which is interpreted as a quasi-particle carrying a conserved energy $E_k$ and a conserved momentum $\theta_k$:
\begin{equation}
 E_k = \frac{1}{\lambda_k^2 + \frac{1}{4}} \qquad\qquad e^{i \theta_k} = \frac{\lambda_k - \frac{i}{2}}{\lambda_k + \frac{i}{2}}.
\label{eq:E,p,of_individual_magnons}
\end{equation}
Furthermore, it can be shown that all the roots have to be different from each other: $\lambda_k \neq \lambda_j$, for $k \neq j$. The total energy and momentum are given by the sum of the contributions from each magnon and can also be expressed in terms of the Baxter operator:
\begin{equation}
 E = \sum_{k=1}^S \frac{1}{\lambda_k^2 + \frac{1}{4}} = i \frac{Q' \left( \frac{i}{2} \right)}{Q \left( \frac{i}{2} \right)} - 
  i \frac{Q' \left( - \frac{i}{2} \right)}{Q \left( - \frac{i}{2} \right)}
   \qquad\qquad e^{i \theta} = \prod_{k=1}^S \frac{\lambda_k - \frac{i}{2}}{\lambda_k + \frac{i}{2}} = \frac{Q \left( \frac{i}{2} \right)}
    {Q \left( - \frac{i}{2} \right)}.
\label{eq:total_E,p_spin_chain}
\end{equation}
The cyclicity of the trace over the colour indices in \eqref{eq:sl2_sector_specific_op} imposes $e^{i \theta} = 1$, while the total energy $E$ yields the 1-loop contribution to the anomalous dimension, through equation \eqref{eq:gamma=H_XXX_-1/2}.

In the general quantum case, the standard approach is then to solve the Bethe equations \eqref{eq:ABAE_spin_chain} and compute the energy of the spin chain from \eqref{eq:total_E,p_spin_chain}.

For the purpose of this thesis, we are mainly going to study the large spin limit $S \to \infty$ with $J$ fixed, in which these equations become increasingly hard to solve, due to the diverging number of magnons. However, as we will see in the next section\footnote{The discussion will be based in particular on sections 2.2 and 3.3 of \cite{Belitsky:2006en}.}, it is possible to determine the spectrum in terms of the roots of the monodromy $t_J (u)$, which are more convenient objects to study, since their number remains constant and equal to $J$. 

\section{Spectrum at large $S$}

In the limit $S \to \infty$, with $J$ fixed, we define the parameter
\begin{equation}
 \eta = \frac{1}{S + \frac{J}{2}} \to 0 \:.
\label{eq:def_eta_BGK}
\end{equation}
In order to determine the spectrum as $S$ becomes infinite, it is possible to construct an asymptotic solution of the Baxter equation \eqref{eq:Baxter_eq}, which is valid in the region $u \sim O (\eta^0)$:
\begin{equation}
 Q^{({\rm as})} (u) = Q_+ (u) Q_- \left( - \frac{i}{2} \right) + Q_- (u) Q_+ \left( \frac{i}{2} \right)
\label{eq:Q(as)_BGK}
\end{equation}
where
\begin{equation}
 Q_\pm (u) = 2^{\mp i u} \prod_{j=1}^J \frac{\Gamma(\mp i u \pm i u_j)}{\Gamma \left( \mp i u + \frac{1}{2} \right)}
\label{eq:Q_pm_BGK}
\end{equation}
and $u_j$, for $j = 1, \ldots, J$, are the roots of the monodromy of the spin chain,
\begin{equation}
 t_J (u) = 2 \prod_{j=1}^J (u - u_j).
\label{eq:factorisation_of_tJ(u)_BGK}
\end{equation}
In the large $S$ limit, the number of magnons becomes infinite. The constraint imposing that the Bethe roots (or magnon rapidities) must all be different from each other acts as an exclusion principle, forcing the rapidities to fill a Dirac sea. It can be shown, however, that there are always $J$ holes in the rapidity distribution. Like magnons, such \emph{holes} can be interpreted as quasi-particles, and the numbers $u_j$ then acquire the meaning of their associated rapidities.

The energy eigenvalues of the spin chain can now be computed by evaluating \eqref{eq:total_E,p_spin_chain} on the asymptotic solution \eqref{eq:Q(as)_BGK}, which yields:
\begin{eqnarray}
 E & = & 2 \log 2 + \sum_{j=1}^J E^{\rm hole}_j \nn \\
 E^{\rm hole}_j & = & \psi \left( \frac{1}{2} + i u_j \right) + \psi \left( \frac{1}{2} - i u_j \right) - 2 \psi (1)
\label{eq:energy_in_terms_of_holes}
\end{eqnarray}
where the contribution $E^{\rm hole}_j$ is interpreted as the energy carried by the $j$-th hole, and $\psi(x) = d [\log \Gamma(x)]/dx$.

In order to link the behaviour of the holes to the behaviour of the conserved charges $q_j$, we need to introduce the semiclassical expansion. First of all, we rescale the spectral parameter by introducing $u = x/ \eta$ and we rewrite the monodromy as
\begin{equation}
 \tau(x) = \left( \frac{\eta}{x} \right)^J t_J \left( \frac{x}{\eta} \right) = 2 + \frac{\hat{q}_2}{x^2} + \frac{\hat{q}_3}{x^3} + \ldots
  + \frac{\hat{q}_J}{x^J},
\label{eq:tau(x)_BGK}
\end{equation}
where $q_k = \hat{q}_k \eta^{-k}$. We then assume that the rescaled charges can be expanded in powers of $\eta$:
\begin{equation}
 \hat{q}_k = \hat{q}_k^{(0)} + \eta \hat{q}_k^{(1)} + \ldots \:.
\label{eq:qhat_semiclassical_exp}
\end{equation}
Note that, in general, some charges may scale slower than $q_k \sim \eta^{-k} \sim S^k$ in the large $S$ limit, which implies that the corresponding coefficient $\hat{q}_k^{(0)}$ will vanish. Hence, we will indicate the highest non-vanishing coefficient as $\hat{q}_{J-M}^{(0)}$, with $M = 0, \ldots, J-2$ (one may check that the charge $\hat{q}^{(0)}_2$ never vanishes). We can now expand $\tau(x)$ as well, and the leading term is given by:
\begin{equation}
 \tau_0 (x) = 2 + \frac{\hat{q}_2^{(0)}}{x^2} + \frac{\hat{q}_3^{(0)}}{x^3} + \ldots + \frac{\hat{q}_{J-M}^{(0)}}{x^{J-M}} \:.
\label{eq:tau0_BGK}
\end{equation}
By rearranging \eqref{eq:tau(x)_BGK}, we obtain:
\begin{equation}
 t_J (u) = \frac{1}{\eta^J} (2 x^J + \hat{q}_2 x^{J-2} + \ldots + \hat{q}_L )
\label{eq:rewrite_tJ(u)_to_get_large/small_holes}
\end{equation}
which implies that the $u_j$ coincide with the zeros of the polynomial on the right-hand side. As $\eta \to 0$, the coefficients $\hat{q}_j = \hat{q}^{(0)}_j + O(\eta)$ vanish for $j = J-M+1, \ldots, J$. Thus, we are left with $J-M$ roots $x_j$, $j = M, \ldots, J$, which remain finite as $\eta \to 0$ and $M$ roots $x_j$, $j = 1, \ldots, M$, which vanish in the same limit. This translates into $u_j = x_j / \eta \sim S$ as $S \to \infty$ for the first set of solutions, and it can also be checked that either $u_j = O(1)$ or $u_j \to 0$ for the second set.

We call the diverging $u_j \sim S$, yielding a contribution $E_{\rm hole} \sim \log S$ to the total energy, ``large'' roots (or ``large'' holes) and the finite $u_j$ ``small'' roots (or ``small'' holes)\footnote{Our definition of ``small'' holes is slightly different from the definition given in \cite{Belitsky:2006en}, according to which ``small'' holes have instead a vanishing rapidity which is $O(1/ \log S)$.}, with energy $O( S^0 )$.

It is only a matter of algebra to expand $E_{\rm hole} (u_j)$ on the ``large'' roots and then eliminate the latter in favour of the charges $\hat{q}^{(0)}_j$. By \eqref{eq:gamma=H_XXX_-1/2}, the spectrum of 1-loop anomalous dimensions is thus given by
\begin{multline}
 \gamma = \frac{\lambda}{4 \pi^2} \left\{ (J-M) \log S + \log \hat{q}^{(0)}_{J-M} + \frac{1}{2} \sum_{j=1}^M
  \left[ \psi \left( \frac{1}{2} + i u_j \right) + \psi \left( \frac{1}{2} - i u_j \right) \right] \right. \\
   \left. \phantom{\sum_{j=1}^M} - J \psi(1) + \ldots \right\} + O(\lambda^2) \:,
\label{eq:1-loop_gamma_sl(2)}
\end{multline}
where the dots indicate terms which vanish in the limit $S \to \infty$ and, up to a moduli-independent constant, the contribution of the ``large'' holes is given by $(J-M) \log S + \log \hat{q}^{(0)}_{J-M}$.

The ``small'' roots $u_j$ are quantised by imposing the condition that the function $Q^{({\rm as})} (u)$ be regular on the real axis in its region of validity $u \sim O( \eta^0 )$. This requires the residues of all the poles of $Q_+ (u)$ and $Q_- (u)$ located in that region to vanish. Note that the poles at $u = u_j$ for $u_j$ a ``large'' root are outside this region, and thus these quantisation conditions do not apply to the large roots.

The result is a set of $M$ Bethe-type equations, one for each ``small'' root, which also receive contributions from the large roots. By approximating these equations as $S \to \infty$ in the case of two ``large'' holes ($J-M = 2$), one obtains
\begin{equation}
 8 u_j \log S = 2 \pi k_j \:, \qquad k_j \in \mathbb{Z} \:,
\label{eq:small_h_quantisation_cond_approx}
\end{equation}
which will be helpful later.
\paragraph{}

The interpretation of these results in terms of the spin chain is the following. As $S \to \infty$, $J-M \geq 2$ spins become highly excited, so that the corresponding spin vectors $\mathcal{L}_k$ become classical variables. These spins are associated with the ``large'' holes and are responsible for the leading diverging $O( \log S)$ contribution to the spectrum of the chain. Their dynamics are governed by the Hamiltonian $H_{J-M} = \log \hat{q}^{(0)}_{J-M}$.

The remaining $M$ spins remain instead in a low-level excited state, and therefore need to be treated quantum mechanically. They are associated with the ``small'' holes, which are in fact subject to the quantum Bethe-type equations mentioned just above, and they yield a subleading $O (S^0)$ contribution to the spectrum.

While the matter is essentially closed concerning the ``small'' holes, we still need to find appropriate semiclassical quantisation conditions for the ``large'' holes (recall that the Bethe equations do not apply to the ``large'' holes). As we will see below, these are described in terms of an algebraic curve.
\paragraph{}

We are now going to focus our attention on the classical spin variables, whose dynamics can be represented in terms of a classical spin chain with $J-M$ sites. The Lax matrix is defined as in \eqref{eq:spin_chain_Lax_matrix}, while the monodromy matrix becomes
\begin{equation}
 \Omega (u) = \mathbb{L}_1 (u) \mathbb{L}_2 (u) \ldots \mathbb{L}_{J-M} (u),
\label{eq:semiclassical_spin_chain_monodromy_matrix}
\end{equation}
so that the corresponding rescaled monodromy is given by $\tau_0 (x)$ as in \eqref{eq:tau0_BGK}. We notice that, as $M$ increases by one unit, we lose a Lax matrix in \eqref{eq:semiclassical_spin_chain_monodromy_matrix} and correspondingly the highest conserved charge in \eqref{eq:tau0_BGK} vanishes.

The quadratic Casimir can be neglected at leading order in this limit, so that the spin components satisfy
\begin{equation}
 \mathcal{L}^+_k \mathcal{L}^-_k + (\mathcal{L}^0_k)^2 = 0,
\label{eq:zero_Casimir_for_classical_spins}
\end{equation}
for $k = 1, \ldots, J$, up to $O(1/S)$ corrections. The quantum commutation relations of the spin chain, arising from the $\mathfrak{sl}(2, \mathbb{R})$ Lie algebra, translate into the following Poisson brackets:
\begin{equation}
 \{ \mathcal{L}^+_k, \mathcal{L}^-_{k'} \} = 2 i \delta_{k k'} \mathcal{L}^0_k \qquad \qquad
  \{ \mathcal{L}^0_k, \mathcal{L}^\pm_{k'} \} = \pm i \delta_{k k'} \mathcal{L}^\pm_k \:.
\label{eq:sl(2,R)_Poisson_brackets}
\end{equation}
As a consequence of these Poisson brackets, it is possible to show that the charges $q_j$ are in involution\footnote{This follows from the general result of \cite{Faddeev:1996iy}, which proves that the charges commute in the quantum case.}, $\{ q_i, q_j \} = 0$, $i,j = 2, \ldots, J$. The existence of this set of conserved charges in involution implies that the classical spin chain is integrable.

In the special case of a highest-weight configuration, which satisfies
\begin{equation}
 \sum_{k=1}^J \mathcal{L}_k^\pm = 0 \:,
\label{eq:highest_weight_condition_spin_chain}
\end{equation}
it is possible to show that
\begin{eqnarray}
 q_2 & = & - \left( S + \frac{J}{2} \right) \left( S + \frac{J}{2} - 1 \right) - \frac{J}{4} \nn\\
     & \simeq & - S^2 \qquad \textrm{as } S \to \infty \:,
\label{eq:q_2_for_highest_weight}
\end{eqnarray}
which implies $\hat{q}_2^{(0)} = -1$. In the following, we will discuss results which hold for both highest- and non-highest-weight states, the only macroscopic difference between them being the value of $q_2$ and $\hat{q}_2^{(0)}$.
\paragraph{}

In order to introduce the algebraic curve, we first express the Baxter operator in terms of the eikonal phase $S(x)$:
\begin{equation}
 Q \left( \frac{x}{\eta} \right) = \eta^{-S} e^{\frac{S(x)}{\eta}} \qquad S(x) = \eta \sum_{k=1}^S \log (x - \eta \lambda_k) \:.
\label{eq:def_S(x)_BGK}
\end{equation}
We then assume that the function $S(x)$ also admits a semiclassical expansion in powers of $\eta$, as the one we saw in equation \eqref{eq:qhat_semiclassical_exp} for the conserved charges:
\begin{equation}
 S(x) = S_0 (x) + \eta S_1 (x) + \ldots \:.
\label{eq:S(x)_semiclassical_exp}
\end{equation}
At this point, we substitute the expansions for $S(x)$ and for the charges into the Baxter equation \eqref{eq:Baxter_eq} and solve it at leading order in $\eta$, obtaining:
\begin{equation}
 2 \cos p(x) = \tau_0 (x) \qquad \qquad p(x) = S'_0 (x).
\label{eq:Baxter_eq_lead_semiclass_order}
\end{equation}
The function $p(x)$ is known as the \emph{quasi-momentum} and we can use the left-hand side of the previous equation in order to study its analytical structure. $p(x)$ has $2(J-M)-2$ square root branch points at $\tau_0 (x) = \pm 2$ and a logarithmic branch point at $x=0$. Consequently, $p(x)$ has an infinite number of possible values at each point on the complex plane, which are related to each other by sign changes and shifts by integer multiples of $2 \pi$.

The differential $dp(x)$,
\begin{equation}
 dp = - \frac{\tau_0' (x)}{\sqrt{4 - \tau_0^2 (x)}} dx,
\label{eq:dp_BGK}
\end{equation}
(or, equivalently, the first derivative $p'(x)$) has a simpler structure, since, while it has the same square root branch points as the quasi-momentum, the singularity at the origin becomes a simple pole. Hence, $dp$ can be made single-valued on a two-sheeted hyper-elliptic Riemann surface $\Gamma_{J-M}$ defined by the following function:
\begin{equation}
 \Gamma_{J-M}: \qquad y = \sqrt{4 - \tau_0^2 (x)} \:.
\label{eq:spectral_curve_BGK}
\end{equation}
This surface is usually called the \emph{spectral curve}.

Because of the fact that the charges $\hat{q}^{(0)}_j$ appearing in \eqref{eq:tau0_BGK} are determined by the monodromy $t_L$ of the spin chain, according to \eqref{eq:spin_chain_tr_Omega}, it can be shown that they only take such values that the $(J-M)-2$ points at which $\tau_0 = 2$ and the $(J-M)$ points at which $\tau_0 = -2$ are all real. Together with the fact that $\tau_0 (x)$ approaches 2 from below as $x \to \pm \infty$, while its only singularity is a pole at the origin, this implies that the square root branch points have to arrange themselves in consecutive pairs on which $\tau_0$ takes the same values, starting with $\tau_0 = -2$ on the two outermost pairs, and then alternating between $\tau_0 = 2$ and $\tau_0 = -2$ on the successive inner pairs, until the origin is reached. This pattern accounts for $2(J-M)-4$ branch points. The two remaining branch points are located at the two sides of the origin $x=0$ and the corresponding values of $\tau_0$ may or may not coincide, depending on the parity of the leading diverging power $(1/x)^{J-M}$.

\begin{figure}
\centering
\psfrag{a1}{\footnotesize{$2 (-1)^\frac{N^+ + 1}{2}$}}
\psfrag{a2}{\footnotesize{$+2$}}
\psfrag{a3}{\footnotesize{$+2$}}
\psfrag{a4}{\footnotesize{$-2$}}
\psfrag{a5}{\footnotesize{$-2$}}
\psfrag{b1}{\footnotesize{$2 (-1)^\frac{N^- + 1}{2}$}}
\psfrag{b2}{\footnotesize{$+2$}}
\psfrag{b3}{\footnotesize{$+2$}}
\psfrag{b4}{\footnotesize{$-2$}}
\psfrag{b5}{\footnotesize{$-2$}}
\psfrag{c1}{\footnotesize{$x_+^{(N^+)}$}}
\psfrag{c2}{\footnotesize{$x_+^4$}}
\psfrag{c3}{\footnotesize{$x_+^3$}}
\psfrag{c4}{\footnotesize{$x_+^2$}}
\psfrag{c5}{\footnotesize{$x_+^1$}}
\psfrag{d1}{\footnotesize{$x_-^{(N^-)}$}}
\psfrag{d2}{\footnotesize{$x_-^4$}}
\psfrag{d3}{\footnotesize{$x_-^3$}}
\psfrag{d4}{\footnotesize{$x_-^2$}}
\psfrag{d5}{\footnotesize{$x_-^1$}}
\includegraphics[width=\columnwidth]{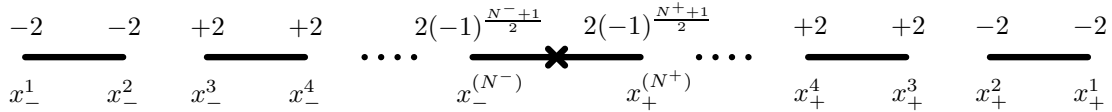}%
\caption{The branch cuts for the quasi-momentum $p(x)$ and the differential $dp(x)$. Above each branch point $x_\pm^{(j)}$ the value $\tau_0 (x_\pm^{(j)})$ is indicated. The cross represents the origin and the dots indicate omitted cuts.}
\label{fig:pm2_pattern_and_cuts}%
\end{figure}

We label the branch points as $x^{(j)}_+$, for $j = 1, \ldots, N^+$, from infinity to the origin on the positive real axis, and similarly as $x^{(j)}_-$, for $j = 1, \ldots, N^-$, on the negative real axis. Note that both $N^+$ and $N^-$ are always odd and that $N^+ + N^- = 2(J-M-1)$. A consistent branch cut structure for $dp(x)$ is obtained by connecting the two branch points belonging to each pair together, defining intervals of the type
\begin{eqnarray}
 I_j^+ & = & [x^{(2j)}_+, x^{(2j-1)}_+] \:, \qquad \textrm{for } j = 1, \ldots, \frac{N^+ - 1}{2} \nn\\
 I_j^- & = & [x^{(2j-1)}_-, x^{(2j)}_-] \:, \qquad \textrm{for } j = 1, \ldots, \frac{N^- - 1}{2} \nn\\
 I_0 & = & [x^{(N^-)}_-, x^{(N^+)}_+]
\label{eq:cuts_BGK}
\end{eqnarray}
so that the genus of the Riemann surface is $(J-M)-2$. This is illustrated in Fig. \ref{fig:pm2_pattern_and_cuts}. The same cut structure is also acceptable for the quasi-momentum itself, although in this case the origin is a logarithmic branch point and the cut $I_0$ must necessarily touch it.

This choice of the branch cuts is peculiar in the sense that it is known that the $S$ Bethe roots $\lambda_k$ are always real and condense along the cuts $I_j^\pm, I_0$ as $S \to \infty$.

Having an infinite number of possible values at each point on the complex plane, $p(x)$ becomes single-valued on a Riemann surface $\tilde{\Gamma}_{J-M}$ with an infinite number of sheets, as opposed to the two-sheeted spectral curve $\Gamma_{J-M}$. These two surfaces should not be confused with each other. We are now going to select an individual branch of the quasi-momentum, specified by the two conditions $p(\infty) = 0$, $p(x^{(1)}_+) = \pi$ and corresponding to a single sheet on $\tilde{\Gamma}_{J-M}$ and then to identify this sheet with the upper sheet of $\Gamma_{J-M}$, defined by the sign choice for the leading behaviour $dp \sim - dx/x^2$ as $x \to \infty$. We call this the \emph{physical sheet}.

Once the reference sheet has been established, $p(x)$ may be computed on its whole Riemann surface as
\begin{equation}
 p(x) = \int_{\infty^+}^x dp \:,
\label{eq:p(x)_abelian_integral_BGK}
\end{equation}
where $\infty^+$ is the point at infinity on the physical sheet and the integral is carried out along a regular contour connecting that point to the other endpoint $x$ lying on $\tilde{\Gamma}_{J-M}$. The notation $x^\pm$ only applies to the two-sheeted surface $\Gamma_{J-M}$, where the points $x^+$ and $x^-$ lie on the upper and lower sheet respectively and correspond to the point $x$ on the complex plane. Due to the definition of the physical sheet, we can identify all the points of the upper sheet of $\Gamma_{J-M}$ with the points of the corresponding sheet of $\tilde{\Gamma}_{J-M}$, so that we may actually say that the above integral starts at $\infty^+$.

\begin{figure}
\centering
\psfrag{a}{\footnotesize{$\tilde{\Gamma}_{J-M}$}}
\psfrag{b}{\footnotesize{$\Gamma_{J-M}$}}
\psfrag{c}{\footnotesize{PS}}
\psfrag{d}{\footnotesize{PS}}
\includegraphics[width=\columnwidth]{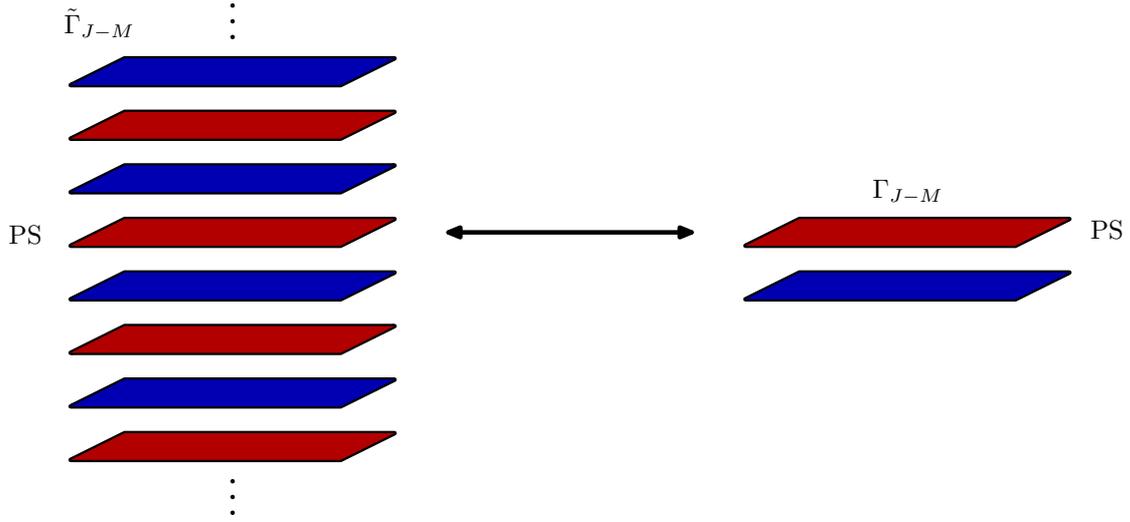}%
\caption{Schematic representation of the Riemann surfaces associated with $p(x)$ and $dp(x)$. Red sheets are dual to the upper sheet of $\Gamma_{J-M}$, while blue sheets are dual to the lower sheet. The physical sheet on both surfaces is marked as ``PS''.}
\label{fig:sheets_identification_BGK}%
\end{figure}

However, there are in general infinitely many possible duals in $\tilde{\Gamma}_{J-M}$ for the two sheets of $\Gamma_{J-M}$ (see Fig. \ref{fig:sheets_identification_BGK}) and therefore, depending on which cuts the contour crosses, the final endpoint of the integral \eqref{eq:p(x)_abelian_integral_BGK} may lie on several different sheets of $\tilde{\Gamma}_{J-M}$, even though it is uniquely specified on 
$\Gamma_{J-M}$. This point will be made clearer shortly.

\begin{figure}
\centering
\psfrag{a}{\footnotesize{$I_0$}}
\psfrag{a2}{\footnotesize{$I_2^+$}}
\psfrag{a1}{\footnotesize{$I_1^+$}}
\psfrag{b2}{\footnotesize{$I_2^-$}}
\psfrag{b1}{\footnotesize{$I_1^-$}}
\psfrag{b}{\footnotesize{$\alpha_0$}}
\psfrag{c1}{\footnotesize{$\alpha_1^+$}}
\psfrag{c2}{\footnotesize{$\alpha_2^+$}}
\psfrag{d1}{\footnotesize{$\alpha_1^-$}}
\psfrag{d2}{\footnotesize{$\alpha_2^-$}}
\psfrag{e1}{\footnotesize{$\gamma_\frac{N^+ + 1}{2}^+$}}
\psfrag{e2}{\footnotesize{$\gamma_2^+$}}
\psfrag{e3}{\footnotesize{$\gamma_1^+$}}
\psfrag{f1}{\footnotesize{$\gamma_\frac{N^- + 1}{2}^-$}}
\psfrag{f2}{\footnotesize{$\gamma_2^-$}}
\psfrag{f3}{\footnotesize{$\gamma_1^-$}}
\includegraphics[width=\columnwidth]{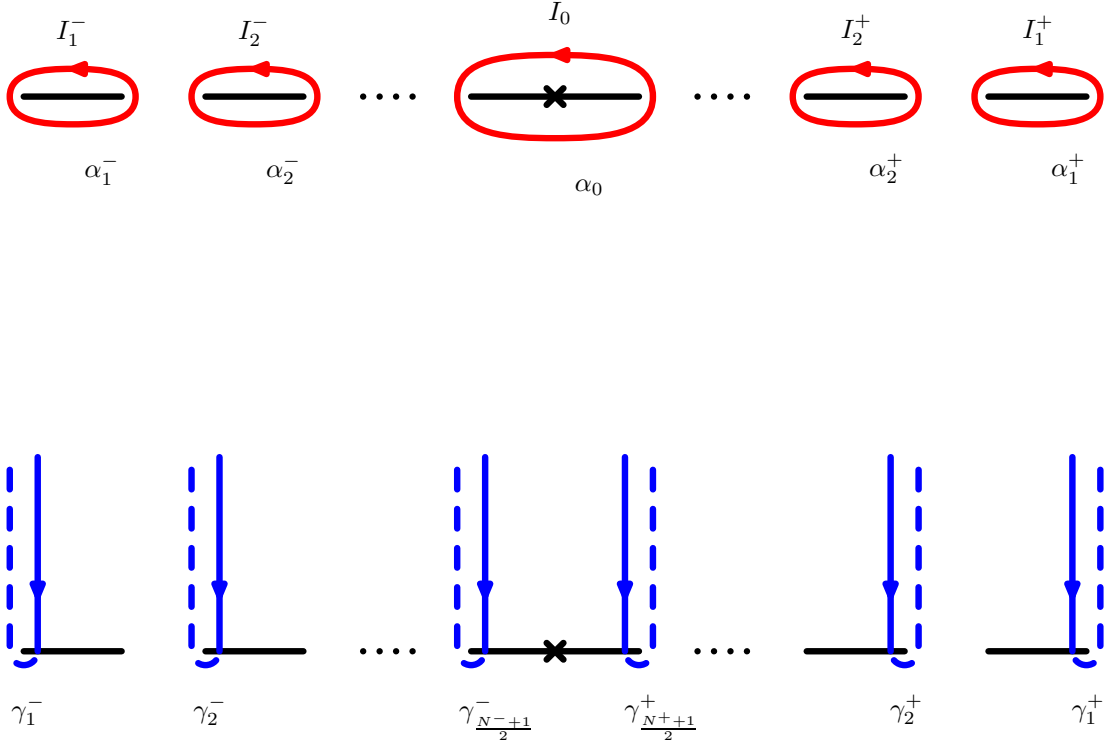}%
\caption{The A-cycles and B-cycles on $\Gamma_{J-M}$.}
\label{fig:Cycles_BGK}%
\end{figure}

By starting, for instance, at a point close to $\infty^+$ near the positive real axis and then following the variation of $p(x)$ from its initial value $p(\infty^+) = 0$ as $x$ moves along a contour leading to the various square root branch points without crossing any cuts, it is possible to show that, on the physical sheet,
\begin{eqnarray}
 p(x^{(2j)}_+) & = & p(x^{2j-1}_+) = \pi j \qquad \textrm{for } j = 1, \ldots, \frac{N^+ - 1}{2} \nn\\
 p(x^{(N^+)}_+) & = & \pi \frac{N^+ + 1}{2} \nn\\
 p(x^{(2j)}_-) & = & p(x^{2(j-1)}_-) = - \pi j \qquad \textrm{for } j = 1, \ldots, \frac{N^- - 1}{2} \nn \\
 p(x^{(N^-)}_-) & = & - \pi \frac{N^- + 1}{2}.
\label{eq:p(x)_at_branch_points_BGK}
\end{eqnarray}
These values determine the A-periods and the B-periods of the differential $dp(x)$: for the former we have
\begin{equation}
 \oint_{\alpha_0} dp = 0 \:, \qquad \oint_{\alpha_j^\pm} dp = 0 \:, \qquad \forall j,
\label{eq:A-periods_BGK}
\end{equation}
where the A-cycles $\alpha_j^\pm$, $\alpha_0$ encircle the cut $I_j^\pm$, $I_0$ on the physical sheet, respectively. For the latter we obtain instead
\begin{eqnarray}
 \int_{\gamma_j^\pm} dp & = & \pm 2 \pi j \qquad \textrm{for } j = 1, \ldots, \frac{N^\pm + 1}{2},
\label{eq:B-periods_BGK}
\end{eqnarray}
where the B-cycles $\gamma_j^\pm$ start at $\infty^+$, cross the cut $I_j^\pm$ and then reach $\infty^-$, on the lower sheet. Note that the two innermost contours $\gamma_{(N^+ + 1)/2}$ and $\gamma_{(N^- + 1)/2}$ cross the cut $I_0$ to the right and to the left of the origin, respectively. (See Fig. \ref{fig:Cycles_BGK}).

Specifically, the above A-periods indicate that the cut structure shown in Fig. \ref{fig:pm2_pattern_and_cuts} is appropriate for $p(x)$, since it does not pick up a monodromy if $x$ follows a contour encircling a cut. The B-periods instead imply that, when $x$ follows a contour starting on the upper sheet at $x^+$ and ending on the lower sheet at the corresponding point $x^-$ and crossing a single cut $I_j^\pm$, the initial and final values of $p(x)$ are related as follows\footnote{For a proof of this fact, see \eqref{eq:monodromy_of_p(x)_FG} and the comments below.}:
\begin{equation}
 p(x^+) + p(x^-) = \pm 2 \pi j \:.
\label{eq:monodromy_of_p(x)_BGK}
\end{equation}
Equivalently, the discontinuity of $p(x)$ across a cut is given by
\begin{equation}
 p(x+ i0) + p(x-i0) = \pm 2 \pi j \:,
\label{eq:discontinuity_of_p(x)_BGK}
\end{equation}
for $x \in I_j^\pm$. As for $I_0$, we have to split it at the origin into two adjacent cuts, $I^-_{(N^- +1)/2} = [x^{(N^-)}_-,0]$ and $I^+_{(N^+ +1)/2} = [0,x^{(N^+)}_-]$, which then behave exactly as all the other cuts.

Due to the fact that $p(x)$ changes sign every time we cross a cut, we may say that the sheets of $\tilde{\Gamma}_{J-M}$ which are reached after crossing an even number of cuts are ``dual'' to the upper sheet of $\Gamma_{J-M}$, while the remaining sheets are ``dual'' to the lower sheet. It is also clear that, because of the cut-dependent constant shift acquired by $p(x)$ after each crossing, there are an infinite number of sheets on $\tilde{\Gamma}_{J-M}$\footnote{For this to be true, it is enough to assume $\hat{q}_2^{(0)} \neq 0$, since then there are always at least the two branch points at the sides of the origin, which introduce two cuts $I^\pm_{(N^\pm +1)/2}$ with different associated periods ($N^\pm = 1$).}. This kind of behaviour is, as we might have expected, a variation of the analytical structure of the inverse cosine function.

As we mentioned above, the Bethe roots condense along the cuts as $S \to \infty$. In particular, from the definition of $S(x)$ we immediately obtain
\begin{equation}
 S'(x) = \eta \sum_{k=1}^N \frac{1}{x - \eta \lambda_k},
\label{eq:S'(x)_from_def_BGK}
\end{equation}
which then implies that the integral of $S'(x)$ along the A-cycles is proportional to the total number $l_j^\pm$, $l_0 \in \mathbb{Z}^+$ of roots condensing along the corresponding cut:
\begin{eqnarray}
 \oint_{\alpha_j^\pm} S'(x) dx & = & 2 \pi i \eta l_j^\pm \simeq 2 \pi i \frac{l_j^\pm}{S} \qquad \textrm{for } j = 1, \ldots, 
  \frac{N^\pm - 1}{2} \nn\\
 \oint_{\alpha_0} S'(x) dx & = & 2 \pi i \eta l_0 \simeq 2 \pi i \frac{l_0}{S}
\label{eq:l_j_from_S'(x)_BGK}
\end{eqnarray}
By exploiting the fact that the sum of the integrals along all the A-cycles is homologous to an integral around $\infty^+$, one can show that the \emph{filling numbers} $l_j^\pm$, $l_0$ have to satisfy
\begin{equation}
 l_0 + \sum_{j=1}^{\frac{N^- - 1}{2}} l_j^- + \sum_{k=1}^{\frac{N^+ - 1}{2}} l_k^+ = S
\label{eq:tot_filling_number=S_BGK}
\end{equation}
which is just a restatement of the fact that the total number of Bethe roots is equal to $S$.

Furthermore, it is possible to show that the following additional constraint must hold:
\begin{equation}
 \sum_{j=1}^{\frac{N^+ + 1}{2}} j l_j^+ - \sum_{j=1}^{\frac{N^- + 1}{2}} j l_j^- = 0,
\label{eq:level_matching_BGK}
\end{equation}
where the filling numbers $l_{(N^\pm + 1)/2}^\pm$ correspond to the contributions to $l_0$ from the two halves of the cut $I_0$ lying at the two sides of the origin ($l_{(N^+ + 1)/2}^+ + l_{(N^- + 1)/2}^- = l_0$).

Finally, by substituting in the leading behaviour of $S'(x)$ in the large $S$ limit from \eqref{eq:Baxter_eq_lead_semiclass_order} and integrating by parts, we may re-express the filling numbers as:
\begin{eqnarray}
 - \frac{1}{2 \pi i} \oint_{\alpha_j^\pm} x dp & \simeq & \frac{l_j^\pm}{S} \qquad \textrm{for } j = 1, \ldots, \frac{N^\pm - 1}{2} \nn\\
 - \frac{1}{2 \pi i} \oint_{\alpha_0} x dp & \simeq & \frac{l_0}{S} \:.
\label{eq:filling_numbers_with_xdp_BGK}
\end{eqnarray}
The importance of these relations lies in the fact that, since the $l_j^\pm$ and $l_0$ have to be integers, they provide the semiclassical quantisation conditions for the moduli $\hat{q}^{(0)}_j$ of the curve $\Gamma_{J-M}$ which we were looking for. Once we impose these conditions, the spectrum of the ``large'' spike Hamiltonian $H_{J-M} (l_j) = \log \hat{q}^{(0)}_{J-M}$ becomes discretised.
\paragraph{}

In summary, equation \eqref{eq:1-loop_gamma_sl(2)}, together with the quantisation conditions for the ``large'' holes \eqref{eq:filling_numbers_with_xdp_BGK} and for the ``small'' holes (given by \eqref{eq:small_h_quantisation_cond_approx} in the case $M = J-2$), yields the semiclassical spectrum of the one-loop anomalous dimensions of operators in the $\mathfrak{sl}(2)$ sector in the limit of large Lorentz spin $S$, with fixed R-charge $J$.

\chapter{Integrability methods in semiclassical string theory on $AdS_3$ and $AdS_3 \times S^1$}
\label{sec:integrability_methods}

In this chapter we will briefly review two techniques which exploit the integrability of the string theory equations of motion, specialising to the cases of $AdS_3$ and $AdS_3 \times S^1$, respectively. Pohlmeyer reduction allows to link solutions of the equations of motion to solutions of simpler two-dimensional models. The finite-gap method allows instead to construct very general classes of solutions to the equations of motion and to determine their spectra in terms of algebraic curves.

\section{String theory basics on $AdS_3 \times S^1$}

$AdS_3$ space is a 3-dimensional hyperboloid embedded in $\mathbb{R}^{2,2}$, defined by the following constraint:
\begin{equation}
X_\mu X^\mu = -X_0^2-X_1^2+X_2^2+X_3^2 = -1
\label{eq:AdS3_constr}
\end{equation}
The bosonic part of the string theory action on $AdS_3 \times S^1$, in conformal gauge, is then defined in terms of the embedding coordinates $X_\mu$ as:
\begin{equation}
I = - \frac{\sqrt{\lambda}}{4\pi}\int d\sigma d\tau \left[ G_{\mu\nu} \partial_a X^\mu \partial^a X^\nu + \Lambda \left( G_{\mu\nu} X^\mu X^\nu + 1 \right) + \partial_a \varphi \partial^a \varphi \right]
\label{eq:AdS3xS1_string_action}
\end{equation}
where $\lambda$ is the t'Hooft coupling, $\varphi$ is the angular coordinate along the $S^1$, $G_{\mu\nu} = \mathrm{diag}(-1,-1,1,1)$ is the $\mathbb{R}^{2,2}$ metric, and the worldsheet indices are contracted with the 2-dimensional Minkowski metric $\eta_{ab} = \mathrm{diag}(-1,1)$.

Once we eliminate the Lagrange multiplier $\Lambda$, the equations of motion for this action become:
\begin{eqnarray}
 && \partial_a \partial^a X_\mu - (\partial_a X^\nu \partial^a X_\nu) X_\mu = 0 \nn\\
 && \partial_a \partial^a \varphi = 0.
\label{eq:AdS3xS1_eom_Xmu}
\end{eqnarray}
We fix the residual gauge symmetry by demanding that the solution to the decoupled equation for the coordinate $\varphi$ take the following form:
\begin{equation}
 \varphi = \frac{J}{\sqrt{\lambda}} \tau.
\label{eq:gauge_fixing_AdS3xS1}
\end{equation}
If we introduce the light-cone worldsheet coordinates $\sigma^\pm = (\tau\pm\sigma)/2$, we can write
the Virasoro constraints as:
\begin{equation}
\partial_\pm X^\mu \partial_\pm X_\mu = - \frac{J^2}{\lambda}.
\label{eq:Virasoro_lc}
\end{equation}
$AdS_3$ can also be parametrised by the global coordinates ($t$,$\rho$,$\phi$), with $\rho \geq 0$, which are related to the embedding coordinates as follows:
\begin{eqnarray}
 X_0 & = & \cosh\rho \cos t \nonumber \\
 X_1 & = & \cosh\rho \sin t \nonumber \\
 X_2 & = & \sinh\rho \cos \phi \nonumber \\
 X_3 & = & \sinh\rho \sin \phi
\label{eq:AdS3_global_to_emb_coord_transf}.
\end{eqnarray}
In terms of these coordinates, the $AdS_3$ line element is:
\begin{equation}
 ds^2 = - \cosh^2\rho \; dt^2 + d\rho^2 + \sinh^2\rho \; d\phi^2.
\label{eq:AdS3_line_element_global_coords}
\end{equation}
Another useful coordinate system on $AdS_3$ is given by the complex coordinates $Z_i$:
\begin{eqnarray}
 Z_1 & = & X_0 + i X_1 = \cosh\rho \; e^{i t} \nonumber \\
 Z_2 & = & X_2 + i X_3 = \sinh\rho \; e^{i \phi}.
\label{eq:AdS3_def_cx_coords}
\end{eqnarray}
It allows us to rewrite the $AdS_3$ constraint, the equations of motion and the Virasoro constraints as:
\begin{eqnarray}
 |Z_1|^2 - |Z_2|^2 & = & 1 \label{eq:AdS3_constr_cx} \\
 \partial_a\partial^a Z_i - (-\partial_b Z_1 \partial^b \bar{Z}_1 + \partial_c Z_2 \partial^c \bar{Z}_2) Z_i & = & 0 \quad i=1,2
   \label{eq:AdS3_eom_cx} \\
 \partial_\pm Z_1 \partial_\pm \bar{Z}_1 - \partial_\pm Z_2 \partial_\pm \bar{Z}_2 & = & \frac{J^2}{\sqrt{\lambda}} \label{eq:AdS3_Virasoro_cx}.
\end{eqnarray}
The string action \eqref{eq:AdS3xS1_string_action} is invariant under global time and angular translations: $t \to t + a$, $\phi \to \phi + b$, $\varphi \to \varphi + c$. By Noether's theorem, the associated conserved charges are the energy $\Delta$\footnote{In the following, we will always indicate the energy of a string solution as $\Delta$, since this quantity is expected to match the scaling dimension $\Delta$ on the gauge side.}, the $AdS_3$ angular momentum $S$ (also referred to as ``spin'') and the $S^1$ angular momentum $J$:
\begin{eqnarray}
 \Delta & = & \frac{\sqrt{\lambda}}{2\pi} \int d\sigma \; \mathrm{Im}(\bar{Z}_1 \partial_\tau Z_1) \nn\\
 S & = & \frac{\sqrt{\lambda}}{2\pi} \int d\sigma \; \mathrm{Im}(\bar{Z}_2 \partial_\tau Z_2) \nn\\
 J & = & \frac{\sqrt{\lambda}}{2\pi} \int d\sigma \; \partial_\tau \varphi
\label{eq:AdS3xS1_charges}
\end{eqnarray}
where the integrals are carried out over the entire range of $\sigma$ (e.g. $\sigma \in [0,2\pi]$ for a closed string).

\section{$AdS_3$ sinh-Gordon connection}

\subsection{Pohlmeyer Reduction on $AdS_3$}

Pohlmeyer's reduction procedure establishes a relation between soutions to complicated systems of second order differential equations and solutions to simpler equations, such as the sine- and sinh-Gordon equations (including their complex versions) or the Liouville equation. In the specific case of $AdS_3$, it connects the $\sigma$-model equations to the sinh-Gordon, cosh-Gordon and Liouville equations. In this brief review, we follow \cite{Larsen:1996gn} and \cite{Jevicki:2007aa}. In order to restrict ourselves to pure $AdS_3$ space, we have to impose $J=0$, but otherwise all the conventions of the previous section still apply\footnote{Pohlmeyer reduction on $AdS_3 \times S^1$ leads to the complex sinh-Gordon equation, but we will not need to consider this case for the purposes of this thesis.}.

The initial system we are going to reduce is given by the $AdS_3$ string equations of motion \eqref{eq:AdS3xS1_eom_Xmu} and the Virasoro constraints \eqref{eq:Virasoro_lc}.

The first step is to define an orthogonal basis for $\mathbb{R}^{2,2}$ in terms of $X$ and its derivatives. Let's first consider the set of vectors $\{X,\partial_+ X,\partial_- X\}$: the first has negative unit norm (by the $AdS_3$ constraint \eqref{eq:AdS3_constr}) with respect to the $\mathbb{R}^{2,2}$ scalar product, whereas the other two have vanishing norm (by the Virasoro constraints). As we can easily see by differentiating the $AdS_3$ constraint, $X$ is orthogonal to the other two vectors:
\begin{equation}
 X_\mu \partial_\pm X^\mu = 0.
\label{eq:X_orthogonal_to_dpm_X}
\end{equation}
Furthermore, the scalar product of the remaining two vectors, $\partial_+ X_\mu \partial_- X^\mu$, can only vanish at isolated points on the worldsheet, since its vanishing turns the $AdS_3$ equations of motion into the flat space equations of motion.
Therefore, the set of vectors $\{X,\partial_+ X,\partial_- X\}$ is linearly independent except possibly at isolated points on the worldsheet. This will be enough for our purposes.

We then add one last element $B$, which we require to be a unit vector orthogonal to all the other vectors in the basis:
\begin{equation}
 B_\mu B^\mu = 1 \:, \qquad B_\mu X^\mu = B_\mu \partial_+ X^\mu = B_\mu \partial_- X^\mu = 0 \:,
\label{eq:def_vector_B_Pohlmeyer}
\end{equation}
and thus we define our basis to be:
\begin{equation}
 e_i = (X,\partial_+ X,\partial_- X,B)_i.
\label{eq:def_R22_basis}
\end{equation}
We also introduce the following functions:
\begin{eqnarray}
 \alpha & = & \ln(-\partial_+ X_\mu \partial_- X^\mu) \label{eq:def_alpha_Pohlmeyer} \\
 u & = & B_\mu \partial_+^2 X^\mu \label{eq:def_u_Pohlmeyer} \\
 v & = & B_\mu \partial_-^2 X^\mu \label{eq:def_v_Pohlmeyer}
\end{eqnarray}
and we then rewrite the equations of motion \eqref{eq:AdS3xS1_eom_Xmu} in terms of $\alpha$:
\begin{equation}
 \partial_+ \partial_- X_\mu = - e^\alpha X_\mu
\label{eq:AdS3_eom_in_terms_of_alpha_Pohlmeyer}
\end{equation}
It is now only a matter of algebra to check that a suitable $B$ is given by the following expression:
\begin{equation}
 B_\mu = e^{-\alpha} \epsilon_{\mu\nu\rho\sigma} X^\nu \partial_- X^\rho \partial_+ X^\sigma
\label{eq:AdS3_expr_for_B_Pohlmeyer}
\end{equation}
where $\epsilon_{\mu\nu\rho\sigma}$ is the anti-symmetric Levi-Civita tensor (and $\epsilon^{0123} = \epsilon_{0123} = 1$ in $\mathbb{R}^{2,2}$).

Some useful identities can be obtained by differentiating the Virasoro constraints \eqref{eq:Virasoro_lc}:
\begin{eqnarray}
 \partial_\pm X_\mu \partial_+ \partial_- X^\mu & = & 0 \label{eq:dpm_X_orthogonal_to_dpdm_X} \\
 \partial_\pm X_\mu \partial_\pm^2 X^\mu & = & 0 \label{eq:dpm_X_orthogonal_to_dpmdpm_X}.
\end{eqnarray}

Now we can start the derivation of the differential equation satisfied by $\alpha$. By directly differentiating its definition \eqref{eq:def_alpha_Pohlmeyer}, we obtain:
\begin{eqnarray}
 \partial_+ \alpha & = & - e^{-\alpha} \partial_+^2 X_\mu \partial_- X^\mu \label{eq:dp_alpha_Pohlmeyer} \\
 \partial_- \alpha & = & - e^{-\alpha} \partial_-^2 X_\mu \partial_+ X^\mu \label{eq:dm_alpha_Pohlmeyer}.
\end{eqnarray}
Then, we can differentiate any of these two equations again and obtain the second derivative of $\alpha$:
\begin{equation}
 \partial_+\partial_-\alpha = - e^{-\alpha}[e^{-\alpha} (\partial_+^2 X_\mu \partial_- X^\mu)(\partial_-^2 X_\mu \partial_+ X^\mu) + \partial_+\partial_-^2 X_\mu \partial_+X^\mu + \partial_-^2 X_\mu \partial_+^2 X^\mu]
\label{eq:derivation_alpha_eom_Pohlmeyer_1}
\end{equation}
where we have rewritten all the first derivatives of $\alpha$ according to \eqref{eq:dp_alpha_Pohlmeyer} and \eqref{eq:dm_alpha_Pohlmeyer}. At this point, we need to decompose $\partial_+^2 X$ and $\partial_-^2 X$ on the basis \eqref{eq:def_R22_basis}. The components can be obtained by evaluating the scalar products of these two second derivatives of $X$ with all the basis vectors. In particular, the component on $X$ must vanish because of \eqref{eq:X_orthogonal_to_dpm_X} and the Virasoro constraints:
\begin{equation}
 \partial_\pm^2 X_\mu X^\mu = \partial_\pm (\partial_\pm X_\mu X^\mu) - \partial_\pm X_\mu \partial_\pm X^\mu = 0
\label{eq:X_orthogonal_to_dpdp_X_and_dmdm_X}
\end{equation}
and the components on $B$ are just $u$ and $v$ by definitions \eqref{eq:def_u_Pohlmeyer} and \eqref{eq:def_v_Pohlmeyer}. Finally, one of the two components on $\partial_+ X$ and $\partial_- X$ also vanishes because it equals the LHS of \eqref{eq:dpm_X_orthogonal_to_dpmdpm_X}. The result is then:
\begin{eqnarray}
 \partial_+^2 X_\mu & = & -(e^{-\alpha}\partial_- X_\mu \partial_+^2 X^\mu) \partial_+ X_\mu + u B_\mu = \partial_+ \alpha \partial_+ X_\mu + u B_\mu \label{eq:decomposition_of_dpdp_X} \\
 \partial_-^2 X_\mu & = & -(e^{-\alpha}\partial_+ X_\mu \partial_-^2 X^\mu) \partial_- X_\mu + v B_\mu = \partial_- \alpha \partial_- X_\mu + v B_\mu\label{eq:decomposition_of_dmdm_X}
\end{eqnarray}
This allows us to compute the scalar product between $\partial_+^2 X$ and $\partial_-^2 X$ in terms of the scalar products of the basis vectors:
\begin{equation}
 \partial_-^2 X_\mu \partial_+^2 X^\mu = - e^{-\alpha}(\partial_- X_\mu \partial_+^2 X^\mu)(\partial_+ X_\mu \partial_-^2 X^\mu) + uv,
\label{eq:scalar_product_of_dpdp_X_and_dmdm_X}
\end{equation}
which we can then substitute back into \eqref{eq:derivation_alpha_eom_Pohlmeyer_1} to find:
\begin{equation}
 \partial_+\partial_-\alpha = - e^{-\alpha}[\partial_+\partial_-^2 X_\mu \partial_+X^\mu + uv ].
\label{eq:derivation_alpha_eom_Pohlmeyer_2}
\end{equation}
We only need to evaluate one last term, and we can do this by differentiating the equations of motion, in the form \eqref{eq:AdS3_eom_in_terms_of_alpha_Pohlmeyer},
\begin{equation}
 \partial_+\partial_-^2 X_\mu = - \partial_- \alpha e^\alpha X_\mu - e^\alpha \partial_- X_\mu,
\label{eq:dpdmdm_X}
\end{equation}
and then taking the scalar product with $\partial_+ X$:
\begin{equation}
 \partial_+\partial_-^2 X_\mu \partial_+ X^\mu = e^{2\alpha}.
\label{eq:scalar_product_of_dpdmdm_X_and_dp_X}
\end{equation}
Thus we have:
\begin{equation}
 \partial_+\partial_-\alpha + e^\alpha + uv e^{-\alpha} = 0,
\label{eq:general_alpha_eom}
\end{equation}
which is the general equation of motion satisfied by $\alpha$. It is now clear that if $uv=0$ for the string solution considered, then $\alpha$ satisfies the Liouville equation. However, in order to understand the connection to the sinh- and cosh-Gordon equations, which appear when $uv \neq 0$, a few more steps are required.

First of all, we calculate $\partial_-\partial_+^2 X$ and $\partial_+\partial_-^2 X$ from equations \eqref{eq:decomposition_of_dpdp_X}, \eqref{eq:decomposition_of_dmdm_X} and \eqref{eq:AdS3_eom_in_terms_of_alpha_Pohlmeyer}, by differentiating once:
\begin{eqnarray}
  \partial_-\partial_+^2 X_\mu & = & \partial_+\partial_-\alpha \partial_+ X_\mu - e^\alpha \partial_+\alpha X_\mu + \partial_- u B_\mu + u 
      \partial_- B_\mu \nonumber \\
  & = & - e^\alpha \partial_+\alpha X_\mu - e^\alpha \partial_+ X_\mu \nonumber \\
  \partial_+\partial_-^2 X_\mu & = & \partial_+\partial_-\alpha \partial_- X_\mu - e^\alpha \partial_-\alpha X_\mu + \partial_+ v B_\mu + v
      \partial_+ B_\mu \nonumber \\
  & = & - e^\alpha \partial_-\alpha X_\mu - e^\alpha \partial_- X_\mu,
\label{eq:dmdpdp_X_and_dpdmdm_X_from_2_sources}
\end{eqnarray}
where we have replaced $\partial_+\partial_- X_\mu$ with $-e^\alpha X_\mu$ in the first and third line, according to \eqref{eq:AdS3_eom_in_terms_of_alpha_Pohlmeyer}. Now, by taking the scalar product of both sides of both equations with $B$, and noticing that:
\begin{equation}
 B_\mu \partial_\pm B^\mu = 0 \:,
\label{eq:B_orthogonal_to_dpm_B}
\end{equation}
as a consequence of the fact that $B$ has a constant norm \eqref{eq:def_vector_B_Pohlmeyer}, we find:
\begin{eqnarray}
 \partial_- u & = & 0 \qquad \mathrm{i.e.} \qquad u=u(\sigma^+) \nn\\ 
 \partial_+ v & = & 0 \qquad \mathrm{i.e.} \qquad v=v(\sigma^-). 
\label{eq:u=u(sigma+)_v=v(sigma-)}
\end{eqnarray}
Now we introduce the following change of variable:
\begin{equation}
 \hat{\alpha} = \alpha - \frac{1}{2} \ln |u||v|,
\label{eq:def_alphahat_Pohlmeyer}
\end{equation}
which, as a consequence of \eqref{eq:u=u(sigma+)_v=v(sigma-)}, implies:
\begin{eqnarray}
 \partial_+ \partial_- \alpha & = & \partial_+ \partial_- \hat{\alpha} \nonumber \\
 e^\alpha & = & e^{\hat{\alpha}} \sqrt{|u||v|} \nonumber \\
 e^{-\alpha} & = & \frac{e^{-\hat{\alpha}}}{\sqrt{|u||v|}}.
  \label{eq:transforming_terms_from_alpha_to_alphahat}
\end{eqnarray}
When we substitute this back into equation \eqref{eq:general_alpha_eom}, we obtain:
\begin{equation}
 \partial_+\partial_- \hat{\alpha} + \sqrt{|u||v|} [e^{\hat{\alpha}} + \mathrm{sgn} (uv) e^{-\hat{\alpha}}] = 0.
\label{eq:general_alphahat_eom_preliminary}
\end{equation}
Finally, we perform the following residual gauge transformation\footnote{Note that, as $\varphi=0$, we do not use the gauge-fixing condition \eqref{eq:gauge_fixing_AdS3xS1} when we work in pure $AdS_3$.} on the worldsheet:
\begin{equation}
 \hat{\sigma}^+ = \int \sqrt{2|u(\sigma^+)|}d\sigma_+ \:, \qquad \hat{\sigma}^- = \int \sqrt{2|v(\sigma^-)|}d\sigma_-,
\label{eq:def_sigmapmhat_Pohlmeyer}
\end{equation}
from which we deduce:
\begin{equation}
 \partial_+ = \sqrt{2|u|} \hat{\partial}_+ \:, \qquad \partial_- = \sqrt{2|v|} \hat{\partial}_-.
\label{eq:transf_from_dpm_to_dpmhat}
\end{equation}
We then express \eqref{eq:general_alphahat_eom_preliminary} in terms of these new worldsheet coordinates and obtain:
\begin{equation}
 \hat{\partial}_+ \hat{\partial}_- \hat{\alpha} + \frac{1}{2} [e^{\hat{\alpha}} + \mathrm{sgn} (uv) e^{-\hat{\alpha}}] = 0,
\label{eq:general_alphahat_eom}
\end{equation}
which reduces to the sinh-Gordon equation when $uv<0$ and to the cosh-Gordon equation when $uv>0$.

In summary, every string solution defines, through \eqref{eq:def_alpha_Pohlmeyer}, a function $\alpha (\tau, \sigma)$ which satisfies the sinh-Gordon, the cosh-Gordon or the Liouville equation. Conversely, starting from a specific $\alpha (\tau, \sigma)$ solving one of these equations, it is possible, through an integrability technique known as inverse scattering transform, to construct a solution $X_\mu (\tau,\sigma)$ to the string equations of motion and Virasoro constraints such that $e^\alpha = -\partial_+ X_\mu \partial_- X^\mu$. While the details of this procedure are beyond the scope of this thesis, we are going to be interested in certain string solutions which were obtained in this way in \cite{Jevicki:2007aa}, starting from some simple solutions of the sinh-Gordon equation that are reviewed in the next section.

\subsection{Review of sinh-Gordon soliton-type solutions}
\label{sec:review_sinhG_soliton_solutions}

We list here, for later convenience, some well-known solutions to the sinh-Gordon equation, all of which have $u=2,v=-2$. It then follows from \eqref{eq:def_alphahat_Pohlmeyer} that:
\begin{equation}
 \hat{\alpha} = \alpha - \ln 2.
\label{eq:relation_alpha_alphahat_sinhG_solitons}
\end{equation}
We will write all the solutions in terms of $\alpha$ and of the original worldsheet variables $(\tau,\sigma)$, without the coordinate transformation \eqref{eq:def_sigmapmhat_Pohlmeyer}, which only amounts to a rescaling of $\sigma^+$, $\sigma^-$. In these coordinates, the equation which $\hat{\alpha}$ satisfies is \eqref{eq:general_alphahat_eom_preliminary}:
\begin{equation}
 \partial_+\partial_- \hat{\alpha} + 4 \sinh \hat{\alpha} = 0.
\label{eq:sinhG-like_eq_satisfied_by_alphahat}
\end{equation}
The simplest solution we are going to consider is the vacuum solution:
\begin{equation}
 \hat{\alpha}_0 = 0 \quad \mathrm{or} \quad \alpha_0 = \ln 2.
\label{eq:def_sinhG-vacuum}
\end{equation}
The one-(anti)soliton solution is:
\begin{equation}
 \alpha_{s,\bar{s}} = \ln 2 \pm \ln[\tanh^2 (\gamma (\sigma - v \tau))]
\label{eq:def_sinhG_1s1a}
\end{equation}
where $\gamma = 1/\sqrt{1-v^2}$, the plus sign is associated with solitons ($s$) and the minus sign is associated with antisolitons ($\bar{s}$). We notice that both these solutions have a single vertical asymptote, located at $\sigma_{s,\bar{s}}(\tau) = v \tau$. We identify the (anti)soliton with the asymptote: $\sigma_{s,\bar{s}}(\tau)$ is then the worldsheet position of the (anti)soliton as a function of worldsheet time and $v$ is its velocity. Therefore, the solution diverges to $(+)-\infty$ in the vicinity of an (anti)soliton.

The two-soliton scattering solutions are:
\begin{eqnarray}
 \alpha_{ss,\bar{s}\bar{s}} & = & \ln 2 \pm \ln \left[ \frac{v \cosh X - \cosh T}{v \cosh X + \cosh T} \right]^2 \nonumber \\
 \alpha_{s\bar{s}} & = & \ln 2 \pm \ln \left[ \frac{v \sinh X - \sinh T}{v \sinh X + \sinh T} \right]^2
\label{eq:def_sinhG_2s2asa_scattering}
\end{eqnarray}
where $X=2\gamma\sigma$, $T=2 v \gamma \tau$, and the (minus)plus sign is associated with two (anti)solitons in the first equation, while both sign choices are associated with one soliton and one antisoliton in the second equation. These solutions are in the centre of mass frame, i.e. the (anti)solitons have equal and opposite velocities $\pm v$. Sinh-Gordon solitons and anti-solitons can be thought of as quasi-particles which undergo factorised scattering.

\section{The finite-gap construction on $AdS_3 \times S^1$}

The finite-gap method relies on the expression of the string equations of motion in terms of a Lax connection and an equivalent linear system. The former can be used to define a Riemann surface which encodes the spectrum of a very general class of string solutions. The latter yields the solutions themselves, although the procedure required to obtain them is rather involved and will not be discussed here, as the results presented in the next chapters only concern the spectrum. Our review mainly follows \cite{Kazakov:2004nh} and \cite{Dorey:2006zj}.

\subsection{Initial setup}

First of all, we introduce the parametrisation of $AdS_3$ in terms of an element $g$ of the group $SU(1,1) \cong SL (2, \mathbb{R})$,
\begin{equation}
 g = \left( \begin{array}{cc}
	Z_1 & Z_2 \\
	\bar{Z}_2 & \bar{Z}_1
     \end{array} \right),
\label{eq:defgSU11elt}
\end{equation}
which is easily seen to satisfy all the $SU(1,1)$ properties:
\begin{eqnarray}
 g^\dag M g & = & M\ , \qquad M = \left( \begin{array}{cc}
                                       1 & 0 \\
                                       0 & -1
                                     \end{array} \right) \nn \\
 \mathrm{det}\: g & = & 1 \label{eq:SU11_properties}
\end{eqnarray}
both of which are a consequence of the fact that $Z_1$ and $Z_2$ satisfy the $AdS_3$ constraint \eqref{eq:AdS3_constr_cx}. For a closed string, $g$ must satisfy the following periodicity condition:
\begin{equation}
 g (\tau, \sigma + 2 \pi) = g (\tau, \sigma).
\label{eq:g_periodicity}
\end{equation}
The Lie algebra $\mathfrak{su}(1,1)$ associated with this group is defined as the space of $2 \times 2$ matrices $m$ satisfying:
\begin{equation}
 \mathrm{Tr} \: m = 0\:, \qquad m^\dag = - M m M.
\label{eq:su11properties}
\end{equation}
We choose the set of generators $s^A = (-i\sigma_3,\sigma_1,-\sigma_2)^A$, $A = 0,1,2$, (where the $\sigma_j$ are the Pauli matrices) as the basis for this vector space, and then introduce the metric $\eta_{AB}/2$ (where $\eta = \mathrm{diag}(-1,1,1)$). The basic properties of the generators are given by
\begin{equation}
 [s^A,s^B] = - 2 \epsilon^{ABC}\eta_{CD}s^D
\label{eq:su11_generators_commutator}
\end{equation}
and
\begin{equation}
 \mathrm{Tr} (s^A s^B) = 2 \eta^{AB}.
\label{eq:su11_generators_trace}
\end{equation}
The decomposition of a vector in the Lie algebra onto the basis is then:
\begin{equation}
 V = V_A s^A = \frac{1}{2}\eta_{AB}V^A s^B = \frac{1}{2} \left( \begin{array}{cc}
														                               iV^0 & V^1 + i V^2 \\
														                               V^1 - i V^2 & -iV^0
                                                         \end{array} \right).
\label{eq:su11_vector_decomposition}
\end{equation}
If we define the $\mathfrak{su}(1,1)$-valued right current $j$,
\begin{equation}
 j_a = g^{-1}\partial_a g,
\label{eq:def_su11_right_current}
\end{equation}
and then express the $\sigma$-model action \eqref{eq:AdS3xS1_string_action} in terms of it (after removing the Lagrange multiplier by imposing the $AdS_3$ constraint on $Z_1$, $Z_2$), we obtain the $SL(2,\mathbb{R})$ Principal Chiral Model action (plus the decoupled term for the $S^1$ angular coordinate):
\begin{equation}
 I = - \frac{\sqrt{\lambda}}{4\pi} \int d^2\sigma \left[ \frac{1}{2} \mathrm{Tr} (j_a j^a) + \partial_a \varphi \partial^a \varphi \right].
\label{eq:SL2R_principal_chiral_model_action}
\end{equation}
In this form, the action is invariant under left and right multiplication by a constant $SU(1,1)$ group element $U_L$/$U_R$:
\begin{equation}
 g \to U_L g\:, \qquad g \to g U_R.
\label{eq:SU11_left_and_right_multiplication}
\end{equation}
The associated conserved currents are
\begin{equation}
 J_L^a = - \frac{\sqrt{\lambda}}{4\pi} l^a\:, \qquad J_R^a = - \frac{\sqrt{\lambda}}{4\pi} j^a,
\label{eq:SU11_multiplication_currents}
\end{equation}
where the right current $j_a$ was defined in \eqref{eq:def_su11_right_current} and the left current $l_a$, given by
\begin{equation}
 l_a = (\partial_a g) g^{-1},
\label{eq:eq:def_su11_left_current}
\end{equation}
is related to the right current by the following transformation:
\begin{equation}
 l_a = g j_a g^{-1}.
\label{eq:su11_l_r_current_transformation}
\end{equation}
The equations of motion are equivalent to the conservation conditions
\begin{equation}
 \partial_+ j_- + \partial_- j_+ = -2 \partial^a j_a = 0 \:, \qquad \partial_+ l_- + \partial_- l_+ = -2 \partial^a l_a = 0,
\label{eq:j_l_conservation}
\end{equation}
and the two currents also satisfy flatness conditions as a direct consequence of their definitions:
\begin{equation}
 \partial_+ j_- - \partial_- j_+ - [j_-,j_+] = 0 \:, \qquad \partial_+ l_- - \partial_- l_+ + [l_-,l_+] = 0.
\label{eq:j_l_flatness}
\end{equation}
The corresponding left and right conserved charges are:
\begin{equation}
 Q_L = \frac{\sqrt{\lambda}}{4\pi} \int d\sigma l_\tau\:, \qquad Q_R = \frac{\sqrt{\lambda}}{4\pi} \int d\sigma j_\tau.
\label{eq:SU11_left_and_right_charges}
\end{equation}
Their components on $s^0$, which generates the Cartan subalgebra, are related to the energy $\Delta$ and $AdS_3$ angular momentum $S$ of the string:
\begin{equation}
 Q_L^0 = \Delta - S \:, \qquad Q_R^0 = \Delta + S.
\label{eq:SU11_charges_first_component}
\end{equation}
A useful property is given by:
\begin{equation}
 -\frac{1}{2} \mathrm{Tr}j^2_a = \mathrm{det} j_a = \partial_a Z_1 \partial_a \bar{Z}_1 - \partial_a Z_2 \partial_a \bar{Z}_2
\label{eq:det_and_tr_of_j}
\end{equation}
(where the first equality is just a consequence of the general $\mathfrak{su}(1,1)$ matrix structure \eqref{eq:su11_vector_decomposition}, and $j^2_a$ indicates the matrix square of $j_a$), which allows to express the Virasoro constraints \eqref{eq:Virasoro_lc} in terms of the current:
\begin{equation}
 -\frac{1}{2} \mathrm{Tr}j^2_\pm = \mathrm{det} \: j_\pm = \frac{J^2}{\lambda}.
\label{eq:Virasoro_j}
\end{equation}

\subsection{Integrability}

The integrability of string theory in $AdS_3$ is a consequence of the existence of a one-parameter family of \emph{Lax currents} (also known as \emph{Lax connections})\footnote{In the following, we will often use the notation $V_\tau \equiv V_0$, $V_\sigma \equiv V_1$ in the $(\tau,\sigma)$ coordinate system.}
\begin{eqnarray}
 \mathcal{J}_\tau (x, \tau, \sigma) & = & \frac{1}{2} \left( \frac{j_+}{1-x} + \frac{j_-}{1+x} \right) \nn\\
 \mathcal{J}_\sigma (x, \tau, \sigma) & = & \frac{1}{2} \left( \frac{j_+}{1-x} - \frac{j_-}{1+x} \right),
\label{eq:Lax_current}
\end{eqnarray}
whose flatness condition
\begin{equation}
 \partial_+ \mathcal{J}_- - \partial_- \mathcal{J}_+ - [ \mathcal{J}_-, \mathcal{J}_+] = 0
\label{eq:Lax_flatness}
\end{equation}
is equivalent to the equations of motion. Specifically, if the equations of motion are verified, then $\mathcal{J} (x,\tau,\sigma)$ is flat for all values of the \emph{spectral parameter} $x$, and vice versa.

The Lax connection allows the introduction of the \emph{auxiliary linear system}:
\begin{eqnarray}
 [\partial_a + \mathcal{J}_a (x, \tau, \sigma)] \Psi (x, \tau, \sigma) = 0 \:, \qquad \textrm{for } a = 0,1
\label{eq:aux_lin_sys}
\end{eqnarray}
where $\Psi (x, \tau, \sigma)$ is a $2 \times 2$ complex matrix.

The consistency condition for such a system, namely $[ \partial_\tau + \mathcal{J}_\tau, \partial_\sigma + \mathcal{J}_\sigma ] = 0$, is equivalent to the flatness of $\mathcal{J}$ and thus the system itself linearises the equations of motion.

The \emph{monodromy matrix} is then defined as:
\begin{equation}
 \Omega (x, \tau, \sigma) = \mathcal{P} \mathrm{exp} \int_{[\gamma (\tau, \sigma)]} \mathcal{J}_a (x, \tau', \sigma') \: d \sigma'^a
\label{eq:def_monodromy_matrix}
\end{equation}
where $\mathcal{P} \mathrm{exp}$ indicates the path-ordered exponential,
\begin{eqnarray}
 \mathcal{P} \mathrm{exp} \int_a^b f(x) dx & = & \lim_{\epsilon \to 0} \: e^{\int_a^{a + \epsilon} f(x) dx} 
  e^{\int_{a + \epsilon}^{a + 2\epsilon} f(x) dx} \ldots e^{\int_{b - \epsilon}^b f(x) dx} \nn\\
  & = & \lim_{\epsilon \to 0} \: [1 + \epsilon f(a) + O(\epsilon^2)] [1 + \epsilon f(a + \epsilon) + O(\epsilon^2)] \ldots \nn \\
  & & [1 + \epsilon f(b - \epsilon) + O(\epsilon^2)],
\label{eq:def_Pexp}
\end{eqnarray}
and $[\gamma (\tau, \sigma)]$ is the homotopy class of a closed path $\gamma(\tau, \sigma)$ with base point $(\tau,\sigma)$, winding once around the closed string worldsheet. By the non-Abelian version of Stokes' theorem, if $\mathcal{J}$ is flat, then, for any simply connected domain $D$, we have
\begin{equation}
 \mathcal{P} \mathrm{exp} \int_{\partial D} \mathcal{J}_a (x, \tau', \sigma') \: d \sigma'^a = 1.
\label{eq:Stokes_thm}
\end{equation}
Hence, the path-ordered exponential of $\mathcal{J}$ along any curve lying on the worldsheet will only depend on its endpoints (even if they coincide) and on its homotopy class (or, equivalently, on its winding number).

\begin{figure}
\centering
\psfrag{a}{\footnotesize{$- \gamma (\tau,\sigma)$}}
\psfrag{b}{\footnotesize{$\gamma (\tau',\sigma')$}}
\psfrag{c}{\footnotesize{$(\tau,\sigma)$}}
\psfrag{d}{\footnotesize{$(\tau',\sigma')$}}
\psfrag{e}{\footnotesize{$- \tilde{\gamma}$}}
\psfrag{f}{\footnotesize{$\tilde{\gamma}$}}
\includegraphics[width=0.40\columnwidth]{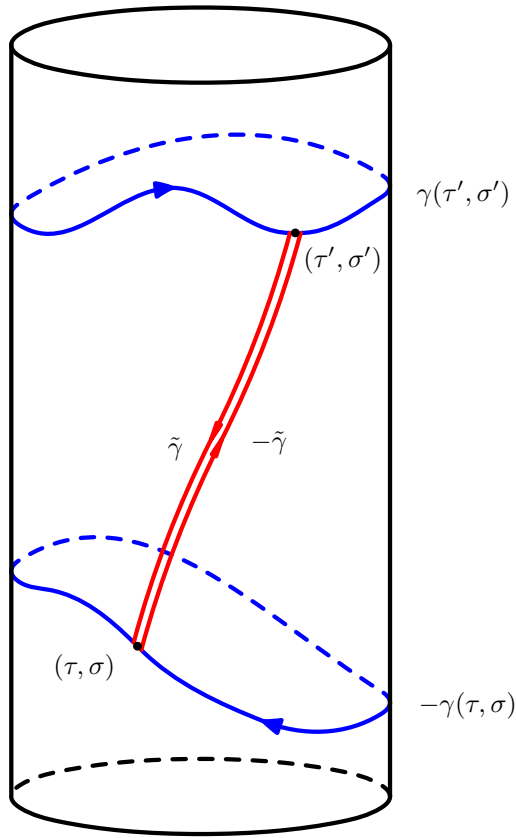}%
\caption{An example of the path $\partial D$ along which Stokes' theorem is applied. Paths with negative signs are run along in the opposite direction with respect to the definition of the path-ordered exponential yielding the corresponding matrix, and hence they generate its inverse.}
\label{fig:Omega_evolution}%
\end{figure}

Another important consequence of Stokes' theorem is the fact that, as we vary the base point $(\tau, \sigma)$ to $(\tau', \sigma')$, the monodromy matrix evolves by conjugation:
\begin{equation}
 \Omega (x, \tau', \sigma') = U \Omega (x, \tau, \sigma) U^{-1} \:, \qquad \textrm{where } U = \mathcal{P} \mathrm{exp} \int_{\tilde{\gamma}}
  \mathcal{J}_a (x, \tau', \sigma') \: d \sigma'^a
\label{eq:Omega_evolution_by_conjugation}
\end{equation}
and $\tilde{\gamma}$ is any curve connecting the two base points $(\tau, \sigma)$ and $(\tau', \sigma')$ with vanishing winding number (see Fig. \ref{fig:Omega_evolution}).

A convenient choice of contour for the monodromy matrix is the following:
\begin{equation}
 \Omega (x, \tau, \sigma) = \mathcal{P} \mathrm{exp} \int_\sigma^{\sigma + 2\pi} \mathcal{J}_\sigma (x, \tau, \sigma') \: d \sigma'
\label{eq:Omega_convenient_choice}
\end{equation}
where we now integrate along a circle of constant $\tau$ on the worldsheet, starting and ending at the point $\sigma$.

The parameter $x$ is complex in general, but $\Omega (x, \tau, \sigma) \in SU(1,1)$ only for real $x$. However, due to the fact that $\mathcal{J}$ is traceless, $\Omega$ will always be unimodular ($\mathrm{det} \Omega (x, \tau, \sigma) = 1$, $\forall x \in \mathbb{C}$). We are now going to study the eigenvalues of $\Omega$, which, by \eqref{eq:Omega_evolution_by_conjugation}, are independent of $(\tau, \sigma)$ and which we will indicate as $t_\pm (x) = e^{\pm i p(x)}$, where the \emph{quasi-momentum} $p(x)$ satisfies
\begin{equation}
 \mathrm{Tr} \: \Omega (x, \tau, \sigma) = 2 \cos p(x)
\label{eq:Tr_Omega=2_cos_p(x)}
\end{equation}
and is thus naturally defined only up to sign changes and shifts by integer multiples of $2 \pi$. Moreover, for $x \in \mathbb{R}$, we have $\cos p(x) \in \mathbb{R}$, but this only implies that, in general, $p(x)$ is either real or pure imaginary (up to shifts by $2 \pi n$) on the real axis.

As a $\tau$-independent function of the spectral parameter, the quasi-momentum represents an infinite tower of conserved charges associated with the classical motion of the string, among which we will find the energy spectrum. But first we will determine some general properties of $p(x)$ in the next section.

\subsection{Analyticity and asymptotics}

The monodromy matrix satisfies the following system of differential equations:
\begin{equation}
 [\partial_a + \mathcal{J}_a, \Omega (x, \tau, \sigma) ] = 0 \:, \qquad \forall a \:,
\label{eq:Omega's_PDEs}
\end{equation}
and hence, by Poincar\'e's theorem on holomorphic differential equations, it is holomorphic for $x \in \mathbb{C} \setminus \{ +1, -1 \}$, while it has essential singularities at $x = \pm 1$, as we can see from the fact that $\mathcal{J}$ has simple poles there. In order to determine the leading behaviour of $p(x)$ near these two points, we first need to approximate the Lax current:
\begin{equation}
 \mathcal{J}_\sigma (x, \tau, \sigma) = - \frac{1}{2} \frac{j_\pm (\tau, \sigma)}{x \mp 1} + O((x \mp 1)^0) \:, \qquad \textrm{as } x \to \pm 1.
\label{eq:J_near_poles}
\end{equation}
Since $\mathrm{Tr} \: j_\pm = 0$, due to the fact that $j_a \in \mathfrak{su}(1,1)$, and $\det j_\pm = J^2/\lambda$ by the Virasoro constraints \eqref{eq:Virasoro_j}, we find that the diagonal form of the components of the right current can be written as
\begin{equation}
 v_\pm (\tau, \sigma) j_\pm (\tau, \sigma) v_\pm^{-1} (\tau, \sigma) = i \frac{J}{\sqrt{\lambda}} \sigma_3.
\label{eq:j_pm_diagonal_form}
\end{equation}
We may then use this result in order to diagonalise the monodromy matrix at leading order, which is achieved by introducing $u_\pm (x, \tau, \sigma) = v_\pm (\tau, \sigma) + O((x \mp 1)^0)$:
\begin{eqnarray}
 u_\pm (x, \tau, \sigma) \Omega (x, \tau, \sigma) u_\pm^{-1} (x, \tau, \sigma) & = & \mathrm{exp} \left[ - \frac{i \pi J}{\sqrt{\lambda}}
  \frac{\sigma_3}{x \mp 1} + O((x \mp 1)^0) \right] \:, \nn\\
   & & \qquad\qquad\qquad\qquad \textrm{as } x \to \pm 1.
\label{eq:Omega_at_singularities}
\end{eqnarray}
Therefore, the quasi-momentum has simple poles at $x = \pm 1$, together with branch points located where the discriminant
\begin{equation}
 D = 4 - [\mathrm{Tr} \: \Omega (x, \tau, \sigma)]^2 = 4 \sin^2 p(x)
\label{eq:def_discriminant}
\end{equation}
has simple zeros\footnote{More generally, odd-multiplicity zeros would also identify branch points, which are however considered unphysical, and hence one usually places the additional constraint that the discriminant may not have zeros of odd multiplicity greater than 1.} and therefore it only becomes single-valued on a Riemann surface which has in general an infinite number of sheets (as follows from the structure of the inverse cosine function). Each sheet will correspond to a particular sign choice for $p(x)$ and to a particular shift by $2 \pi n$, $n \in \mathbb{Z}$. Following the usual procedure, we identify a reference sheet, or \emph{physical sheet}, by making a specific choice:
\begin{equation}
 p(x) = \frac{\pi J}{\sqrt{\lambda}} \frac{1}{x \mp 1} + O((x \mp 1)^0) \:, \qquad \textrm{as } x \to \pm 1.
\label{eq:p(x)_near_poles}
\end{equation}
Note that, since the discriminant has essential singularities at $x = \pm 1$, it must have an infinite number of zeros accumulating at these points. This, however, does not necessarily imply that there is an infinite number of branch points, since most of these zeros may have even multiplicity; we will refer to multiplicity two zeros as ``double points'' (zeros of higher even multiplicity can be thought of as multiple coincident double points).

Another important remark is the fact that, since $\cos p(x) = \pm 1$ at the branch points, the latter are necessarily of the square-root type, in the sense that, if $x$ follows a path that circles around one of them twice, there is no variation of $p(x)$ between the initial and final points.

We now move on to considering the asymptotic behaviour of the quasi\hyp{}momentum at infinity. As before, we start by expanding the Lax current:
\begin{equation}
 \mathcal{J}_\sigma (x, \tau, \sigma) = - \frac{1}{x} j_0 (\tau, \sigma) + O \left( \frac{1}{x^2} \right) \:, \qquad \textrm{as } x \to \infty,
\label{eq:J_at_infty}
\end{equation}
and then proceed to expand the monodromy matrix as well:
\begin{eqnarray}
 \Omega (x, \tau, \sigma) & = & 1 - \frac{1}{x} \int_{\sigma}^{\sigma + 2 \pi} j_0 (\tau, \sigma') d \sigma' + O \left( \frac{1}{x^2} \right)
  \nn\\
                          & = & 1 - \frac{1}{x} \frac{4 \pi}{\sqrt{\lambda}} Q_R + O \left( \frac{1}{x^2} \right) \:, 
                           \qquad \textrm{as } x \to \infty.
\label{eq:Omega_near_infty}
\end{eqnarray}
In the case of a highest-weight string configuration, the vector of the components of $Q_R$ on the $\mathfrak{su}(1,1)$ generators is parallel to $s^0$:
\begin{equation}
 Q_R = \frac{i}{2} (\Delta + S) \sigma_3 \:,
\label{eq:Q_R_highest_weight}
\end{equation}
whereas all the other equivalent solutions can be obtained by applying arbitrary left and right $SU(1,1)$ rotations: $g \to U_L g U_R$, with $\partial_a U_R = \partial_a U_L = 0$, $\forall a$. In the following, we will restrict ourselves to highest-weight solutions. 
We therefore obtain
\begin{equation}
 \Omega (x, \tau, \sigma) = 1 - \frac{2 \pi i}{\sqrt{\lambda}} \frac{\Delta + S}{x} \sigma_3 + O \left( \frac{1}{x^2} \right) \:, 
                             \qquad \textrm{as } x \to \infty
\label{eq:Omega_infty_Delta+S}
\end{equation}
which yields
\begin{equation}
 p(x) = \frac{2 \pi}{\sqrt{\lambda}} (\Delta + S) \frac{1}{x} + O \left( \frac{1}{x^2} \right) \:, \qquad \textrm{as } x \to \infty \:,
\label{eq:p(x)_near_infty}
\end{equation}
where we have made conventional choices for the branch of the quasi-momentum which define the physical sheet (e.g. $p(\infty) = 2 \pi n \pm 2 \pi (\Delta + S) /(x \sqrt{\lambda})$ on other sheets).

Finally, the asymptotic behaviour of the Lax current near the origin $x = 0$ is given by:
\begin{equation}
 \mathcal{J}_\sigma (x, \tau, \sigma) = j_\sigma (\tau, \sigma) - x j_\tau (\tau, \sigma) + O(x^2) \:, \qquad \textrm{as } x \to 0 \:,
\label{eq:J_near_0}
\end{equation}
which we can now substitute into the usual definition of $\Omega$ to obtain:
\begin{equation}
 \Omega (x, \tau, \sigma) = g^{-1} (\tau, \sigma) \mathcal{P} \mathrm{exp} \left[ - x \int_{\sigma}^{\sigma + 2 \pi} l_\tau (\tau, \sigma')
  d \sigma' + O (x^2) \right] g (\tau, \sigma + 2 \pi).
\label{eq:Omega_near_0}
\end{equation}
If we use the periodicity of the $SU(1,1)$ group element \eqref{eq:g_periodicity} and then expand the remaining path-ordered exponential for small $x$, we find:
\begin{eqnarray}
 g (\tau, \sigma) \Omega (x, \tau, \sigma) g^{-1} (\tau, \sigma) & = & 1 - x \int_{\sigma}^{\sigma + 2 \pi} l_\tau (\tau, \sigma')
                                                                        d \sigma' + O (x^2) \nn\\
                                                                 & = & 1 - x \frac{4 \pi}{\sqrt{\lambda}} Q_L + O(x^2) \nn\\
                                                                 & = & 1 - x \frac{2 \pi i}{\sqrt{\lambda}} (\Delta - S) \sigma_3
                                                                        + O(x^2) \:, \qquad \textrm{as } x \to 0 \:, \nn\\
                                                                 & &
\label{eq:gOmega(g^-1)_near_0}
\end{eqnarray}
where again we are considering a highest-weight solution.

Thus, we have
\begin{equation}
 p(x) = - \frac{2 \pi}{\sqrt{\lambda}} (\Delta - S) x + O(x^2)  \:, \qquad \textrm{as } x \to 0 \:,
\label{eq:p(x)_near_0}
\end{equation}
which is used as one of the defining conditions of the physical sheet.

\subsection{The spectral curve}

The Riemann surface associated with the eigenvalues of $\Omega$, $t_\pm (x) = e^{\pm i p(x)}$, has two properties which make it complicated to study: firstly, it has an infinite number of sheets and, secondly, it has essential singularities located at $x = \pm 1$ on all its sheets. In fact, it is more convenient to consider the Riemann surface $\Sigma$ corresponding to $p'(x)$, which instead only has double poles at $x = \pm 1$ and two sheets, since it only maintans the sign ambiguity of $p(x)$, while it loses the arbitrary shift by $2 \pi n$.

It has been proved (see section 3.1 of \cite{Dorey:2006zj} and references therein) that this surface, which is called the \emph{spectral curve}, can be written as:
\begin{equation}
 \Sigma : \quad y^2 = \prod_{j=1}^\infty (x - x_j) \:,
\label{eq:Sigma_arbitrary_genus}
\end{equation}
where the branch points $x_j$ correspond to the infinite zeros of the discriminant $D$ \eqref{eq:def_discriminant}, which, as we saw in the previous discussion, accumulate at $x = \pm 1$. From now on, we are going to restrict ourselves to the case in which only $2K$ of these zeros have multiplicity one, while all the others have multiplicty two, becoming double points, so that the spectral curve takes the following hyper-elliptic form:
\begin{equation}
 \Sigma : \quad y^2 (x) = \prod_{j=1}^{2K} (x - x_j) \:,
\label{eq:Sigma_hyper-elliptic_form}
\end{equation}
having $K$ branch cuts and therefore genus $K-1$. String solutions associated with this class of spectral curves are known as \emph{finite-gap solutions}. Strictly speaking, with this restriction in place, we are unable to describe directly surfaces with an infinite genus. However, it is believed that these should be recovered as the $K \to \infty$ limit of a $K$-gap solution, and that therefore finite-gap solutions are completely generic. Hence, the following analysis will yield results which should apply to all the possible string solutions in $AdS_3 \times S^1$ in the large $S$ limit.

We now introduce the meromorphic differential $dp = p'(x) dx$, which inherits all its analytic and asymptotic properties from $p(x)$. Such properties were studied in the previous section, and determine the behaviour of the differential near the poles, near the origin and at infinity. As we did in the case of the gauge theory spin chain (see Fig. \ref{fig:sheets_identification_BGK}), we choose the physical sheet (or upper sheet) on $\Sigma$ so that $dp$ corresponds to $p(x)$ on the physical sheet of its own Riemann surface, as it was defined above. Thus we have:
\begin{eqnarray}
 dp & = & - \frac{\pi J}{\sqrt{\lambda}} \frac{dx}{(x \mp 1)^2} + O((x \mp 1)^0) \:, \qquad \textrm{as } x \to \pm 1 \:, \nn\\
 dp & = & - \frac{2 \pi}{\sqrt{\lambda}} (\Delta + S) \frac{dx}{x^2} + O \left( \frac{1}{x^3} \right)  \:, \qquad \textrm{as } x \to \infty
         \:, \nn\\
 dp & = & - \frac{2 \pi}{\sqrt{\lambda}} (\Delta - S) dx + O(x) \:, \qquad \textrm{as } x \to 0 \:,
\label{eq:def_properties_of_dp_FG}
\end{eqnarray}
on the upper sheet of $\Sigma$, while all the signs flip on the lower sheet. We are also going to use the usual notation, according to which $x^+$ and $x^-$ identify the point above $x$ on the upper and, respectively, lower sheet of the spectral curve.

Apart from the double poles at $x = \pm 1$, the only other singularities of the differential are the branch points $x_j$, which are of the square root type.

The most general meromorphic differential with the required branch points and double poles is given in terms of the function $y(x)$, of a generic holomorphic part $f(x)$ and of a particular singular part $g(x)$:
\begin{eqnarray}
 dp  & = & dp_{1}\,\,+\,\,dp_{2}\,\,=\,\,-\frac{dx}{y}\left[f(x)\,\,+\,\, g(x)\right]  \nn\\ 
f(x) & = & \sum_{\ell=0}^{K-2}\,\,C_{\ell}\,x^{\ell} \nn \\
g(x) & = & \frac{\pi J}{\sqrt{\lambda}} \left[ \frac{y_{+}}{(x-1)^{2}}\,\,+\,\,\frac{y_{-}}{(x+1)^{2}} 
            + \frac{y'_{+}}{(x-1)}\,\,+\,\,\frac{y'_{-}}{(x+1)} \right] \:,
\label{eq:most_general_dp}
\end{eqnarray}
where
\begin{equation}
 y_\pm = y(\pm 1) \qquad \textrm{and} \qquad y'_\pm = \left. \frac{dy}{dx} \right|_{x=\pm 1} \:.
\label{eq:def_y_pm_and_y'_pm}
\end{equation}
The corresponding branch cut structure is the same that applies to $y(x)$: the points $x_j$ are divided into pairs and each pair is then joined by a single cut $\mathcal{C}_I$, for $I = 1, \ldots, K$, such that
\begin{equation}
 dp (x + \epsilon) + dp (x - \epsilon) = 0 \:,
\label{eq:discontinuity_of_dp(x)_FG}
\end{equation}
where $x \in \mathcal{C}_I$ and $\epsilon$ is an infinitesimal shift orthogonal to the cut. We may also equivalently say that $dp (x^+) = - dp(x^-)$.

This choice of cuts is also appropriate for $p(x)$, since, as we saw above, its branch points coincide with the $x_j$ and all of them are of the square root type.

As in the case of the $SL(2, \mathbb{R})$ spin chain, the pair $(\Sigma, dp)$ allows us to reconstruct $p(x)$ as the Abelian integral
\begin{equation}
 p(x) = \int_{\infty^+}^x dp \:,
\label{eq:p(x)_abelian_integral_FG}
\end{equation}
which starts on the physical sheet of $\Sigma$. Again, by choosing an integration path which crosses the appropriate cuts, we may place the point $x$ at which $p$ is evaluated on any sheet of its Riemann surface, although, from the point of view of $dp$, the path always lies on the two-sheeted spectral curve.

\begin{figure}%
\centering
\psfrag{a}{\footnotesize{$(a)$}}
\psfrag{b}{\footnotesize{$(b)$}}
\psfrag{a1}{\footnotesize{$-dp(x)$}}
\psfrag{a2}{\footnotesize{$\phantom{-}dp(x)$}}
\psfrag{b1}{\footnotesize{$\mathcal{B}_I$}}
\psfrag{b2}{\footnotesize{$\mathcal{A}_I$}}
\psfrag{b3}{\footnotesize{$\mathcal{C}_I$}}
\psfrag{c1}{\footnotesize{$\infty^+$}}
\psfrag{c2}{\footnotesize{$\infty^-$}}
\psfrag{d1}{\footnotesize{$\mathcal{A}_I$}}
\psfrag{d2}{\footnotesize{$\mathcal{B}_I$}}
\includegraphics[width=\columnwidth]{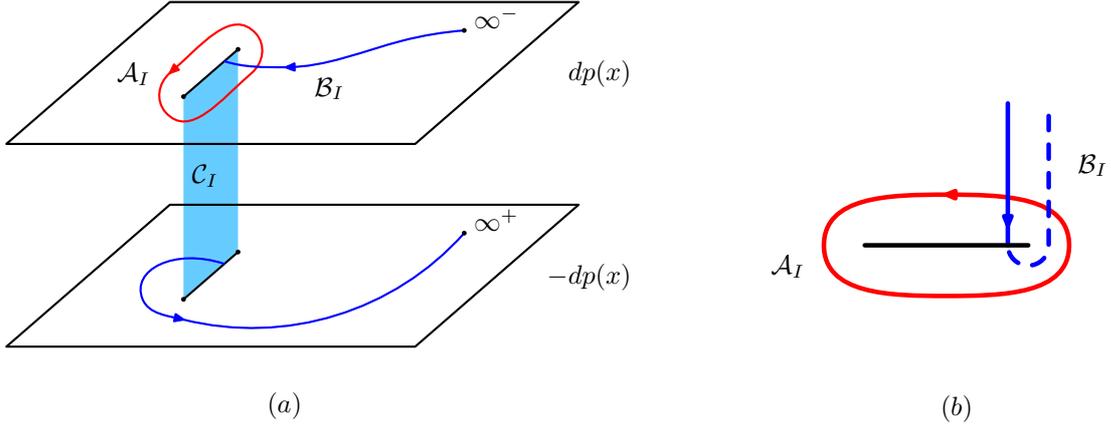}%
\caption{(a) Two-sheeted representation of the A-cycle $\mathcal{A}_I$ and of the B-cycle $\mathcal{B}_I$ associated with a given cut $\mathcal{C}_I$ on $\Sigma$. (b) Equivalent representation of the cycles on the complex plane; the dashed part of the contour lies on the lower sheet.}%
\label{fig:Cycles_FG}%
\end{figure}

The set of conditions \eqref{eq:def_properties_of_dp_FG} must be supplemented with period conditions for each cut, which are once again determined by the behaviour of $p(x)$. For each cut $\mathcal{C}_I$, we define an A-cycle $A_I$, which encircles the cut and no other singularities, and a B-cycle $\mathcal{B}_I$, which starts at $\infty^+$, crosses the cut, and ends at $\infty^-$ (see Fig. \ref{fig:Cycles_FG}).

\begin{figure}%
\centering
\psfrag{a}{\footnotesize{$(a)$}}
\psfrag{b}{\footnotesize{$(b)$}}
\psfrag{a1}{\footnotesize{$-dp(x)$}}
\psfrag{a2}{\footnotesize{$\phantom{-}dp(x)$}}
\psfrag{b1}{\footnotesize{$x^+$}}
\psfrag{b2}{\footnotesize{$x^-$}}
\psfrag{c1}{\footnotesize{$\infty^-$}}
\psfrag{c2}{\footnotesize{$\infty^+$}}
\psfrag{d1}{\footnotesize{$\mathcal{B}_I$}}
\psfrag{d2}{\footnotesize{$\mathcal{C}_I$}}
\includegraphics[width=\columnwidth]{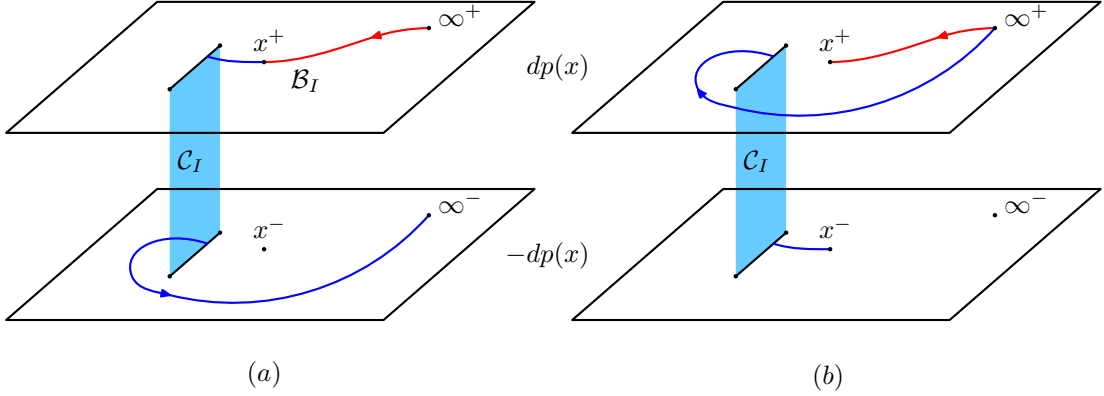}%
\caption{(a) The standard B-cycle for the cut $\mathcal{C}_I$, chosen so that it touches the point $x^+$. (b) The rearranged version of the B-cycle, where its second half has been swapped from one sheet to the other and has undergone a direction flip.}%
\label{fig:p_monodromy_proof}%
\end{figure}

Due to the fact that the variation of $p(x)$ is zero along an A-cycle, since the cycle does not cross any cut, all the A-periods must vanish:
\begin{equation}
 \oint_{\mathcal{A}_I} dp = 0 \:, \qquad \textrm{for } I = 1, \ldots, K.
\label{eq:A_periods_FG}
\end{equation}
On the other hand, while $p(\infty^+) = 0$ since $\infty^+$ is always chosen to be located on the physical sheet of the Riemann surface of $p(x)$, $\infty^-$ may lie on different sheets, depending on the cut we crossed, and hence, by \eqref{eq:Omega_infty_Delta+S}, $p(\infty^-) \in 2 \pi \mathbb{Z}$ in general. Thus the B-periods are given by
\begin{equation}
 \int_{\mathcal{B}_I} dp = p(\infty^+) - p(\infty^-) = 2 \pi n_I \:, \qquad n_I \in \mathbb{Z},
\label{eq:B_periods_FG}
\end{equation}
where the integers $n_I$ are called \emph{mode numbers}. This also implies that $p(x_j) = p(x_{j+1}) = \pi n_I$ at the two branch points connected by $\mathcal{C_I}$, and that the discontinuity of $p(x)$ across a cut is given by:
\begin{equation}
 p (x + \epsilon) + p (x - \epsilon) = 2 \pi n_I \:,
\label{eq:discontinuity_of_p(x)_FG}
\end{equation}
for $x \in \mathcal{C}_I$, or equivalently that, for any contour starting at $x^+$ on the physical sheet and ending at $x^-$ after crossing a single cut, we have:
\begin{equation}
 p(x^+) + p(x^-) = 2 \pi n_I \:.
\label{eq:monodromy_of_p(x)_FG}
\end{equation}
This may be seen, for instance, by choosing the contour so that it touches the point $x^+$ and then splitting the B-period into two contributions:
\begin{equation}
 2 \pi n_I = \int_{\infty^+}^{x^+} dp + \int_{x^+}^{\infty^-} dp \:,
\label{eq:proof_monodromy_p(x)_FG_1}
\end{equation}
where the contour corresponding to the first term lies completely on the upper sheet (drawn in red in Fig. \ref{fig:p_monodromy_proof}(a)), while the contour of the second term lies on both sheets (drawn in blue). We then bring the part of the second contour which lies on the upper sheet to the lower sheet and vice versa, which, by \eqref{eq:discontinuity_of_dp(x)_FG}, requires us to change the sign of $dp$ (first step of the next equation). Equivalently, we flip the direction of the contour (second step of the next equation, also see Fig. \ref{fig:p_monodromy_proof}(b)) and finally apply \eqref{eq:p(x)_abelian_integral_FG}:
\begin{equation}
 2 \pi n_I = \int_{\infty^+}^{x^+} dp - \int_{x^-}^{\infty^+} dp = \int_{\infty^+}^{x^+} dp + \int_{\infty^+}^{x^-} dp = p(x^+) + p(x^-) \:.
\label{eq:proof_monodromy_p(x)_FG_2}
\end{equation}

At the double points $\hat{x}_k$ we also have $p( \hat{x}_k ) = n_k \pi$, $n_k \in \mathbb{Z}$, which is exactly what happens at the branch points, with the difference that $p(x)$ does not pick up a monodromy if $x$ circles once around $\hat{x}_k$. In fact, double points can be thought of as collapsed cuts, whose branch points have become coincident.

It is known \cite{Beisert:2005bm} that the mode numbers are in a 1-1 correspondence with the elements of the set $\{ \mathcal{C}_I, \hat{x}_k \}$, i.e. that each mode number is either represented by a single cut or by a single double point and this accounts for all the zeros of the discriminant, accumulating at $x = \pm 1$. Double points correspond to mode numbers which are ``turned off'', while cuts represent mode numbers which are ``turned on''. As more and more mode numbers are excited, the genus of $\Sigma$ increases, and the corresponding family of string solutions acquires additional moduli, as will shortly become apparent.

In fact, we see from \eqref{eq:most_general_dp} and \eqref{eq:Sigma_hyper-elliptic_form} that the differential has $3K - 1$ parameters $\{ x_j, C_l \}$, for $j = 1, \ldots, 2K$ and $l = 0, \ldots, K-2$. Once we impose the $2K$ constraints \eqref{eq:A_periods_FG} and \eqref{eq:B_periods_FG}, we are left with $K-1$ independent parameters, which represent the moduli.

The spectrum $\Delta - S$ of these string solutions is encoded in $(\Sigma, dp)$ through \eqref{eq:def_properties_of_dp_FG} and is a function of the moduli, which, at the classical level, vary in a continuum. In order to move on to the semiclassical level, appropriate Bohr-Sommerfeld quantisation conditions must be imposed. These were studied in \cite{Dorey:2006mx} and can be formulated as integrality constraints on the so-called \emph{filling fractions} associated with the cuts: $\mathcal{S}_I \in \mathbb{Z}^+$, for $I = 1, \ldots, K$, with
\begin{equation}
 \mathcal{S}_{I} = - \frac{1}{2\pi i} \frac{\sqrt{\lambda}}{4\pi}\,\,\oint_{\mathcal{A}_{I}} \,
  \left( x \, + \, \frac{1}{x} \right) \, dp \:.
\label{eq:def_filling_fractions_FG}
\end{equation}
The $K$ filling fractions, subject to the level-matching constraint,
\begin{equation}
 \sum_{I=1}^K \,\, n_I \mathcal{S}_I = 0 \:,
\label{eq:level_matching_FG}
\end{equation}
were also shown to constitute a valid parametrisation of the moduli space, which therefore has dimension $K-1$. The total angular momentum in $AdS_3$ is also related to the filling fractions, according to
\begin{equation}
 S = \sum_{I=1}^K \, \mathcal{S}_I \:.
\label{eq:S_in_terms_of_filling_fractions}
\end{equation}
In summary, the semiclassical spectrum of the family of $K$-gap solutions,
\begin{equation}
 \Delta = \Delta\left[\mathcal{S}^{+}_{1},\mathcal{S}^{-}_{1}, \ldots, \mathcal{S}^{+}_{K/2},\mathcal{S}^{-}_{K/2}\right] \:,
\label{eq:finite-gap_spectrum_as_f-n_of_moduli}
\end{equation}
is determined by a genus $K-1$ hyper-elliptic spectral curve $\Sigma$, equipped with a differential $dp$ which has two double poles and $K$ branch cuts on the complex plane, together with specific analytical and asymptotic properties \eqref{eq:def_properties_of_dp_FG}, and must satisfy the period conditions \eqref{eq:A_periods_FG} and \eqref{eq:B_periods_FG}. The moduli $\mathcal{S}^\pm_I$ have to be integers due to the semiclassical quantisation conditions.

For this semiclassical description to be valid, we need the conserved charges $\Delta$, $S$ and $J$ to be $O(\sqrt{\lambda})$, with $\lambda \gg 1$. Note, however, that, when we will later consider the $S \to \infty$ limit of the finite gap construction, $J$ will be kept fixed and finite, albeit large.
\paragraph{}
Finally, it is interesting to compare the differential of the quasi-momentum in gauge theory to the same object in string theory. There are several similarities between the two versions of $dp$: their associated Riemann surface is hyper-elliptic in both cases and the cut structure is  identical, including the fact that cuts can collapse into double points, although the branch points have to lie on the real axis in the spin chain case, while there is in general no constraint on their position in the string case, as long as they do not overlap with other singularities. Furthermore, all the A- and B-periods coincide, with the exception of those associated with the cut $I_0$.

The singularities are however different: while $dp$ has a simple pole at the origin on each sheet on the gauge side, it has two double poles at $x = \pm 1$ on both sheets on the string side. Consequently, the quasi-momentum develops a logarithmic branch point along $I_0$ in the first case, and it has no other singularities other than the square root branch points; the number of double points is also finite for finite $J$. Logarithmic branch points are instead absent in the second case, which is the reason why there are no cuts with the properties of $I_0$ on the string side. Furthermore, two simple poles appear, acting as accumulation points of a numerable infinite set of double points.

Most importantly, in the case of the semiclassical spin chain, the quasi-momentum $p(x)$ is known from the start, and all the properties of the corresponding spectral curve can be directly inferred from it. On the string side, instead, the general theory yields a generic form for the spectral curve and then places analytical, asymptotic and period constraints on the differential, leaving us with the task of determining it. Solving this so-called \emph{spectral problem} is usually very hard, due to the transcendental nature of the period constraints.

\chapter{The large spin limit of the finite-gap spectrum}
\label{ch:FGA}

In this chapter, we will consider the limit $S \to \infty$ with $J = O(\lambda) \gg 1$ of the finite-gap spectrum discussed in the previous chapter, with \cite{Dorey:2008zy} and \cite{Dorey:2010id} as our main references. The main result of \cite{Dorey:2008zy} is the derivation of the fact that the spectrum of finite-gap strings in a fairly general large $S$ limit reproduces the ``large'' hole contribution to the spectrum of the $SL (2, \mathbb{R})$ spin chain. The limit considered in \cite{Dorey:2010id} is more general and yields a second branch of the spectrum, which may be associated with a family of quasi-particles. The corresponding dispersion relation matches the one associated with  ``small'' holes, in the limit of very large momentum.

\section{Setup}
\label{sec:FGA_setup}

We will now focus on $K$-gap solutions with all the branch cuts lying on the real axis, outside the interval $[-1,1]$, which means that only classical transverse oscillator modes with positive spin $S = + 1$ are activated. Therefore, such string solutions are expected to correspond to gauge theory operators in the $\mathfrak{sl}(2)$ sector with only covariant derivatives of the type $D_+$ appearing inside the trace, as in \eqref{eq:sl2_sector_specific_op}.

For simplicity, we are going to consider $K$ to be even and the cuts to be equally distributed between the two regions $x < - 1$ and $x > 1$, although the results still apply if these restrictions are lifted.

\begin{figure}
\centering
\psfrag{a}{\footnotesize{$(a)$}}
\psfrag{b}{\footnotesize{$(b)$}}
\psfrag{c}{\footnotesize{$(c)$}}
\psfrag{a1}{\footnotesize{$a^{(K-M-1)}_-$}}
\psfrag{a2}{\footnotesize{}}
\psfrag{a3}{\footnotesize{}}
\psfrag{a4}{\footnotesize{}}
\psfrag{a5}{\footnotesize{$a^{(1)}_-$}}
\psfrag{a6}{\footnotesize{$b^{(M)}_-$}}
\psfrag{a7}{\footnotesize{}}
\psfrag{a8}{\footnotesize{}}
\psfrag{a9}{\footnotesize{$b^{(1)}_-$}}
\psfrag{a10}{\footnotesize{$-b$}}
\psfrag{a11}{\footnotesize{$b$}}
\psfrag{a12}{\footnotesize{$b^{(1)}_+$}}
\psfrag{a13}{\footnotesize{}}
\psfrag{a14}{\footnotesize{}}
\psfrag{a15}{\footnotesize{$b^{(M)}_+$}}
\psfrag{a16}{\footnotesize{$a^{(1)}_+$}}
\psfrag{a17}{\footnotesize{}}
\psfrag{a18}{\footnotesize{}}
\psfrag{a19}{\footnotesize{}}
\psfrag{a20}{\footnotesize{$a^{(K-M-1)}_+$}}
\psfrag{b1}{\footnotesize{$\mathcal{C}_1^-$}}
\psfrag{b2}{\footnotesize{}}
\psfrag{b3}{\footnotesize{$\mathcal{C}_{\frac{K-M}{2}}^-$}}
\psfrag{b4}{\footnotesize{}}
\psfrag{b5}{\footnotesize{$\mathcal{C}_{\frac{K}{2}}^-$}}
\psfrag{b6}{\footnotesize{$\mathcal{C}_{\frac{K}{2}}^+$}}
\psfrag{b7}{\footnotesize{}}
\psfrag{b8}{\footnotesize{$\mathcal{C}_{\frac{K-M}{2}}^+$}}
\psfrag{b9}{\footnotesize{}}
\psfrag{b10}{\footnotesize{$\mathcal{C}_1^+$}}
\psfrag{c1}{\footnotesize{$\mathcal{A}_1^-$}}
\psfrag{c2}{\footnotesize{}}
\psfrag{c3}{\footnotesize{$\mathcal{A}_{\frac{K-M}{2}}^-$}}
\psfrag{c4}{\footnotesize{}}
\psfrag{c5}{\footnotesize{$\mathcal{A}_{\frac{K}{2}}^-$}}
\psfrag{c6}{\footnotesize{$\mathcal{A}_{\frac{K}{2}}^+$}}
\psfrag{c7}{\footnotesize{}}
\psfrag{c8}{\footnotesize{$\mathcal{A}_{\frac{K-M}{2}}^+$}}
\psfrag{c9}{\footnotesize{}}
\psfrag{c10}{\footnotesize{$\mathcal{A}_1^+$}}
\psfrag{d1}{\footnotesize{$\mathcal{B}_1^-$}}
\psfrag{d2}{\footnotesize{}}
\psfrag{d3}{\footnotesize{$\mathcal{B}_{\frac{K-M}{2}}^-$}}
\psfrag{d4}{\footnotesize{}}
\psfrag{d5}{\footnotesize{$\mathcal{B}_{\frac{K}{2}}^-$}}
\psfrag{d6}{\footnotesize{$\mathcal{B}_{\frac{K}{2}}^+$}}
\psfrag{d7}{\footnotesize{}}
\psfrag{d8}{\footnotesize{$\mathcal{B}_{\frac{K-M}{2}}^+$}}
\psfrag{d9}{\footnotesize{}}
\psfrag{d10}{\footnotesize{$\mathcal{B}_1^+$}}
\includegraphics[width=\columnwidth]{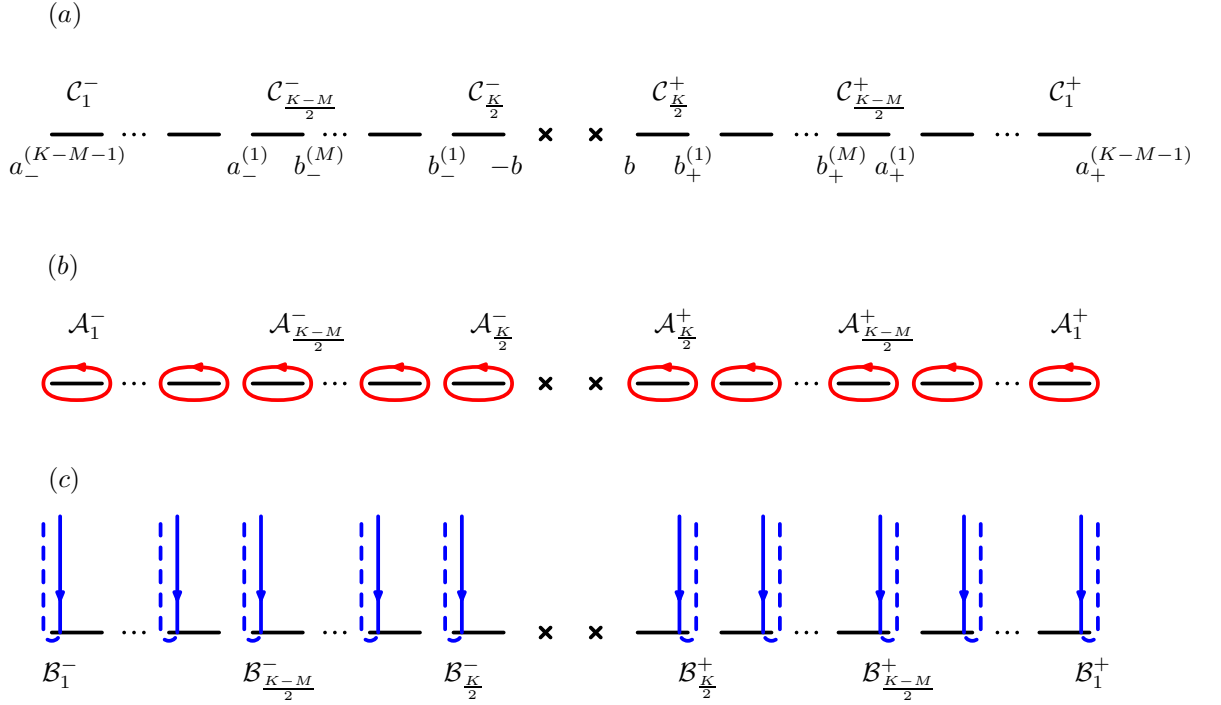}%
\caption{Initial configuration for the surface $\Sigma$. (a) Branch points and cuts. (b) A-cycles. (c) B-cycles. The double poles at $x = \pm 1$ are indicated as crosses.}
\label{fig:init_cuts_bps_Sigma}%
\end{figure}

We label the $2K$ branch points as follows:
\begin{eqnarray}
 a_{-}^{(K-M-1)} \,\, \leq \,\, a_{-}^{(K-M-2)} \,\, \leq \,\, \ldots \,\, \leq \,\, a_{-}^{(1)} \,\, \leq \,\, b_{-}^{(M)}
   \,\, \leq & \ldots & \leq \,\, b_{-}^{(0)}\,\,\leq -1 \nn \\ 
 a_{+}^{(K-M-1)} \,\, \geq \,\, a_{+}^{(K-M-2)} \,\, \geq \,\, \ldots \,\, \geq \,\, a_{+}^{(1)} \,\, \geq \,\, b_{+}^{(M)}
  \,\, \geq & \ldots & \geq \,\, b_{+}^{(0)}\,\,\geq +1 \qquad
\label{eq:branch_point_ordering_FGA}
\end{eqnarray}
where $M = 0, \ldots, K-2$ must be \emph{even}\footnote{As we will see shortly, $M$ acquires the meaning of the number of cuts whose endpoints both migrate onto the Riemann surface $\tilde{\Sigma}_2$ and has to be even in our symmetric setup. However, our results also hold when this number is odd.} and we also assume the two innermost branch points to be symmetric with respect to the origin, i.e. $b^{(0)}_\pm = \pm b$, with $b \geq 1$. The spectral curve is then expressed as
\begin{equation}
 \Sigma: \quad y^2 = (x^2 - b^2) \prod_{j=1}^M (x - b^{(j)}_+) (x - b^{(j)}_-) \prod_{k=1}^{K-M-1} (x - a^{(k)}_+) (x - a^{(k)}_-)
\label{eq:def_Sigma_FGA}
\end{equation}
and the cuts and their corresponding standard A- and B-cycles are in turn respectively relabelled as $\mathcal{C}^\pm_I, \mathcal{A}^\pm_I$ and $\mathcal{B}^\pm_I$, for $I = 1, \ldots, K/2$, as shown in Fig. \ref{fig:init_cuts_bps_Sigma}.

Note that, assuming that the double points $\hat{x}_k$ also lie on the real axis\footnote{This is actually the case when we consider the $SL(2, \mathbb{R})$ spin chain, and hence it seems reasonable to impose the same condition while we try to reobtain the same spectrum from string theory.}, accumulating at $x = \pm 1$ from both sides, we recover a pattern which is very similar to what we saw on the gauge side. In fact, by essentially the same reasoning, relying on the fact that $\cos p(x)$ is real for $x \in \mathbb{R}$, one may check that the mode numbers associated with the cuts and the double points have to start at $-1$ on the leftmost branch point or cut and at $+1$ at the rightmost object. Then, as we move inwards, approaching the double poles, the mode numbers respectively decrease and increase by one unit when we move from an object to the next, eventually diverging towards $-\infty$ and $+\infty$ due to the infinite number of double points. As in the case of the semiclassical spin chain, there is an underlying conventional choice in the initial definition of the branch of the inverse cosine: $p(x_0) = -\pi$, where $x_0$ represents the leftmost object in $\{ a^{(K-M-1)}_-, \hat{x}_k \}$. With the opposite convention, $p(x_0) = +\pi$ all the signs of the mode numbers would be reversed.

The important point here is that there is a general restriction on the way in which we may choose the mode numbers, demanding that they be monotonically decreasing (respectively, increasing) as we approach the pole at $x = -1$ ($x = +1$) starting from $\mathcal{C}^-_1$ ($\mathcal{C}^+_1$). This has an impact on the period conditions, which are relabelled as
\begin{equation}
 \oint_{\mathcal{A}^{\pm}_{I}} dp = 0   \qquad{} \oint_{\mathcal{B}^{\pm}_{I}} dp = 2\pi n^{\pm}_{I}.
\label{eq:period_conditions_FGA}
\end{equation}
Taking the restriction into account, we choose to activate only the lowest possible oscillator modes, which is achieved by fixing $n_I = \pm I$, $I = 1, \ldots, K/2$. This also implies that all the double points are located inside the interval $x \in (-b,b)$.

The filling fractions are rewritten as
\begin{equation}
 \mathcal{S}^{\pm}_{I} = - \frac{1}{2\pi i} \frac{\sqrt{\lambda}}{4\pi} \oint_{\mathcal{A}^{\pm}_{I}} \left(x + \frac{1}{x} \right) dp \:,
\label{eq:filling_fractions_FGA}
\end{equation}
for $I = 1, \ldots, K/2$. They are subject to the level-matching constraint
\begin{equation}
 \sum_{I=1}^{K/2} \left( n^{+}_{I} \mathcal{S}^{+}_{I} + n^{-}_{I} \mathcal{S}^{-}_{I} \right) = 0 
\label{eq:level_matching_FGA}
\end{equation}
and they are related to the total $AdS_3$ angular momentum:
\begin{equation}
 S = \sum_{I=1}^{K/2} \left( \mathcal{S}^{+}_{I} + \mathcal{S}^{-}_{I} \right) \:.
\label{eq:S_filling_fracts_FGA}
\end{equation}

\section{The large $S$ limit}

\subsection{General discussion of the limit}

\begin{figure}
\centering
\psfrag{a}{\footnotesize{(a)}}
\psfrag{b}{\footnotesize{(b)}}
\psfrag{c}{\footnotesize{(c)}}
\psfrag{a1}{\footnotesize{$\mathcal{A}_{\frac{K-M}{2}-1}^+$}}
\psfrag{a2}{\footnotesize{$\mathcal{A}_1^+$}}
\psfrag{a3}{\footnotesize{$\mathcal{A}_{\frac{K-M}{2}-1}^-$}}
\psfrag{a4}{\footnotesize{$\mathcal{A}_1^-$}}
\psfrag{a5}{\footnotesize{$\tilde{\mathcal{A}}_0$}}
\psfrag{a6}{\footnotesize{$\hat{\mathcal{A}}_{\frac{K}{2}-1}^-$}}
\psfrag{a7}{\footnotesize{$\hat{\mathcal{A}}_{\frac{K}{2}}^-$}}
\psfrag{a8}{\footnotesize{$\hat{\mathcal{A}}_{\frac{K-M}{2}}^+$}}
\psfrag{a9}{\footnotesize{$\hat{\mathcal{A}}_{\frac{K}{2}-1}^+$}}
\psfrag{a10}{\footnotesize{$\hat{\mathcal{A}}_{\frac{K}{2}}^+$}}
\includegraphics[width=\columnwidth]{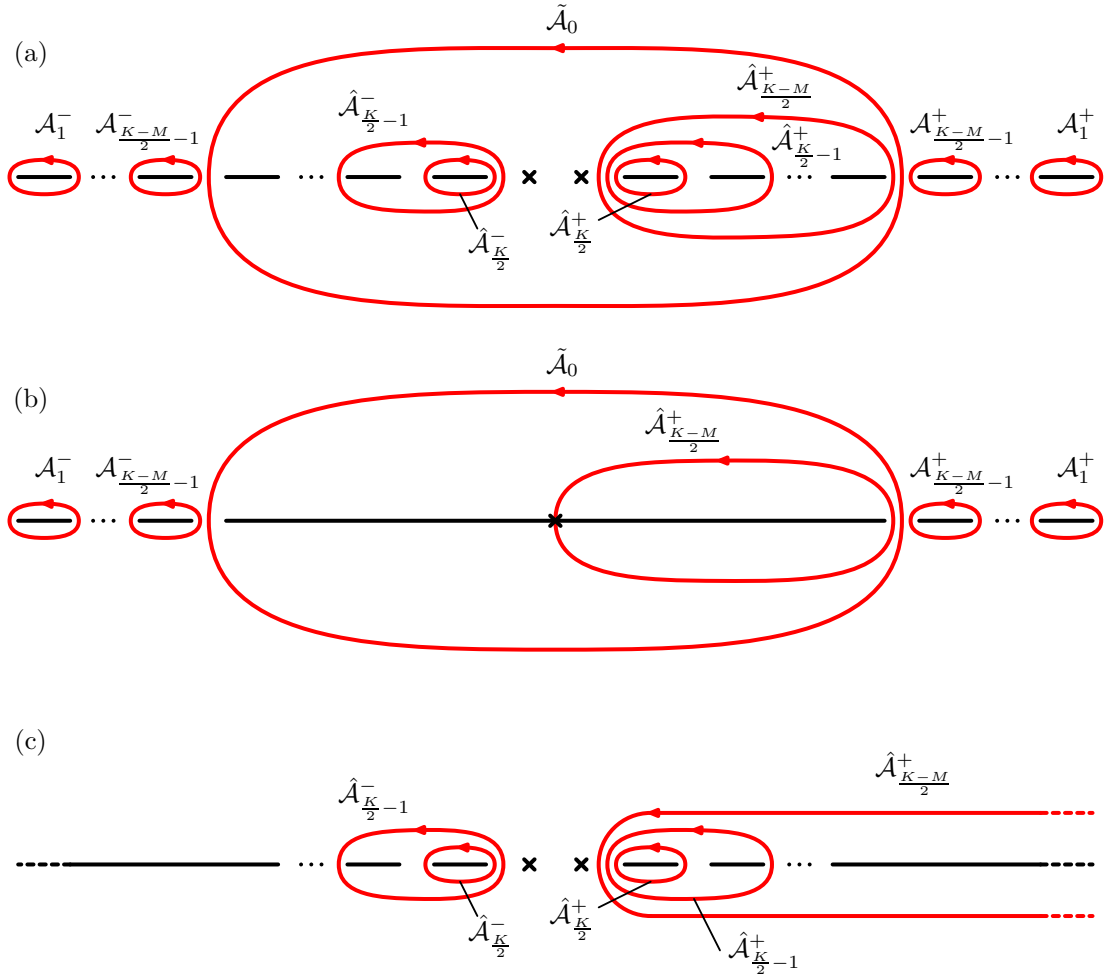}%
\caption{(a) The rearranged A-cycle configuration on $\Sigma$. (b) The A-cycles on $\tilde{\Sigma}_1$; the marked point indicates the simple pole at $x=0$. (c) The A-cycles on $\tilde{\Sigma}_2$; the marked points indicate the double poles at $x = \pm 1$. All dashed lines extend to infinity along the real axis.}
\label{fig:Acycles}%
\end{figure}

\begin{figure}
\centering
\psfrag{a}{\footnotesize{(a)}}
\psfrag{b}{\footnotesize{(b)}}
\psfrag{c}{\footnotesize{(c)}}
\psfrag{b1}{\footnotesize{$\mathcal{B}_1^-$}}
\psfrag{b2}{\footnotesize{$\mathcal{B}_{\frac{K-M}{2}-1}^-$}}
\psfrag{b3}{\footnotesize{$\mathcal{B}_1^+$}}
\psfrag{b4}{\footnotesize{$\mathcal{B}_{\frac{K-M}{2}-1}^+$}}
\psfrag{b5}{\footnotesize{$\mathcal{B}_{\frac{K-M}{2}}^+$}}
\psfrag{b6}{\footnotesize{$\hat{\mathcal{B}}$}}
\psfrag{b7}{\footnotesize{$\hat{\mathcal{B}}_{\frac{K}{2}}^-$}}
\psfrag{b8}{\footnotesize{$\hat{\mathcal{B}}_{\frac{K}{2}-1}^-$}}
\psfrag{b9}{\footnotesize{$\hat{\mathcal{B}}_{\frac{K-M}{2}+1}^-$}}
\psfrag{b10}{\footnotesize{$\hat{\mathcal{B}}_{\frac{K}{2}}^+$}}
\psfrag{b11}{\footnotesize{$\hat{\mathcal{B}}_{\frac{K}{2}-1}^+$}}
\psfrag{b12}{\footnotesize{$\hat{\mathcal{B}}_{\frac{K-M}{2}+1}^+$}}
\includegraphics[width=\columnwidth]{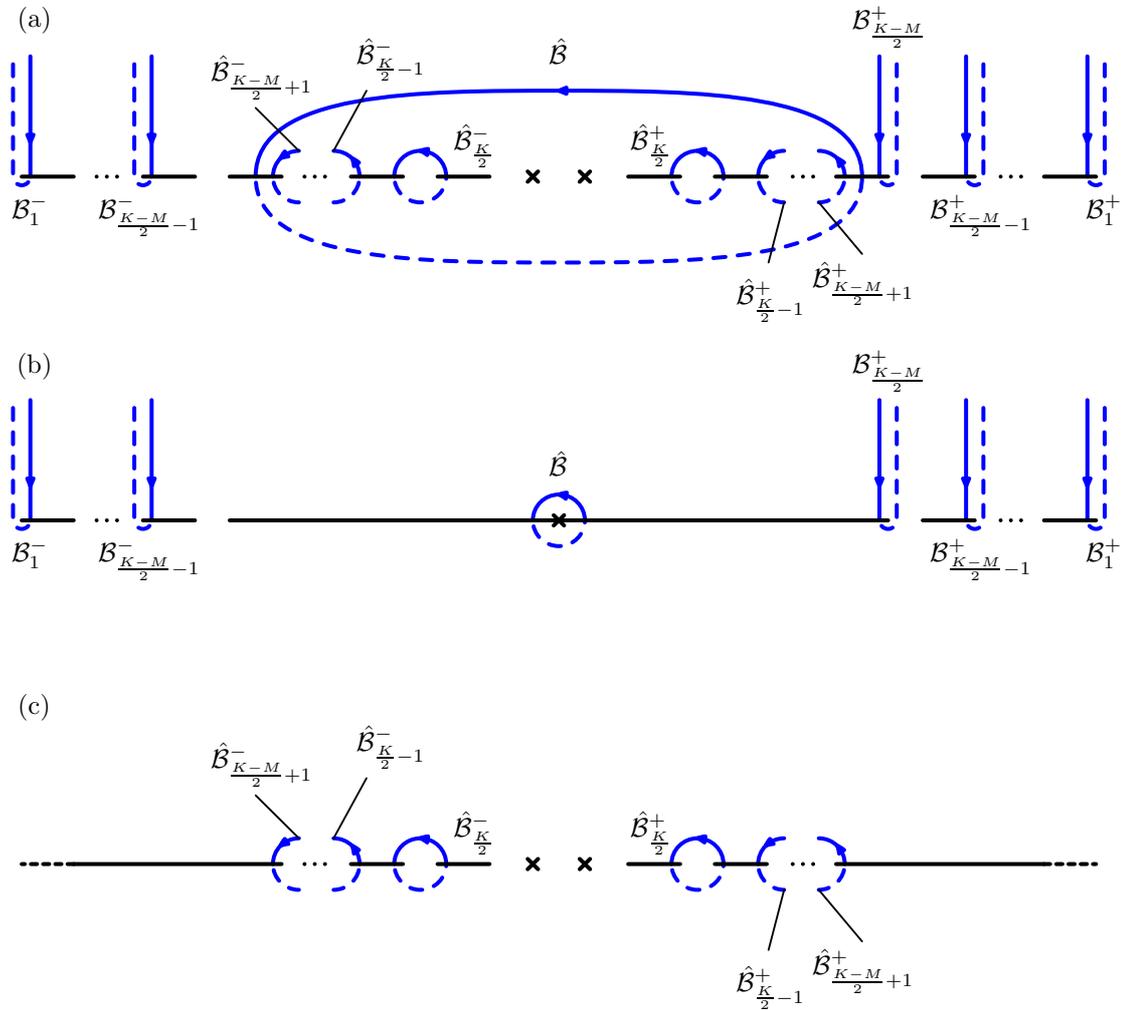}%
\caption{(a) The rearranged B-cycle configuration on $\Sigma$. Dashed blue lines indicate paths on the lower sheet throughout the picture. (b) The B-cycles on $\tilde{\Sigma}_1$; the marked point indicates the simple pole at $x=0$. (c) The B-cycles on $\tilde{\Sigma}_2$; the marked points indicate the double poles at $x = \pm 1$. The black dashed lines extend to infinity along the real axis.}
\label{fig:Bcycles}%
\end{figure}

In the following we are going to consider a specific behaviour of the branch points which results in a factorisation of the spectral curve $\Sigma$ into two separate Riemann surfaces, joined by two contact points. For the purposes of the subsequent analysis, it is convenient to rearrange the A- and B-cycles in a non-canonical configuration before we take the limit $S \to \infty$. The new equivalent configurations of the A-cycles and of the B-cycles are shown in Fig. \ref{fig:Acycles}(a) and \ref{fig:Bcycles}(a) respectively. In particular, we define $\hat{\mathcal{A}}_I^\pm = \sum_{J=I}^{K/2} \mathcal{A}_J^\pm$, for $I = (K-M)/2, \ldots, K/2$, and $\hat{\mathcal{A}}_0 = \sum_{J=(K-M)/2}^{K/2}(\mathcal{A}_J^+ + \mathcal{A}_J^-)$, together with 
$\hat{\mathcal{B}} = \mathcal{B}_{(K-M)/2}^- - \mathcal{B}_{(K-M)/2}^+$, $\hat{\mathcal{B}}_I^- = \mathcal{B}_{I-1}^- - \mathcal{B}_I^-$ and $\hat{\mathcal{B}}_I^+ = \mathcal{B}_I^+ - \mathcal{B}_{I-1}^+$, for $I = (K-M)/2 +1, \ldots, K/2$.

The new period conditions for $dp$ on $\Sigma$ associated with this new configuration are given by
\begin{eqnarray}
  \oint_{\mathcal{A}_I^\pm} dp = 0 & \qquad & \oint_{\mathcal{B}_I^\pm} dp = 2 \pi n_I^\pm = \pm 2 \pi I 
   \qquad \textrm{for } I = 1, \ldots, \frac{K-M}{2} - 1 \nn \\
  \oint_{\tilde{\mathcal{A}}_0} dp = 0 & \qquad & \oint_{\hat{\mathcal{B}}} dp = - 2\pi (K-M) \nn \\
  \oint_{\hat{\mathcal{A}}_\frac{K-M}{2}^+} dp = 0 & \qquad & \oint_{\mathcal{B}_\frac{K-M}{2}^+} dp = \pi (K-M) \nn \\
  \oint_{\hat{\mathcal{A}}_J^\pm} dp = 0 & \qquad & \oint_{\hat{\mathcal{B}}_J^\pm} dp = 2 \pi \qquad \textrm{for } I = \frac{K-M}{2} +1, \ldots, K/2 \:.
\label{eq:new_period_conditions_S}
\end{eqnarray}
\paragraph{}
We now introduce a scaling parameter $\rho$, and impose
\begin{equation}
 a^{(j)}_{\pm} = \rho \tilde{a}_{\pm}^{(j)}
\label{eq:def_rho_FGA}
\end{equation}
for $j=1,2,\ldots, K-M-1$. We then take the limit $\rho \rightarrow \infty$ with $\tilde{a}^{(j)}_{\pm}$,  $b^{(j)}_{\pm}$ and $b$ held fixed. Thus we are dividing the branch points into ``large'' ($a^{(j)}_\pm$) and ``small'' ($b^{(j)}_\pm$). This is a generalisation of the limit considered in \cite{Dorey:2008zy}, which in the present notation corresponds to $M = 0$.

\begin{figure}
\centering
\psfrag{a}{\footnotesize{$\Sigma$}}
\psfrag{b}{\footnotesize{$\tilde{\Sigma}_1$}}
\psfrag{c}{\footnotesize{$\tilde{\Sigma}_2$}}
\psfrag{d}{\footnotesize{contact points}}
\includegraphics[width=\columnwidth]{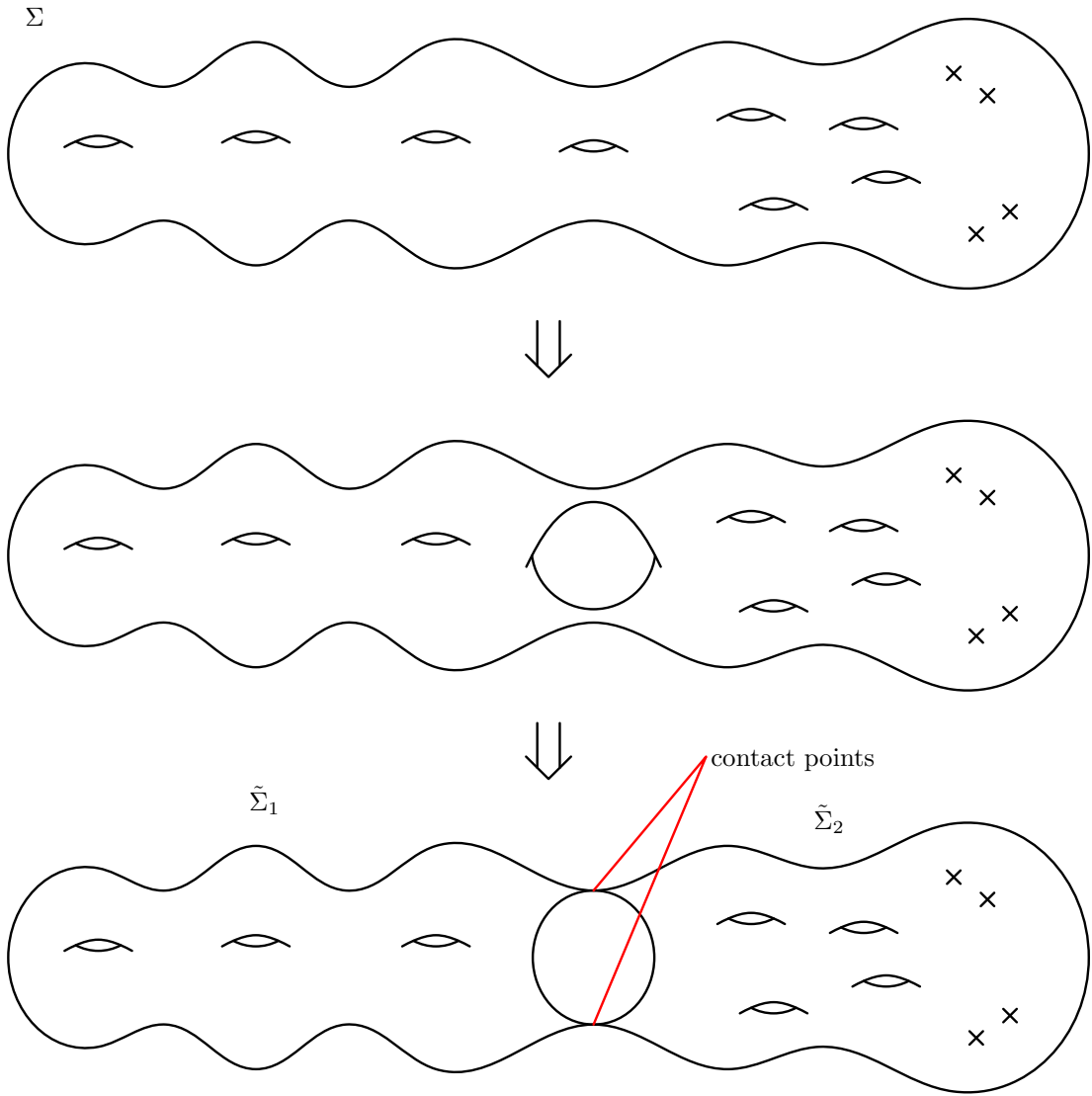}%
\caption{The factorisation of the finite-gap spectral curve: $\Sigma \to \tilde{\Sigma}_1 \cup \tilde{\Sigma}_2$, in the special case $K=9$, $M=4$. Note that the behaviour of $\tilde{\Sigma}_2$ as shown here is only conjectured for the time being, and in fact it will turn out to be different (see Fig. \ref{fig:exact_factor_Sigma}).}
\label{fig:Sigma_factor_1}%
\end{figure}

The ``large'' branch points move towards infinity on both sheets, while the ``small'' branch points remain close to the origin and to the double poles. As the regions in the proximity of the two groups of singularities become infinitely separated, the spectral curve factorises:
\begin{equation}
 \Sigma \longrightarrow \tilde{\Sigma}_{1} \cup \tilde{\Sigma}_{2}
\label{eq:Sigma_factorisation_FGA}
\end{equation}
where $\tilde{\Sigma}_1$ is the area containing the ``large'' branch points and the two points at infinity ($\infty^\pm$), while $\tilde{\Sigma}_2$ contains the ``small'' branch points, the four double poles ($x = 1^\pm$, $x = -1^\pm$), the two origins ($x=0^\pm$) and all the double points. This process can be visualised by ``blowing up'' one of the handles of the original surface until the tubes at its sides, connecting the two regions, squeeze into two contact points, as shown in Fig. \ref{fig:Sigma_factor_1}.

If we introduce the rescaled spectral parameter $\tilde{x} = x/ \rho$, $\tilde{\Sigma}_1$ then corresponds to the region $\tilde{x} \neq 0$, whereas $\tilde{\Sigma}_2$ is represented by all the finite values of $x$, i.e. $x \neq \infty$. The two variables $x$ and $\tilde{x}$ represent the coordinate systems which we will use to parametrise the two surfaces. Some notable points after the factorisation are the two contact points (respectively located at $\tilde{x} = 0^\pm$ and $x = \infty^\pm$) the two points at infinity on $\Sigma$ (lying on $\tilde{\Sigma}_1$ at $\tilde{x} = \infty^\pm$), the four double poles on $\Sigma$ (lying on $\tilde{\Sigma}_2$ at $x = 1^\pm$ and at $x = -1^\pm$) and finally the two origins on $\Sigma$ (lying on $\tilde{\Sigma}_2$ at $x = 0^\pm$).

\begin{figure}
\centering
\psfrag{a}{\footnotesize{(a)}}
\psfrag{b}{\footnotesize{(b)}}
\psfrag{a1}{\footnotesize{$\mathcal{C}_{\frac{K-M}{2}-1}^+$}}
\psfrag{a2}{\footnotesize{$\mathcal{C}_1^+$}}
\psfrag{a3}{\footnotesize{$\mathcal{C}_{\frac{K-M}{2}-1}^-$}}
\psfrag{a4}{\footnotesize{$\mathcal{C}_1^-$}}
\psfrag{a5}{\footnotesize{$\tilde{\mathcal{C}}_0$}}
\psfrag{a6}{\footnotesize{$\tilde{a}^{(K-M-1)}_-$}}
\psfrag{a7}{\footnotesize{$\tilde{a}^{(1)}_-$}}
\psfrag{a8}{\footnotesize{$\tilde{a}^{(K-M-1)}_+$}}
\psfrag{a9}{\footnotesize{$\tilde{a}^{(1)}_+$}}
\psfrag{b1}{\footnotesize{$\mathcal{C}_{\frac{K}{2}}^-$}}
\psfrag{b2}{\footnotesize{$\mathcal{C}_{\frac{K}{2}}^+$}}
\psfrag{b3}{\footnotesize{$b^{(M)}_-$}}
\psfrag{b4}{\footnotesize{$b^{(M)}_+$}}
\psfrag{b5}{\footnotesize{$+b$}}
\psfrag{b6}{\footnotesize{$b^{(1)}_+$}}
\psfrag{b7}{\footnotesize{$-b$}}
\psfrag{b8}{\footnotesize{$b^{(1)}_-$}}
\includegraphics[width=\columnwidth]{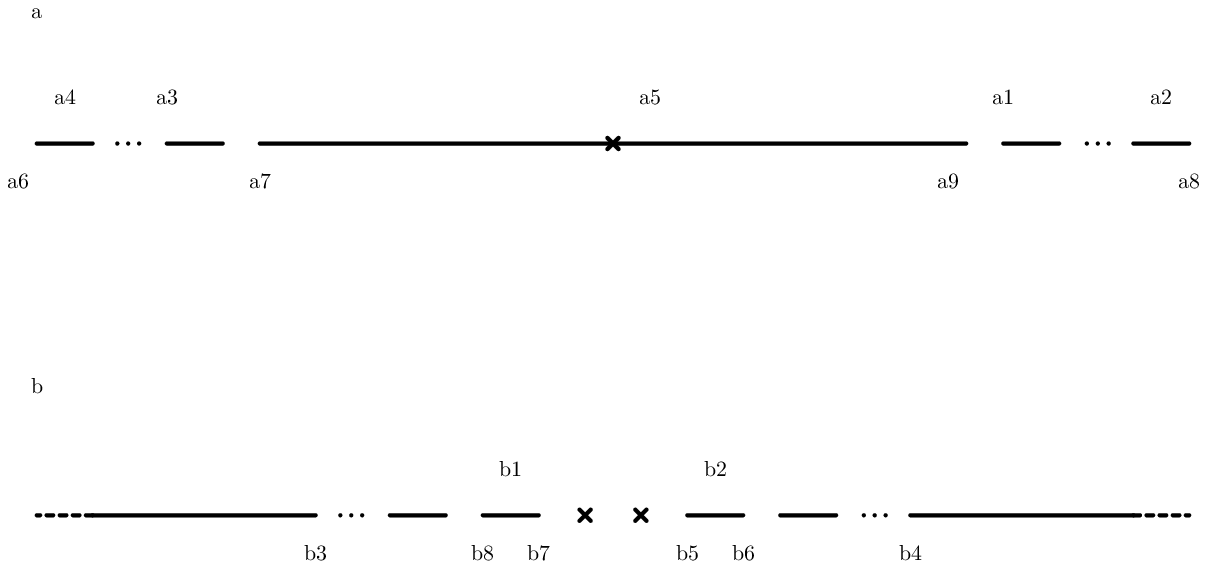}%
\caption{The surviving branch cuts and branch points on $\tilde{\Sigma}_1$ (a) and on $\tilde{\Sigma}_2$ (b).}
\label{fig:bps_and_cuts_denominations_Sigma12tilde}%
\end{figure}

The key point is that the periods of the differential $dp$ must be preserved during this process. This requirement will determine all of the crucial properties of the two new surfaces emerging after the factorisation. The situation on $\tilde{\Sigma}_1$ will turn out to be almost identical to the one discussed in \cite{Dorey:2008zy}, and hence the derivation will proceed in the same fashion. The behaviour on $\tilde{\Sigma}_2$ will instead become rather complicated due to the larger number of ``small'' branch points migrating onto that surface with respect to the $M=0$ case. In fact, we previously modified all of the cycles involving at least one of the ``small'' branch points in order to deal with the period conditions on the second surface more effectively.
\paragraph{}
As $\rho$ increases, due to the progressive separation of the two groups of branch points, only a subset of the initial cycles remains on each of the two resulting surfaces after the factorisation. Fig. \ref{fig:Acycles}(b) and \ref{fig:Bcycles}(b) show which cycles survive on $\tilde{\Sigma}_1$ (which contains the $a^{(j)}_\pm$), while Fig. \ref{fig:Acycles}(c) and \ref{fig:Bcycles}(c) show the configuration for $\tilde{\Sigma}_2$ (which contains the $b^{(j)}_\pm$). Fig. \ref{fig:bps_and_cuts_denominations_Sigma12tilde} shows the labelling of cuts and branch points on the two surfaces.

The limit $\rho \to \infty$ and the associated migration of the branch points also modify the expressions for $y$ and $dp$ on $\tilde{\Sigma}_1$ and $\tilde{\Sigma}_2$. On the first surface,
\begin{equation}
 \tilde{\Sigma}_1 \:\: : \qquad \tilde{y}_1^2 = \prod_{k=1}^{K-M-1} (\tilde{x} - \tilde{a}^{(k)}_+) (\tilde{x} - \tilde{a}^{(k)}_-) \:,
\label{eq:y1tilde_Mn0}
\end{equation}
we have the following limiting form of the differential:
\begin{equation}
 d \tilde{p}_1 = - \frac{d \tilde{x}}{\tilde{y}_1} \sum_{l=M}^{K-2} \tilde{C}_l \tilde{x}^{l-M-1} \:.
\label{eq:def_dp1tilde_Mn0}
\end{equation}
The $C_l$ have been rescaled as follows:
\begin{equation}
 C_l = \left\{
  \begin{array}{ll}
	 \tilde{C}_l \, \rho^{K-l-1} \:, & \textrm{for } l \geq M \\
	 \tilde{C}_l \, \rho^{K-M-1} \:, & \textrm{for } l < M
	\end{array}
	 \right. \:.
\label{eq:rescaling_C_l_Mn0}
\end{equation}
This ensures that $d \tilde{p}_1$ has a simple pole at $\tilde{x} = 0$ and that none of these parameters disappears from both $d \tilde{p}_1$ and $d \tilde{p}_2$ due to suppression by negative powers of $\rho$. The first property is crucial to the construction of an explicit solution $\tilde{p}_1(\tilde{x})$ in closed form, while the second simply ensures full generality in that we do not lose any free parameters, while we are still free to set the rescaled $\tilde{C}_l$ equal to zero if we want.

As we can see from Figures \ref{fig:Acycles}(b) and \ref{fig:Bcycles}(b), the differential is subject to the following period conditions, which are inherited from $\Sigma$ \eqref{eq:new_period_conditions_S}:
\begin{eqnarray}
  \oint_{\mathcal{A}_I^\pm} d \tilde{p}_1 = 0 & \qquad & \oint_{\mathcal{B}_I^\pm} d \tilde{p}_1 = 2 \pi n_I^\pm = \pm 2 \pi I 
   \qquad \textrm{for } I = 1, \ldots, \frac{K-M}{2} - 1 \nn \\
  \oint_{\tilde{\mathcal{A}}_0} d \tilde{p}_1 = 0 & \qquad & \oint_{\hat{\mathcal{B}}} d \tilde{p}_1 = - 2\pi (K-M) \nn \\
  & & \oint_{\mathcal{B}_\frac{K-M}{2}^+} d \tilde{p}_1 = \pi (K-M) \:.
\label{eq:period_conditions_S1}
\end{eqnarray}
On $\tilde{\Sigma}_2$,
\begin{equation}
 \tilde{\Sigma}_2 \:\: : \qquad \tilde{y}_2^2 = (x^2 - b^2) \prod_{i=1}^M (x - b^{(i)}_+) (x - b^{(i)}_-) \:,
\label{eq:eq:Sigma_2_tilde_Mn0}
\end{equation}
we find instead
\begin{equation}
 d \tilde{p}_2 = - \frac{dx}{\tilde{Q} \tilde{y}_2} \sum_{l=0}^M \tilde{C}_l x^l - \frac{\pi J}{\sqrt{\lambda}} \left[ 
  \frac{\tilde{y}_2 (1)}{(x-1)^2} + \frac{\tilde{y}_2 (-1)}{(x+1)^2} + 
   \frac{\tilde{y}'_2 (1)}{x-1} + \frac{\tilde{y}'_2 (-1)}{x+1} \right] \frac{dx}{\tilde{y}_2}
\label{eq:def_dp2tilde_Mn0}
\end{equation}
with $\tilde{Q}^2 = \tilde{y}_1^2(0)$. The corresponding period conditions are given by:
\begin{eqnarray}
  \oint_{\hat{\mathcal{A}}_J^\pm} d \tilde{p}_2 = 0 & \qquad & \oint_{\hat{\mathcal{B}}_J^\pm} d \tilde{p}_2 = 2 \pi \qquad \textrm{for } I = \frac{K-M}{2} +1, \ldots, K/2 \:,
\label{eq:period_conditions_S2}
\end{eqnarray}
as shown in Figures \ref{fig:Acycles}(c) and \ref{fig:Bcycles}(c).

\begin{figure}
\centering
\psfrag{a}{\footnotesize{$\Sigma$}}
\psfrag{b}{\footnotesize{$\tilde{\Sigma}_1$}}
\psfrag{c}{\footnotesize{$\tilde{\Sigma}_2$}}
\includegraphics[width=\columnwidth]{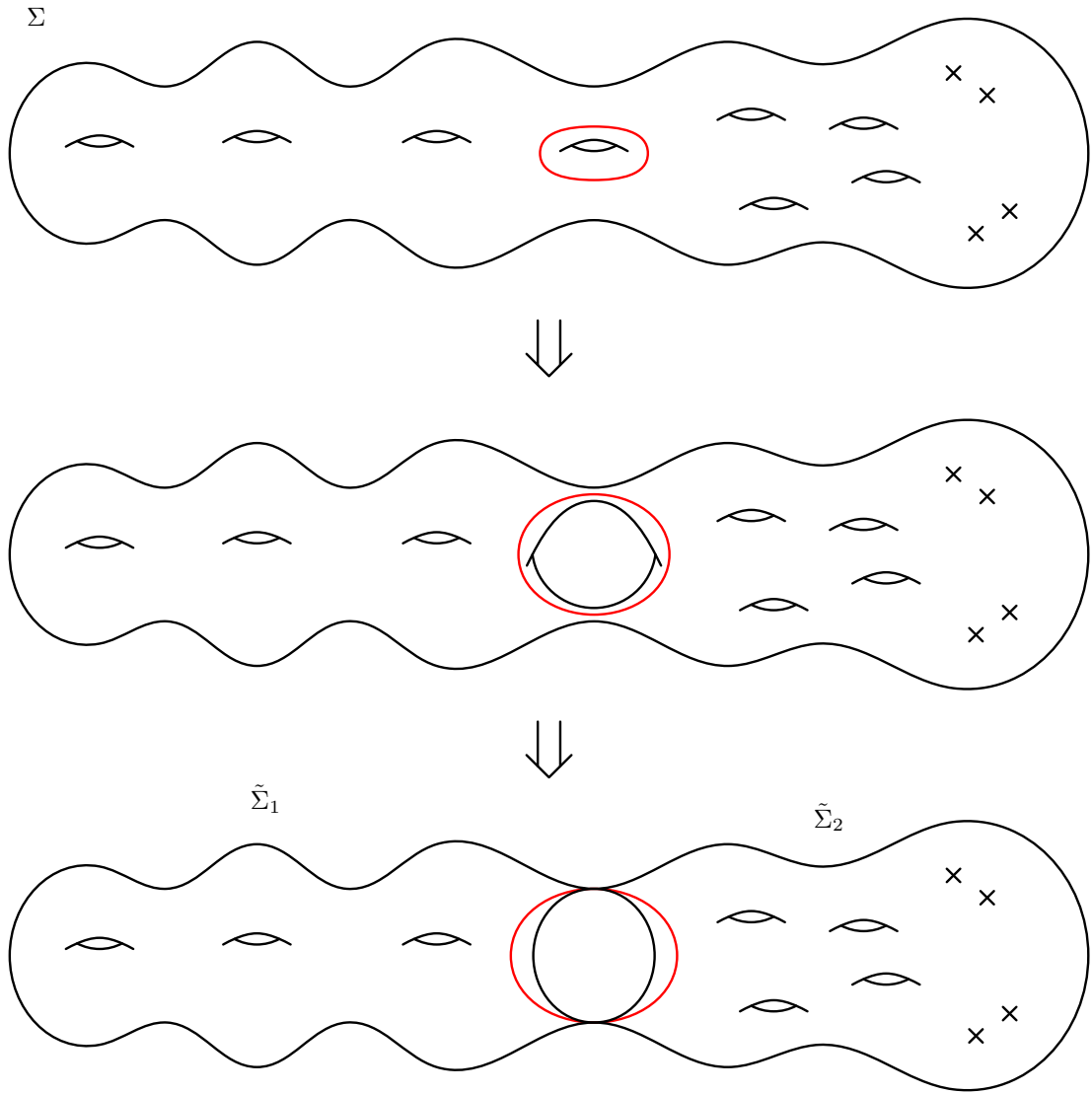}%
\caption{The A-cycle which is pinched at the contact points, providing the only period constraint which links the two otherwise separate spectral problems on $\tilde{\Sigma}_1$ and $\tilde{\Sigma}_2$ after factorisation.}
\label{fig:match_cond_cycle}%
\end{figure}

Only one period condition from the original set \eqref{eq:new_period_conditions_S} remains, namely
\begin{equation}
 \oint_{\hat{\mathcal{A}}_\frac{K-M}{2}^+} dp = 0 \:,
\label{eq:match_cond}
\end{equation}
which involves the only cycle that survives both on $\tilde{\Sigma}_1$ and on $\tilde{\Sigma}_2$. As we have just seen, the original spectral problem for $(\Sigma, dp)$ has reduced to two almost separate spectral problems $(\tilde{\Sigma}_1, d \tilde{p}_1)$ and $(\tilde{\Sigma}_2, d \tilde{p}_2)$. The constraint \eqref{eq:match_cond} then provides the only relationship between these two problems and we will therefore refer to it as the ``matching condition''. The associated A-cycle, which is pinched by the expanding handle on the surface, is shown in Fig. \ref{fig:match_cond_cycle}.

\subsection{Explicit solution on $\tilde{\Sigma}_1$}

The key difference with respect to the case of $dp$ on $\Sigma$ is that $d \tilde{p}_1$ only has a simple pole at $\tilde{x}=0$ on both sheets of $\tilde{\Sigma}_1$, instead of two double poles on each sheet:
\begin{equation}
 d \tilde{p}_1 \to \pm \frac{K-M}{i} \frac{d \tilde{x}}{\tilde{x}} \qquad \textrm{as } \tilde{x} \to 0 \:,
\label{eq:dptilde1_near_0}
\end{equation}
where the residue of $\tilde{p}'_1 (\tilde{x})$ at $\tilde{x} = 0$ is determined by the period condition \eqref{eq:period_conditions_S1} on the cycle $\hat{\mathcal{B}}$, which imposes $\tilde{C}_M = - i \tilde{Q} (K-M)$. If we integrate this relation, we obtain the asymptotic behaviour of $\tilde{p}_1 ( \tilde{x} )$ near the origin:
\begin{equation}
 \tilde{p}_1 ( \tilde{x} ) \to \pm \frac{K-M}{i} \log \tilde{x} \qquad \textrm{as } \tilde{x} \to 0
\label{eq:ptilde1_near_0}
\end{equation}
and thus, on one sheet, we have
\begin{equation}
 \mathrm{exp} (\pm i \tilde{p}_1 (\tilde{x})) \to (\tilde{x})^{\pm (K-M)} \qquad \textrm{as } \tilde{x} \to 0 \:,
\label{eq:exp(p1tilde)_near_0}
\end{equation}
while one of the $\pm$ must be replaced with $\mp$ on the other sheet. Hence, the function
\begin{equation}
 f (\tilde{x}) = 2 \cos \tilde{p}_1 (\tilde{x}) = \mathrm{exp} (i \tilde{p}_1 (\tilde{x})) + \mathrm{exp} (-i \tilde{p}_1 (\tilde{x}))
\label{eq:def_f(xtilde)}
\end{equation}
has a pole of order $K-M$ at $\tilde{x} = 0$ on both sheets. Furthermore, since the periods of $d \tilde{p}_1$ are normalised in integer units (which means that $\tilde{p}_1$ changes sign and is shifted by an integer multiple of $2 \pi$ whenever we cross a cut), $f (\tilde{x})$ has no branch cuts, and it is therefore analytic on the complex plane, with its only singularity given by the pole. The behaviour of $\tilde{p}_1 (\tilde{x})$ at infinity can be obtained by integration from that of $d \tilde{p}_1$:
\begin{equation}
 \tilde{p}_1 (\tilde{x}) \simeq \frac{\tilde{C}_{K-2}}{\tilde{x}} \:, \qquad \textrm{as } \tilde{x} \to \infty
\label{eq:p1tilde_near_infinity}
\end{equation}
and then determines the behaviour of $f (\tilde{x})$:
\begin{equation}
 f (\tilde{x}) = 2 \cos \tilde{p}_1 (\tilde{x}) \simeq 2 + \frac{\tilde{q}_2}{\tilde{x}^2} \:, \qquad \textrm{as } \tilde{x} \to \infty
\label{eq:f_near_infinity}
\end{equation}
where we have defined $\tilde{q}_2 \equiv - \tilde{C}_{K-2}^2$.

The most general function $f (\tilde{x})$ satisfying these analytic constraints is given by the following expression, parametrised in terms of $K-M-1$ undetermined coefficients $\tilde{q}_j$, $j = 2, \ldots, K-M$:
\begin{equation}
 f (\tilde{x}) \equiv \mathbb{P}_{K-M} \left( \frac{1}{\tilde{x}} \right) = 2 + \frac{\tilde{q}_2}{\tilde{x}^2} + \frac{\tilde{q}_3}{\tilde{x}^3}
  + \ldots + \frac{\tilde{q}_{K-M}}{\tilde{x}^{K-M}} \:.
\label{eq:most_general_form_of_f}
\end{equation}
This yields an explicit form for $\tilde{p}_1 (\tilde{x}) = \cos^{-1} (f/2)$ and its differential:
\begin{equation}
 d \tilde{p}_1 = -i \frac{d \tilde{x}}{\tilde{x}^2} \frac{\mathbb{P}'_{K-M} \left( \frac{1}{\tilde{x}} \right)}{\sqrt{\mathbb{P}_{K-M}^2
  \left( \frac{1}{\tilde{x}} \right) -4}} \:.
\label{eq:explicit_dp1tilde_Mn0}
\end{equation}
It is now only a matter of algebra to cast this expression into the previous form \eqref{eq:def_dp1tilde_Mn0} in order to read off the equation defining the spectral curve:
\begin{equation}
 \tilde{\Sigma}_1 \:\: : \qquad \tilde{y}_1^2 = \frac{\tilde{x}^{2(K-M)}}{4 \tilde{q}_2} \left[ \mathbb{P}_{K-M}^2 
  \left( \frac{1}{\tilde{x}} \right) - 4 \right]
\label{eq:explicit_y1tilde}
\end{equation}
and of the rescaled coefficients:
\begin{equation}
 \tilde{C}_l = - \frac{(K-l) \tilde{q}_{K-l}}{2 \sqrt{- \tilde{q}_2}}
\label{eq:Ctilde(qtilde)_Mn0}
\end{equation}
for $l = M, \ldots, K-2$. In this way, we have expressed the solution in terms of the $K-M-1$ new parameters $\tilde{q}_j$, $j = 2, \ldots, K-M$, which represent the moduli on $\tilde{\Sigma}_1$. One can also check that the period conditions \eqref{eq:period_conditions_S1} are indeed satisfied, so that this is a genuine solution to the spectral problem on this surface.

With reference to Fig. \ref{fig:Sigma_factor_1}, $\tilde{\Sigma}_1$ is the surface on the left-hand side of the bottom picture. Its main features are the $K-M-1$ branch cuts and the two singular contact points with $\tilde{\Sigma}_2$, which are located at $\tilde{x}=0^\pm$.

\subsection{Explicit solution on $\tilde{\Sigma}_2$}

The following argument relies on the evaluation of the matching condition, at first only at leading order and using the explicit solution \eqref{eq:explicit_dp1tilde_Mn0}, which allows us to obtain a simpler expression for the differential $d \tilde{p}_2$ in terms of a single undetermined coefficient. Then this new expression is used to evaluate the constraint up to the first subleading order so as to fix the last coefficient. The required calculations lead to cumbersome expressions and involve several technical problems. Therefore, they have been relegated to appendix \ref{sec:matching_condition}.

By imposing the matching condition \eqref{eq:match_cond} at leading order, it is possible to show that the two innermost branch points on $\tilde{\Sigma}_2$ have to collide with the neighbouring double poles, i.e. that $b \to 1$ so that\footnote{The result \eqref{eq:def_dp2tilde_Mn0} still holds even though we now have a diverging factor $1/ \sqrt{1 - b^2}$ coming from $\tilde{y}'_2 (\pm 1)$, since corrections to that limit are suppressed by inverse powers of $\rho$ and hence a logarithmic divergence is too weak to make them $O(\rho^0)$.}:
\begin{equation}
 \frac{1}{\sqrt{1 - b^2}} \sim i \log \epsilon
\label{eq:sqrt(1-b^2)_lead_ord_Mn0}
\end{equation}
where $\epsilon = 1/\rho$. 

\begin{figure}
\centering
\psfrag{a}{\footnotesize{(a)}}
\psfrag{b}{\footnotesize{(b)}}
\psfrag{a1}{\footnotesize{$\hat{\mathcal{A}}_{\frac{K}{2}}^+$}}
\psfrag{a2}{\footnotesize{$\hat{\mathcal{A}}_{\frac{K}{2}-1}^+$}}
\psfrag{a3}{\footnotesize{$\hat{\mathcal{A}}_{\frac{K-M}{2}+1}^+$}}
\psfrag{a4}{\footnotesize{$\hat{\mathcal{A}}_{\frac{K-M}{2}}^+$}}
\psfrag{a5}{\footnotesize{$\hat{\mathcal{A}}_{\frac{K}{2}}^-$}}
\psfrag{a6}{\footnotesize{$\hat{\mathcal{A}}_{\frac{K}{2}-1}^-$}}
\psfrag{a7}{\footnotesize{$\hat{\mathcal{A}}_{\frac{K-M}{2}+1}^-$}}
\psfrag{c1}{\footnotesize{$c^{\left( 1 \right)}_+$}}
\psfrag{c2}{\footnotesize{$c^{\left( 2 \right)}_+$}}
\psfrag{c3}{\footnotesize{$c^{\left( \frac{M}{2} \right)}_+$}}
\psfrag{c4}{\footnotesize{$c^{\left( 1 \right)}_-$}}
\psfrag{c5}{\footnotesize{$c^{\left( 2 \right)}_-$}}
\psfrag{c6}{\footnotesize{$c^{\left( \frac{M}{2} \right)}_-$}}
\includegraphics[width=\columnwidth]{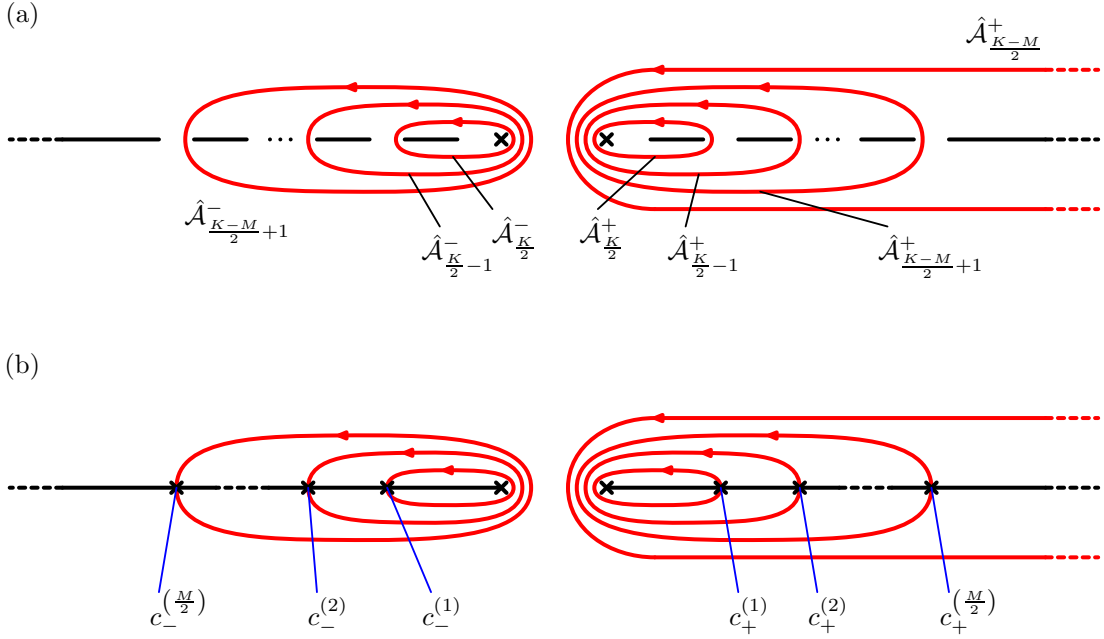}%
\caption{(a) The A-cycles on $\tilde{\Sigma}_2$ with the double poles inside them. (b) As $\rho \to \infty$, the A-cycles become pinched at the points $c^{(j)}_\pm$, for $j = 1, \ldots, M/2$.}
\label{fig:S2pinch}%
\end{figure}

We now focus our attention on the A-cycle conditions on $\tilde{\Sigma}_2$. First of all, we notice that, due to the fact that the double poles at $x = \pm 1$ have vanishing residues, $b \to 1$ does not imply that the contours are pinched at these singularities. In fact, even before the limit $\rho \to \infty$ is taken, we are free to rearrange the cycles so that they cross the real axis along the interval $-1 < x < 1$, where clearly there can be no pinching (see Fig. \ref{fig:S2pinch}(a)). It then follows that, even in the $\rho \to \infty$ limit, the contours $\hat{\mathcal{A}}_I^\pm$ do not touch any of the singularities of the differential. Therefore, the only diverging contribution the integrals receive comes from the factor $1/ \sqrt{1 - b^2}$ inside the integrand. For them to vanish, a second infinite contribution must arise in order to compensate.

\begin{figure}
\centering
\psfrag{a}{\footnotesize{$\Sigma$}}
\psfrag{b}{\footnotesize{$\tilde{\Sigma}_1$}}
\psfrag{c}{\footnotesize{$\tilde{\Sigma}_2$}}
\psfrag{d}{\footnotesize{contact points}}
\includegraphics[width=\columnwidth]{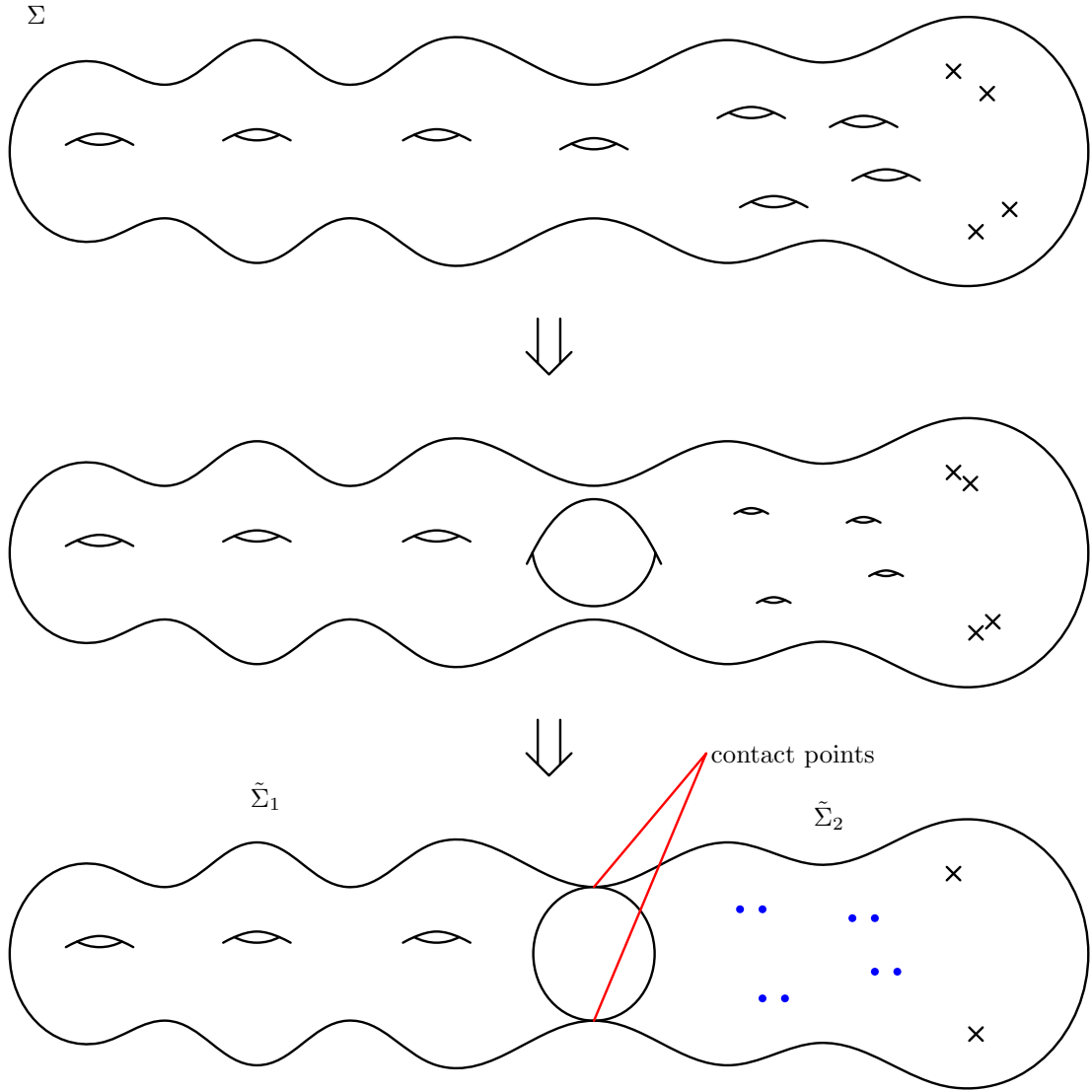}%
\caption{The complete description of the factorisation $\Sigma \to \tilde{\Sigma}_1 \cup \tilde{\Sigma}_2$, in the special case $K=9$, $M=4$. As $\rho \to \infty$, the handles on the second surface collapse and are replaced by pairs of simple poles, indicated as blue dots. Meanwhile, the double poles collide in pairs with two branch points, and thus we are left with two double poles overlapping with branch points in the final picture.}
\label{fig:exact_factor_Sigma}%
\end{figure}

This can only happen if all the branch points on $\tilde{\Sigma}_2$ coalesce in pairs as $\rho \to \infty$:
\begin{eqnarray}
 b^{(1)}_\pm , b^{(2)}_\pm & \to & c^{(1)}_\pm \nonumber\\
             & \vdots & \nonumber\\
 b^{(2j-1)}_\pm , b^{(2j)}_\pm & \to & c^{(j)}_\pm \nonumber\\
             & \vdots & \nonumber\\
 b^{(M-1)}_\pm , b^{(M)}_\pm & \to & c^{ \left( \frac{M}{2} \right) }_\pm \:.
\label{eq:bps_coalesce}
\end{eqnarray}
The differential then develops a simple pole at each collision site $c^{(j)}_\pm$ and all the A-cycles are pinched at one (and only one) of these poles, as shown in Fig. \ref{fig:S2pinch}(b). Correspondingly, the genus of $\tilde{\Sigma}_2$ reduces from $M$ to 0:
\begin{eqnarray}
 \tilde{y}_2 (x) & \to & \sqrt{x^2 - b^2} \prod_{j=1}^\frac{M}{2} ( x - c^{(j)}_+ ) ( x - c^{(j)}_- )  \qquad \textrm{as } \rho \to \infty
  \nonumber\\
                 & \equiv & \sqrt{x^2 - b^2} \: \hat{y}_2 (x) \:.
\label{eq:simplification_of_y2tilde_after_collisions}
\end{eqnarray}
We can visualise the effect of this process with the help of Figure \ref{fig:exact_factor_Sigma}. As the two surfaces separate from each other, the ``handles'' on $\tilde{\Sigma}_2$ (which lies on the right-hand side) collapse and each of them is replaced by a simple pole on each sheet, represented as a pair of blue dots in the picture.

This qualitative reasoning is already sufficient to determine the explicit form of $d \tilde{p}_2$. In particular, if we take into account the behaviour of all the branch points on this surface, we can write:
\begin{equation}
 d \tilde{p}_2 = \frac{h(x)}{\sqrt{x^2 - b^2}} dx + \ldots \qquad \textrm{as } \rho \to \infty
\label{eq:remove_sqrt(x^2-b^2)_from_dp2tilde_by_introducing_h(x)}
\end{equation}
where the dots denote terms which vanish in the limit considered\footnote{Strictly speaking, this is only true if corrections to \eqref{eq:bps_coalesce} are $O (\epsilon^\alpha)$ for some $\alpha > 0$, so that the logarithmically diverging factor $1/ \sqrt{1 - b^2}$ cannot generate $O(1)$ terms. It is possible to use the final explicit form of $d \tilde{p}_2$ in order to retrospectively check that this is the case. We discuss this in \ref{sec:behav_branch_pts_Sigma2tilde}.} and
\begin{equation}
 h(x) = - \frac{1}{\hat{y}_2 (x)} \left[ \frac{1}{\tilde{Q}} \sum_{l=0}^M \tilde{C}_l x^l + \frac{\pi J}{\sqrt{\lambda}} \frac{1}{\sqrt{1-b^2}}
  \left( \frac{\hat{y}_2 (1)}{x-1} - \frac{\hat{y}_2 (-1)}{x+1} \right) \right]
\label{eq:def_h(x)_for_limit_of_dp2tilde}
\end{equation}
is an analytic function which has simple poles at $x = \pm 1$ and $x = c^{(j)}_\pm$, for $j = 1, \ldots, M/2$. The limit of $h(x)$ as $x \to \infty$ and its residues at $x = \pm 1$ can be computed directly, while the residues at $x = c^{(j)}_\pm$ are determined by the B-period conditions in equation \eqref{eq:period_conditions_S2}. $h(x)$ can then be determined by analyticity constraints, yielding an explicit form for the differential of the quasi-momentum:
\begin{equation}
 d \tilde{p}_2 = dw_0 + d \hat{w} + \sum_{j=1}^\frac{M}{2} (dw_j^+ + dw_j^-)
\label{eq:explicit_dp2tilde_no_match_cond}
\end{equation}
where we have defined
\begin{eqnarray}
 dw_0 & = & - \frac{2 \pi}{\sqrt{\lambda}} \frac{J}{\sqrt{1-b^2}} \frac{dx}{(x^2-1)^\frac{3}{2}} \nonumber\\
 dw_j^\pm & = & \frac{1}{i} \frac{\sqrt{(c^{(j)}_\pm)^2 - 1}}{x-c^{(j)}_\pm} \frac{dx}{\sqrt{x^2-1}} \nonumber\\
 d \hat{w} & = & - \frac{K-M}{i} \frac{dx}{\sqrt{x^2-1}} \:.
\label{eq:def_dw0_dwhat_dwj^pm}
\end{eqnarray}
Finally, we can use this result in order to impose the matching condition up to $O(\rho^0)$ (see appendix \ref{sec:matching_condition}), which allows us to obtain the value of the only remaining unknown coefficient:
\begin{equation}
 \frac{2 \pi}{\sqrt{\lambda}} \frac{J}{\sqrt{1-b^2}} = - i (K-M) \log \rho - i \log (\tilde{q}_{K-M}) - \frac{1}{2i} \sum_{j=1}^\frac{M}{2} \left[ T(c^{(j)}_+) + T(c^{(j)}_-) \right] - R
\label{eq:match_cond_without_infinite_sums}
\end{equation}
where $R$ is an undetermined moduli-independent constant and
\begin{equation}
 T(c) = \log \left( \frac{c - \sqrt{c^2-1}}{c + \sqrt{c^2-1}} \right) \:.
\label{eq:def_T(c)}
\end{equation}
Now that $b$ has been eliminated by the matching condition, we observe that the only free parameters left in the final form of $d \tilde{p}_2$ are the $M$ positions of the poles, $c^{(j)}_\pm$, for $j = 1, \ldots, M/2$, which thus represent the  moduli on $\tilde{\Sigma}_2$. Adding these to the $K-M-1$ moduli from $\tilde{\Sigma}_1$, we obtain a $(K-1)$-dimensional moduli space of solutions.

As we can see in Fig. \ref{fig:exact_factor_Sigma}, the final configuration of $\tilde{\Sigma}_2$ after the factorisation is characterised by two singular contact points with $\tilde{\Sigma}_1$, located at $x = \infty^\pm$, and $M$ simple poles on each sheet (represented as blue dots). As the branch points approach $x = \pm 1$ the four double poles above these points collide in pairs and the resulting differential $dw_{0}$ exhibits two double poles which coincide with branch points at $x = \pm 1$.

\subsection{Semiclassical spectrum}

Now that we have the explicit form of the differential of the quasi-momentum on both surfaces, we can apply the asymptotic relations \eqref{eq:def_properties_of_dp_FG} to \eqref{eq:explicit_dp1tilde_Mn0} and \eqref{eq:explicit_dp2tilde_no_match_cond} respectively\footnote{We recall that the point $x=\infty$ on $\Sigma$ lies on $\tilde{\Sigma}_1$ after the factorisation, while $x=0$ on $\Sigma$ migrates to $\tilde{\Sigma}_2$.}, obtaining
\begin{equation}
 \Delta + S \simeq \frac{\sqrt{\lambda}}{2 \pi} \sqrt{- \tilde{q}_2} \,\, \rho
\label{eq:Delta+S_Mn0}
\end{equation}
and
\begin{equation}
 \Delta - S \simeq \frac{\sqrt{\lambda}}{2 \pi} \left[ (K-M) \log \rho + \log (\tilde{q}_{K-M}) 
  + \sum_{j=1}^\frac{M}{2} \left( G( c^{(j)}_+ ) + G( c^{(j)}_- ) \right) + \textrm{const.} \right]
\label{eq:Delta-S_Mn0}
\end{equation}
where the constant is moduli-independent and we have introduced
\begin{equation}
 G(c) \equiv \frac{1}{2} \log \left( \frac{c + \sqrt{c^2-1}}{c - \sqrt{c^2-1}} \right) - \frac{\sqrt{c^2-1}}{c} \:.
\label{eq:def_G(c)}
\end{equation}
We notice that $\Delta + S$ diverges faster than $\Delta - S$ and hence
\begin{equation}
 \Delta \simeq S \simeq \frac{\sqrt{\lambda}}{4 \pi} \sqrt{- \tilde{q}_2} \,\, \rho \:,
\label{eq:S_as_rho_to_infty}
\end{equation}
which then implies
\begin{multline}
 \Delta - S \simeq \frac{\sqrt{\lambda}}{2 \pi} \left[ (K-M) \log S + \log \left( \frac{\tilde{q}_{K-M}}{(- \tilde{q}_2)^\frac{K-M}{2}} \right)
  \phantom{\sum_{j=1}^\frac{M}{2}} \right. \\
 \left. + \sum_{j=1}^\frac{M}{2} \left( G( c^{(j)}_+ ) + G( c^{(j)}_- ) \right) + \textrm{const.} \right]
\label{eq:Delta-S_Mn0_final}
\end{multline}
where the constant is again independent of the moduli.

\paragraph{}

In order to complete the picture, we need to implement the semiclassical quantisation conditions, which require the filling fractions, defined in equation \eqref{eq:filling_fractions_FGA}, to be integers.

On $\tilde{\Sigma}_1$, we consider the filling fractions which are associated with the cuts $\mathcal{C}_I^\pm$, for $I = 1, \ldots, (K-M)/2 - 1$, and $\tilde{\mathcal{C}}_0$. After changing variables to $\tilde{x} = x/ \rho$ and approximating the integrand at leading order as $\rho \to \infty$, we impose
\begin{eqnarray}
 - \frac{1}{2 \pi i} \frac{S}{\sqrt{- \tilde{q}_2}} \oint_{\mathcal{A}_I^\pm} \tilde{x} \, d \tilde{p}_1 & = & l_I^\pm
  \qquad \textrm{for } I = 1, \ldots, \frac{K-M}{2} - 1 \nonumber\\
 - \frac{1}{2 \pi i} \frac{S}{\sqrt{- \tilde{q}_2}} \oint_{\tilde{\mathcal{A}}_0} \tilde{x} \, d \tilde{p}_1 & = & l_0
\label{eq:filling_fractions_S1}
\end{eqnarray}
with $l_I^\pm, l_0 \in \mathbb{Z}^+$, which leads to the discretisation of the moduli of that surface: $\tilde{q}_j = \tilde{q}_j (l_I^\pm, l_0)$, for $j = 2, \ldots, K-M-1$. In particular, note that the filling fraction $l_0$ is actually the sum of two separate contributions from the cuts $\mathcal{C}_{(K-M)/2}^+$ and $\mathcal{C}_{(K-M)/2}^-$ which collide at $\tilde{x}=0$ in the large $S$ limit: $l_0 = l_{(K-M)/2}^+ + l_{(K-M)/2}^-$, with
\begin{equation}
 - \frac{1}{2 \pi i} \frac{S}{\sqrt{- \tilde{q}_2}} \oint_{\mathcal{A}_{\frac{K-M}{2}}^\pm} \tilde{x} \, d \tilde{p}_1 = l_{\frac{K-M}{2}}^\pm
  \in \mathbb{Z}^+ \:.
\label{eq:filling_fractions_l_0^pm_S1}
\end{equation}
However, if we want to have as many filling fractions as moduli, we have to combine $l_{(K-M)/2}^+$ and $l_{(K-M)/2}^-$ into $l_0$.

On $\tilde{\Sigma}_2$, we redefine the remaining filling fractions by replacing the contours $\mathcal{A}_I^\pm$ with $\hat{\mathcal{A}}_I^\pm$, for $I = (K-M)/2 + 1, \ldots, K/2$. We may compute the relevant contour integrals at leading order by using the explicit expression for $d \tilde{p}_2$ \eqref{eq:explicit_dp2tilde_no_match_cond}\footnote{This step requires particular care: the contours are pinched at the poles $c^{(j)}_\pm$, and hence the filling fractions must be regulated. For this purpose, before we take $\rho \to \infty$, we convert the integral along $\hat{\mathcal{A}}_{K/2 - (j - 1)}^\pm$ into an open chain starting at $b^{(2j-1)}_\pm$ on one side of the cut, intersecting the real axis between the double poles, and ending at $b^{(2j-1)}_\pm$ on the other side of the cut, for $j = 1, \ldots, M/2$. We then write $b^{(2j-1)}_\pm = c^{(j)}_\pm \mp \eta^{(j)}_\pm$, according to \eqref{eq:bps_coalesce}, and use \eqref{eq:explicit_dp2tilde_no_match_cond} to compute the integral. Finally, the behaviour of the regulator $\eta^{(j)}_\pm$ can be determined by imposing, at leading order, the A-period condition involving the same contour $\hat{\mathcal{A}}_{K/2 - (j - 1)}^\pm$, again by turning the latter into an open chain and using \eqref{eq:explicit_dp2tilde_no_match_cond}. In fact, this last step is carried out in \eqref{eq:A-periods_on_Sigma2tilde_written_as_one-endpoint_integrals_Mn0} and \eqref{eq:lead_behaviour_of_b_j-branch_points_j_odd_Mn0}.}:
\begin{equation}
 \hat{\mathcal{S}}_j^\pm = - \frac{\sqrt{\lambda}}{4 \pi} \frac{1}{2 \pi i} \oint_{\hat{\mathcal{A}}^\pm_{\frac{K}{2}-(j-1)}}
  \left( x + \frac{1}{x} \right) dp = \mathcal{S} (c^{(j)}_\pm) \in \mathbb{Z}
\label{eq:filling_fractions_S2}
\end{equation}
for $j = 1, \ldots, M/2$, where
\begin{equation}
 \mathcal{S} (c) = \frac{\sqrt{\lambda}}{4 \pi^2} \left[ \sqrt{c^2-1} + \tan^{-1} \left( \frac{1}{\sqrt{c^2-1}} \right) \right] (K-M) \log \rho
  \:.
\label{eq:def_S(c)}
\end{equation}
We now analyse the relation between the filling fractions and the total angular momentum $S$ \eqref{eq:S_filling_fracts_FGA} and the level-matching condition \eqref{eq:level_matching_FGA}. Each term in the various sums involved is a contour integral, which we have already evaluated above at leading order. In particular, we have changed variables to $\tilde{x} = x / \rho$ for the filling fractions $\mathcal{S}_I$, for $I = 1, \ldots, (K-M)/2$, while we have rearranged the remaining filling fractions according to: $\hat{\mathcal{S}}_I^\pm = \sum_{J=I}^{K/2} \mathcal{S}_J^\pm$, for $I = (K-M)/2 + 1, \ldots, K/2$, without changing variables.

At this point, we notice that, in the large $S$ limit, the first set of contributions, corresponding to \eqref{eq:filling_fractions_S1}, diverges as $\mathcal{S}_I^\pm \sim S$, for $I = 1, \ldots, (K-M)/2$, while the second set of contributions \eqref{eq:filling_fractions_S2} diverges slower, $\hat{\mathcal{S}}_I^\pm \sim \log S$, for $I = (K-M)/2 + 1, \ldots, K/2$. Therefore, we may neglect the latter at leading order, and thus \eqref{eq:S_filling_fracts_FGA} yields:
\begin{equation}
 S = \sum_{I=1}^\frac{K-M}{2} \left( \mathcal{S}^{+}_{I} + \mathcal{S}^{-}_{I} \right) \:.
\label{eq:S_filling_fracts_S1}
\end{equation}
Similarly, \eqref{eq:level_matching_FGA} can be written as:
\begin{equation}
 \sum_{I=1}^\frac{K-M}{2} \left( I \mathcal{S}^{+}_{I} - I \mathcal{S}^{-}_{I} \right) = 0 \:,
\label{eq:level_matching_FGA_S1}
\end{equation}
where we have substituted $n_I^\pm = \pm I$ for the mode numbers. Hence, the total spin and the level-matching condition only involve the filling fractions associated with $\tilde{\Sigma}_1$ after the factorisation. In particular, notice that those filling fractions may only cancel at leading order, leaving an arbitrary subleading $O (\log S)$ contribution which would then allow the filling fractions on $\tilde{\Sigma}_2$ to violate the level-matching condition.

In conclusion, the semiclassical spectrum of this class of finite-gap solutions in the large $S$ limit,
\begin{equation}
 \Delta = \Delta [l_I^\pm, l_0, \mathcal{S}_j^\pm] \:,
\label{eq:semiclassical_spectrum_FGA}
\end{equation}
for $I = 1, \ldots, (K-M)/2 - 1$ and $j = 1, \ldots, M/2$, is controlled by two Riemann surfaces $\tilde{\Sigma}_1$ of genus $K-M-2$ and  $\tilde{\Sigma}_2$ of genus 0, respectively parametrised by the moduli $\tilde{q}_j$, $j = 2, \ldots, K-M$ and $c^{(k)}_\pm$, $k = 1, \ldots, M/2$. The spectrum is given by \eqref{eq:Delta-S_Mn0_final} as a function of the moduli, which are discretised by the quantisation conditions \eqref{eq:filling_fractions_S1} and \eqref{eq:filling_fractions_S2}, associated with the filling fractions on both surfaces.

As a final remark on the derivation, the case we have considered here is that of equal total numbers of branch cuts on the two halves of the real axis, and of equal numbers of cuts moving to $\tilde{\Sigma}_2$ from these two regions. We have restricted ourselves to this case for simplicity, but the above reasoning still applies when these restrictions are lifted. Therefore, the result generalises to
\begin{equation}
 \Delta - S \simeq \frac{\sqrt{\lambda}}{2 \pi} \left[ (K-M) \log S + \log \left( \frac{\tilde{q}_{K-M}}{(- \tilde{q}_2)^\frac{K-M}{2}} \right)
  + \sum_{j=1}^M G( c^{(j)} ) + \textrm{const.} \right]
\label{eq:Delta-S_FGA_general_K_M}
\end{equation}
with no parity restrictions on either $K$ or $M$ ($M = 0, 1, \ldots, K-2$). The quantisation conditions become
\begin{eqnarray}
 \mathcal{S}_I & = & - \frac{1}{2 \pi i} \frac{S}{\sqrt{- \tilde{q}_2}} \oint_{\mathcal{A}_I} \tilde{x} \, d \tilde{p}_1 = l_I
  \in \mathbb{Z}^+ \qquad \textrm{for } I = 1, \ldots, K-M-1 \nonumber\\
 \hat{\mathcal{S}}_j & = & \mathcal{S} (c^{(j)}) \in \mathbb{Z} \:, \qquad \textrm{for } j = 1, \ldots, M \:,
\label{eq:quantisation_conditions_FGA_general_K_M}
\end{eqnarray}
where, with this labelling, the cut $\mathcal{C}_1$ contains the origin and thus is the result of the collision of two separate cuts, $\mathcal{C}_1^+$ and $\mathcal{C}_1^-$, so that $\mathcal{S}_1 = \mathcal{S}_1^+ + \mathcal{S}_1^-$. Lastly, the total $AdS$ angular momentum is given by
\begin{equation}
 S = \sum_{I=1}^{K-M-1} l_I \:,
\label{eq:S_filling_fracts_S1_general_K_M}
\end{equation}
and the level-matching condition reads:
\begin{equation}
 \sum_{I=2}^{K-M-1} n_I l_I + n_1^+ l^+_1 + n_1^- l^-_1 \:,
\label{eq:level_matching_FGA_general_K_M}
\end{equation}
where the mode numbers are chosen according to the usual criterion (i.e. $n_I = \pm 1$ on the two outermost cuts and then, as we move inwards, the number increases in absolute value by one unit for each consecutive cut).

\subsection{Interpretation}
\label{sec:FGA_interpretation}

The interpretation of the first two terms in \eqref{eq:Delta-S_FGA_general_K_M}, which are associated with $\tilde{\Sigma}_1$, was given in \cite{Dorey:2008zy}. The main idea is that strings in this family should develop $K-M$ spikes, which approach the boundary of $AdS_3$ as $S$ increases towards infinity. Each section of the string containing a spike becomes infinitely long and hence yields an infinite contribution to $\Delta - S$, given by $(\sqrt{\lambda}/ 2\pi) \log S$ at leading order, while the $O (\sqrt{\lambda},S^0)$ corrections are represented by a function of $\tilde{q}_2$ and $\tilde{q}_{K-M}$. Spikes behaving in this way were thus called ``large'' spikes and they account for the leading branch of the spectrum, which corresponds to the surface $\tilde{\Sigma}_1$.

\begin{figure}
\centering
\includegraphics[width=100mm]{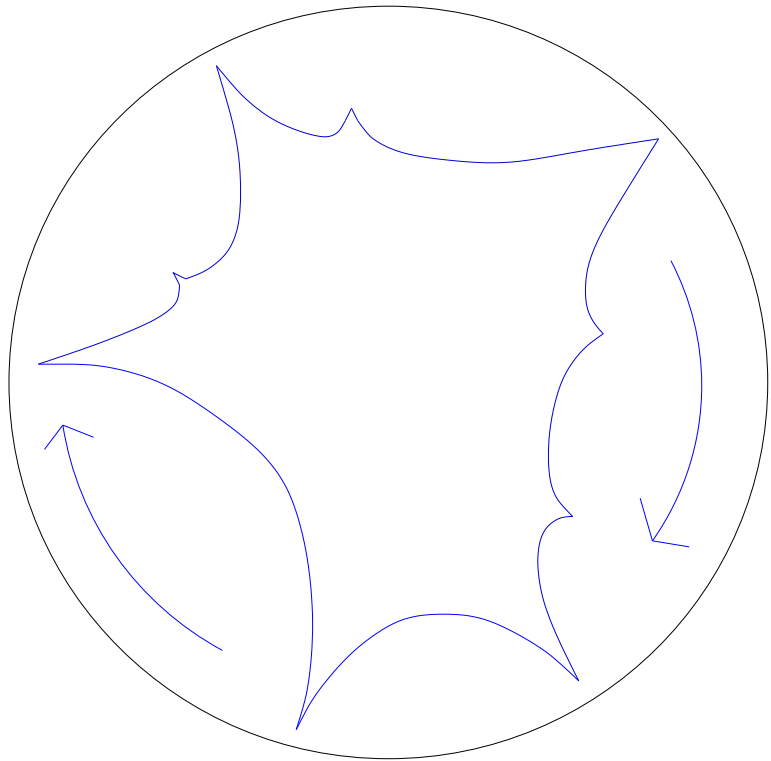}
\caption{``Small'' spikes propagating on a background of ``large'' spikes. The
  specific solution illustrated corresponds to the case $K=9$, $M=4$ 
and to the curve shown in Figure \ref{fig:exact_factor_Sigma}.}
\label{fig:large_small_spikes}
\end{figure}

The extended analysis of \cite{Dorey:2010id}, which we have discussed above, found the second branch of the spectrum represented by the third term in \eqref{eq:Delta-S_FGA_general_K_M} and corresponding to $\tilde{\Sigma}_2$. It consists of excitations of order $O(\sqrt{\lambda},S^0)$, each of which is associated with a simple pole in the differential of the quasi-momentum on $\tilde{\Sigma}_2$ located at $x = c^{(j)}$. The proposed interpretation is that such excitations should correspond to solitonic objects moving along the string with worldsheet velocity $v = 1/c$. They should appear as ``small'' spikes, in the sense that they do not stretch up to the boundary in the large $S$ limit, propagating on the background of the ``large'' spikes, as shown in Fig. \ref{fig:large_small_spikes}.

Moreover, both ``large'' and ``small'' spikes should correspond to solitons of the complex sinh-Gordon equation\footnote{This conjecture is based on the properties of some explicit solutions displaying this type of behaviour. We will discuss this point in the next chapter.}, which emerges from Pohlmeyer reduction on $AdS_3 \times S^1$, and should therefore undergo factorised scattering, also due to the fact that classical string theory on this background is integrable.

\paragraph{}

We are now going to make the quasi-particle interpretation of the ``small'' spikes more precise. From the spectrum \eqref{eq:Delta-S_FGA_general_K_M}, we immediately see that each solitonic excitation carries an amount of energy
\begin{equation}
 E_{\rm sol} (v) = \frac{\sqrt{\lambda}}{2 \pi} \left[ \frac{1}{2} \log \left( \frac{1 + \sqrt{1-v^2}}{1 - \sqrt{1-v^2}} \right) - \sqrt{1-v^2} \right] \:,
\label{eq:E(v)}
\end{equation}
where, due to the original ordering restrictions placed on the branch points of $\Sigma$, we have $|c^{(j)}| > 1$, $\forall j$, and hence $-1 < v < 1$. When $v = \pm 1$, the energy vanishes and thus the spike disappears: this corresponds to $|c| \to 1$, so that the associated simple pole on $\tilde{\Sigma}_2$ collides with one of the branch points (which are already coincident with the double poles).

In the opposite limiting case, $v \to 0$, the energy diverges and the simple pole at $x = c \to \infty$ leaves $\tilde{\Sigma}_2$ and migrates onto $\tilde{\Sigma}_1$. It is hard however to look at this phenomenon after the surface has factorised, since the contact points are singular. If we instead consider the surface $\Sigma$ before the limit $S \to \infty$ is taken, the process $v \to 0$ should make two more branch points move towards infinity as $\rho$ diverges, with the effect of transferring one of the moduli from $\tilde{\Sigma}_2$ to $\tilde{\Sigma}_1$, thereby increasing the genus of the latter by 1. Hence, the ``small'' spike becomes a ``large'' spike when its velocity vanishes. Consequently, ``large'' spikes should always correspond to stationary solitons on the worlsheet.

The conserved momentum carried by the excitations may be extracted from the quantisation conditions \eqref{eq:filling_fractions_S2}, but we first need a leading order estimate of the length of the string. Since we have no explicit expression for this class of finite-gap solutions to work with, we are going to rely on the results of \cite{Dorey:2008vp} and \cite{Dorey:2010iy}, which will be discussed in the next chapter. Here, it will be sufficient to say that several explicit solutions that show the behaviour we predicted above have a length that equals, at leading order, $N \log S$, where $N$ is the number of ``large'' spikes present. We will therefore assume that this is also the case for a generic finite-gap string in the family we are considering.

We now wish to interpret the constraint $\mathcal{S} (c) \in \mathbb{Z}$ as the Bohr-Sommerfeld quantisation condition for a particle of momentum $P(c)$ in a box of length $L \simeq (K-M) \log \rho$. Such a condition would read
\begin{equation}
 P(c) L / (2 \pi) \in \mathbb{Z} \:,
\label{eq:semiclassical_quant_condition_P(c)}
\end{equation}
leading us to the identification
\begin{equation}
 \frac{P_{\rm sol}(c) L}{2 \pi} = \mathcal{S} (c) \:.
\label{eq:identify_P(c)_and_S(c)}
\end{equation}
This allows us to introduce a conserved momentum associated with each excitation, which we now express as a function of the velocity $v$:
\begin{equation}
 P_{\rm sol}(v) = \frac{\sqrt{\lambda}}{2 \pi} \left[ \frac{\sqrt{1-v^2}}{v} - \mathrm{Tan}^{-1} \left( \frac{\sqrt{1-v^2}}{v} \right) \right]
\label{eq:P(v)}
\end{equation}
where $\mathrm{Tan}^{-1}$ is the principal branch of the inverse tangent and we have chosen the origin of the scale of momenta so that $P_{\rm sol}(\pm 1) = 0$, which is justified since, as we have just seen, $E_{\rm sol}(\pm 1) = 0$ and hence the excitations disappear for these extremal values of the velocity.

\paragraph{}

Another important point is the comparison of the string theory result \eqref{eq:Delta-S_FGA_general_K_M} with the semiclassical spectrum of the $SL (2, \mathbb{R})$ spin chain \eqref{eq:1-loop_gamma_sl(2)}. The main result of \cite{Dorey:2008zy} is the fact that, if $J-M = K-M$, i.e. $J = K$, and we identify the charges of the spin chain with the moduli on $\tilde{\Sigma}_1$ in the following way:
\begin{equation}
 \hat{q}_j^{(0)} = \frac{\tilde{q}_j}{(- \tilde{q}_2)^\frac{j}{2}} \:,
\label{eq:identification_charges_spin_chain_Sigma1}
\end{equation}
then the ``large'' hole branch of the gauge spectrum precisely matches the ``large'' spike branch of the string spectrum, modulo the usual replacement of the prefactor $\lambda/ 4 \pi^2$ with $\sqrt{\lambda} / 2\pi$, according to the leading behaviour of the cusp anomalous dimension \eqref{eq:cuspanomdim_weak/strong_coupling}, and up to the moduli-independent constant.

Assuming the finite-gap solution satisfies the highest-weight condition for both the left and the right current, with the identifications \eqref{eq:identification_charges_spin_chain_Sigma1} and if we rescale the spectral parameter as
\begin{equation}
 \tilde{x}' = \frac{\tilde{x}}{\sqrt{- \tilde{q}_2}} = \frac{\sqrt{\lambda}}{4 \pi S} x \:,
\label{eq:rescale_x_gauge-string}
\end{equation}
$\tilde{\Sigma}_1$ \eqref{eq:explicit_y1tilde}, expressed as a function of $\tilde{x}'$, is identical to the spectral curve of the spin chain \eqref{eq:spectral_curve_BGK}, provided we identify $\tilde{x}'$ with the gauge theory spectral parameter\footnote{Basically this follows from the fact that $\tau_0 (x)$ from the spin chain coincides with $\mathbb{P} (1 / \tilde{x}')$ from $\tilde{\Sigma}_1$ if identify $x$ with $\tilde{x}'$. The two curves \eqref{eq:spectral_curve_BGK} and \eqref{eq:explicit_y1tilde} look slightly different since they are written in different reference forms. In order to see that they match, one should compare \eqref{eq:spectral_curve_BGK} with $\sqrt{4 - \mathbb{P}^2 (1 / \tilde{x}')}$.} and we define
\begin{equation}
 d \tilde{p}_1 (\tilde{x}) = dp_{\rm gauge} (\tilde{x}') \:, \qquad \tilde{p}_1 (\tilde{x}) = p_{\rm gauge} (\tilde{x}') \:.
\label{eq:p,dp_identification_FGA_plus_sign}
\end{equation}
Note that we expect $\hat{q}^{(0)}_2 = -1$ for a highest-weight configuration of the spin chain, due to \eqref{eq:q_2_for_highest_weight}, and that this is correctly reproduced by \eqref{eq:identification_charges_spin_chain_Sigma1}. On the other hand, this also means that, while the gauge theory spectral curve only has $K-M-2$ moduli in the highest-weight case, the string theory curve maintains the full set of $K-M-1$ moduli, with the extra modulus $\tilde{q}_2$ essentially not being mapped to the gauge theory side due to the above rescaling.

\begin{figure}
\centering
\psfrag{a}{\footnotesize{(a)}}
\psfrag{b}{\footnotesize{(b)}}
\psfrag{a1}{\footnotesize{$\mathcal{B}_{2}^-$}}
\psfrag{a2}{\footnotesize{$\mathcal{B}_{1}^-$}}
\psfrag{a3}{\footnotesize{$\mathcal{B}_{1}^+$}}
\psfrag{b1}{\footnotesize{$\gamma_1^-$}}
\psfrag{b2}{\footnotesize{$\gamma_2^+$}}
\psfrag{b3}{\footnotesize{$\gamma_1^+$}}
\includegraphics{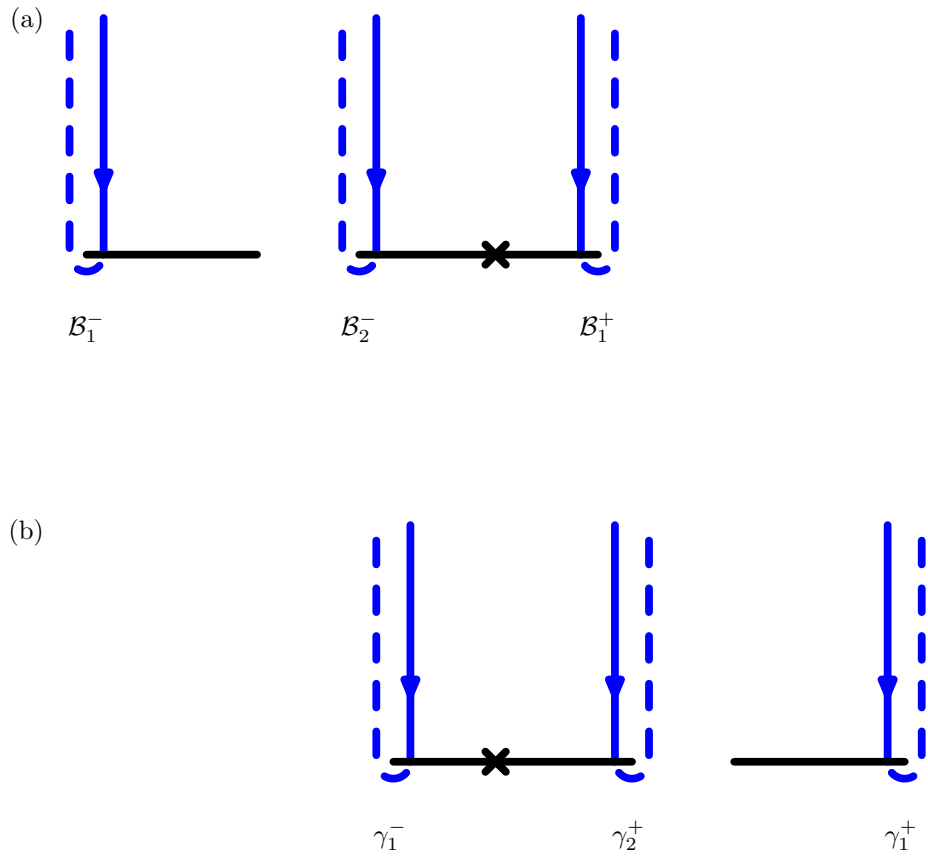}%
\caption{An example of the reflection occurring between $\tilde{\Sigma}_1$ and $\Gamma_{K-M}$ when a minus sign is introduced in the rescaling of the spectral parameter. (a) Cut configuration on $\tilde{\Sigma}_1$. (b) Corresponding cut configuration on $\Gamma_{K-M}$. The origin is marked on both curves.}
\label{fig:periods_reflection_gauge_string}%
\end{figure}

Moreover, the quantisation conditions \eqref{eq:filling_numbers_with_xdp_BGK} and \eqref{eq:quantisation_conditions_FGA_general_K_M} (upper equation only), with the associated relations to the total angular momentum \eqref{eq:tot_filling_number=S_BGK} and \eqref{eq:S_filling_fracts_S1_general_K_M} and level-matching constraints \eqref{eq:level_matching_BGK} and \eqref{eq:level_matching_FGA_general_K_M}, also coincide due to the above rescaling of the spectral parameter.
\paragraph{}
As far as the other branch of the spectrum is concerned, the remaining terms of \eqref{eq:1-loop_gamma_sl(2)} are certainly different from the last term of \eqref{eq:Delta-S_FGA_general_K_M}. However, we may compare the ``small'' spike dispersion relation $E_{\rm sol}(P_{\rm sol})$ from \eqref{eq:E(v)} and \eqref{eq:P(v)} to the ``small'' hole dispersion relation. The latter is obtained by identifying the momentum carried by each hole, interpreted as a quasi-particle, from the semiclassical quantisation condition, i.e. the Bethe-type equations. While these are hard to solve in general, in the special case $J-M = 2$ (i.e. two ``large'' holes), they take the form \eqref{eq:small_h_quantisation_cond_approx}:
\begin{equation}
 8 u_j \log S = 2 \pi k_j \:, \qquad k_j \in \mathbb{Z} \:.
\label{eq:small_h_quantisation_cond_approx_repeat}
\end{equation}
If we now assume that the holes are propagating along an object of length $2 \log S$, then the momentum is given by:
\begin{equation}
 P_{\rm hole} (u) = 4 u \:,
\label{eq:small_h_momentum_2_large_h}
\end{equation}
while the energy is given by \eqref{eq:energy_in_terms_of_holes}:
\begin{equation}
 E_{\rm hole} (u) = \psi \left( \frac{1}{2} + i u \right) + \psi \left( \frac{1}{2} - i u \right) - 2 \psi (1) \:.
\label{eq:small_h_energy}
\end{equation}
Then the dispersion relation $E_{\rm hole} (P_{\rm hole})$ has the following behaviour in the limit $P_{\rm hole} = 4 u \gg 1$:
\begin{equation}
 E_{\rm hole} = \frac{\lambda}{4\pi^{2}} \log |P_{\rm hole}| + O \left( P_{\rm hole}^0 \right) \:.
\label{eq:small_h_disprel_large_P}
\end{equation}
With the standard replacement of $\lambda/4\pi^{2}$ by the cusp anomalous dimension $\Gamma(\lambda)$, this precisely matches the large momentum ($P \gg 1$) form of the ``small'' spike dispersion relation $E_{\rm sol}(P_{\rm sol})$. It is therefore reasonable to conjecture that these solitonic excitations of the string worldsheet are dual to the gauge theory holes, and that the generic object that reduces to ``small'' holes at $\lambda \ll 1$ and to ``small'' spikes at $\lambda \gg 1$ should have the above large-momentum dispersion relation with an interpolating prefactor given by $\Gamma(\lambda)$.
\paragraph{}
Finally, it is important to mention that the gauge theory spectrum may also be recovered from the finite-gap spectrum through the rescaling
\begin{equation}
 \tilde{x}' = - \frac{\tilde{x}}{\sqrt{- \tilde{q}_2}} = - \frac{\sqrt{\lambda}}{4 \pi S} x
\label{eq:rescale_x_gauge-string_minus}
\end{equation}
and the corresponding identifications
\begin{eqnarray}
 \hat{q}_j^{(0)} & = & (-1)^j \frac{\tilde{q}_j}{(- \tilde{q}_2)^\frac{j}{2}} \nn\\
 d \tilde{p}_1 (\tilde{x}) & = & - dp_{\rm gauge} (\tilde{x}') \nn\\
 \tilde{p}_1 (\tilde{x}) & = & - p_{\rm gauge} (\tilde{x}') \:.
\label{eq:identifications_Gamma_Sigma1tilde_minus}
\end{eqnarray}
This introduces an extra $\log (-1)^{K-M}$ term in $\Delta - S$, which can however be absorbed into the definition of the moduli-independent constant. More importantly, the cut configurations of the string theory and gauge theory spectral curves are now related by a reflection with respect to the origin (and of course by the usual rescaling by $\sqrt{- \tilde{q}_2}$). For example, in order to reproduce a gauge curve $\Gamma_{K-M}$ with the central cut containing the origin plus a single additional cut on the positive real axis, with mode numbers
\begin{equation}
 \int_{\gamma_1^-} dp_{\rm gauge} = - 2\pi \:, \qquad \int_{\gamma_1^+} dp_{\rm gauge} = 2\pi \:,
  \qquad \int_{\gamma_2^+} dp_{\rm gauge} = 4\pi \:,
\label{eq:reflect_example_periods_gauge_curve}
\end{equation}
we need to define $\tilde{\Sigma}_1$ with the central cut and one extra cut on the negative real axis, with mode numbers
\begin{equation}
 \int_{\mathcal{B}_1^-} d \tilde{p}_1 = - 2 \pi \:, \qquad \int_{\mathcal{B}_2^-} d \tilde{p}_1 = - 4 \pi \:, \qquad
  \int_{\mathcal{B}_1^+} d \tilde{p}_1 = 2 \pi \:.
\label{eq:reflect_example_periods_string_curve}
\end{equation}
This is shown in Fig. \ref{fig:periods_reflection_gauge_string}.

The extra minus sign in the last two equations \eqref{eq:identifications_Gamma_Sigma1tilde_minus} has the effect of swapping the two sheets while mapping $\tilde{\Sigma}_1$ to $\Gamma_{K-M}$: this ensures that, despite the reflection, the filling fractions remain positive and that the B-periods are canonically defined (i.e. positive periods for cuts lying on the positive real axis) on both surfaces.

The sign change $x \to - x$, together with $dp \to - dp$, is equivalent to the parity transformation $\sigma \to - \sigma$: $(x,dp)$ corresponds to performing the finite-gap construction in the coordinates $(\tau,\sigma)$, while $(-x,-dp)$ is obtained by working with $(\tau,-\sigma)$. \eqref{eq:identifications_Gamma_Sigma1tilde_minus} may in fact be derived by comparing the results of the procedure in these two coordinate frames. Therefore, the rescaling \eqref{eq:rescale_x_gauge-string_minus} only amounts to choosing a different gauge in string theory, which has no physical implications on the finite-gap solutions considered.

Although the choice \eqref{eq:rescale_x_gauge-string} is the most intuitive in the framework of the previous analysis, we will see in the next chapter that the alternative \eqref{eq:rescale_x_gauge-string_minus} allows a smoother identification between the string theory and the gauge theory degrees of freedom.

\subsection{Non-highest-weight extension}
\label{sec:non-hw_extension}

Another point worth mentioning is the relationship between the highest-weight conditions on the two sides. 

On the string side there are two highest-weight conditions, associated with the right and left current. The former should be identified with the highest-weight condition \eqref{eq:highest_weight_condition_spin_chain} for the spin chain. The reason for this comes from the fact that, as we will see in section \ref{sec:eneral_large_spike_behaviour}, the string counterparts of the spin vectors $\mathcal{L}_k$ may be extracted from the large $S$ behaviour of the right current and hence its highest-weight constraint becomes equivalent to \eqref{eq:highest_weight_condition_spin_chain}. 


As some of the explicit string solutions which we will consider in chapter \ref{sec:explicit_solutions} do not satisfy one or both of the highest-weight conditions, it is necessary to study the corresponding modifications to the finite-gap spectrum. In fact, the only changes with respect to our previous analysis occur at the level of the asymptotics of the monodromy matrix \eqref{eq:Omega_near_infty} and \eqref{eq:gOmega(g^-1)_near_0}:
\begin{eqnarray}
 \Omega (x, \tau, \sigma) & = & 1 - \frac{1}{x} \frac{4 \pi}{\sqrt{\lambda}} Q_R + O \left( \frac{1}{x^2} \right) \:, 
  \qquad \textrm{as } x \to \infty \nn\\
 g (\tau, \sigma) \Omega (x, \tau, \sigma) g^{-1} (\tau, \sigma) & = & 1 - x \frac{4 \pi}{\sqrt{\lambda}} Q_L + O(x^2) \:,
  \qquad \textrm{as } x \to 0 \:,
\label{eq:non-hw-Omega_asymptotics}
\end{eqnarray}
where $Q_R$ and $Q_L$ are no longer diagonal. However, a non-highest-weight solution is always related to a highest-weight solution by an appropriate global $SU(1,1)$ rotation, so that the non-diagonal charges of the former may be expressed as the matrix conjugates of the diagonal charges associated with the latter. In particular, the transformation rules
\begin{eqnarray}
 g & \to & g' = U_R^{-1} \: g \: U_L \nn\\
 j & \to & j' = U_R \: j \: U_R^{-1} \nn \\
 l & \to & l' = U_L \: l \: U_L^{-1}
\label{eq:j_l_SU(1,1)_transf_rules}
\end{eqnarray}
imply
\begin{equation}
 Q_R = U_R^{-1} Q'_R U_R \qquad \textrm{and} \qquad Q_L = U_L^{-1} Q'_L U_L \:,
\label{eq:Q_R_L_SU(1,1)_transf_rules}
\end{equation}
where
\begin{equation}
 Q'_R = \frac{i}{2} (\Delta' + S') \sigma_3 \qquad \textrm{and} \qquad Q'_L = \frac{i}{2} (\Delta' - S') \sigma_3
\label{eq:Q_R_L_relation_Q'_R_L_non_hw}
\end{equation}
are the charges of the highest-weight solution. These are related to the charges of the non-highest-weight state by a rescaling
\begin{equation}
 \Delta + S = (\Delta' + S') (|\alpha|^2 + |\beta|^2) \qquad \Delta - S = (\Delta' - S') (|\gamma|^2 + |\delta|^2)
\label{eq:Delta_pm_S_non_hw_rescaling}
\end{equation}
which depends on the parameters of the rotations performed:
\begin{equation}
 U_R =
  \begin{pmatrix}
	 \alpha & \beta \\
	 \bar{\beta} & \bar{\alpha}
  \end{pmatrix}
   \qquad U_L =
    \begin{pmatrix}
     \gamma & \delta \\
     \bar{\delta} & \bar{\gamma}
    \end{pmatrix} \:,
\label{eq:SU(1,1)_rot_matrices_non_hw}
\end{equation}
with $|\alpha|^2 - |\beta|^2 = 1$ and $|\gamma|^2 - |\delta|^2 = 1$. We may therefore write
\begin{equation}
 Q_R = \frac{i}{2} (\Delta + S)
  \begin{pmatrix}
	 1 & \frac{2 \bar{\alpha} \beta}{|\alpha|^2 + |\beta|^2} \\
	 - \frac{2 \alpha \bar{\beta}}{|\alpha|^2 + |\beta|^2} & -1
	\end{pmatrix} \qquad
 Q_L = \frac{i}{2} (\Delta - S)
  \begin{pmatrix}
	 1 & \frac{2 \bar{\gamma} \delta}{|\gamma|^2 + |\delta|^2} \\
	 - \frac{2 \gamma \bar{\delta}}{|\gamma|^2 + |\delta|^2} & -1
	\end{pmatrix} \:.
\label{eq:Q_R_L_non-hw_in_terms_of_Delta-S}
\end{equation}
We can then immediately see that the eigenvalues of the monodromy matrix in the two asymptotic limits \eqref{eq:non-hw-Omega_asymptotics} are given by
\begin{eqnarray}
  p(x) & = & \frac{2 \pi}{\sqrt{\lambda}} (\Delta' + S') \frac{1}{x} + O \left( \frac{1}{x^2} \right) \:, \qquad \textrm{as } x \to \infty \nn\\
  p(x) & = & - \frac{2 \pi}{\sqrt{\lambda}} (\Delta' - S') x + O(x^2)  \:, \qquad \textrm{as } x \to 0 \:,
\label{eq:p(x)_asymptotics_non_hw}
\end{eqnarray}
so that \eqref{eq:Delta+S_Mn0} becomes
\begin{equation}
 \Delta + S \simeq \frac{\sqrt{\lambda}}{2 \pi} \sqrt{- \tilde{q}_2} (|\alpha|^2 + |\beta|^2) \rho \:,
\label{eq:Delta+S_non_hw}
\end{equation}
yielding the following result for the spectrum:
\begin{multline}
 \Delta - S \simeq \frac{\sqrt{\lambda}}{2 \pi} (1 + 2|\delta|^2) \left\{ (K-M) \log S + \log \tilde{q}_{K-M}
  + \sum_{j=1}^M G( c^{(j)} ) \right. \\ 
  \left. \phantom{\sum_{j=1}^M} - (K-M) \log \left[ \sqrt{- \tilde{q}_2} (1 + 2|\beta|^2) \right] + \textrm{const.} \right\} \:,
\label{eq:Delta-S_FGA_general_K_M_non_hw}
\end{multline}
where the moduli-independent constant is unchanged with respect to \eqref{eq:Delta-S_FGA_general_K_M}.

\paragraph{Case 1: subleading violation of the left highest-weight condition.} Suppose $Q_L^+$ is subleading with respect to $Q_L^0 \sim \Delta - S$. This requires
\begin{equation}
 \frac{|\bar{\gamma} \delta|}{|\gamma|^2 + |\delta|^2} \to 0 \quad \Leftrightarrow \quad | \delta | \to 0
\label{eq:Q_L^+_is_subleading}
\end{equation}
as $S \to \infty$. This implies that the extra factor appearing in \eqref{eq:Delta-S_FGA_general_K_M_non_hw} may be Taylor-expanded and thus any corrections to \eqref{eq:Delta-S_FGA_general_K_M_non_hw} will be additive, and not multiplicative, so that the leading behaviour of the spectrum is preserved. By \eqref{eq:Delta_pm_S_non_hw_rescaling}, we obtain
\begin{equation}
 \Delta' - S' = \frac{\Delta - S}{1 + 2 |\delta|^2} = (\Delta - S) [1 - 2 |\delta|^2 + O (|\delta|^4)] \:,
\label{eq:Delta'-S'_with_delta_to_0}
\end{equation}
which implies that all corrections vanish in the large $S$ limit if
\begin{equation}
 |\delta|^2 (\Delta - S) \to 0 \:, \qquad \textrm{as } S \to \infty \:.
\label{eq:no_corrections_condition_left_hw_violation}
\end{equation}
Note that the violation of the left highest-weight condition at leading order (i.e. $|\delta| \nrightarrow 0$) has potentially dramatic consequences on the spectrum, since it changes the leading logarithmic behaviour.

\paragraph{Case 2: subleading violation of the right highest-weight condition.} Similarly to the previous case, if $Q_R^+$ is subleading with respect to $Q_R^0 \sim \Delta + S$, we have
\begin{equation}
 \frac{|\bar{\alpha} \beta|}{|\alpha|^2 + |\beta|^2} \to 0 \quad \Leftrightarrow \quad | \beta | \to 0
\label{eq:Q_R^+_is_subleading}
\end{equation}
as $S \to \infty$. However, we are only interested in the logarithm of $\Delta + S$, and we find
\begin{equation}
 \log \Delta' + S' = \log \left( \frac{\Delta + S}{1 + 2 |\beta|^2} \right) = \log(\Delta + S) - \log [1 - 2 |\beta|^2 + O (|\beta|^4)] \:,
\label{eq:log_Delta'+S'_with_beta_to_0}
\end{equation}
so that
\begin{equation}
 \log \rho \simeq \log S + \log \left( \frac{4 \pi}{\sqrt{\lambda}} \right) - \log \sqrt{- \tilde{q}_2} - \log [1 - 2 |\beta|^2 + O (|\beta|^4)] \:,
\label{eq:log_rho_with_beta_to_0}
\end{equation}
in the large $S$ limit. Thus, the condition $|\beta| \to 0$ is sufficient to guarantee that all the corrections to \eqref{eq:Delta-S_FGA_general_K_M_non_hw} vanish. Hence, a subleading violation of the highest-weight condition for the right current does not modify the spectrum.

\paragraph{Case 3: violation of the right highest-weight condition at leading order.} Let us now consider the case $Q_R^+ \sim Q_R^0 \sim \Delta + S$. This implies
\begin{equation}
 \frac{|\bar{\alpha} \beta|}{|\alpha|^2 + |\beta|^2} = O(1) \quad \Leftrightarrow \quad | \beta | \nrightarrow 0 \:.
\label{eq:Q_R^+_is_leading}
\end{equation}
Note that this exhausts all the possible cases for the right current, since, as we can easily see from the above expression, $Q_R^+/(\Delta + S)$ may not diverge (i.e. the $+$ component may not dominate over the $0$ component of the right charge).

From \eqref{eq:Delta+S_non_hw} and \eqref{eq:Delta-S_FGA_general_K_M_non_hw}, we then obtain (assuming \eqref{eq:no_corrections_condition_left_hw_violation} is satisfied)
\begin{multline}
 \Delta - S \simeq \frac{\sqrt{\lambda}}{2 \pi} \left\{ (K-M) \log S + \log \left( 
  \frac{\tilde{q}_{K-M}}{(\sqrt{- \tilde{q}_2})^{K-M} (1 + 2|\beta|^2)^{K-M}} \right) \phantom{\sum_{j=1}^M} \right. \\
   + \left. \sum_{j=1}^M G( c^{(j)} ) + \textrm{const.} \right\} \:.
\label{eq:Delta+S_lead_violation_right_current_hw}
\end{multline}
The above spectrum, together with the spectral curve and the filling fractions, matches the corresponding gauge theory results if we identify
\begin{eqnarray}
 \hat{q}^{(0)}_j & = & \frac{(\pm 1)^j \tilde{q}_j}{(\sqrt{- \tilde{q}_2})^j (1 + 2 |\beta|^2)^j} \nn\\
 \tilde{x}' & = & \pm \frac{\tilde{x}}{\sqrt{- \tilde{q}_2} (1 + 2 |\beta|^2)} = \pm \frac{\sqrt{\lambda}}{4 \pi S} x \nn\\
 dp_{\rm gauge} (\tilde{x}') & = & \pm d \tilde{p}_1 (\tilde{x}) \nn\\
 p_{\rm gauge} (\tilde{x}') & = & \pm \tilde{p}_1 (\tilde{x}) \:.
\label{eq:identifications_Gamma_Sigma1tilde_non-hw_R}
\end{eqnarray}
Note that the relationship between the original spectral parameter $x$ and the final rescaled parameter $\tilde{x}'$ expressed in \eqref{eq:identifications_Gamma_Sigma1tilde_non-hw_R} is in fact the same as we found in the highest-weight case (see \eqref{eq:rescale_x_gauge-string} and \eqref{eq:rescale_x_gauge-string_minus}; only the relationship with $\tilde{x}$ is different).

Furthermore, we observe that in this way we have introduced an extra modulus $\hat{q}^{(0)}_2$ on the gauge side, which is determined by the parameter $\beta$ of the $SU(1,1)$ rotation,
\begin{equation}
 \hat{q}^{(0)}_2 = - \frac{1}{(1 + 2 |\beta|^2)^2} \:.
\label{eq:q2hat_extra_modulus}
\end{equation}
We also correctly find $\hat{q}^{(0)}_2 = - 1$ if $|\beta| \to 0$ as $S \to \infty$, i.e. when the violation of the highest-weight condition is negligible.

It is important to notice that, despite the similar notation, $\hat{q}^{(0)}_2$ and $\tilde{q}_2$ are not related to each other. $\tilde{q}_2$ is an extra modulus of the string theory finite-gap solution which is not mapped to the gauge side, regardless of whether the highest-weight condition for the right current is satisfied or not. $\hat{q}^{(0)}_2$ is instead the gauge theory equivalent of the rotation parameter $\beta$.
\paragraph{}
In summary, at the cost of these extra complications, the gauge theory semiclassical spectrum may also be reproduced from the finite-gap construction when the two highest-weight conditions are violated, although this must happen at subleading order for the left charge.

\chapter{Explicit solutions in $AdS_3$ in the large $S$ limit}
\label{sec:explicit_solutions}

This chapter is devoted to the analysis of some families of explicit solutions living in $AdS_3$ which exhibit the behaviour corresponding to the ``large'' and ``small'' spikes discussed in the previous chapter. Although these strings have no motion on $S^5$, they may still be compared to the finite-gap result, due to the fact that we are going to work in the limit of large $AdS_3$ spin $S$, with a constant angular momentum $J$ on $S^1$. Since $J$ remains finite, albeit large ($J = O( \sqrt{\lambda} ) \gg 1$), the motion along $S^1$ is negligible at leading order when $S$ becomes infinite, and thus the semiclassical spectra of strings in $AdS_3 \times S^1$ and in pure $AdS_3$ become degenerate in the limit considered.

\section{``Large'' spikes in $AdS_3$}

In this section, based on the results of \cite{Dorey:2008vp}, we will study the large $S$ behaviour of some well-known explicit solutions in $AdS_3$, namely the GKP string \cite{Gubser:2002tv} and the Kruczenski spiky string \cite{Kruczenski:2004wg}, together with an approximate family of solutions, which however becomes exact when $S$ diverges. All these objects have ``large'' spikes, which approach the boundary of $AdS_3$ in the large angular momentum limit, and their spectra reproduce the ``large'' spike contribution (associated with the surface $\tilde{\Sigma}_1$) to the finite-gap spectrum \eqref{eq:Delta-S_FGA_general_K_M}, thus providing evidence in support of the conjecture of \cite{Dorey:2008zy}. None of these solutions describes ``small'' spikes.

\subsection{General ``large'' spike behaviour in $AdS_3 \times S^1$}
\label{sec:eneral_large_spike_behaviour}

In this section, we will assume to be able to work in a gauge such that the $AdS_3$ global time $t$ coincides with the worldsheet time $\tau$. This will be the case with the simpler $N$-folded GKP solution, but not with the Kruczenski solution. We will deal with the ensuing complications in due course, but for now we will restrict ourselves to the simplest case in order to better clarify the key ideas involved.
\paragraph{}
A spike (or cusp) is defined as a discontinuity in the spacelike unit tangent vector to the string, which may in fact occur on a smooth worldsheet, provided that all the components of the tangent vector vanish at the cusp:
\begin{equation}
 \partial_\sigma \left. \rho (\tau, \sigma) \right|_{\sigma = \sigma_0} = 0 = \partial_\sigma \left. \phi (\tau, \sigma)
  \right|_{\sigma = \sigma_0} \:,
\label{eq:spike_condition_tau}
\end{equation}
for a spike located at $\sigma = \sigma_0 (\tau)$ on the worldsheet. This condition will always be satisfied in the gauge we will use, but it may be spoiled by a coordinate transformation on the worldsheet if it is singular at the cusp and it may therefore not hold for other gauge choices.

As we have already seen in section \ref{sec:FGA_interpretation}, the ``large'' spikes should approach the boundary of $AdS_3$ (i.e. their radial coordinate should diverge) in the limit $S \to \infty$, each yielding a contribution of $(\sqrt{\lambda}/ 2 \pi)$ $\log S$ to the quantity $\Delta - S$. In addition to postulating this, the conjecture of \cite{Dorey:2008zy} also proposes an explanation for the fact that the spectrum of a system with a continuum of degrees of freedom such as string theory coincides, as $S \to \infty$, with the spectrum of the gauge theory spin chain, which has a discrete set of degrees of freedom, given by the components $\mathcal{L}^A_k$ of the spin vector at each site. Note that here we are referring to the semiclassical spin chain which only involves the highly excited spins and is discussed in terms of equations \eqref{eq:semiclassical_spin_chain_monodromy_matrix}-\eqref{eq:q_2_for_highest_weight}.

The key idea is that, as $S$ becomes infinite, the charge density for $\Delta + S$, corresponding to the timelike component of the right current $j_\tau (\tau, \sigma)$, should become $\delta$-function localised at the spikes (for a normalised spacelike coordinate $\sigma$ such that $g(\sigma) = g (\sigma + 2 \pi)$):
\begin{equation}
 j^A_\tau (\tau,\sigma) \rightarrow  \frac{8\pi}{\sqrt{\lambda}} \sum_{m=0}^{K-1} L_m^A \delta(\sigma - \sigma_m) \:, \qquad \textrm{as }
  S \to \infty \:,
\label{eq:def_spin_vector_at_each_cusp}
\end{equation}
where $L_m^A$ represents the component on the generator $s^A$ of the $\mathfrak{su}(1,1)$-valued quantity $L_m$ and we have labelled the positions of the spikes as $\sigma_m$, with $m = 0, \ldots, K-1$ for a generic string with $K$ ``large'' spikes (and $M=0$ ``small'' spikes). We also expect the component $j_\sigma (\tau,\sigma)$ to be subleading with respect to $j_\tau (\tau, \sigma)$. Note that \eqref{eq:def_spin_vector_at_each_cusp}, together with the fact that
\begin{equation}
 \Delta + S = \frac{\sqrt{\lambda}}{4 \pi} \int_0^{2 \pi} d \sigma j_\tau^0 (\tau, \sigma) \sim 2 S \:, \qquad \textrm{as } S \to \infty \:,
\label{eq:Delta+S_sim_2S_as_Sinfty}
\end{equation}
implies that $j_\tau (\tau, \sigma) \sim S$.

The vectors $L_m$ represent a discrete set of degrees of freedom emerging from the string worldsheet at large $S$. They should be identified with the gauge theory spin vectors according to
\begin{equation}
 L^0_k \leftrightarrow \mathcal{L}^0_{k} \qquad\qquad
  L^\pm_k \leftrightarrow i \mathcal{L}^\pm_{k} \:,
\label{eq:L_spin_string_identification}
\end{equation}
for $k = 0, \ldots, K-1$. This identification is mainly supported by the fact that it allows to reproduce the spin chain spectral curve from the string side, as we will see shortly.
\paragraph{}
First of all, we notice that, for a highest-weight string solution, since $\Delta \sim S \to \infty$ in our limit, the following properties must be satisfied\footnote{Here we use a column vector notation for the components of the Lie algebra-valued quantity $L_m$ on the three generators $s^A$.}:
\begin{equation}
 \sum_{m=0}^{K-1} L_m = \begin{pmatrix}
	                         S \\
	                         0 \\
	                         0
                          \end{pmatrix} \:, \qquad \eta_{AB} L^A_m L^B_m = 0 \:,
\label{eq:properties_of_spin_vectors}
\end{equation}
for $m=0,\ldots,K-1$. The first equation, taking the normalisation of $L_m$ into account, is just the highest-weight condition \eqref{eq:Q_R_highest_weight} for the right charge, which, through \eqref{eq:L_spin_string_identification}, turns into the highest-weight condition \eqref{eq:highest_weight_condition_spin_chain} for the spin chain. The second equation instead reproduces the quadratic Casimir constraint \eqref{eq:zero_Casimir_for_classical_spins} and can be seen as a consequence of the Virasoro constraints \eqref{eq:Virasoro_j}, which, in $AdS_3 \times S^1$, read:
\begin{equation}
 -\frac{1}{2} \mathrm{Tr} \: j^2_\pm = \mathrm{det} \: j_\pm = \frac{J^2}{\sqrt{\lambda}} \:.
\label{eq:Virasoro_j_AdS3}
\end{equation}
If we now take the limit $S \to \infty$, due to the fact that we expect $j_\tau (\tau, \sigma) \sim S$ while $J$ and $\lambda$ are held fixed, we can approximate the right-hand side of the equation with zero, obtaining:
\begin{equation}
 \mathrm{Tr} \: [ j^2_\tau (\tau,\sigma_m) ] = 0 \:,
\label{eq:Virasoro_near_spikes_large_S}
\end{equation}
where we have used the fact that $j_\sigma$ is subleading with respect to $j_\tau$, or alternatively the spike condition \eqref{eq:spike_condition_tau}, which implies $j_\sigma (\tau, \sigma_m) = 0$.

By substituting in equation \eqref{eq:def_spin_vector_at_each_cusp}, decomposing $j_\tau$ onto the generators $s^A$ as in \eqref{eq:su11_vector_decomposition} and using the property \eqref{eq:su11_generators_trace}, we see that the only way in which this condition can hold is that the spin vectors satisfy the second equation \eqref{eq:properties_of_spin_vectors}.

We now proceed to compute the monodromy matrix of the string solution. For definiteness, we are going to choose a base point with $\sigma = 0$:
\begin{eqnarray}
 \Omega (x, \tau) & = & \mathcal{P} \mathrm{exp} \int_0^{2 \pi} \mathcal{J}_\sigma (x, \tau, \sigma') \: d \sigma' \nn\\
                  & = & \mathcal{P} \mathrm{exp} \left[ \frac{1}{2} \int_0^{2 \pi} \left( \frac{j_+ (x, \tau, \sigma')}{1-x}
                         - \frac{j_- (x, \tau, \sigma')}{1+x} \right) d \sigma' \right] \:.
\label{eq:Omega_0_2pi}
\end{eqnarray}
Since $j_\tau \sim S \to \infty$ in the limit considered, we have to rescale the spectral parameter as $x \sim S$ in order to keep the exponent finite. The approximate form of the monodromy matrix as $S \to \infty$ then becomes:
\begin{equation}
 \Omega(x,\tau) \simeq \mathcal{P} \mathrm{exp} \left[ - \frac{1}{x} \int_0^{2\pi} d\sigma j_\tau (\tau,\sigma) \right] \:,
\label{eq:approx_monodromy_matrix_large_S_1}
\end{equation}
where it is important to notice that the exponent is in general $O(1)$, while we have neglected vanishing corrections. This prevents us from expanding the path-ordered exponential as we did when we studied the large-$x$ asymptotics of the monodromy matrix (this is a large $S$ limit, not a large-$x$ limit). We can, however, replace $j_\tau$ by its limiting form \eqref{eq:def_spin_vector_at_each_cusp} as $S \to \infty$. The resulting sum of $\delta$-functions in the integrand converts the path-ordered exponential into a finite ordered product of exponentials: 
\begin{equation}
 \Omega(x,\tau) \simeq \prod_{m=0}^{K-1} \mathrm{exp} \left[ - \frac{4\pi}{\sqrt{\lambda}} \frac{1}{x} L^A_m s^B \eta_{AB} \right] \:,
\label{eq:approx_monodromy_matrix_large_S_2}
\end{equation}
where we have also expressed $L_m$ in terms of the $\mathfrak{su}(1,1)$ generators, according to \eqref{eq:su11_vector_decomposition}. We then observe that:
\begin{equation}
 (\eta_{AB} L^A_m s^B)^2 = \frac{1}{2} \mathbb{I} \: \eta_{AB} L^A_m L^B_m = 0
\label{eq:square_of_matrix_in_exp_vanishes_large_S_monodromy}
\end{equation}
as a consequence of the fact that the generators $s^A = (-i\sigma_3,\sigma_1,-\sigma_2)^A$ satisfy
\begin{equation}
 \{ s^A, s^B \} = 2 \eta^{AB}
\label{eq:anticommutator_of_our_su11_generators}
\end{equation}
and of the second property \eqref{eq:properties_of_spin_vectors} of the spin vectors. Therefore, the series expansion for the exponential in \eqref{eq:approx_monodromy_matrix_large_S_2} actually truncates at the linear term:
\begin{equation}
 \Omega(x,\tau) \simeq \prod_{m=0}^{K-1} \left[ \mathbb{I} + \frac{1}{u S} \eta_{AB} L^A_m s^B \right] \:,
\label{eq:approx_monodromy_matrix_large_S_3}
\end{equation}
where we have defined a rescaled spectral parameter $u = - x \sqrt{\lambda} / (4\pi S)$. Note that the rescaling coincides with \eqref{eq:rescale_x_gauge-string_minus}, so that $u \equiv \tilde{x}'$ should be identified with the rescaled gauge theory spectral parameter (indicated as $x$ in chapter \ref{sec:gauge_theory_spin_chain}).

Finally, we rewrite \eqref{eq:approx_monodromy_matrix_large_S_3} as
\begin{equation}
 \Omega(u) \simeq \frac{1}{u^K} \prod_{m=0}^{K-1} \mathbb{L}_m (u)
\label{eq:monodromy_matrix_large_S_L-product}
\end{equation}
where:
\begin{equation}
 \mathbb{L}_m (u) = \begin{pmatrix}
	                       u + i \frac{L^0_m}{S}    & \frac{L^+_m}{S} \\
	                       \frac{L^-_m}{S}          & u - i \frac{L^0_m}{S}
                        \end{pmatrix} \:.
\label{eq:L_m(u)}
\end{equation}
We now notice that $\mathbb{L}_m (u)$ and $\Omega (x, \tau)$, as defined just above, are respectively identical to the Lax matrix \eqref{eq:spin_chain_Lax_matrix} and to the monodromy matrix \eqref{eq:semiclassical_spin_chain_monodromy_matrix} of the semiclassical spin chain (up to an overall factor $u^{-K}$ and up to a rescaling by a factor of $S$), provided that $K=J$ (with $M=0$ in \eqref{eq:semiclassical_spin_chain_monodromy_matrix}, since there are no ``small'' spikes and thus no ``small'' holes) and that the spin vectors on the two sides are identified as in \eqref{eq:L_spin_string_identification}. In particular, by comparing with \eqref{eq:spin_chain_tr_Omega}, \eqref{eq:tau(x)_BGK} and \eqref{eq:tau0_BGK}, we see that the trace of $\Omega (u)$ yields precisely the rescaled monodromy $\tau_0 (x)$ of the semiclassical spin chain\footnote{Recall that $u$ in this chapter is equivalent to the $x$ of chapter \ref{sec:gauge_theory_spin_chain}, while the $u$ of chapter \ref{sec:gauge_theory_spin_chain} has no equivalent here.} (which is where the extra factor of $S$ becomes important). Therefore, by working at leading order in $S$, we can compute the gauge theory quasi-momentum $p(u)$:
\begin{equation}
 \mathrm{Tr} \: \Omega (u) = 2 \cos p(u) \:,
\label{eq:rel_Tr_Omega_p(u)}
\end{equation}
from which we obtain the semiclassical spectrum and the spectral curve.

Finally, as mentioned in \cite{Dorey:2008zy}, the Hamiltonian formalism for the Principal Chiral Model yields the following Poisson brackets for the spin vectors $L_k$ (which are in turn derived from the brackets for $j_\tau (\tau,\sigma)$):
\begin{equation}
 \{ L^A_j, L^B_k \} = - 2 \epsilon^{ABC} \delta_{jk} L_{C \, k} \:.
\label{eq:Poisson_brackets_L_k}
\end{equation}
These reproduce the corresponding brackets \eqref{eq:sl(2,R)_Poisson_brackets} for the gauge theory spin vectors\footnote{As explained in \cite{Dorey:2008zy}, the analysis of the Hamiltonian structure is actually rather complicated and requires several assumptions, but the previous discussion at least provides some evidence that the proposed identification \eqref{eq:L_spin_string_identification} may be correct. Further evidence comes from the consistent results obtained by the direct analysis of explicit string solutions, which will be the object of the next sections.}.

Hence, the identification \eqref{eq:L_spin_string_identification} allows to reproduce the degrees of freedom of the semiclassical spin chain and their behaviour on the string side in the large $AdS_3$ angular momentum limit.
\paragraph{}
In the next sections we will study three families of spiky strings in $AdS_3$ in the large $S$ limit. The strings have ``large'' spikes only and hence should correspond to finite-gap solutions with $M=0$. They all display the $\delta$-function localisation of the right current in the form \eqref{eq:def_spin_vector_at_each_cusp}, which we will exploit in order to compute the monodromy matrix of the spin chain, through the rescaling
\begin{equation}
 u = - \frac{\sqrt{\lambda}}{4 \pi S} x
\label{eq:rescaling_x_for_explicit_solutions} \:.
\end{equation}
From the monodromy matrix, we will obtain the quasi-momentum $p(u)$, the spectral curve $\Gamma_K$, including all its moduli $\hat{q}^{(0)}_k$, for $k = 2, \ldots, K$, and the semiclassical spectrum, including the filling fractions for the first two families. We will then compute the quantity $\Delta - S$ through elementary string theory techniques and compare it to the gauge theory spectrum \eqref{eq:1-loop_gamma_sl(2)}, finding agreement up to $O(\sqrt{\lambda}, S^0)$ and up to a moduli-independent constant. The latter should therefore coincide with the constant appearing in the finite-gap spectrum \eqref{eq:Delta-S_FGA_general_K_M} and its value is consistent for all the string solutions considered.

As a final observation, note that, ideally, we would want to compute the general monodromy matrix, quasi-momentum and spectral curve for the explicit solution and for any value of $S$. Then, we would be able to take the large $S$ limit, observe the factorisation and only at this point, by introducing the appropriate rescaling \eqref{eq:rescale_x_gauge-string_minus} and identifications, see that $\tilde{\Sigma}_1$ reproduces the dual gauge theory results (up to the cusp anomalous dimension and the extra moduli-independent constant appearing in the spectrum).

Unfortunately, the standard finite-gap computation method would require us to directly solve the auxiliary linear system \eqref{eq:aux_lin_sys}, which is a very hard task to accomplish. What we will instead do through the procedure we described above, is to introduce the rescaling \eqref{eq:rescale_x_gauge-string_minus} right from the start, in the form \eqref{eq:rescaling_x_for_explicit_solutions}, and then compute the monodromy matrix, quasi-momentum and spectral curve by approximating them in the large $S$ limit. This will effectively allow us to circumvent the computational problems associated with the general procedure, at the cost of restricting ourselves to studying $\tilde{\Sigma}_1$ only, and only after the final rescaling \eqref{eq:rescale_x_gauge-string_minus}.

Therefore, all the results will be obtained in the form which reproduces their gauge theory duals. In particular, the spectral curve will be given in the form
\begin{equation}
 \tilde{\Sigma}_1 \,\,: \qquad t + \frac{1}{t} = \mathbb{P}_K \left( \frac{1}{u} \right) = 
  2 + \frac{\hat{q}_2^{(0)}}{u^2} + \frac{\hat{q}_3^{(0)}}{u^3} + \ldots + \frac{\hat{q}_{K}^{(0)}}{u^{K}}
\label{eq:spectral_curve_explicit_sol-s}
\end{equation}
(recall that the eigenvalues of the monodromy matrix are $t_\pm = \exp (\pm i p(u))$). Similarly, the filling fractions, which are obtained from \eqref{eq:quantisation_conditions_FGA_general_K_M}, will reduce to the gauge theory expression \eqref{eq:filling_numbers_with_xdp_BGK}:
\begin{equation}
 l_I = - \frac{S}{2 \pi i} \oint_{\mathcal{A}_I} u \: dp \qquad \textrm{for } I = 1, \ldots, K-1 \:.
\label{eq:filling_fractions_explicit_sol-s}
\end{equation}
Moreover, the spectrum \eqref{eq:Delta-S_FGA_general_K_M} will become
\begin{equation}
 \Delta - S = \frac{\sqrt{\lambda}}{2 \pi} \left[ K \log S + \log \hat{q}^{(0)}_{K} + C_{\rm string} + \ldots \right]
  + O \left( (\sqrt{\lambda})^0 \right) \:,
\label{eq:general_Delta-S_finite-gap_explicit_sol-s}
\end{equation}
where the dots indicate vanishing subleading corrections and $C_{\rm string}$ is the moduli-independent constant (after absorbing the extra $\log(-1)^K$ due to the identification \eqref{eq:identifications_Gamma_Sigma1tilde_minus}). We will thus be unable to compute the extra string modulus $\tilde{q}_2$ through this procedure, which will prevent us from computing the complete expression for $\tilde{\Sigma}_1$.

Another feature of this technique is that, since we are bypassing the intermediate steps connected to $\tilde{x}$ by jumping directly to $\tilde{x}' \equiv u$, the rescaling \eqref{eq:rescaling_x_for_explicit_solutions} is the same both when the solution is a highest-weight state and when the highest-weight condition for the right charge is violated at leading order.

\subsection{The $N$-folded GKP string}
\label{sec:NGKP}

\subsubsection{General properties}

We will now discuss a generalisation of the string solution which was first presented in \cite{Gubser:2002tv}. We start with the following ansatz\footnote{In the following we reserve the notation $(\tau,\sigma)$ for worldsheet coordinates with periodicity $\sigma \to \sigma + 2 \pi$. These are related to the present worldsheet coordinates $(\tilde{\tau},\tilde{\sigma})$ by a rescaling we will describe below.}: $t = \tilde{\tau}$, $\rho = \rho(\tilde{\sigma})$, $\phi = \phi_0 + \omega \tilde{\tau}$. The corresponding solution to the equations of motion and Virasoro constraints is
\begin{equation}
 \rho(\tilde{\sigma}) = - i \: \mathrm{am} (i \tilde{\sigma}|\sqrt{1-\omega^2}) \:,
\label{eq:GKP_rho_of_sigmatilde}
\end{equation}
which lies in the interval $\rho(\tilde{\sigma}) \in [0,\rho_1]$, where $\coth{\rho_1} = \omega$ with $\omega > 1$. In the limit $\omega\to 1$, we have $\rho_1 \to + \infty$ and the upper bound is lifted. In order to study the shape of the solution, we consider
\begin{eqnarray}
  \coth\rho(\tilde{\sigma}) & = & \frac{\omega}{\mathrm{sn} \left( \omega\tilde{\sigma} \left| \frac{1}{\omega} \right. \right)} \:,
     \label{eq:coth_rho_GKP_preliminary}
\end{eqnarray}
which is periodic in $\tilde{\sigma}$ with period\footnote{In this chapter we will widely use the notation $\mathbb{K} (k)$ and $\mathbb{E} (k)$ for the elliptic integral of the first and second kind respectively. The shorthand notation $\mathbb{K}$ and $\mathbb{E}$ will instead represent specific values of these integrals for particular choices of the elliptic modulus $k$, which will be different for each family of string solutions discussed. See appendix \ref{sec:elliptic_integrals} for conventions associated with elliptic integrals and elliptic functions, such as the elliptic sine $\mathrm{sn} (z|k)$.} $(4/\omega)\mathbb{K}(1/\omega) \equiv 4\tilde{L}$, but does not have a definite sign. If we restrict ourselves to the first half-period $\tilde{\sigma} \in [0,2\tilde{L}]$, we see that $\rho$ increases from zero to its maximum value $\rho_1$ at $\tilde{\sigma}=\tilde{L}$ and then returns to zero at $\tilde{\sigma}=2\tilde{L}$. Since $\phi$ does not depend on $\tilde{\sigma}$, the corresponding snapshot of the string at constant global time $t$ is given by two identical overlapping straight line segments stretching out of the origin of $AdS_3$ in the radial direction.

\begin{figure}%
\begin{center}
\includegraphics{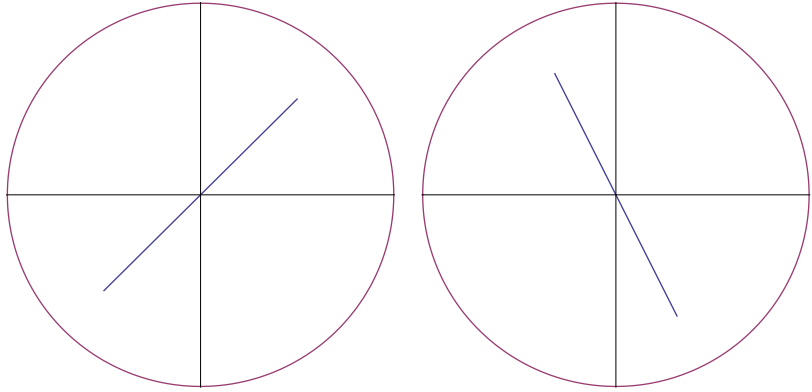}%
\end{center}
\caption{The GKP string appears as a straight line segment of finite length, rigidly rotating around its midpoint, which coincides with the centre of $AdS_3$. The spikes are located at the endpoints of the segment.}%
\label{fig:plot_GKP_finite}%
\end{figure}

We can then glue two more overlapping segments extending in the opposite direction by adding another half-period (with $\tilde{\sigma}$ shifted backwards by $2 \tilde{L}$ in order to keep $\rho \geq 0$) and shifting $\phi$ by $\pi$. This describes the original folded rotating GKP string which consists of two overlapping straight segments rigidly rotating around the centre of $AdS_3$ and has two spikes, located at $\tilde{\sigma} = \tilde{L}$ and $\tilde{\sigma} = 3 \tilde{L}$. The $N$-folded version has $N$ overlapping pairs of segments (two for each ``fold''), and is obtained by allowing $\tilde{\sigma}$ to range over $N$ periods\footnote{Here $[X]$ denotes the greatest integer less than $X$.}:
\begin{eqnarray}
 \rho & = & \rho \left( \tilde{\sigma}- l(\tilde{\sigma}) (2\tilde{L})
     \right) \:, \quad \textrm{where}\,\,\, l(\tilde{\sigma}) = \left[
     \frac{\tilde{\sigma}}{2\tilde{L}} \right]\nonumber\\
 \phi & = & \phi_1 + \omega \tilde{\tau} = \begin{cases}
              \phi_0 + \omega \tilde{\tau} & \text{if } l(\tilde{\sigma}) \textrm{ is even} \\
              \phi_0 + \omega \tilde{\tau}+\pi & \text{if } l(\tilde{\sigma}) \textrm{ is odd}
            \end{cases}
\label{eq:GKP_Nfolded_solution}
\end{eqnarray}
with $\tilde{\sigma}\in[0,4N \tilde{L}]$. The plot is identical in both cases (since additional folds overlap with the initial string) and is given in Fig. \ref{fig:plot_GKP_finite}.

We define:
\begin{equation}
 L = 4 N \tilde{L} = \frac{4 N}{\omega} \mathbb{K}
\label{eq:def_L_max_value_of_sigmatilde_GKP_Nfolded} \nonumber
\end{equation}
as the period of the coordinate $\tilde{\sigma}$, together with the shorthand notations
\begin{equation}
 \tilde{\sigma}_* = \tilde{\sigma}- l(\tilde{\sigma}) (2\tilde{L})
\label{eq:def_sigmatilde_star} \nonumber
\end{equation}
and
\begin{equation}
 \mathbb{E} \equiv \mathbb{E} \left( \frac{1}{\omega} \right) \qquad \mathbb{K} \equiv \mathbb{K} \left( \frac{1}{\omega} \right) \:,
\label{eq:shorthand_EK_GKP}
\end{equation}
which we will use throughout the rest of the discussion of the GKP solution.

The $N$-folded string has $K=2N$ cusps, that is two for each fold of the string, located at the tips $\rho = \rho_1$ of the line segment, which we can identify with the following worldsheet positions:
\begin{equation}
 \tilde{\sigma}_m = (2m+1) \tilde{L} \:, \quad m = 0,\ldots,K-1 \:.
\label{eq:cusp_positions_Nfolded_GKP_sigmatilde}
\end{equation}
One may check that the associated sinh-Gordon field \eqref{eq:def_alpha_Pohlmeyer} through Pohlmeyer reduction is in a $K$-soliton periodic configuration, where the worldsheet positions of the solitons coincide with those of the spikes.

As $\omega\to 1$, $\rho_1$ tends to $+\infty$, and hence the spikes touch the boundary of $AdS_3$ and the string becomes infinitely long. As we will see in section \ref{sec:small_spikes}, the solitons and the spikes are ``pushed away'' at infinity in this process, so that the exact solution of the equations of motion and Virasoro constraints at $\omega = 1$ corresponds to the sinh-Gordon vacuum \eqref{eq:def_sinhG-vacuum}. The next string solution we will consider, the symmetric spiky string, will also display this type of behaviour.

\subsubsection{Conserved charges}

The energy and the angular momentum of the solution \eqref{eq:GKP_Nfolded_solution} can straightforwardly be computed from \eqref{eq:AdS3xS1_charges}:
\begin{eqnarray}
 \Delta & = & 2N \frac{\sqrt{\lambda}}{2\pi} \int_0^{2 \tilde{L}} 
d\tilde{\sigma} \frac{1}{\mathrm{dn}^2 \left(\omega \tilde{\sigma}_* \left|
    \frac{1}{\omega} \right) \right.} = K \frac{\sqrt{\lambda}}{\pi} \frac{\omega}{\omega^2 - 1} \mathbb{E} \nn\\
 S & = & 2N \frac{\sqrt{\lambda}}{2\pi} \frac{1}{\omega} 
\int_0^{2 \tilde{L}} d\tilde{\sigma} \frac{\mathrm{sn}^2 \left(\omega \tilde{\sigma}_* \left|
    \frac{1}{\omega} \right) \right.}{\mathrm{dn}^2 \left(\omega
    \tilde{\sigma}_* \left| \frac{1}{\omega} \right) \right.} 
\nonumber \\
   & = & K \frac{\sqrt{\lambda}}{\pi} \left[ \frac{\omega^2}{\omega^2 - 1} \mathbb{E} - \mathbb{K} \right]
\label{eq:E_S_Nfolded_GKP}
\end{eqnarray}
As we could expect due to the periodicity of the integrands, the values are just N times the original GKP values, as they appear in \cite{Jevicki:2008mm}.

The large $S$ limit corresponds to $\omega\to 1$. Both $\Delta$ and $S$ diverge in this limit and, if we define
\begin{equation}
 \omega = 1 + \eta \:,
\label{eq:def_eta_omega_goes_to_1} \nonumber
\end{equation}
with $\eta > 0$, then their leading behaviour as $\eta \to 0$ is given by
\begin{equation}
 \Delta \simeq S = K \frac{\sqrt{\lambda}}{2 \pi} \frac{1}{\eta} + O ( \log \eta ) \:,
\label{eq:ES_lead_behaviour_omega_to_1_GKP}
\end{equation}
so that each spike contributes $\sqrt{\lambda}/(2\pi\eta)$ to the leading order term. By expanding up to $O(\eta^0)$, we obtain the spectrum
\begin{equation}
 \Delta - S = \frac{K\sqrt{\lambda}}{2\pi} \log \left( \frac{2\pi
 S}{K\sqrt{\lambda}} \right) + \frac{K\sqrt{\lambda}}{2\pi} (3\log 2-1) + O(\eta\log\eta) \:,
\label{eq:E_S_omega_to_1_behaviour}
\end{equation}
which exhibits the usual logarithmic growth with the angular momentum that is also found for the operators lying in the $\mathfrak{sl}(2)$ sector of Super Yang-Mills theory in the large conformal spin limit.

\subsubsection{Spectral curve for large $S$}
\label{sec:spectral_curve_NGKP}

We now proceed to compute the monodromy matrix and the dual gauge theory spectral curve for the $N$-folded string solution discussed above. As we saw in the previous section, our starting point is the time-like component of the right current $j$, which we are going to compute in the rescaled coordinate system $(\tau, \sigma)$,
\begin{equation}
 (\tau,\sigma) = \frac{2\pi}{L} (\tilde{\tau},\tilde{\sigma}) =  \frac{\pi\omega}{K \mathbb{K}}(\tilde{\tau},\tilde{\sigma}) \:,
\label{eq:def_rescaled_coords_Nfolded_GKP} \nonumber
\end{equation}
such that now the periodicity is the standard $\sigma \to \sigma + 2 \pi$. This is important in order to keep the period of the spatial worldsheet coordinate finite as $\omega \to 1$, which then allows us to see the expected $\delta$-function localisation. We find
\begin{eqnarray}
 j^0_\tau (\tau,\sigma) & = & \frac{2 K \mathbb{K}}{\pi\omega} \frac{\left[ 1 + \frac{1}{\omega} \mathrm{sn}^2 \left( \omega
        \sigma_* \left| \frac{1}{\omega} \right) \right] \right.}{\mathrm{dn}^2 \left( \omega \sigma_* \left| \frac{1}{\omega} \right) \right.}
         \nonumber \\
 j^1_\tau (\tau,\sigma) + i j^2_\tau (\tau,\sigma) & = & i
        \frac{2 K \mathbb{K}}{\pi} \frac{\omega + 1}{\omega^2}
        \frac{\mathrm{sn} 
\left(\omega \sigma_* \left| \frac{1}{\omega} \right) \right.}{\mathrm{dn}^2 \left( \omega \sigma_* \left| \frac{1}{\omega}
          \right) \right.} e^{i \left[ \phi_1 + (\omega - 1) \frac{K \mathbb{K}}{\pi\omega} \tau \right]} \:,
\label{eq:j_tau_in_terms_of_tau_sigma_Nfolded_GKP}
\end{eqnarray}
where $\sigma_*$ is just $\tilde{\sigma}_*$ written as a function of $\sigma$. In the new coordinates, the spikes are located at
\begin{equation}
 \sigma_m = (2m+1) \frac{\pi}{K} \:, \qquad m=0,\ldots,K-1 \:.
\label{eq:cusp_positions_Nfolded_GKP_sigma}
\end{equation}
\paragraph{}
The next step is to take the limit $\omega\rightarrow 1$ so that the angular momentum $S$ of the solution diverges. As explained earlier, the key point here is that, as $S \to \infty$, the charge density is dominated by the vicinity of the cusp points $\sigma = \sigma_{m}$. To demonstrate this we expand around the $m$-th cusp point setting:
\begin{equation}
 \sigma = \sigma_m + \hat{\sigma} \:, \qquad \textrm{with } |\hat{\sigma}| < \frac{\pi}{K}
\label{eq:def_sigma_near_cusps_Nfolded_GKP}
\end{equation}
which is equivalent to  $\sigma_* = \tilde{L} + \hat{\sigma}
K \mathbb{K}/ (\pi\omega)$ for each $m$. 
We can then use the quarter-period transformation formulae for the elliptic functions appearing in \eqref{eq:j_tau_in_terms_of_tau_sigma_Nfolded_GKP} to get:
\begin{eqnarray}
 \mathrm{sn} \left( \omega \sigma_* \left| \frac{1}{\omega} \right)
 \right. & = & \mathrm{sn} \left( \mathbb{K} + 
\frac{K \mathbb{K}}{\pi}
  \hat{\sigma} \left| \frac{1}{\omega} \right) \right. = 
\mathrm{cd} \left( \frac{K \mathbb{K}}{\pi} \hat{\sigma} \left| \frac{1}{\omega} \right)
   \right. \nonumber\\
 \mathrm{dn} \left( \omega \sigma_* \left| \frac{1}{\omega} \right)
  \right. & = & \mathrm{dn} \left( \mathbb{K} + \frac{K \mathbb{K}}
{\pi}
  \hat{\sigma} \left| \frac{1}{\omega} \right) \right. = \sqrt{1
 -\frac{1}{\omega^2}} \; \mathrm{nd} \left( \frac{K \mathbb{K}}{\pi} 
\hat{\sigma}
   \left| \frac{1}{\omega} \right) \right.
\label{eq:half_period_transf_for_sn_dn_Nfolded_GKP}
\end{eqnarray}
As we are interested in the limit $\omega \to 1$, we can 
use the standard series expansions for $\mathrm{cd}(z|k)$ and
$\mathrm{nd}(z|k)$ in powers of $(k-1)$ (see for instance \cite{Byrd::1971}), with   
\begin{equation}
 z = \frac{K \mathbb{K}}{\pi} \hat{\sigma}
\label{eq:z_elliptic_fn_series_GKP}
\end{equation}and $k=1/\omega$. Note however that 
$\mathbb{K}\simeq - (1/2) \ln (\omega-1)$ for 
$\omega\rightarrow 1$. This implies we must consider a limit where 
not only $k\to 1$, but also $z\to\pm\infty$, depending on the sign of
$\hat{\sigma}$. However, since $|z|=K \mathbb{K}|\hat{\sigma}|/\pi <
\mathbb{K}$, i.e. $z$ is always within the first quarter-period in
both directions, one can check that higher order terms remain
suppressed. We thus consider only the lowest order terms in these series,
which give the leading behaviour of 
$j_{\tau}$ near each spike as:
\begin{eqnarray}
 j_\tau^0(\tau,\sigma) & \simeq & \frac{2 K \mathbb{K}}{\pi}\frac{1}{\eta \cosh^2 \left( \frac{K \mathbb{K}}{\pi} \hat{\sigma} \right)} \nonumber
  \\
 j_\tau^1(\tau,\sigma) + i j_\tau^2(\tau,\sigma) & \simeq & (-1)^m i \frac{2 K \mathbb{K}}{\pi} \frac{1}{\eta \cosh^2 \left(
  \frac{K \mathbb{K}}{\pi} \hat{\sigma} \right)}
\label{eq:leading_behaviour_for_cpts_of_j_tau_GKP_Nfolded}
\end{eqnarray}
Here the factor $(-1)^m$ comes from the extra $\pi$ which is added to $\phi$ every other period, as specified in \eqref{eq:GKP_Nfolded_solution}, and consequently affects the contribution of every other cusp.

If we now use the identity
\begin{equation}
 \lim_{\epsilon\to 0} \frac{1}{2 \epsilon} \frac{1}{\cosh^2 \left( \frac{x}{\epsilon} \right)} = \delta(x)
\label{eq:identity_for_delta_function}
\end{equation}
and eliminate $\hat{\sigma}$ in favour of $\sigma$ according to \eqref{eq:def_sigma_near_cusps_Nfolded_GKP}, we find the $\delta$-function localisation of the right charge density exactly as defined in \eqref{eq:def_spin_vector_at_each_cusp}, with
\begin{equation}
 L_m = \frac{S}{K} \begin{pmatrix}
	                        1 \\
	                        (-1)^{m+1} \\
	                        (-1)^m
                         \end{pmatrix} \qquad m=0,\ldots,K-1 \:.
\label{eq:spin_vector_at_mth_cusp_Nfolded_GKP}
\end{equation}
It is now easy to check that the spin vectors satisfy the required properties \eqref{eq:properties_of_spin_vectors}. This also implies that the highest-weight condition for the right charge is satisfied at leading order. There may be non-vanishing subleading corrections, but, as discussed in section \ref{sec:non-hw_extension}, these have no effect on the leading semiclassical spectrum, up to the order $O(\sqrt{\lambda},S^0)$. By similarly approximating the left current $l (\tau, \sigma)$, it is also possible to check that $Q_L^+$ vanishes as $\eta \to 0$, so that the GKP solution is a highest-weight state with respect to the left charge as well.
\paragraph{}
We then proceed to calculate the Lax matrices from \eqref{eq:L_m(u)},
\begin{equation}
 \mathbb{L}_m (u) = \begin{pmatrix}
	                       u + \frac{i}{K}                    & (-1)^m \frac{i}{K} \\
	                       -(-1)^m \frac{i}{K}                & u - \frac{i}{K}
                        \end{pmatrix} \:,
\label{eq:L_m(u)_for_Nfolded_GKP}
\end{equation}
and the monodromy matrix from \eqref{eq:monodromy_matrix_large_S_L-product},
\begin{equation}
 \Omega(u) \simeq \frac{1}{u^K} [\mathbb{L} (u)]^{\frac{K}{2}} \:,
\label{eq:monodromy_matrix_for_Nfolded_GKP}
\end{equation}
where we have exploited the fact that $\mathbb{L}_m (u)$ only depends on the parity of $m$ and we have defined $\mathbb{L} (u) = \mathbb{L}_0 (u) \mathbb{L}_1 (u)$.

\begin{figure}
\centering
\psfrag{a}{\footnotesize{$u=\frac{1}{K}$}}
\psfrag{b}{\footnotesize{$u=-\frac{1}{K}$}}
\psfrag{c}{\footnotesize{$0$}}
\includegraphics[width=\columnwidth]{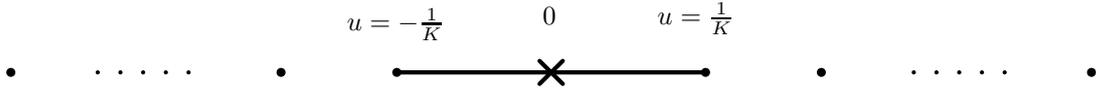}%
\caption{The cut and branch points configuration on the surface $\tilde{\Sigma}_1$ for the $N$-folded GKP string. The distribution of the branch points is symmetric with respect to the origin.}
\label{fig:bps_and_cuts_NGKP}%
\end{figure}

It follows that the eigenvalues of $\Omega$ can be expressed in terms of the eigenvalues of the matrix $\mathbb{L} (u)$, which can be
evaluated explicitly as:
\begin{equation}
 \kappa_{\pm} = u^2 - \frac{2}{K^2} \pm 2 \sqrt{\frac{1}{K^4} - \frac{u^2}{K^2}} \:.
\label{eq:evalues_of_L(utilde)_Nfolded_GKP}
\end{equation}
Finally we can write the trace of the monodromy matrix as
\begin{eqnarray}
 \mathrm{Tr} \: \Omega (u) & = & \frac{\kappa_+^\frac{K}{2}+\kappa_-^\frac{K}{2}}{u^K} \nonumber \\
                           & = & 2 T_K \left( \sqrt{1 - \frac{1}{K^2 u^2}} \right) = 
                                  2 \cos \left[ K \sin^{-1} \left( \frac{1}{K u} \right) \right] \:,
\label{eq:tr_Omega_Nfolded_GKP}
\end{eqnarray}
where $\mathrm{T}_k (y)$ is the Chebyshev polynomial of the first kind:
\begin{equation}
 T_k (y) = \frac{1}{2} \left[ \left( y + i \sqrt{1-y^2} \right)^k + \left( y - i \sqrt{1-y^2} \right)^k \right] = \cos (k \arccos y) \:.
\label{eq:def_Tk(y)_Chebyshev}
\end{equation}
This is a polynomial in $y$ of degree $k$. The above result yields the quasi-momentum through \eqref{eq:rel_Tr_Omega_p(u)}:
\begin{equation}
 p(u) = K \sin^{-1} \left( \frac{1}{K u} \right) \:.
\label{eq:quasi-momentum_N-folded_GKP}
\end{equation}
A similar result was derived using different methods in \cite{Casteill:2007ct}.
\paragraph{}
We may rewrite the string theory spectral curve as follows:
\begin{equation}
 \tilde{\Sigma}_1 \,\,: \qquad t + \frac{1}{t} = \mathbb{P}_K \left( \frac{1}{u} \right) = 
  2 \cos \left[ K \sin^{-1} \left( \frac{1}{K u} \right)\right] \:.
\label{eq:spectral_curve_NGKP}
\end{equation}
As anticipated at the end of section \ref{sec:eneral_large_spike_behaviour}, the above expression is already in the appropriate form \eqref{eq:spectral_curve_explicit_sol-s} for comparision with the gauge theory spectral curve $\Gamma_K$. We then find that the limiting string theory curve for the $N$-folded GKP solution corresponds to a particular point in the moduli space of the gauge theory curve where the conserved charges $\hat{q}^{(0)}_{k}$ take the particular values:
\begin{equation}
 \hat{q}^{(0)}_k = \left( \frac{2}{K} \right)^k \sum_{1\leq j_1 < j_2 < \ldots < j_k \leq K} \prod_{l=1}^k \sin \left[ \frac{\pi}{2}
  (j_{l+1} - j_l) \right] \:, \quad k=2, 4, \ldots, K \:,
\label{eq:qktilde_from_spin_chain_N-folded_GKP}
\end{equation}
where $j_{k+1} \equiv j_1$ and we notice that $\hat{q}^{(0)}_k = 0$ for odd $k$ (the normalisation $\hat{q}^{(0)}_2 = - 1$ is checked in appendix \ref{sec:app_computing_qtilde_2_for_N-folded_GKP_and_K}).
\paragraph{}
To characterise more precisely the spectral curve associated with this string solution it is useful to determine the pattern of the branch points, which coincide with the simple zeros of the discriminant \eqref{eq:def_discriminant} $D = 4 \sin^2 p(u)$:
\begin{equation}
 D \left(u \right) = 4 \left( \frac{4}{K^2 u^2} - \frac{4}{K^4 u^4} \right) U_{\frac{K}{2}-1}^2 \left( 1 - \frac{2}{K^2 u^2} \right) \:,
\label{eq:discriminant_NGKP}
\end{equation}
where $U_k (y)$ is the Chebyshev polynomial of the second kind:
\begin{eqnarray}
 U_k (y) & = & \frac{1}{2i\sqrt{1-y^2}} \left[ \left( y + i \sqrt{1-y^2} \right)^{k+1} - \left( y - i \sqrt{1-y^2} \right)^{k+1} \right]
  \nonumber \\
         & = & \frac{\sin[(k+1)\arccos y]}{\sin \arccos y} \:.
\label{eq:def_Uk(y)_Chebyshev}
\end{eqnarray}
The zeros of the discriminant can then be determined as:
\begin{equation}
 \begin{array}{ll}
  u = \pm \frac{1}{K}  & \qquad \textrm{simple} \\
  u = \infty           & \qquad \textrm{double} \\
  u = \pm \frac{1}{K \sin \left( \frac{k \pi}{K} \right)} \:, \qquad k=1,\ldots,\frac{K}{2}-1 & \qquad
   \textrm{double} \:.
 \end{array}
\label{eq:pattern_of_zeros_of_Q_n_for_Nfolded_GKP}
\end{equation}
The double zero at infinity is a common feature of spectral curves of the form \eqref{eq:spectral_curve_explicit_sol-s}, while the presence of the remaining $K-2$ double zeros together with the two branch points indicate that the curve has degenerated to genus zero. In particular, all of the outer cuts have collapsed into double points and the only cut left is the one extending from $u = -1/K$ to $u = 1/K$ along the real axis, touching the origin (see Fig. \ref{fig:bps_and_cuts_NGKP}). Accordingly, the quasi-momentum has a logarithmic branch point at $u = 0$ and two square root branch points at $u = \pm 1/K$ and is analytic everywhere else. Correspondingly, we have
\begin{equation}
 dp(u) = - \frac{d u}{ u \sqrt{u^2 - \frac{1}{K^2}}} \:,
\label{eq:dp(x)_N-folded_GKP}
\end{equation}
which displays a simple pole at $u=0$ and two square root branch points at $u = \pm 1/K$ and is made single-valued by the same cut as we just introduced for $p(u)$.

We also calculate the filling fraction associated with the cut: this is given by \eqref{eq:filling_fractions_explicit_sol-s}. As we saw earlier, the filling fraction corresponding to the cut touching the origin should be split into separate contributions from the two parts of it lying at the sides of the origin, which coincide with the intervals $[-1/K,0]$ and $[0,1/K]$:
\begin{equation}
 l_1^- = \frac{S}{\pi i} \int_{- \frac{1}{K}}^{0} \frac{du}{\sqrt{u^2 - \frac{1}{K^2}}} = \frac{S}{2} \qquad
  l_1^+ = \frac{S}{\pi i} \int_{0}^{\frac{1}{K}} \frac{du}{\sqrt{u^2 - \frac{1}{K^2}}} = \frac{S}{2} \:.
\label{eq:filling_fractions_NGKP}
\end{equation}
We notice that $l_1^+ + l_1^- = S$ and that, since the double points are symmetric with respect to the origin, the mode numbers associated with the two halves of the cut are $n_1^\pm = \pm K/2$, which implies $n_1^+ l_1^+ + n_1^- l_1^- = 0$. Hence, the properties \eqref{eq:S_filling_fracts_S1_general_K_M} and \eqref{eq:level_matching_FGA_general_K_M} are satisfied.
\paragraph{}
Finally, in order to compute the finite-gap spectrum \eqref{eq:general_Delta-S_finite-gap_explicit_sol-s}, we obtain the highest conserved charge 
$\hat{q}^{(0)}_K$ from (\ref{eq:qktilde_from_spin_chain_N-folded_GKP}),
\begin{equation}
 \hat{q}^{(0)}_K = (-1)^\frac{K}{2} \left( \frac{2}{K} \right)^K \:,
\label{eq:highest_conserved_charge_Nfolded_GKP}
\end{equation}
which then yields
\begin{equation}
 \Delta - S = \frac{K \sqrt{\lambda}}{2\pi} \log S + \frac{\sqrt{\lambda}}{2\pi} \left( K \log 2 - K \log K + \log (-1)^\frac{K}{2} 
  + C_{\mathrm{string}}(K) \right) \:.
\label{eq:gauge_theory_prediction_for_E-S_Nfolded_GKP}
\end{equation}
Comparison with the direct calculation \eqref{eq:E_S_omega_to_1_behaviour} suggests
\begin{equation}
 C_{\mathrm{string}}(K) = K \left[ \log \left( \frac{8 \pi}{\sqrt{\lambda}} \right) - 1 \right] - \log(-1)^\frac{K}{2}
\label{eq:C(n)_from_Nfolded_GKP}
\end{equation}
We would also like to remark that all the calculations concerning the N-folded GKP string reduce to the standard GKP results for $N=1$ (i.e. $K=2$), as listed in \cite{Dorey:2008zy,Gubser:2002tv,Jevicki:2008mm}.

\subsection{The symmetric spiky string}
\label{sec:JJ}

\subsubsection{Gauge considerations}

\begin{figure}
\includegraphics[width=\columnwidth]{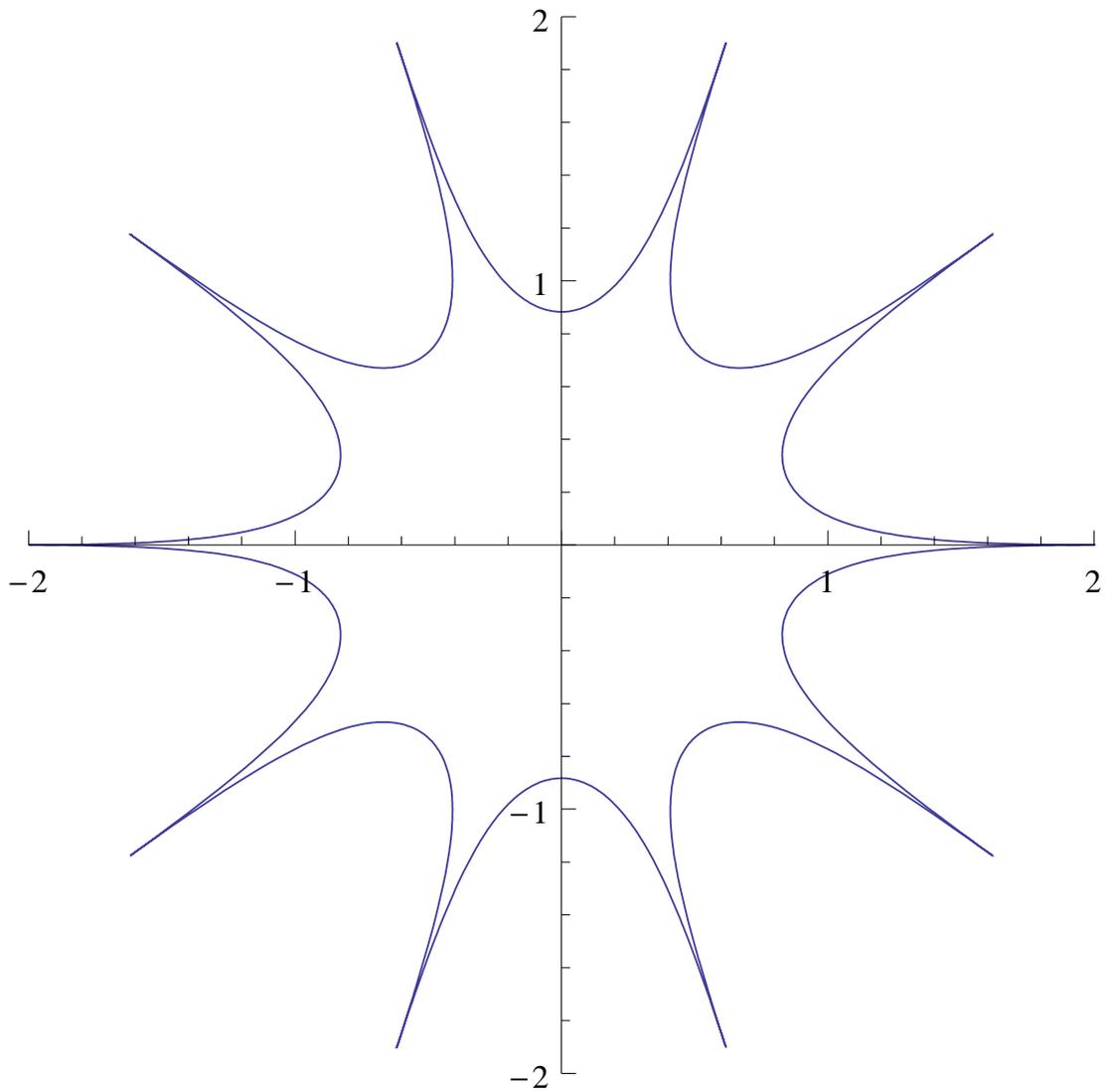}%
\caption{The Kruczenski spiky string in the $(\rho,\phi)$-plane at $t=0$, with $\rho_0 = 0.882663$ and $\rho_1 = 2$.}%
\label{fig:Kruc_10_spikes}%
\end{figure}

The Kruczenski spiky string was first discovered \cite{Kruczenski:2004wg} as a solution to the equations of motion generated by the Nambu-Goto action. It is given, in Kruczenski's gauge, which we label as $(\bar{\tau},\bar{\sigma})$, by $t = \bar{\tau}$, $\rho = \rho(\bar{\sigma})$, $\phi = \omega \bar{\tau} + \bar{\sigma}$, where the function $\rho(\bar{\sigma})$ is only known implicitly, while we have an explicit expression for its inverse:
\begin{equation}
 \bar{\sigma} = \pm \frac{\sinh 2\rho_0}{\sqrt{2}\sqrt{w_0 + w_1} \sinh\rho_1} \left\{ \Pi \left( \frac{w_1-w_0}{w_1-1}, \beta, p \right)
  - \Pi \left( \frac{w_1-w_0}{w_1+1}, \beta, p \right) \right\} \:,
\label{eq:sigma_of_rho_Kruc_Nambu-Goto_1}
\end{equation}
where
\begin{equation}
 p \equiv \sqrt{\frac{w_1-w_0}{w_1+w_0}} \:, \qquad \sin\beta \equiv \sqrt{\frac{w_1-w(\hat{\rho})}{w_1-w_0}}
\label{eq:p_and_beta_in_Kruc_Nambu-Goto_1}
\end{equation}
($\beta \in [0,\pi/2]$) and we define $w(x) \equiv \cosh (2x)$, $w_0 \equiv \cosh 2\rho_0$ and $w_1 \equiv \cosh 2\rho_1$. Its plot is shown in Fig. \ref{fig:Kruc_10_spikes}. While the coincidence of worldsheet time and global time is a desirable feature of Kruczenski's gauge, the fact that $\rho(\bar{\sigma})$ is only known implicitly makes our calculations harder.

We will therefore work with the conformal gauge version, which was more recently found in \cite{Jevicki:2008mm}. In appendix \ref{sec:app_gauge_transf_for_K}, we verify directly that the two solutions are indeed gauge equivalent, by deriving the explicit transformation from Kruczenski's gauge, represented by the coordinates $(\bar{\tau},\bar{\sigma})$, to conformal gauge, with our usual coordinates $(\tilde{\tau}, \tilde{\sigma})$ having non-standard periodicity. The problematic feature of the solution in conformal gauge is the inequivalence of the two time coordinates, $t \neq \tilde{\tau}$, which, as previously anticipated, causes additional complications.

We will postpone the discussion of the solution and the analysis of its plot to the next section, while we are now going to deal with these extra issues.
\paragraph{}
First of all, since we identify the ``real'' time coordinate with the global time $t$, every time we think of a snapshot of the string frozen in its motion, we mean that $t$ is constant. Hence, starting from a generic string solution in global coordinates $t (\tilde{\tau}, \tilde{\sigma}), \rho (\tilde{\tau}, \tilde{\sigma}), \phi (\tilde{\tau}, \tilde{\sigma})$, when discussing its time evolution, we will always eliminate the worldsheet time through the constraint $t (\tilde{\tau}, \tilde{\sigma}) = \textrm{constant}$, which is solved by a function $\tilde{\tau} (t, \tilde{\sigma})$, and then parametrise the two remaining coordinates as $\rho (t, \tilde{\sigma}), \phi (t, \tilde{\sigma})$. The plot of the string is then obtained by letting $\tilde{\sigma}$ run over its full range while keeping $t$ fixed.
\paragraph{}
Coordinate changes may also have an impact on the spike condition. We will assume the spikes to exist at constant $t$, not at constant $\tilde{\tau}$, which means that the spike condition \eqref{eq:spike_condition_tau} becomes
\begin{equation}
 \partial_{\tilde{\sigma}} \rho \left. \Big( \tilde{\tau}(t, \tilde{\sigma}), \tilde{\sigma} \Big) \right|_{\tilde{\sigma}_0 (t)} = 0
  = \partial_{\tilde{\sigma}} \phi \left. \Big( \tilde{\tau}(t, \tilde{\sigma}), \tilde{\sigma} \Big) \right|_{\tilde{\sigma}_0 (t)} \:,
\label{eq:spike_conditions_constant_t}
\end{equation}
for a spike located at $\tilde{\sigma} = \tilde{\sigma}_0(t)$. This equation is in fact the spike condition formulated in a different coordinate system $(\tau', \sigma') = (t(\tilde{\tau},\tilde{\sigma}), \tilde{\sigma})$ and, as such, it does not in general imply that the same condition should be satisfied in the frame $(\tilde{\tau}, \tilde{\sigma})$, so that we may well have
\begin{equation}
 \partial_{\tilde{\sigma}} \rho \left. ( \tilde{\tau}, \tilde{\sigma} ) \right|_{\tilde{\sigma}_0 (t)} \neq 0 \neq
   \partial_{\tilde{\sigma}} \phi \left. ( \tilde{\tau}, \tilde{\sigma} ) \right|_{\tilde{\sigma}_0 (t)}
\label{eq:spike_conditions_violated_conf_gauge}
\end{equation}
in conformal gauge.
\paragraph{}
Another important consequence of the gauge choice concerns the closedness condition. Assuming the string is closed at constant $t$, i.e. in the gauge $(\tau', \sigma')$ from above:
\begin{equation}
 \phi( \tau', \sigma' ) = \phi( \tau', \sigma' + L ) \qquad
  \rho( \tau', \sigma' ) = \rho( \tau', \sigma' + L ) \:,
\label{eq:t_closedness}
\end{equation}
where $L$ is some period, the same may not be true in conformal gauge. Specifically, the fact that, while $t = \tau'$ is automatically constant along the whole string in the first coordinate system, this may no longer be true in the $(\tilde{\tau}, \tilde{\sigma})$ coordinates is a potential source of trouble.

In general, in order to define a closed contour along the worldsheet in conformal gauge, it is necessary to let $\tilde{\tau}$ vary together with $\tilde{\sigma}$. In our case, it will be enough to split the contour into two parts: $\gamma = \gamma_1 \cup \gamma_2$, with $\tilde{\tau}$ remaining constant along the first and $\tilde{\sigma}$ remaining constant along the second.

This extra timelike segment becomes crucial when calculating the conserved charges and the monodromy matrix of the string, since the gauge-invariant definition of such objects involves integrals along closed contours $\gamma (\tilde{\tau}, \tilde{\sigma})$ starting at a generic base point $(\tilde{\tau}, \tilde{\sigma})$ and wrapping once around the worldsheet. We have already seen that the monodromy matrix is defined as \eqref{eq:def_monodromy_matrix}, which should only be specialised to \eqref{eq:Omega_convenient_choice} when the string is closed at constant $\tilde{\tau}$. The invariant definition of a conserved charge $Q$ associated with a current $J_a$ is instead:
\begin{equation}
 Q = - \int_{[\gamma (\tau, \sigma)]} (* J)_a (\tau', \sigma') \: d \sigma'^a \:,
\label{eq:gauge_inv_charge}
\end{equation}
where $*J$ is the Hodge dual of the current $J$,
\begin{equation}
 (*J)_a = \epsilon_{ba} V^b \:,
\label{eq:def_Hodge_dual}
\end{equation}
with the anti-symmetric tensor satisfying $\epsilon_{01} = 1$. In the following analysis, we will need to use such invariant definitions, together with appropriate closed contours involving timelike segments.

It is important to observe that, if $j_\sigma$ is subleading with respect to $j_\tau \sim S$, when we scale $x \sim S$ in the definition of the monodromy matrix the timelike segment does not contribute:
\begin{eqnarray}
 \Omega (x, \tau) & = & \mathcal{P} \mathrm{exp} \left[ \int \mathcal{J}_\sigma (x, \tau, \sigma') \: d \sigma' \right]
                         \left[ \mathcal{P} \mathrm{exp} \int \mathcal{J}_\tau (x, \tau', \sigma) \: d \tau' \right] \nn\\
                  & = & \mathcal{P} \mathrm{exp} \left[ \frac{1}{2} \int \left( \frac{j_\tau (x, \tau, \sigma')}{x} + \ldots \right) d \sigma'
                         \right] \nn\\
                  &   & \times \mathcal{P} \mathrm{exp} \left[ \frac{1}{2} \int \left( - \frac{j_\sigma (x, \tau', \sigma)}{x}
                           - \frac{j_\tau (x, \tau', \sigma)}{x^2} + \ldots \right) d \tau' \right] \nn\\
                  & \simeq & \mathcal{P} \mathrm{exp} \left[ \frac{1}{2} \int \left( \frac{j_\tau (x, \tau, \sigma')}{x} \right)
                              d \sigma' \right] \:,
\label{eq:Omega_gamma1_gamma2}
\end{eqnarray}
where the dots denote subleading terms. Therefore, we may continue to use the result \eqref{eq:monodromy_matrix_large_S_L-product}, \eqref{eq:L_m(u)} when computing the monodromy matrix.
\paragraph{}
Lastly, as a marginal note, we recall that, while discussing the argument leading to the derivation of the Casimir constraint for the spin vectors identified on the string side (see \eqref{eq:properties_of_spin_vectors}, right-hand equation), we saw that, instead of using the fact that $j_\sigma$ is subleading with respect to $j_\tau$ in the large $S$ limit, we may exploit the spike condition \eqref{eq:spike_condition_tau} if it is satisfied in our gauge. In fact, this property extends to the current situation, assuming that such a condition holds in the $(\tau', \sigma')$ coordinate system: $j_{\tilde{\sigma}} (\tilde{\tau}, \tilde{\sigma}_m) \neq 0$, while $j_{\sigma'} (\tau', \sigma'_m) = 0$ for a spike located at $\tilde{\sigma} = \tilde{\sigma}_m$ or equivalently $\sigma' = \sigma'_m$.

In order to see this, we start by using the tensor transformation rule to express $j_{\tilde{\sigma}}$ in terms of $j_{\tau'}$ and $j_{\sigma'}$, the latter of which vanishes at the cusp. Therefore, $j_{\tilde{\sigma}}$ at the spike is a function of $j_{\tau'}$ only, which in turn is proportional to $j_{\tilde{\tau}}$. Hence, we may write the Virasoro constraints \eqref{eq:Virasoro_j_AdS3} at the cusp as:
\begin{equation}
 \left( 1 \pm \frac{t' (\tilde{\tau},\tilde{\sigma}_m)}{\dot{t} (\tilde{\tau},\tilde{\sigma}_m)} \right)^2 \mathrm{Tr} \: [ j^2_{\tilde{\tau}}
  (\tilde{\tau},\tilde{\sigma}_m) ] = \frac{J^2}{\sqrt{\lambda}} \:,
\label{eq:Virasoro_near_spikes}
\end{equation}
where $\dot{t} (\tilde{\tau},\tilde{\sigma}) = \partial_{\tilde{\tau}} t (\tilde{\tau},\tilde{\sigma})$ and $t' (\tilde{\tau},\tilde{\sigma}) = \partial_{\tilde{\sigma}} t (\tilde{\tau},\tilde{\sigma})$. Now, even if $t'/\dot{t}$ was to equal either $+1$ or $-1$ at the spike (or approach these values in the large $S$ limit), one of the two equations would still have a non-vanishing pre-factor. We can then assume that the left-hand side dominates over the right-hand side and proceed as before.

\subsubsection{General properties}

The initial ansatz for the conformal gauge version of this solution is\footnote{As usual $\tilde{\sigma}$ denotes the worldsheet coordinate prior to a rescaling which normalises its periodicity to $2\pi$.} $t = \tilde{\tau} + f(\tilde{\sigma})$, $\phi = \phi_0 + \omega \tilde{\tau} + g(\tilde{\sigma})$, $\rho = \rho(\tilde{\sigma})$, and it leads to the following solution to the equations of motion and Virasoro constraints:
\begin{eqnarray}
 \partial_{\tilde{\sigma}} f(\tilde{\sigma}) & = & \frac{\omega \sinh 2 \rho_0}{2 \cosh^2 \rho} \:, \qquad
  \partial_{\tilde{\sigma}} g(\tilde{\sigma}) = \frac{\sinh 2 \rho_0}{2 \sinh^2 \rho} \nonumber \\
 \left[ \partial_{\tilde{\sigma}} \rho(\tilde{\sigma}) \right]^2 & = & \frac{(\cosh^2 \rho - \omega^2 \sinh^2\rho)(\sinh^2 2\rho - \sinh^2 
  2 \rho_0)}{\sinh^2 2\rho} \:.
\label{eq:derivatives_for_JJ_solution}
\end{eqnarray}
We impose $\rho_0 \leq \rho \leq \rho_1$, with $\coth\rho_1 = \omega$, so that both factors in the numerator of $[\partial_{\tilde{\sigma}} \rho(\tilde{\sigma})]^2$ are positive. These equations can be integrated to give:
\begin{equation}
 \rho (\tilde{\sigma}) = \frac{1}{2} \cosh^{-1} \: [ \cosh 2\rho_1 \: \mathrm{cn}^2 (v|k) + \cosh 2\rho_0 \: \mathrm{sn}^2 (v|k)] \:,
\label{eq:rho_of_sigmatilde_JJ}
\end{equation}
where
\begin{equation}
 v \equiv \sqrt{\frac{\cosh 2\rho_1 + \cosh 2\rho_0}{\cosh 2 \rho_1 - 1}} \tilde{\sigma} \:, \qquad
  k \equiv \sqrt{\frac{\cosh 2\rho_1 - \cosh 2\rho_0}{\cosh 2 \rho_1 + \cosh 2 \rho_0}}
\label{eq:def_v_and_k_JJ}
\end{equation}
and
\begin{eqnarray}
 f (\tilde{\sigma}) & = & \frac{\sqrt{2} \omega \sinh 2 \rho_0 \sinh \rho_1}{(\cosh 2 \rho_1 + 1)\sqrt{\cosh 2 \rho_1 + \cosh 2 \rho_0}}
  \Pi \left( \frac{\cosh 2\rho_1 - \cosh 2\rho_0}{\cosh 2\rho_1 + 1}, x, k \right) \nonumber \\
 g (\tilde{\sigma}) & = & \frac{\sqrt{2} \sinh 2 \rho_0 \sinh \rho_1}{(\cosh 2 \rho_1 - 1)\sqrt{\cosh 2 \rho_1 + \cosh 2 \rho_0}}
  \Pi \left( \frac{\cosh 2\rho_1 - \cosh 2\rho_0}{\cosh 2\rho_1 - 1}, x, k \right) \:, \nonumber \\
\label{eq:f_and_g_of_sigmatilde_JJ}
\end{eqnarray}
with $x = \mathrm{am}(v|k)$ ($0 \leq k \leq 1$). For simpler notation, we introduce $w \equiv \cosh 2\rho$, $w_0 \equiv \cosh 2\rho_0$, $w_1 \equiv \cosh 2\rho_1$ and define:
\begin{equation}
 n_\pm \equiv \frac{w_1 - w_0}{w_1 \pm 1} \:, \qquad \mathbb{K} \equiv \mathbb{K}(k) \:, \qquad \mathbb{E} \equiv \mathbb{E}(k) \:,
\label{eq:def_np_nm_JJ}
\end{equation}
which will be used throughout the rest of this section.
\paragraph{}
In order to understand the shape of this solution, we first need to observe that $\rho(v(\tilde{\sigma}))$ is periodic of period $2\mathbb{K}$, starting off at $\rho(0) = \rho_1$, then decreasing to $\rho(\mathbb{K}) = \rho_0$ at half the period and finally going back to  $\rho(2\mathbb{K}) = \rho_1$. Therefore, for the string to be closed (whether at constant $t$ or at constant $\tilde{\tau}$ makes no difference at this stage, since $\rho$ only depends on $\tilde{\sigma}$), we impose $v(\tilde{\sigma}) \in [0,2 K \mathbb{K}]$, which corresponds to $\tilde{\sigma} \in [0,L]$, with:
\begin{equation}
 L = 2 K \mathbb{K}\sqrt{\frac{w_1 - 1}{w_1 + w_0}} \equiv 2 K \tilde{L} \:.
\label{eq:range_of_sigmatilde_in_JJ}
\end{equation}
The functions $f$ and $g$ are instead pseudo-periodic of pseudo-period $2\tilde{L}$:
\begin{eqnarray}
 f(\tilde{\sigma} + 2 m \tilde{L}) & = & f(\tilde{\sigma}) + \frac{\sqrt{2} \omega \sinh 2 \rho_0 \sinh \rho_1}{(w_1 + 1)\sqrt{w_1 + w_0}} 
  2 m \Pi (n_+,k) \nonumber \\
 g(\tilde{\sigma} + 2 m \tilde{L}) & = & g(\tilde{\sigma}) + \frac{\sqrt{2} \sinh 2 \rho_0 \sinh \rho_1}{(w_1 - 1)\sqrt{w_1 + w_0}} 
  2 m \Pi (n_-,k) \:,
\label{eq:pseudo-periodicities_of_f_and_g_JJ}
\end{eqnarray}
due to the pseudo-periodicities of the amplitude function and of the incomplete elliptic integral of the third kind.

Now, in order to have a closed string at constant global time $t$, we need to substitute $\tau = t - f(\tilde{\sigma})$ into the original ansatz for $\phi$, thus finding $\phi(t,\tilde{\sigma}) = \omega t + g(\tilde{\sigma}) - \omega f(\tilde{\sigma})$, and then to impose $\phi(t,L) = \phi(t,0) + 2 n \pi$, for $n\in\mathbb{Z}$. By the pseudo-periodicity, we can easily see that $\phi(t,L) - \phi(t,0) = 2 K \Delta\phi$, where:
\begin{eqnarray}
 \Delta\phi & = & \frac{\sqrt{2}\sinh 2\rho_0 \sinh \rho_1}{\sqrt{w_1+w_0}} \left[ \frac{\Pi(n_-,k)}{w_1 - 1} - \frac{\omega^2\Pi(n_+,k)}
  {w_1 + 1} \right] \nonumber\\
  					& = & \frac{\sinh 2\rho_0}{\sqrt{2} \sinh \rho_1 \sqrt{w_1+w_0}} \left[ \Pi(n_-,k) - \Pi(n_+,k) \right]
\label{eq:deltaphi_for_JJ}
\end{eqnarray}
The closedness constraint then becomes:
\begin{equation}
 \Delta\phi = \frac{n}{K} \pi \:.
\label{eq:closedness_constraint_JJ}
\end{equation}
On the other hand, in order to define a closed contour winding once around the worldsheet to be used when implementing the definitions of the monodromy matrix and the conserved charges, we may use the following path $\gamma (\tilde{\tau}_0) = \gamma_1 (\tilde{\tau}_0) \cup \gamma_2 (\tilde{\tau}_0)$, with base point $(\tilde{\tau}_0,0)$:
\begin{eqnarray}
 \gamma_1 (\tilde{\tau}_0) & : & \tilde{\tau} = \tilde{\tau}_0 \:, \qquad \tilde{\sigma} \in [0, L] \nn\\
 \gamma_2 (\tilde{\tau}_0) & : & \tilde{\tau} \in [\tilde{\tau}_0, \tilde{\tau}_0 - f(L)] \:, \qquad \tilde{\sigma} = L \:,
\label{eq:closed_contour_worldsheet_JJ}
\end{eqnarray}
where $\gamma_1 (\tilde{\tau}_0)$ and $\gamma_2 (\tilde{\tau}_0)$ are respectively the spacelike and the timelike segment. One may straightforwardly check that, at the endpoints of this contour, the global coordinates $t, \rho, \phi$ take the values $(\tilde{\tau}_0, \rho_1, \omega \tilde{\tau}_0)$ and $(\tilde{\tau}_0, \rho_1, \omega \tilde{\tau}_0 + 2 K \Delta \phi)$. Closedness is therefore ensured by \eqref{eq:closedness_constraint_JJ}.

The resulting plot at constant $t$ is shown in Fig. \ref{fig:Kruc_10_spikes}, and consists of $K$ arcs of equal angular separation $\Delta\theta = 2\Delta\phi$; a cusp is present at the joining point between each pair of consecutive arcs, where $\rho=\rho_1$. As global time varies, the string rigidly rotates.

The $K$ cusps are located at:
\begin{equation}
 \tilde{\sigma}_m = 2 m \tilde{L} \:, \qquad m = 0,\ldots,K-1 \:.
\label{eq:cusp_positions_JJ_sigmatilde}
\end{equation}
In particular, for this string solution, the spike condition \eqref{eq:spike_condition_tau} is satisfied in the coordinate system $(\tau', \sigma') = (t,\tilde{\sigma})$ we introduced earlier, while this is no longer true in conformal gauge $(\tilde{\tau}, \tilde{\sigma})$ (and in Kruczenski's gauge, as one may easily check from the corresponding ansatz in appendix \ref{sec:app_gauge_transf_for_K}).

Moreover, in \cite{Jevicki:2008mm} it is also shown that each spike corresponds to a static soliton of the sinh-Gordon field associated with this string solution by Pohlmeyer reduction. In particular, the worldsheet positions of the spikes coincide with those of the solitons and the sinh-Gordon solution describes a periodic $K$-soliton configuration.

\begin{figure}%
\includegraphics[width=\columnwidth]{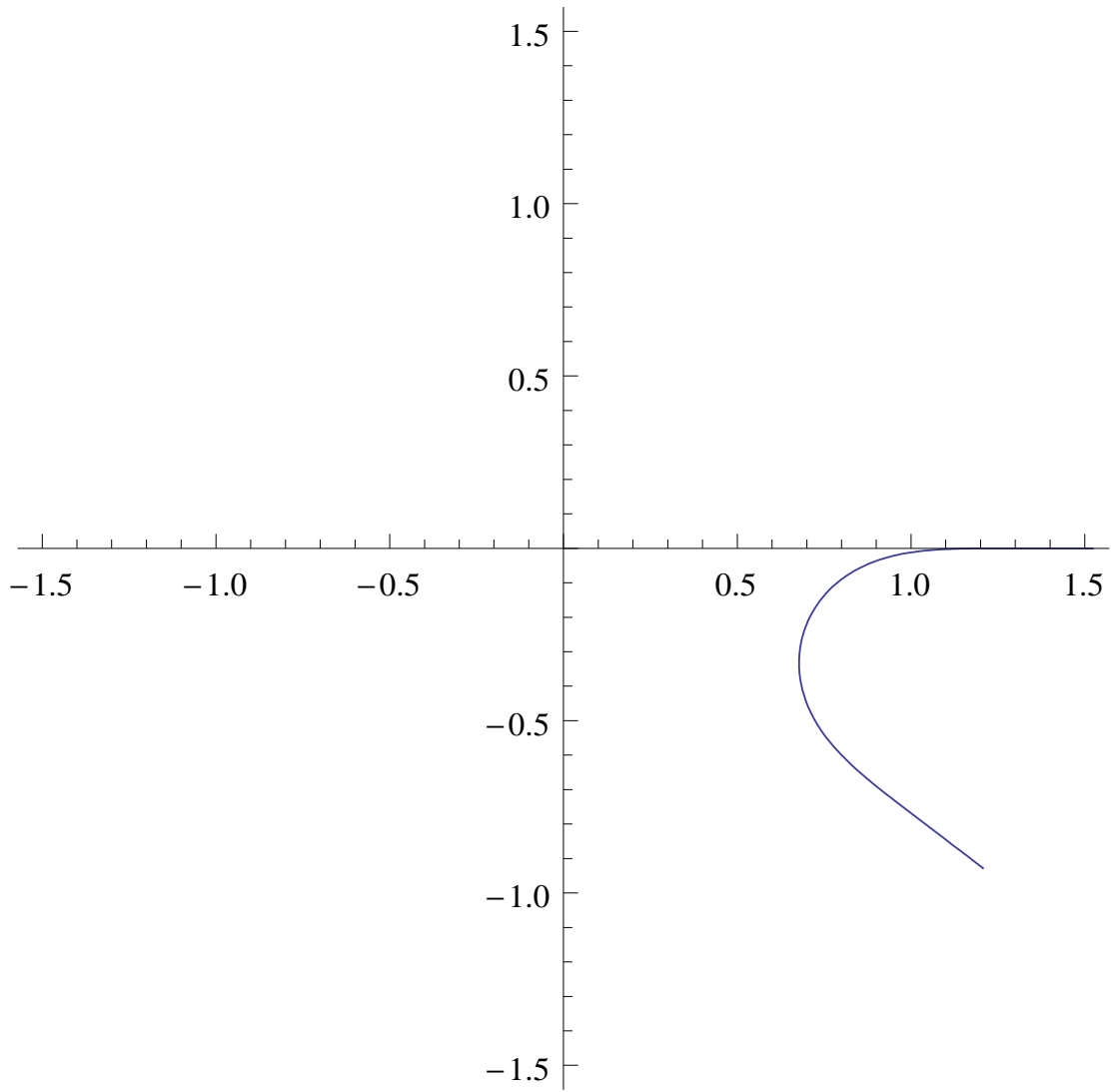}%
\caption{The Kruczenski spiky string in conformal gauge, for $\omega = 1$ and $\rho_0 = 0.9$.}%
\label{fig:JJ_single_arc_rho0=0.9}%
\end{figure}

As in the GKP case, for $\omega\to 1$ we find that $\rho_1 \to \infty$, the spikes touch the boundary and, as we will see shortly, the energy and angular momentum diverge. Note that, when considering this limit, $\rho_0$ is not fixed, since it depends on $\rho_1$ through equations \eqref{eq:deltaphi_for_JJ} and \eqref{eq:closedness_constraint_JJ}. Instead, it changes so that $\Delta\phi$ remains constant.
\paragraph{}
Furthermore, it is possible to compute a solution to the equations of motion and Virasoro constraints which holds for $\omega = 1$:
\begin{eqnarray}
 \rho(\tilde{\sigma}) & = & \frac{1}{2} \cosh^{-1} (w_0 \cosh 2\tilde{\sigma}) \nonumber \\
 t(\tilde{\tau},\tilde{\sigma}) & = & \tilde{\tau} + \arctan \left[ \coth 2\rho_0 \: e^{2 \tilde{\sigma}} + \frac{1}{\sinh 2\rho_0} \right]
  \nonumber \\
 \phi(\tilde{\tau},\tilde{\sigma}) & = & \tilde{\tau} + \arctan \left[ \coth 2\rho_0 \: e^{2 \tilde{\sigma}} - \frac{1}{\sinh 2\rho_0} \right]
  \:,
\label{eq:JJ_single_arc_solution_omega=1}
\end{eqnarray}
which is obtained simply by integrating \eqref{eq:derivatives_for_JJ_solution} after substituting in $\omega = 1$. It describes a single arc which has its endpoints on the boundary of $AdS$, reached for $\tilde{\sigma} \to \pm \infty$ (Fig. \ref{fig:JJ_single_arc_rho0=0.9}). What we see is the result of ``blowing up'' one of the interconnecting arcs located between two consecutive cusps in the original $\omega > 1$ solution. In the process, the spikes are ``pushed away'' into the region in which $\tilde{\sigma}$ becomes infinite, ultimately disappearing from the worldsheet, exactly as in the case of the infinite GKP string discussed in section \ref{sec:small_spikes}. Other than from the plot, we can also see this from the fact that now we have $\left[ \partial_{\tilde{\sigma}} \rho(\tilde{\sigma}) \right]^2 = (\sinh^2 2\rho - \sinh^2 2 \rho_0)/\sinh^2 2\rho$, and thus the first derivative of $\rho(\tilde{\sigma})$ does not vanish any longer at the endpoints $\rho \to +\infty$. This is also the case with the sinh-Gordon field, which for this simplified solution reduces to the vacuum solution \eqref{eq:def_sinhG-vacuum} due to the fact that the two static solitons at the spikes have disappeared at infinity, together with the spikes themselves.

The reason why this solution is particularly helpful is that it allows us to obtain the relationship between the angular separation $\Delta\theta = 2 \Delta\phi$ at constant $t$ of the arcs in the original solution and the parameter $\rho_0$ in the limit $\omega \to 1$. Since \eqref{eq:JJ_single_arc_solution_omega=1} describes one of these arcs at $\omega =1$, all we have to do is to compute $\Delta\theta$ from it:
\begin{equation}
 \Delta\theta = \left( \lim_{\tilde{\sigma} \to +\infty} - \lim_{\tilde{\sigma} \to -\infty} \right) \phi(t,\tilde{\sigma}) = 2 \mathrm{Arctan}
  \frac{1}{\sinh 2\rho_0} \:.
\label{eq:Deltatheta_single_arc_JJ_omega=1}
\end{equation}
Therefore, we deduce that the expression defined in \eqref{eq:deltaphi_for_JJ} has the following behaviour:
\begin{equation}
 \Delta\theta \simeq 2 \mathrm{Arctan} \frac{1}{\sinh 2\rho_0} \:, \qquad \textrm{as } \omega \to 1 \:.
\label{eq:Deltatheta_single_arc_JJ_omega_to_1}
\end{equation}
Note that $\Delta\theta \in (0,\pi)$, since $0<\rho_0<+\infty$; this is true for any $\omega>1$, due to the fact that, as we remarked earlier, $\Delta\theta$ is always fixed at a constant value by the closedness constraint \eqref{eq:closedness_constraint_JJ}. This implies that $n \leq K/2$ (the value $n = K/2$ may be accessed as a limiting case). \eqref{eq:Deltatheta_single_arc_JJ_omega_to_1} will be useful to us later, when computing the conserved charges and the monodromy matrix for large $S$, since it shows that $\rho_0$ always approaches a constant non-zero value as $\rho_1$ diverges, and therefore it always behaves as $O(1)$ in the limit $\omega \to 1$.

We can recognise the GKP $N$-folded string solution as a special case of the symmetric spiky string. In particular, \eqref{eq:Deltatheta_single_arc_JJ_omega_to_1} shows that, when $\Delta\theta\to \pi$, we have $\rho_0 \to 0$ and we recover a folded string solution, which passes through the origin $\rho=0$. This is the only case in which $n = K/2$.

\subsubsection{Conserved charges}

The energy and angular momentum may be computed by applying the invariant definition \eqref{eq:gauge_inv_charge} to the corresponding charge densities, appearing in \eqref{eq:AdS3xS1_charges}, and to the closed contour \eqref{eq:closed_contour_worldsheet_JJ}:
\begin{eqnarray}
 \Delta & = & \frac{\sqrt{\lambda}}{2 \pi} \left( \int_{\gamma_1 (\tilde{\tau}_0)} \mathrm{Im} (\bar{Z}_1 \partial_{\tilde{\tau}} Z_1)
               d \tilde{\sigma} + \int_{\gamma_2 (\tilde{\tau}_0)} \mathrm{Im} (\bar{Z}_1 \partial_{\tilde{\sigma}} Z_1) d \tilde{\tau} \right)
                \nn\\
        & = & \frac{\sqrt{\lambda}}{2\pi} \left( \int_0^L d\tilde{\sigma} \cosh^2 \rho (\tilde{\sigma})
               + \frac{\omega \sinh 2 \rho_0}{2} \int_{\tilde{\tau}_0}^{\tilde{\tau}_0 - f(L)} d \tilde{\tau} \right) \nn\\
        & = & K \frac{\sqrt{\lambda}}{\pi} \sqrt{\frac{w_1-1}{w_1+w_0}} \left[ \frac{1}{2} (w_1+w_0) \mathbb{E} - \sinh^2 \rho_0
               \mathbb{K} \right] \nn\\
        &   & \phantom{++} - K \frac{\sqrt{\lambda}}{2 \pi} \frac{\sqrt{2} \omega^2 \sinh^2 2 \rho_0 \sinh \rho_1}{\sqrt{w_1 + w_0} (w_1 + 1)}
               \Pi (n_+, k) \nn\\
 S & = & \frac{\sqrt{\lambda}}{2 \pi} \left( \int_{\gamma_1 (\tilde{\tau}_0)} \mathrm{Im} (\bar{Z}_2 \partial_{\tilde{\tau}} Z_2)
          d \tilde{\sigma} + \int_{\gamma_2 (\tilde{\tau}_0)} \mathrm{Im} (\bar{Z}_2 \partial_{\tilde{\sigma}} Z_2) d \tilde{\tau} \right) \nn\\
        & = & \frac{\sqrt{\lambda}}{2\pi} \left( \omega \int_0^L d\tilde{\sigma} \sinh^2 \rho (\tilde{\sigma})
               + \frac{\sinh 2 \rho_0}{2} \int_{\tilde{\tau}_0}^{\tilde{\tau}_0 - f(L)} d \tilde{\tau} \right) \nn\\
        & = & K \frac{\omega\sqrt{\lambda}}{\pi} \sqrt{\frac{w_1-1}{w_1+w_0}} \left[ \frac{1}{2} (w_1+w_0) \mathbb{E} - \cosh^2 \rho_0
               \mathbb{K} \right] \nn\\
        &   & \phantom{++} - K \frac{\sqrt{\lambda}}{2 \pi} \frac{\sqrt{2} \omega \sinh^2 2 \rho_0 \sinh \rho_1}{\sqrt{w_1 + w_0} (w_1 + 1)}
               \Pi (n_+, k) \:.
\label{eq:E_S_JJ}
\end{eqnarray}
These expressions agree with the results calculated in Kruczenski's gauge in \cite{Kruczenski:2004wg}. Again, we consider the limit as $\omega \to 1$, with $\omega = 1+\eta$, $\eta \gtrsim 0$, and compute the leading behaviour of these two quantities, finding that, as before, each spike contributes with an amount $\sqrt{\lambda}/(2 \pi \eta)$ to both charges:
\begin{equation}
 \Delta \simeq S = K \frac{\sqrt{\lambda}}{2 \pi} \frac{1}{\eta} + O ( \log \eta ) \:.
\label{eq:ES_lead_behaviour_omega_to_1_JJ}
\end{equation}
We also compute the $O(1)$ correction to the anomalous dimension:
\begin{equation}
 \Delta - S = \frac{K\sqrt{\lambda}}{2\pi} \log \left( \frac{2\pi S}{K\sqrt{\lambda}} \right) + \frac{K\sqrt{\lambda}}{2\pi} \left[ 3\log 2 - 1 +
  \log \left( \sin \frac{\Delta\theta}{2} \right) \right] + \ldots \:,
\label{eq:E-S_omega_to_1_behaviour}
\end{equation}
where we have used the relation:
\begin{equation}
 \frac{1}{w_0} \simeq \sin \frac{\Delta\theta}{2} \qquad \textrm{as } \omega \to 1 \:,
\label{eq:relation_between_u0_and_sin_Deltatheta/2}
\end{equation}
which is easily obtained from \eqref{eq:Deltatheta_single_arc_JJ_omega_to_1}. Again we find the usual logarithmic growth $\Delta-S\sim K\log S$ which is characteristic of the gauge theory anomalous dimensions for operators of twist $K$.

\subsubsection{Spectral curve for large $S$}

We will now repeat the calculation of section \ref{sec:spectral_curve_NGKP}, which will yield the spectral curve associated with the Kruczenski solution.

As before, we start by computing the components of the right current, expressed in terms of the rescaled worldsheet coordinates $(\tau,\sigma)$, defined so that $\sigma \in [0,2\pi]$:
\begin{equation}
 (\tau,\sigma) = \frac{2\pi}{L} (\tilde{\tau},\tilde{\sigma}) = \frac{\pi}{K \mathbb{K}}
  \sqrt{\frac{w_1+w_0}{w_1-1}}(\tilde{\tau},\tilde{\sigma}) \:.
\label{eq:def_rescaled_coords_JJ}
\end{equation}
The charge density is then given by
\begin{eqnarray}
 j^0_\tau(\tau,\sigma) & = & \frac{K \mathbb{K}}{\pi}\sqrt{\frac{w_1-1}{w_1+w_0}} \nonumber \\
  & & \times \{ (\omega+1) [w_1 \mathrm{cn}^2 (v|k) + w_0 \mathrm{sn}^2 (v|k)] + 1-\omega \} \nonumber \\
 j^1_\tau(\tau,\sigma) + i j^2_\tau(\tau,\sigma) & = & i \frac{K \mathbb{K}}{\pi} \sqrt{\frac{w_1-1}{w_1+w_0}} (\omega+1) e^{i(\phi-t)}  
  \nonumber \\
  & & \times \sqrt{[w_1 \mathrm{cn}^2 (v|k) + w_0 \mathrm{sn}^2 (v|k)]^2 - 1} \:,
\label{eq:eq:j_tau_in_terms_of_tau_sigma_JJ}
\end{eqnarray}
where now
\begin{equation}
 v \equiv \sqrt{\frac{w_1 + w_0}{w_1 - 1}} \tilde{\sigma} = \frac{K \mathbb{K}}{\pi} \sigma \:.
\label{eq:v_in_terms_of_sigma_JJ}
\end{equation}
In terms of the new coordinates, the cusps are located at:
\begin{equation}
 \sigma_m = 2m \frac{\pi}{K} \:, \qquad m=0,\ldots,K-1
\label{eq:cusp_positions_JJ_sigma}
\end{equation}
As in the GKP case, the leading order of the charge density is dominated by the contributions coming from the cusps, which we compute individually by setting $\sigma = \sigma_m + \hat{\sigma}$, with $|\hat{\sigma}| < \pi/K$, and then expanding \eqref{eq:eq:j_tau_in_terms_of_tau_sigma_JJ} as $\omega \to 1$, obtaining:
\begin{eqnarray}
 j_\tau^0 (\tau,\sigma) & \simeq & \frac{2 K \mathbb{K}}{\pi} \frac{1}{\eta \cosh^2 \left( \frac{K \mathbb{K}}{\pi} \hat{\sigma} \right)} \nn\\
 j^1_\tau(\tau,\sigma) + i j^2_\tau(\tau,\sigma) & \simeq & i \frac{2 K \mathbb{K}}{\pi} \frac{1}{\eta \cosh^2 \left( \frac{K \mathbb{K}}{\pi}
  \hat{\sigma} \right)} e^{i m \frac{2n\pi}{K}} \:.
\label{eq:leading_behaviour_of_jtau_JJ}
\end{eqnarray}
Note that the component $j_\sigma$ of the right current is $O( \eta^0)$ in the limit $\eta \to 0$, and is therefore subleading with respect to $j_\tau$, as we had previously assumed. Hence, it does not contribute to the spin vectors, which are still defined according to \eqref{eq:def_spin_vector_at_each_cusp}:
\begin{equation}
 L_m = \frac{S}{K} \begin{pmatrix}
	                  1 \\
	                  -\sin \left(m \frac{2 n \pi}{K} \right) \\
	                  \cos \left(m \frac{2 n \pi}{K} \right)
                   \end{pmatrix} \:.
\label{eq:spin_vector_at_mth_cusp_JJ}
\end{equation}
As in the GKP case, these vectors satisfy the properties \eqref{eq:properties_of_spin_vectors}, so that the highest-weight condition for the right-charge is satisfied at least at leading order for large $S$ (as usual, subleading corrections to this statement do not modify the spectrum). Moving on to the left charge, we find that $Q_L^+ = O(1)$, which is subleading with respect to $Q_L^0 = \Delta - S = O( \log \eta )$ and in particular satisfies the condition \eqref{eq:no_corrections_condition_left_hw_violation}\footnote{The only way in which we were able to calculate $Q_L^+$ is by approximating the left current in a similar fashion to what we have done with the right charge in \eqref{eq:leading_behaviour_of_jtau_JJ}, and only then evaluating the integral. Due to the leading order approximations introduced, the current conservation condition $\partial_a l^a = 0$ is no longer satisfied and hence the component $Q_L^+$ becomes time-dependent. Nevertheless, the estimate $Q_L^+ = O(1)$ remains acceptable.}, implying that there are no modifications of the spectrum due to the violation of the left charge highest-weight condition.

Then, \eqref{eq:L_m(u)} and \eqref{eq:monodromy_matrix_large_S_L-product} respectively yield the Lax matrix
\begin{equation}
 \mathbb{L}_m (u) = \begin{pmatrix}
	                           u + \frac{i}{K}   & \frac{i}{K} e^{i m \frac{2 n \pi}{K} } \\
	                           -\frac{i}{K} e^{-i m \frac{2 n \pi}{K} } & u - \frac{i}{K}
                            \end{pmatrix}
\label{eq:L_m(utilde)_for_JJ_1}
\end{equation}
and the monodromy matrix. We notice that this time no simplification occurs, i.e. the product of two consecutive matrices $\mathbb{L}_m (u)$ and $\mathbb{L}_{m+1} (u)$ still depends on $m$, and therefore we can't proceed as we did earlier. Instead, we introduce the following sequence of matrices:
\begin{equation}
 S_m = \begin{pmatrix}
	      c e^{i m \frac{n \pi}{K}} & d e^{i m \frac{n \pi}{K}} \\
	      \bar{d} e^{-i m \frac{n \pi}{K}} & \bar{c} e^{-i m \frac{n \pi}{K}}
       \end{pmatrix} \:, \qquad \textrm{with } |c|^2 - |d|^2 = 1 \:,
\label{eq:acceptable_choice_for_S_m_JJ}
\end{equation}
for $m=0,\ldots,K$ (where $c$ and $d$ are arbitrary, apart from the constraint on their absolute values), and notice that it makes the product $S_m^{-1} \mathbb{L}_m (u) S_{m+1} \equiv \mathbb{M} (u)$ independent of $m$. We also observe that:
\begin{equation}
 S_K = \begin{pmatrix}
	      c e^{i n \pi} & d e^{i n \pi} \\
	      \bar{d} e^{-i n \pi} & \bar{c} e^{-i n \pi}
       \end{pmatrix} = (-1)^{n} \begin{pmatrix}
	                                       c & d \\
	                                       \bar{d} & \bar{c}
                                        \end{pmatrix} = (-1)^{n} S_0 \:.
\label{eq:relation_between_S_n_and_S_0_JJ}
\end{equation}
We can now compute the trace of the monodromy matrix by inserting copies of the identity matrix, in the form of the products $S_m S_m^{-1}$, between consecutive matrices $\mathbb{L}_m (u)$:
\begin{eqnarray}
 \mathrm{Tr} \, \Omega(u) & = & \frac{1}{u^K} \mathrm{Tr} \prod_{m=0}^{K-1} \mathbb{L}_m(u) \nonumber\\
  & = & \frac{1}{u^K} \mathrm{Tr} \left[ S_0 S_0^{-1} \mathbb{L}_0 (u) S_1 S_1^{-1} \mathbb{L}_1(u) S_2 S_2^{-1}
   \ldots S_{K-1} S_{K-1}^{-1} \mathbb{L}_{K-1}(u) \right] \nonumber \\
  & = &  \frac{(-1)^{n}}{u^K} \mathrm{Tr} \left[ S_0^{-1} \mathbb{L}_0 (u) S_1 S_1^{-1} \mathbb{L}_1(u) S_2
   S_2^{-1} \ldots S_{K-1} S_{K-1}^{-1} \mathbb{L}_{K-1}(u) S_K \right] \nonumber \\
  & = & (-1)^{n} \frac{1}{u^K} \mathrm{Tr} \left[ \mathbb{M} (u)^K \right] \:,
\label{eq:trace_of_monodromy_matrix_in_terms_of_M_utilde_JJ}
\end{eqnarray}
where in obtaining the third line we have used \eqref{eq:relation_between_S_n_and_S_0_JJ} and the cyclicity property of the trace.

\begin{figure}
\centering
\psfrag{a}{\footnotesize{$u_+$}}
\psfrag{b}{\footnotesize{$u_-$}}
\psfrag{c}{\footnotesize{$0$}}
\psfrag{d}{\footnotesize{$u_{n+1}$}}
\psfrag{e}{\footnotesize{$u_{K-1}$}}
\psfrag{f}{\footnotesize{$u_{n-1}$}}
\psfrag{g}{\footnotesize{$u_1$}}
\includegraphics[width=\columnwidth]{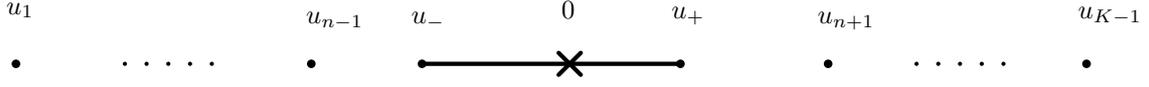}%
\caption{The cut and branch points configuration on the surface $\tilde{\Sigma}_1$ for the Kruczenski symmetric spiky string. The distribution of the branch points is not symmetric with respect to the origin.}
\label{fig:bps_and_cuts_JJ}%
\end{figure}

The rest of the calculation proceeds as in the GKP case. We first determine the eigenvalues of $\mathbb{M} (u)$,
\begin{equation}
 \kappa_\pm = u \cos \frac{n\pi}{K} - \frac{1}{K} \sin \frac{n\pi}{K} \pm \sqrt{- \frac{2}{K} u 
  \sin \frac{n\pi}{K} \cos \frac{n\pi}{K} + \left( \frac{1}{K^2} - u^2 \right) \sin^2 \frac{n\pi}{K}} \:,
\label{eq:evalues_of_M(utilde)_JJ}
\end{equation}
and then deduce
\begin{eqnarray}
 \mathrm{Tr}\: \Omega(u) & = & (-1)^{n} \frac{1}{u^K} 
  ( \kappa_+^K + \kappa_-^K ) \nonumber \\
  & = & (-1)^{n} 2 T_K \left( \cos \frac{n\pi}{K}  - \frac{\sin \frac{n\pi}{K}}{K u} \right) \nonumber \\
  & = & 2 \cos \left[ n \pi + K \cos^{-1} \left( \cos \frac{n\pi}{K}  - \frac{\sin \frac{n\pi}{K}}{K u} \right)
   \right] \:,
\label{eq:tr_Omega_JJ}
\end{eqnarray}
where again we have used \eqref{eq:def_Tk(y)_Chebyshev}. Hence, we obtain the following expression for the quasi-momentum:
\begin{equation}
 p(u) = n \pi + K \cos^{-1} \left( \cos \frac{n\pi}{K}  - \frac{\sin \frac{n\pi}{K}}{K u} \right) \:,
\label{eq:quasi-momentum_K}
\end{equation}
and for the spectral curve associated with the Kruczenski solution:
\begin{equation}
 \tilde{\Sigma}_1 \: : \qquad t + \frac{1}{t} = \mathbb{P}_K \left( \frac{1}{u} \right) = 2 \cos \left[
  n \pi + K \cos^{-1} \left( \cos \frac{n\pi}{K}  - \frac{\sin \frac{n\pi}{K}}{K u} \right) \right] \:,
\label{eq:string_spectral_curve_K}
\end{equation}
which, as expected, is already in the form \eqref{eq:spectral_curve_explicit_sol-s}. We see that again this corresponds to a point in the moduli space of the gauge theory curve $\Gamma_K$, where the conserved charges take the following values:
\begin{equation}
 \hat{q}^{(0)}_k = \left( - \frac{2}{K} \right)^k \sum_{1\leq j_1 < j_2 < \ldots < j_k \leq K} \prod_{l=1}^k \sin \left[ \frac{n\pi}{K}
  (j_{l+1} - j_l) \right] \:, \quad k=2, \ldots, K \:,
\label{eq:qktilde_from_spin_chain_K}
\end{equation}
where $j_{k+1} \equiv j_1$ (the normalisation $\hat{q}^{(0)}_2 = -S^2$ is checked in appendix \ref{sec:app_computing_qtilde_2_for_N-folded_GKP_and_K}).
\paragraph{}
We can now compute the discriminant $D = 4 \sin^2 p(u)$:
\begin{eqnarray}
 D (u) & = & 4 \left[1 - T_K^2 \left( \cos \frac{n\pi}{K} - \frac{\sin
  \frac{n\pi}{K}}{K u} \right) \right] = \nonumber \\
 & = & - 4 \sin \frac{n\pi}{K} \left(- \sin \frac{n\pi}{K} - \frac{2}{K u} \cos \frac{n\pi}{K} +
  \frac{1}{K^2 u^2} \sin \frac{n\pi}{K} \right) \nonumber \\
 & & \times U_{K-1}^2 \left( \cos \frac{n\pi}{K} - \frac{\sin
  \frac{n\pi}{K}}{K u} \right) \:,
\label{eq:rewriting_D(utilde)_JJ}
\end{eqnarray}
which we use to determine the pattern of branch points for the spectral curve \eqref{eq:string_spectral_curve_K}:
\begin{equation}
 \begin{array}{ll}
  u  =  \frac{\sin \frac{n\pi}{K}}{K} \left( \cos \frac{n\pi}{K} \pm 1 \right)^{-1} \equiv u_\pm  & \qquad \textrm{simple} \\
  u  =  \frac{\sin \frac{n\pi}{K}}{K} \left( \cos \frac{n\pi}{K} - \cos \frac{j\pi}{K} \right)^{-1} \equiv u_j
   \:, \qquad \textrm{for } j=1,\ldots,K-1 & \qquad \textrm{double} \:.
 \end{array}
\label{eq:pattern_of_zeros_of_Q_n_for_JJ}
\end{equation}
One may check that $u_n = \infty$ reproduces the usual double point at infinity\footnote{Recall that $n<K/2$ (or $n = K/2$ in the exceptional GKP case), and thus $n \leq K-1$ for $K \geq 2$, which is required in order to reproduce the minimal structure of branch points for the finite-gap spectral curves $\tilde{\Sigma}_1$ (i.e. two branch points at the sides of the origin and a double point at infinity).}, while the other points satisfy $u_j < u_- < 0 < u_+ < u_k$, for $j = 1, \ldots, n-1$ and $k = n+1, \ldots, K-1$, so that again the curve has degenerated to genus zero, with only the central cut $[u_-, u_+]$ surviving and all the outer cuts collapsing into double points (see Fig. \ref{fig:bps_and_cuts_JJ}). Therefore, the quasi-momentum has a logarithmic branch point at $u=0$ and two square root branch points at $u = u_\pm$ and is analytic away from these points. The differential
\begin{equation}
 dp(u) = - \frac{K \sin \frac{n\pi}{K} du}{u \sqrt{\left(K u \sin \frac{n\pi}{K} + \cos \frac{n\pi}{K} \right)^2 - 1}}
\label{eq:dp(x)_K}
\end{equation}
displays a simple pole at $u=0$, two square root branch points at $u=u_\pm$. As in the GKP case, the same cut we introduced for $p(u)$ also makes $dp(u)$ single-valued.

\paragraph{}
We also compute the filling fractions associated with the two parts of the cut:
\begin{eqnarray}
 l_1^- & = & \frac{S}{\pi i} \int_{u_-}^{0} \frac{K \sin \frac{n\pi}{K} du}{\sqrt{\left( K u
  \sin \frac{n\pi}{K} + \cos \frac{n\pi}{K} \right)^2 - 1}} = S \left( 1 - \frac{n}{K} \right) \nonumber \\
 l_1^+ & = & \frac{S}{\pi i} \int_{0}^{u_+} \frac{K \sin \frac{n\pi}{K} du}{\sqrt{\left( K u
  \sin \frac{n\pi}{K} + \cos \frac{n\pi}{K} \right)^2 - 1}} = S \frac{n}{K} \:.
\label{eq:filling_fractions_original_cuts_K}
\end{eqnarray}
We observe that, differently from the GKP case, there is an asymmetry in the filling fractions, which is clearly due to the fact that the two square root branch points in $p(u)$ are no longer symmetric with respect to the origin. The symmetry is restored in the limiting GKP case $n = K/2$.

As before, we have $l_1^+ + l_1^- = S$ and $n_1^+ l_1^+ + n_1^- l_1^- = 0$, in agreement with \eqref{eq:S_filling_fracts_S1_general_K_M} and \eqref{eq:level_matching_FGA_general_K_M}.
\paragraph{}
Finally, as we did in the previous case, we derive the highest conserved charge:
\begin{equation}
 \hat{q}^{(0)}_K = (-1)^{K + n} \left( \frac{2}{K} \right)^K \left( \sin \frac{n\pi}{K} \right)^K \:.
\label{eq:highest_conserved_charge_JJ}
\end{equation}
Thus, the finite-gap expression \eqref{eq:general_Delta-S_finite-gap_explicit_sol-s} yields
\begin{multline}
 \Delta - S = \frac{K\sqrt{\lambda}}{2\pi} \log S + \frac{\sqrt{\lambda}}{2\pi} \left[ K \log 2 
  - K \log K + \log (-1)^{K+n} \phantom{\left( \sin \frac{n\pi}{K} \right)} \right. \\
 \left. + K \log \left( \sin \frac{n\pi}{K} \right)  + C_{\mathrm{string}}(K) \right] \:,
\label{eq:gauge_theory_prediction_for_E-S_JJ}
\end{multline}
where again we have omitted subleading terms in the limit $\omega \to 1$. Comparison with \eqref{eq:E-S_omega_to_1_behaviour} (we recall that $\Delta\theta = 2 \Delta\phi = 2 n \pi /K$) results in following value for the moduli-independent constant:
\begin{equation}
 C_{\mathrm{string}}(K) = K \left[ \log \left( \frac{8 \pi}{\sqrt{\lambda}} \right) - 1 \right] - \log (-1)^{K + n} \:.
\label{eq:C(n)_from_JJ}
\end{equation}
\paragraph{}
Furthermore, we would like to observe that it is possible to obtain all results for the GKP $N$-folded string from the Kruczenski string in conformal gauge, if we assume, of course, that the two solutions have the same number of cusps. This is done by interpreting the GKP configuration as a set of $K=2N$ spikes with angular separation between consecutive cusps equal to $\pi$, or, in other words, a set of 2 spikes for each turn around the origin in $AdS$ space, which can then be described by a Kruczenski-type solution with $K$ even and $n = K/2$ (which implies $\Delta\theta = \pi$).


\subsection{The general patched solution}
\label{sec:patching_K}

\subsubsection{General properties}

In this section, we are going to discuss a generalised version of Kruczenski's solution in conformal gauge, which describes arcs with arbitrary individual angular separations $\Delta\theta_j$, $j=1,\ldots,K$ (still subject to the constraint $\Delta\theta_j \in (0,\pi)$ which is an intrinsic property of the symmetric spiky string discussed in the previous section). The main objective of this exercise is to analyse the spectral curve associated with a larger family of strings. In particular, the GKP string only has one discrete parameter, $N = K/2$ (i.e. the number of folds), and the Kruczenski spiky string only has two discrete parameters, the total number of spikes $K$ and the $AdS_3$ winding number $n$. On the other hand, in addition to $K$ and $n$, this solution has $K-1$ continous parameters $\Delta \theta_j$ corresponding to the angular separations (one of which is always eliminated by the constraint $\sum_j \Delta \theta_j = 2 n \pi$). This will lead to considerably more complicated expressions for the moduli $\hat{q}^{(0)}_j$, allowing us to test the spectrum \eqref{eq:general_Delta-S_finite-gap_explicit_sol-s} and the spin vector identification \eqref{eq:def_spin_vector_at_each_cusp} on a much larger moduli space of solutions, which in fact has the same dimension as the moduli space of the finite-gap curves discussed in chapter \ref{ch:FGA}.

\begin{figure}
\centering
\psfrag{a}{\footnotesize{$\Delta\theta_1$}}
\psfrag{b}{\footnotesize{$\Delta\theta_2$}}
\psfrag{c}{\footnotesize{$\Delta\theta_3$}}
\psfrag{f}{\footnotesize{$\Delta\theta_K$}}
\includegraphics[width=100mm]{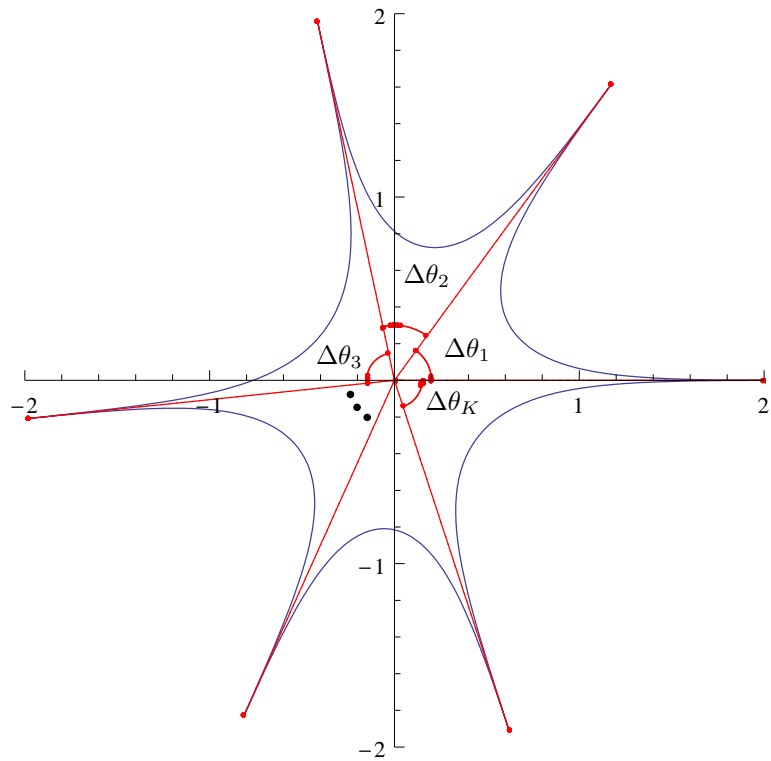}
\caption{The general multi-spike solution with an arbitrary choice of angular separations.}
\label{fig:patched_K_generic}
\end{figure}

The idea is to use different versions of \eqref{eq:rho_of_sigmatilde_JJ} and \eqref{eq:f_and_g_of_sigmatilde_JJ} to describe each single arc, and then to patch all the arcs together by gluing them at the endpoints. In this way, we will construct an approximate solution, which becomes exact in the large angular momentum limit $\omega \to 1$. A plot for a possibile configuration of this string is given in Fig. \ref{fig:patched_K_generic}.

We start by considering equation \eqref{eq:deltaphi_for_JJ}, which determines the angular separation between two consecutive cusps, as a function of the two parameters $\rho_0$ and $\rho_1$. We keep $\rho_1$ fixed, and define $K$ parameters $\rho_0^{(j)}$, $j=1,\ldots,K$, by imposing the following constraints:
\begin{equation}
 \Delta\theta_j = \frac{\sqrt{2}\sinh 2\rho_0^{(j)}}{\sinh \rho_1 \sqrt{w_1+w_0^{(j)}}} \left[ \Pi(n_-^{(j)},k^{(j)}) - \Pi(n_+^{(j)},k^{(j)})
  \right] \:, \qquad \textrm{for } j=1,\ldots,K \:,
\label{eq:def_rho_0^j_patched_K}
\end{equation}
where $n_\pm^{(j)}$ and $k^{(j)}$ (and also $v^{(j)}$, which we will use later) are defined in \eqref{eq:def_v_and_k_JJ} and \eqref{eq:def_np_nm_JJ}, with $\rho_0$ replaced by $\rho_0^{(j)}$. Each pair $(\rho_0^{(j)},\rho_1)$ defines a different version of the solution given in \eqref{eq:rho_of_sigmatilde_JJ} and \eqref{eq:f_and_g_of_sigmatilde_JJ}, with different fundamental half-period $\mathbb{K}(k^{(j)})$ and angular separation $\Delta\theta_j$, but with the same radial position of the spikes $\rho=\rho_1$. We introduce the shorthand notation
\begin{equation}
 \mathbb{E}_j \equiv \mathbb{E}(k^{(j)}) \qquad \mathbb{K}_j \equiv \mathbb{K}(k^{(j)})
\label{eq:E,K_shorthand_pK}
\end{equation}
together with the cusp positions $\tilde{\sigma}_j$ and the overall period $L$ of the coordinate $\tilde{\sigma}$:
\begin{equation}
 \tilde{\sigma}_j = 2 \sum_{k=1}^j \tilde{L}_k 
  \:, \qquad L = 2 \sum_{j=1}^K \tilde{L}_j \:, \qquad \tilde{L}_j = \mathbb{K}_j \sqrt{\frac{w_1 - 1}{w_1+w_0^{(j)}}} \:.
\label{eq:def_Ltilde_j_sigmatilde_j_and_L_patched_K}
\end{equation}
In order to glue these different solutions together, we let $\tilde{\sigma}$ run in the interval $[0,L]$: for $0 = \tilde{\sigma}_0 \leq \tilde{\sigma} \leq \tilde{\sigma}_1$ we want the patched solution to describe the first period of the $(\rho_0^{(1)},\rho_1)$ spiky string, for $\tilde{\sigma}_1 \leq \tilde{\sigma} \leq \tilde{\sigma}_2$ we want it to describe the first period of the $(\rho_0^{(2)},\rho_1)$ spiky string, and so on until we see the first period of the $(\rho_0^{(K)},\rho_1)$ spiky string for $\tilde{\sigma}_{K-1} \leq \tilde{\sigma} \leq L$. This is achieved by the following definition:
\begin{eqnarray}
 \rho (\tilde{\sigma}) & = & \rho ( \tilde{\sigma}- \tilde{\sigma}_{j-1}, \rho_0^{(j)} ) \nonumber\\
 f (\tilde{\sigma}) & = & f ( \tilde{\sigma}- \tilde{\sigma}_{j-1}, \rho_0^{(j)} ) + \sum_{k=1}^{j-1} f(2\tilde{L}_k,\rho_0^{(k)}) \nonumber\\
 g (\tilde{\sigma}) & = & g ( \tilde{\sigma}- \tilde{\sigma}_{j-1}, \rho_0^{(j)} ) + \sum_{k=1}^{j-1} g(2\tilde{L}_k,\rho_0^{(k)}) \:,
  \nonumber\\
 & & \qquad\qquad\qquad\qquad\qquad \textrm{for } \tilde{\sigma}_{j-1} \leq \tilde{\sigma} \leq \tilde{\sigma}_j \:,
\label{eq:def_patched_K_solution}
\end{eqnarray}
where $\rho (\tilde{\sigma},\rho_0)$, $f (\tilde{\sigma},\rho_0)$ and $g (\tilde{\sigma},\rho_0)$ are given by equations \eqref{eq:rho_of_sigmatilde_JJ} and \eqref{eq:f_and_g_of_sigmatilde_JJ}. In obtaining this, we have used our freedom to shift $\tilde{\sigma}$, $f$ and $g$ by a constant (these are all symmetries of the equations \eqref{eq:derivatives_for_JJ_solution}).
\paragraph{}
We note that, although \eqref{eq:def_patched_K_solution} clearly satisfies the equations of motion and Virasoro constraints in each interval $\tilde{\sigma}_{j-1} < \tilde{\sigma} < \tilde{\sigma}_j$, it is not smooth at the junction points. In particular, $\rho (\tilde{\sigma})$ is $C^1$, whereas $f (\tilde{\sigma})$ and $g (\tilde{\sigma})$ are only $C$. In the case of $\rho$, we can see this by considering equation \eqref{eq:derivatives_for_JJ_solution}:
\begin{eqnarray}
 && \partial_{\tilde{\sigma}} \rho (\tilde{\sigma}) = \pm \sqrt{h(\rho)} \:, \qquad h(\rho) = h_1 (\rho) h_2(\rho) \nonumber \\
 && h_1(\rho) = \cosh^2 \rho - \omega^2 \sinh^2 \rho \:, \qquad h_2 (\rho) = 1 - \frac{\sinh^2 2 \rho_0^{(j)}}{\sinh^2 2 \rho} \nonumber \\
 && \qquad \qquad \qquad \qquad \qquad \qquad \textrm{for } \tilde{\sigma}_{j-1} \leq \tilde{\sigma} \leq \tilde{\sigma}_j \:,
\label{eq:def_h_h1_h2_studying_derivs_of_rho_patched_K}
\end{eqnarray}
where the sign is plus or minus depending on which half of the $j$-th arc we are considering ($\rho$ is an increasing function of $\tilde{\sigma}$ along one half of every arc and it is instead decreasing on the other half). Clearly, at the junction points $\rho = \rho_1$ we have $\partial_{\tilde{\sigma}} \rho (\tilde{\sigma}) = 0$, independently of $j$, due to the fact that $h_1(\rho_1) = 0$. However, when we turn our attention to the second derivative of $\rho$, we obtain:
\begin{equation}
 \partial_{\tilde{\sigma}}^2 \rho (\tilde{\sigma}) = \frac{h'(\rho)}{2} \:, \qquad \textrm{where } ' \equiv \partial_\rho
\label{eq:rho''_patched_K}
\end{equation}
and it is very easy to see that $h'(\rho)$ contains a term which does not vanish at $\rho = \rho_1$ and which depends on $\rho_0^{(j)}$:
\begin{equation}
 h'(\rho_1) = h_1' (\rho_1) h_2(\rho_1) = (1- \omega^2) \sinh 2 \rho_1 \left( 1 - \frac{\sinh^2 2 \rho_0^{(j)}}{\sinh^2 2 \rho_1} \right) \:.
\label{eq:discontinuity_term_in_rho''_patched_K}
\end{equation}
This term generates a discontinuity in $\partial_{\tilde{\sigma}}^2 \rho (\tilde{\sigma})$ at the junction points, since the value of $\rho_0$ jumps from $\rho_0^{(j)}$ to $\rho_0^{(j+1)}$ there.

Similarly,
\begin{equation}
 \partial_{\tilde{\sigma}} g (\tilde{\sigma}) = \frac{\sinh 2 \rho_0^{(j)}}{2} l(\rho) \:, \qquad l(\rho) = \frac{1}{\sinh^2 \rho}  
\label{eq:def_l_in_g'_patched_K} \nonumber
\end{equation}
explicitly depends on $\rho_0^{(j)}$ at $\rho = \rho_1$ and thus is discontinuous at the junction points. The same clearly applies to $\partial_{\tilde{\sigma}} f (\tilde{\sigma})$.

Therefore, the patched version of Kruczsenki's solution is not a proper closed string solution for fixed $\omega>1$. However, the situation changes as $\omega \to 1$. In fact, in this limit, the function $\rho(\tilde{\sigma})$ from equation \eqref{eq:rho_of_sigmatilde_JJ} displays the following leading behaviour near the cusp located at $\tilde{\sigma}=0$:
\begin{equation}
 \rho(\tilde{\sigma}) = - \frac{1}{2} \log \eta + \frac{1}{2} \log ( 2 \mathrm{sech}^2 \tilde{\sigma} ) + O(\eta) \:,
\label{eq:leading_universal_behaviour_of_rho_near_cusps_patched_K}
\end{equation}
where, in our usual notation, $\omega = 1 + \eta$. Clearly, the situation is identical near any other cusp, due to the periodicity of $\rho(\tilde{\sigma})$. Thus, $\rho (\tilde{\sigma})$ has a universal profile near the cusps, which is independent of $\rho_0$, so that it is no longer sensitive to jumps in that parameter as we move across the junction points.

We now study $h_1 (\rho)$ in more detail:
\begin{eqnarray}
 h_1 (\rho_1) & = & 0 \nonumber \\
 \partial_\rho^k h_1 (\rho_1) & = & 2^{k-1} (1-\omega^2)
  \begin{cases}
   \sinh 2 \rho_1 & k \textrm{ even} \\
   \cosh 2 \rho_1 & k \textrm{ odd}
  \end{cases} \:.
\label{eq:h1_and_derivs_at_rho_1_patched_K}
\end{eqnarray}
It is easy to check that $w_1 = \cosh 2 \rho_1 = 1/\eta + O(1)$ and $\sinh 2 \rho_1 = 1/\eta + O(1)$, which then implies $\partial_\rho^k h_1 (\rho_1) = O(1)$, $\forall k > 0$. Of course, no discontinuities arise from this factor. Next, we consider the troublesome function $h_2(\rho)$:
\begin{eqnarray}
 && h_2 (\rho_1) = 1 - \frac{\sinh^2 2 \rho_0^{(j)}}{\sinh^2 2 \rho_1} \nonumber \\
 && \partial_\rho h_2 (\rho_1) = 2 \sinh^2 2 \rho_0^{(j)} \frac{\cosh 2 \rho_1}{\sinh^3 2 \rho_1} \nonumber \\
 && \partial_\rho^k h_2 (\rho_1) = 2^k \sinh^2 2 \rho_0^{(j)} \frac{P_k (2\rho_1)}{\sinh^{k+2} 2 \rho_1} \:, \qquad \textrm{for } \tilde{\sigma}_{j-1} \leq
  \tilde{\sigma} \leq \tilde{\sigma}_j \:,
\label{eq:h2_and_derivs_at_rho_1_patched_K}
\end{eqnarray}
where $P_k (2 \rho_1)$ is a polynomial of degree $k$ in $\cosh 2\rho_1$ and $\sinh 2\rho_1$. It is now easy to deduce that $h_2 (\rho_1) \to 1$ and $\partial_\rho^k h_2 (\rho_1) \to 0$ as $\omega \to 1$, which means that $h_2 (\rho)$ becomes smooth at the junction points in this limit. Taking the behaviour of $h_1 (\rho)$ into account we can then deduce
\begin{equation}
 \partial_\rho^k h (\rho_1) = \sum_{m=0}^k \partial_\rho^m h_1 (\rho_1) \partial_\rho^{k-m} h_2 (\rho_1) \to \partial_\rho^k h_1 (\rho_1) = O(1) \:, \qquad \textrm{as } \omega \to 1
\label{eq:behaviour_of_derivs_of_h_at_rho1_as_omega_to_1_patched_K}
\end{equation} 
and therefore $h(\rho)$ also becomes smooth as $\omega \to 1$. This immediately shows that $\partial_{\tilde{\sigma}}^2 \rho (\tilde{\sigma})$ from \eqref{eq:rho''_patched_K} is continuous in the same limit. Now, working from that equation, we see that, in general, $\partial_{\tilde{\sigma}}^k \rho (\tilde{\sigma})$ is a sum of products of derivatives of $h (\rho)$, up to order $k-1$, and of derivatives of $\rho$, up to order $k-2$. The latter can all be re-expressed in terms of lower derivatives of $h (\rho)$ through \eqref{eq:rho''_patched_K} and \eqref{eq:def_h_h1_h2_studying_derivs_of_rho_patched_K}, so that, in the end, we're only left with derivatives of $h (\rho)$ which all become smooth in the limit considered (in particular, notice that there are never any diverging factors involved, so that the exponential suppression of the discontinuities as $\omega \to 1$ is never undone). Hence, $\partial_{\tilde{\sigma}} \rho (\tilde{\sigma})$ becomes a smooth function as $\omega \to 1$.

If we now consider $l(\rho)$, we easily see that:
\begin{equation}
 \partial_\rho^k l (\rho_1) = \frac{P_k (\rho_1)}{\sinh^{k+2} \rho_1} \to 0 \:, \qquad \textrm{as } \omega \to 1 \:,
\label{eq:l_and_derivs_patched_K}
\end{equation}
which then implies that all derivatives of $g(\tilde{\sigma})$ at the junction points vanish as these points approach the boundary, since they are given by sums of products of derivatives of $l(\rho)$ and of $\rho(\tilde{\sigma})$ (these then reduce to derivatives of $h(\rho)$ through $(\partial_{\tilde{\sigma}} \rho )^2 = h(\rho)$ and $\partial_{\tilde{\sigma}}^2 \rho = h'(\rho)/2$); the former vanish, while the latter do not diverge. Thus, $g(\tilde{\sigma})$ and, similarly, $f(\tilde{\sigma})$ both become smooth as $\omega \to 1$.

Consequently, the patched Kruczenski string is an approximate solution which only becomes exact in the limit of large angular momentum, as the spikes approach the boundary of $AdS_3$. At $\omega = 1$, it reduces to a collection of arcs with endpoints on the boundary, such as the one shown in Fig. \ref{fig:JJ_single_arc_rho0=0.9} and represented by \eqref{eq:JJ_single_arc_solution_omega=1}, glued together to form a closed string. This is another indication of the fact that the solution is acceptable in the large $S$ limit.

\begin{figure}%
\includegraphics[width=\columnwidth]{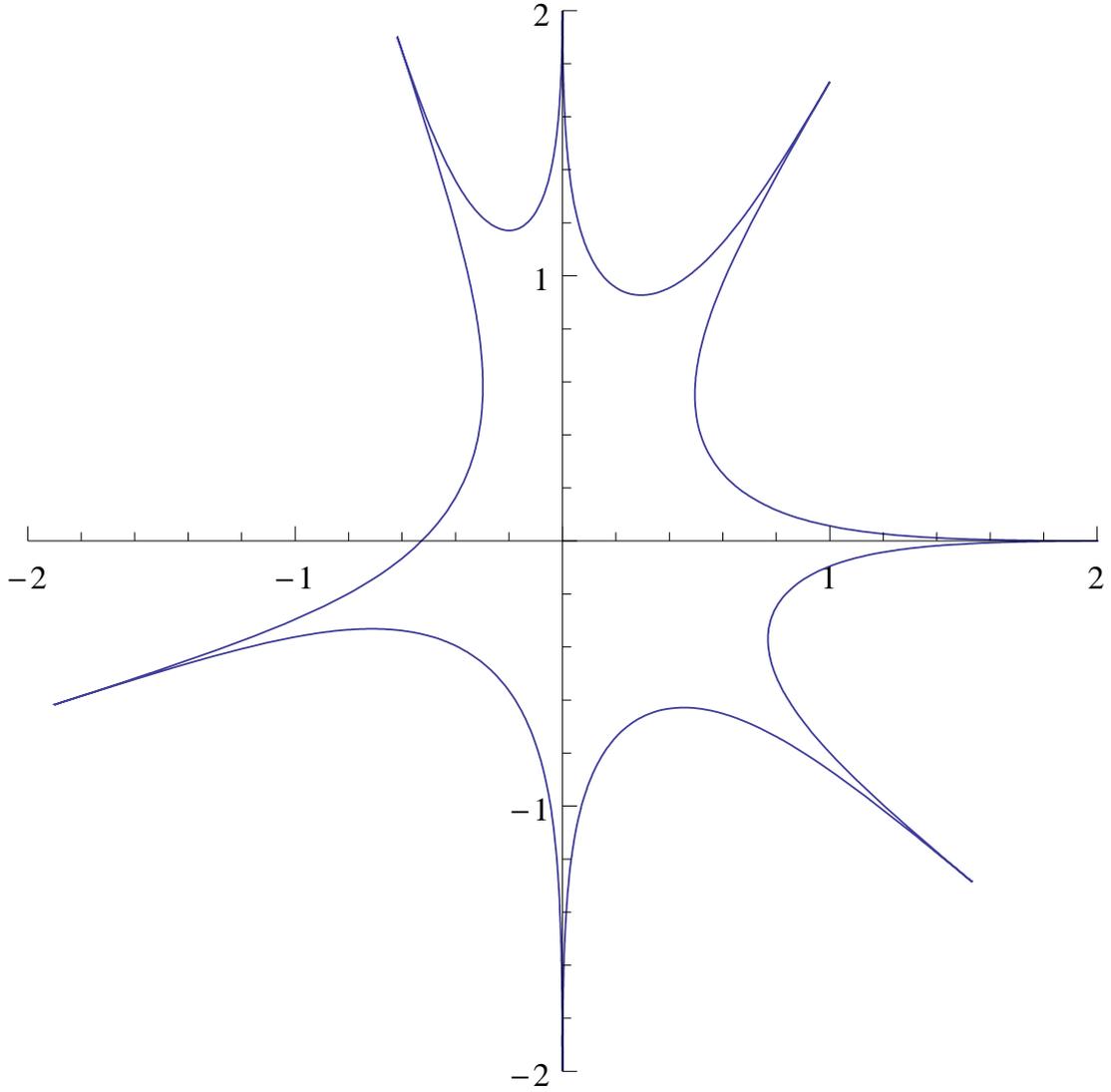}
\caption{The patched Kruczenski spiky string, with 7 spikes, angular separations equal to $\pi/3$, $\pi/6$, $\pi/10$, $\pi/2$, $2\pi/5$, $5\pi/18$, $2\pi/9$ and corresponding parameters given by $\rho_1 = 2$, $\rho_0^{(1)} = 0.638475$, $\rho_0^{(2)} = 0.965792$, $\rho_0^{(3)} = 1.18657$, $\rho_0^{(4)} = 0.43007$, $\rho_0^{(5)} = 0.546767$, $\rho_0^{(6)} = 0.727608$, $\rho_0^{(7)} = 0.833672$.}
\label{fig:patched_K_7_cusps}
\end{figure}

However, it is important to notice that the solution \eqref{eq:JJ_single_arc_solution_omega=1}, displays a different type of behaviour, since, by definition, it satisfies $\partial_{\tilde{\sigma}} \rho (\tilde{\sigma}) = h_2 (\rho)$, and thus now the first derivative of $\rho(\tilde{\sigma})$ no longer vanishes at the endpoints, where the spikes should be located. The reason is that in this solution we see the extreme consequences of the $\omega \to 1$ limit: the cusps are ``pushed away'' at infinity and eventually disappear from the worldsheet, which is now infinitely long. It is hence necessary to maintain $\omega > 1$ and then to study the type of limit we used just above, in order to keep track of the spikes and to be able to approximate the monodromy matrix as we have been doing in the previous sections.
\paragraph{}
We now return to the analysis of the general properties of the patched string. By construction, this solution, when plotted at constant $t$, has $K$ arcs of angular separation $\Delta\theta_j$, for $j=1,\ldots,K$, and hence the closedness condition is
\begin{equation}
 \sum_{j=1}^K \Delta\theta_j = 2 n \pi \:.
\label{eq:closedness_constraint_patched_K}
\end{equation}
As previously mentioned, we have $K$ cusps located at $\tilde{\sigma} = \tilde{\sigma}_j$, for $j=0,\ldots,K-1$ (the analysis carried out for the symmetric spiky string still applies, and thus we have $\partial_{\tilde{\sigma}} (\rho,\phi) = (0,0)$ at each cusp). We denote their angular positions by $\phi_j \equiv \omega \tilde{\tau} + \theta_j$, where, without loss of generality, we can assume $\theta_0 = 0$. We also define $\theta_{K} \equiv 2 n \pi - \theta_0 = 2 n \pi$, so that $\Delta\theta_j = \theta_j - \theta_{j-1}$ and $\sum_{j=1}^m \Delta\theta_j = \theta_m$. Another plot at constant time $t$ is shown in Fig. \ref{fig:patched_K_7_cusps}.

As usual, we will be interested in the limit of this solution as the spikes touch the boundary of $AdS_3$, i.e. $\omega\to 1$, in which $\rho_1 \to + \infty$ and $\rho_0^{(j)}$ approaches the value which satisfies the following equation, coming from \eqref{eq:Deltatheta_single_arc_JJ_omega_to_1}:
\begin{equation}
 \Delta\theta_j \simeq 2 \mathrm{Arctan} \frac{1}{\sinh 2\rho_0^{(j)}} \:, \qquad \textrm{as } \omega \to 1 \:.
\label{eq:Deltatheta_single_arc_patched_K_omega_to_1}
\end{equation}

\subsubsection{Conserved charges}

The contour defined in \eqref{eq:closed_contour_worldsheet_JJ} is still closed and winds once around the worldsheet. The energy and angular momentum can thus be calculated as in the Kruczenski case, and reduce to the sum of the usual Kruczenski-type contributions from each individual arc:
\begin{eqnarray}
 \Delta & = & \frac{\sqrt{\lambda}}{2\pi} \left( \sum_{j=1}^K \int_{2 \tilde{L}_{j-1}}^{2 \tilde{L}_j} d \tilde{\sigma}
               \cosh^2 \rho (\tilde{\sigma}, \rho_0^{(j)})
                + \frac{\omega \sinh 2 \rho_0^{(K)}}{2} \int_{\tilde{\tau}_0}^{\tilde{\tau}_0 - f(L)} d \tilde{\tau} \right) \nn\\
        & = & \sum_{j=1}^K \frac{\sqrt{\lambda}}{\pi} \sqrt{\frac{w_1-1}{w_1+w_0^{(j)}}} \left[ \frac{1}{2} (w_1+w_0^{(j)}) \mathbb{E}_j - 
               \sinh^2 \rho_0^{(j)} \mathbb{K}_j \right] \nn\\
        &   & \phantom{++} - \frac{\sqrt{\lambda}}{2 \pi} \sinh 2 \rho_0^{(K)} \sum_{j=1}^K \frac{\sqrt{2} \omega^2 \sinh 2 \rho_0^{(j)}
               \sinh \rho_1}{\sqrt{w_1 + w_0^{(j)}} (w_1 + 1)} \Pi (n_+^{(j)}, k^{(j)}) \nn\\
 S & = & \frac{\sqrt{\lambda}}{2\pi} \left( \omega \sum_{j=1}^K \int_{2 \tilde{L}_{j-1}}^{2 \tilde{L}_j} d\tilde{\sigma} 
          \sinh^2 \rho (\tilde{\sigma}, \rho_0^{(j)})
           + \frac{\sinh 2 \rho_0^{(K)}}{2} \int_{\tilde{\tau}_0}^{\tilde{\tau}_0 - f(L)} d \tilde{\tau} \right) \nn\\
   & = & \sum_{j=1}^K \frac{\omega\sqrt{\lambda}}{\pi} \sqrt{\frac{w_1-1}{w_1+w_0^{(j)}}} \left[ \frac{1}{2} (w_1+w_0^{(j)}) \mathbb{E}_j - 
          \cosh^2 \rho_0^{(j)} \mathbb{K}_j \right] \nn\\
   &   & \phantom{++} - \frac{\sqrt{\lambda}}{2 \pi} \sinh 2 \rho_0^{(K)} \sum_{j=1}^K \frac{\sqrt{2} \omega \sinh^2 2 \rho_0^{(j)}
          \sinh \rho_1}{\sqrt{w_1 + w_0^{(j)}} (w_1 + 1)} \Pi (n_+^{(j)}, k^{(j)}) \:.
\label{eq:E_S_pK}
\end{eqnarray}
where we have introduced $\tilde{L}_0 \equiv 0$.

As $\omega\to 1$, with $\omega = 1+\eta$, the leading behaviour of $\Delta$ and $S$ is unchanged with respect to the symmetric spiky string:
\begin{eqnarray}
 \Delta \simeq S = K \frac{\sqrt{\lambda}}{2 \pi} \frac{1}{\eta} + O ( \log \eta ) \:,
\label{eq:ES_lead_behaviour_omega_to_1_pK}
\end{eqnarray}
while the $O(1)$ correction to $\Delta - S$ is now dependent on the $K-1$ new moduli:
\begin{equation}
 \Delta - S = \frac{K\sqrt{\lambda}}{2\pi} \log \left( \frac{2\pi S}{K\sqrt{\lambda}} \right)
  + \frac{\sqrt{\lambda}}{2\pi} \left[ K (3\log 2 - 1) +
   \sum_{j=1}^K \log \left( \sin \frac{\Delta\theta_j}{2} \right) \right] + \ldots \:,
\label{eq:E-S_omega_to_1_behaviour_patched_K}
\end{equation}
where we have used:
\begin{equation}
 \frac{1}{w_0^{(j)}} \simeq \sin \frac{\Delta\theta_j}{2} \qquad \textrm{as } \omega \to 1
\label{eq:relation_between_u0j_and_sin_Deltathetaj/2}
\end{equation}
which generalises \eqref{eq:relation_between_u0_and_sin_Deltatheta/2}.

\subsubsection{Spectral curve for large S}

As always, we introduce the rescaled worldsheet coordinates, defined as
\begin{equation}
 (\tau,\sigma) = \frac{2\pi}{L} (\tilde{\tau},\tilde{\sigma}) 
\label{eq:def_rescaled_coords_patched_K}
\end{equation}
in terms of which the cusp positions become
\begin{equation}
 \sigma_j = \frac{4\pi}{L} \sum_{k=1}^j \tilde{L}_k 
   \:, \qquad \textrm{for } j=0,\ldots,n-1 \:.
\label{eq:cusp_positions_patched_K_sigma}
\end{equation}
The rescaled charge density is then given by
\begin{eqnarray}
 j^0_\tau(\tau,\sigma) & = & \frac{L}{2\pi} \left\{ (\omega+1) \left[ w_1 \mathrm{cn}^2 \left(
  \left. v^{(j)} \left(\frac{L}{2\pi}(\sigma-\sigma_{j-1})\right) \right| k^{(j)} \right) \right. \right. \nonumber \\
  & & \left. + w_0^{(j)} \left. \mathrm{sn}^2 \left( \left. v^{(j)}\left(\frac{L}{2\pi}(\sigma-\sigma_{j-1})\right) \right|k^{(j)} \right)
   \right] + 1-\omega  \right\} \nonumber \\
 j^1_\tau(\tau,\sigma) + i j^2_\tau(\tau,\sigma) & = & i \frac{L}{2\pi} (\omega+1) e^{i(\phi-t)} \nonumber \\
  & & \times \left\{ \left[ w_1 \mathrm{cn}^2 \left( \left. v^{(j)}\left(\frac{L}{2\pi}(\sigma-\sigma_{j-1})\right) \right|k^{(j)} \right)
   \right. \right. \nonumber \\
  & & + \left. \left. w_0^{(j)} \mathrm{sn}^2 \left( \left. v^{(j)}\left(\frac{L}{2\pi}(\sigma-\sigma_{j-1})\right) \right|k^{(j)} \right)
   \right]^2 - 1\right\}^\frac{1}{2}\nonumber \\
  & & \qquad\qquad\qquad\qquad\qquad\qquad \textrm{for } \sigma \in [\sigma_{j-1},\sigma_j] \:.
\label{eq:eq:j_tau_in_terms_of_tau_sigma_patched_K}
\end{eqnarray}
As before, we express $\sigma$ near the $m$-th spike as $\sigma = \sigma_m + \hat{\sigma}$, where $\hat{\sigma}$ is never allowed to reach one half of the distance to the nearest cusp in both directions. For the patched Kruczenski solution, this translates into an asymmetric condition on $\hat{\sigma}$ (since the fundamental periods $\tilde{L}_j$ are in general different from each other), which, however, reduces to the usual $|\hat{\sigma}| \leq \pi/K$ as $\omega \to 1$ (since all $\tilde{L}_j$ become identical in this limit).

At this point, it is only a matter of tedious algebra to carry out the usual expansion of the elliptic functions and integrals as $\omega \to 1$ and obtain:
\begin{eqnarray}
 j_\tau^0 (\tau,\sigma) & \simeq & \frac{2 \mathbb{K}}{\pi} \frac{1}{\eta \cosh^2 \left( \frac{\mathbb{K}}{\pi} \hat{\sigma} \right)} \nonumber\\
 j^1_\tau(\tau,\sigma) + i j^2_\tau(\tau,\sigma) & \simeq & i \frac{2 \mathbb{K}}{\pi} \frac{1}{\eta \cosh^2 \left( \frac{\mathbb{K}}{\pi}
  \hat{\sigma} \right)} e^{i \sum_{j=1}^m \Delta\theta_j} 
\label{eq:leading_behaviour_of_jtau_patched_K}
\end{eqnarray}
where $\mathbb{K} \equiv \sum_{j=1}^K \mathbb{K}_j$. As before, we recover the $\delta$-function localisation as in \eqref{eq:def_spin_vector_at_each_cusp}, which we use to compute the spin vectors at each cusp:
\begin{equation}
 L_m = \frac{S}{K} \begin{pmatrix}
	                  1 \\
	                  -\sin \left( \sum_{j=1}^m \Delta\theta_j \right) \\
	                  \cos \left( \sum_{j=1}^m \Delta\theta_j \right)
                   \end{pmatrix} = \frac{S}{K} \begin{pmatrix}
	                                                   1 \\
	                                                   -\sin \theta_m \\
	                                                   \cos \theta_m
                                                    \end{pmatrix}
\label{eq:spin_vector_at_mth_cusp_patched_K}
\end{equation}
(we recall that $\theta_0 = 0$). This time, when testing the properties \eqref{eq:properties_of_spin_vectors}, we find that the highest-weight condition is not satisfied at leading order, namely that $Q_R^\pm \simeq \sum_{j=0}^{K-1} L^\pm_j \neq 0$. As we discussed in section \ref{sec:non-hw_extension}, this leads to the appearance of an extra modulus $\hat{q}^{(0)}_2 \neq -1$ on the gauge side, which is determined by the $SU(1,1)$ rotation linking the string solution to the corresponding highest-weight state. In particular, from \eqref{eq:Q_R_L_non-hw_in_terms_of_Delta-S}, we obtain
\begin{equation}
 i \frac{\bar{\alpha} \beta}{|\alpha|^2 + |\beta|^2} 4 S \simeq \frac{\sqrt{\lambda}}{\pi} \frac{i}{\eta} \sum_{m=0}^{K-1} e^{i \theta_m} \:.
\label{eq:explicit_Q_R_+_rel_with_beta_pK}
\end{equation}
As far as the left charge is concerned, we have a subleading non-vanishing contribution to $Q_L^+$, which, as usual, does not modify the spectrum.

We proceed along the usual path and calculate the matrix $\mathbb{L}_m (u)$ from \eqref{eq:spin_vector_at_mth_cusp_patched_K}:
\begin{equation}
 \mathbb{L}_m (u) = \begin{pmatrix}
	                           u + \frac{i}{K} & \frac{i}{K} e^{i \theta_m} \\
	                           - \frac{i}{K} e^{-i \theta_m} & u - \frac{i}{K}
                            \end{pmatrix} 
\label{eq:L_m(utilde)_for_patched_K}
\end{equation}
As expected, this matrix coincides with the one computed for the original Kruczenski solution if we choose $\Delta\theta_j = 2 n \pi/K$, $\forall j$, and therefore, under this condition, all the subsequent results will reduce to those we obtained for that solution.

Due to the arbitrariness of the angular separations $\Delta\theta_j$, the procedure we used in order to calculate $\mathrm{Tr} \: \Omega(u)$ for the symmetric Kruczenski string is no longer effective. Nonetheless, the trace of the monodromy matrix is clearly still in the form \eqref{eq:spectral_curve_explicit_sol-s} and it is possible, through a tedious calculation which however only involves elementary reasoning, to show that the conserved charges $\hat{q}^{(0)}_k$ can be expressed as
\begin{multline}
 \hat{q}^{(0)}_k = 2 \left( \frac{1}{K} \right)^k \sum_{r=0}^{\left[ \frac{k}{2} \right]} (-1)^r \sum_{\begin{array}{l}
	                                                                                 d_1,\ldots,d_r = 0,\ldots,K-2r \\
	                                                                                 D \leq K-2r
                                                                                  \end{array}} C(k,K,r,D)\\
    \times \sum_{\begin{array}{l}
	          j_1,\ldots,j_r = 2,\ldots,K \\
	          j_{l+1} > j_l + d_l + 1
           \end{array}}
    \mathrm{Re} \left[ i^k e^{-i \sum_{l=1}^r (\theta_{j_l+d_l} - \theta_{j_l - 1})} \right] \:,
\label{eq:qtilde_k_from_trace_patched_K}
\end{multline}
where we define
\begin{equation}
 \begin{gathered}
  D \equiv \sum_{l=1}^r d_l \:, \\
  C(k,K,r,D) \equiv \sum_{j=\max\{k-K+D,0\},\ldots,\min\{D,k-2r\}} (-1)^j \binom{K-2r-D}{k-2r-j} \binom{D}{j} \:.
 \end{gathered}
\label{eq:def_of_D_Deltatheta_jd_and_C(knrD)_patched_K}
\end{equation}
Since the method we used in obtaining \eqref{eq:qtilde_k_from_trace_patched_K} defines $\hat{q}^{(0)}_k$ as the coefficient of $1/u^k$ in $\mathrm{Tr} \: \Omega(u)$, this expression also correctly reproduces the known term, $\hat{q}^{(0)}_0 = 2$, and the missing linear term in $1/u$, $\hat{q}^{(0)}_1 = 0$.

A less unwieldy expression for $\hat{q}^{(0)}_k$ can be obtained by identifying the string theory equivalents of the spin chain variables $z_k$ and $p_k$ introduced in \cite{Korchemsky:1997yy}, for $k = 1, \ldots, K$ (where $K$ is the number of spins in the chain, which matches the number of cusps). As described in \cite{Korchemsky:1997yy}\footnote{In their notation, $L_3 \equiv \sum_{k=1}^N \mathcal{L}_k^0$, $L_+ \equiv \sum_{k=1}^N \mathcal{L}_k^+$ and $L_- \equiv \sum_{k=1}^N \mathcal{L}_k^-$.}, these parameters are related to the individual spin vectors $\mathcal{L}_k$ at each site of the chain:
\begin{equation}
 \mathcal{L}_k^0 = i z_k p_k \:, \quad \mathcal{L}_k^+ = i z_k^2 p_k \:, \quad \mathcal{L}_k^- = -i p_k \:.
\label{eq:spin_chain_spin_vectors_at_each_site_in_terms_of_zk,pk_patched_K}
\end{equation}
We relate these to the spin vectors at each cusp \eqref{eq:spin_vector_at_mth_cusp_patched_K} according to \eqref{eq:L_spin_string_identification}. It is now straightforward to obtain:
\begin{equation}
 p_k = - \frac{i S}{K} e^{-i \theta_k} \:, \qquad z_k = e^{i \theta_k} \:.
\label{eq:identification_of_spin_chain_parameters_in_stringy_calculation_patched_K}
\end{equation}
Then, the $k$-th conserved charge is given by
\begin{equation}
 \hat{q}^{(0)}_k = \sum_{1\leq j_1 < j_2 < \ldots < j_k \leq K} z_{j_1 j_2} z_{j_2 j_3} \ldots z_{j_{k-1} j_k} z_{j_k j_1} p_{j_1} p_{j_2}
  \ldots p_{j_k} \:, \quad k=2, \ldots, K \:,
\label{eq:q_k_from_spin_chain_patched_K}
\end{equation}
where $z_{ab} = z_a - z_b$. After some algebra, we can recast this expression into the following form:
\begin{equation}
 \hat{q}^{(0)}_k =  \left( - \frac{2}{K} \right)^k \sum_{1\leq j_1 < j_2 < \ldots < j_k \leq K} \prod_{l=1}^k \sin \left(
  \frac{\theta_{j_{l+1}} - \theta_{j_l}}{2} \right) \:, \quad k=2, \ldots, K
\label{eq:qktilde_from_spin_chain_patched_K}
\end{equation}
where we have defined $j_{k+1} \equiv j_1$. A lengthy calculation shows that this expression equals \eqref{eq:qtilde_k_from_trace_patched_K} for $k\geq 2$, and thus yields the conserved charges from $\mathrm{Tr} \: \Omega$. We can hence use this result, together with \eqref{eq:explicit_Q_R_+_rel_with_beta_pK}, to check that the relationship \eqref{eq:q2hat_extra_modulus} between $\hat{q}^{(0)}_2$ and rotation parameter $\beta$ is satisfied. In appendix \ref{sec:app_computing_qtilde_2_for_N-folded_GKP_and_K}, we show that $\hat{q}^{(0)}_2$ has a complicated dependence on the angular separations $\Delta \theta_j$ and that, in general, $\hat{q}^{(0)}_2 \neq - 1$. The parameter is however equal to $-1$ when all the $\Delta\theta_j$ are equal, which, as we can see from \eqref{eq:spin_vector_at_mth_cusp_patched_K}, makes the patched solution a highest-weight state. This is exactly the behaviour we expected from the general finite-gap picture.

As usual, we are interested in the highest conserved charge, which is given by:
\begin{eqnarray}
 \hat{q}^{(0)}_K & = & \left( - \frac{2}{K} \right)^K \prod_{l=1}^K \sin \left( \frac{\theta_{l+1} - \theta_l}{2} \right) \nonumber\\
             & = & \left( - \frac{2}{K} \right)^K \sin \left( \frac{\Delta\theta_1 - 2 n \pi}{2} \right) \prod_{l=1}^{K-1} \sin \left(
              \frac{\Delta\theta_{l+1}}{2} \right) \nonumber \\
             & = & \left( - \frac{2}{K} \right)^K (-1)^{n} \prod_{l=1}^{K} \sin \left( \frac{\Delta\theta_{l}}{2} \right) \:,
\label{eq:qtilde_n_for_patched_K}
\end{eqnarray}
where we have used $\theta_1 - \theta_K = - \sum_{m=2}^K \Delta\theta_m = - 2 n \pi + \Delta\theta_1$ in deriving the second line. We can now substitute this into \eqref{eq:general_Delta-S_finite-gap_explicit_sol-s} in order to obtain the finite-gap spectrum $\Delta-S$:
\begin{eqnarray}
 \Delta - S & = & \frac{K\sqrt{\lambda}}{2\pi} \log S + \frac{\sqrt{\lambda}}{2\pi} \left[ K \log 2 - K \log K + \log (-1)^{K + n}
   \phantom{\sum_{j=1}^K} \right. \nonumber\\
 & & \left. + \sum_{j=1}^K \log \left( \sin \frac{\Delta\theta_j}{2} \right) + C_{\mathrm{string}}(K) \right]
\label{eq:gauge_theory_prediction_for_E-S_patched_K} \nonumber
\end{eqnarray}
(where, as always, we have omitted subleading terms as $\omega \to 1$) which, by comparison with the direct result \eqref{eq:E-S_omega_to_1_behaviour_patched_K}, implies:
\begin{equation}
 C_{\mathrm{string}}(K) = K \left[ \log \left( \frac{8 \pi}{\sqrt{\lambda}} \right) - 1 \right] - \log (-1)^{K + n} \:.
\label{eq:C(n)_from_patched_K} \nonumber
\end{equation}
This expression agrees with those obtained in the previous two cases, \eqref{eq:C(n)_from_JJ} and \eqref{eq:C(n)_from_Nfolded_GKP} (in order to obtain the latter, we must set $K$ even and $n = K/2$, as we saw before).

Finally, we observe that, as we noticed earlier in this section, it is possible to obtain all the previous results concerning the N-folded GKP string and the Kruczenski string from this generalised version. For instance, it is easy to check that the highest conserved charge \eqref{eq:qtilde_n_for_patched_K} reduces to the expression \eqref{eq:highest_conserved_charge_JJ} valid in the Kruczenski case if we set $\Delta\theta_j = 2 n \pi /K$, $\forall j$.

\section{``Small'' spikes in $AdS_3$}
\label{sec:small_spikes}

This section is based on the results of \cite{Dorey:2010iy} and is devoted to the discussion of explicit string solutions exhibiting, in the large angular momentum limit, the ``small'' spikes which we introduced in section \ref{sec:FGA_interpretation}, where we also computed their dispersion relation \eqref{eq:E(v)}, \eqref{eq:P(v)},
\begin{eqnarray}
 E_{\rm sol} (v) & = & \frac{\sqrt{\lambda}}{2 \pi} \left[ \frac{1}{2} \log \left( \frac{1 + \sqrt{1-v^2}}{1 - \sqrt{1-v^2}} \right) -
  \sqrt{1-v^2} \right] \nn\\
 P_{\rm sol}(v) & = & \frac{\sqrt{\lambda}}{2 \pi} \left[ \frac{\sqrt{1-v^2}}{v} - \mathrm{Tan}^{-1} \left( \frac{\sqrt{1-v^2}}{v} \right)
  \right] \:.
\label{eq:small_spike_disprel}
\end{eqnarray}
Together with the dispersion relation, these solutions will also reproduce the finite-gap semiclassical quantisation condition \eqref{eq:semiclassical_quant_condition_P(c)} in the form
\begin{equation}
 P_{\rm sol} (v) \: (K-M) \log S \in 2 \pi \mathbb{Z} \:.
\label{eq:small_spike_P_quantisation_condition}
\end{equation}
We recall that we were originally able to restate the integrality constraint for the filling fractions in this way by assuming that the length of the string along which the ``small'' spikes were propagating was
\begin{equation}
 L = (\Delta - S) (2 \pi / \sqrt{\lambda}) \simeq (K-M) \log S
\label{eq:def_L_string_and_relation_with_Delta-S}
\end{equation}
at leading order as $S \to \infty$, where $K-M$ is the number of ``large'' spikes on the string. This was in fact the case with all the previously considered explicit solutions displaying ``large'' spikes (see sections \ref{sec:NGKP}, \ref{sec:JJ} and \ref{sec:patching_K}), which include the GKP string, and will be true also for the solutions which we will study in this section.

The semiclassical dispersion relation \eqref{eq:small_spike_disprel}, \eqref{eq:small_spike_P_quantisation_condition} was then found to agree with the ``small'' hole dispersion relation from gauge theory, in the limit of large momentum $P_{\rm sol}$ and only in the special case $K-M=2$, which will apply to all the solutions considered here.
\paragraph{}
Apart from the fact that they constitute relatively simple explicit examples of the string behaviour predicted by the finite-gap analysis, these objects are also interesting as the $AdS_3$ version of the Giant Magnons discussed in \cite{Hofman:2006xt}.

In particular, all the solutions we will consider describe ``small'' spikes moving along the infinite GKP string, which is just the solution discussed in section \ref{sec:NGKP}, in the special case $N=1$ and $\omega = 1$, and therefore has $K-M=2$ ``large'' spikes and $M=0$ ``small'' spikes. The GKP string is interpreted as the reference vacuum state and its spectrum of small quadratic fluctuations was computed in \cite{Alday:2007mf}, where these were found to correspond to a single transverse mode with a relativistic dispersion relation, $O((\sqrt{\lambda})^0)$ energy and mass $m^2 = 4$.

The ``small'' spikes can also be thought of as excitations over the GKP vacuum, but they carry a large, albeit finite, $O (\sqrt{\lambda})$ energy and their dispersion relation \eqref{eq:small_spike_disprel} is not relativistic, due to the fact that the Lorentz invariance of the string action and Virasoro constraints in conformal gauge is broken by the fixing of the residual gauge symmetry. The relativistic behaviour is however restored in the low-momentum regime, where we recover the dispersion relation for a massless particle:
\begin{equation}
 E_{\rm sol} \simeq |P_{\rm sol}| + {\rm O} \left( P^{\frac{5}{3}}_{\rm sol} \right)  \:.
\label{eq:E(P)_small_spikes_at_low_P}
\end{equation}
This also happens in the case of Giant Magnons and allows their identification with the small fluctuations over the BMN vacuum. In fact, since a string in $AdS_3$ only has one transverse mode, it is natural to conjecture that ``small'' spikes should be continuously related to the small fluctuations of \cite{Alday:2007mf}, representing their large energy limit. This picture is consistent at the semiclassical level, since the $O(1)$ mass of these excitations becomes negligible at $O( \sqrt{\lambda} )$, and it is therefore acceptable to have a massless dispersion relation at leading order. A more accurate test of this proposal would require the computation of the first quantum correction, $O(1)$, which should then introduce the appropriate mass term for the quadratic fluctuations. However, such a test is no longer necessary, as the above conjecture has been found true in \cite{Basso:2010in}.

``Small'' spikes also correspond to sinh-Gordon solitons propagating along the infinitely long GKP string and undergoing factorised scattering, as a consequence of the integrability of semiclassical string theory. This is in perfect analogy with the case of Giant Magnons, which were found to represent sine-Gordon solitons.

Due to all these similarities between the Giant Magnons and the ``small'' spikes, and due to their conjectured duality to the ``small'' holes of gauge theory, we proposed the name ``Giant Holes'' for these objects.
\paragraph{}
Before we begin the analysis of the two string solutions containing Giant Holes, we will briefly summarise the facts we are interested in concerning the GKP vacuum.

\begin{figure}%
\begin{center}
\includegraphics{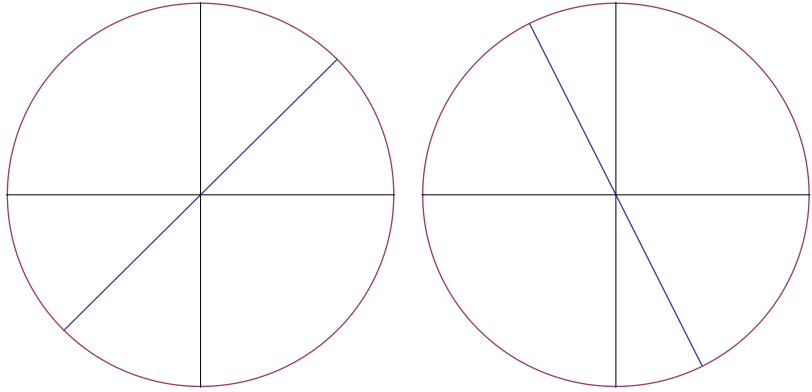}%
\end{center}
\caption{The infinite GKP string is represented as a straight line extending up to the boundary of $AdS_3$, rigidly rotating around the centre.}%
\label{fig:plot_GKP_infinite}%
\end{figure}

For $\omega = 1$, the solution \eqref{eq:GKP_rho_of_sigmatilde} simplifies to:
\begin{eqnarray}
 Z_1 & = & e^{i\tau} \cosh \sigma \nonumber \\
 Z_2 & = & e^{i\tau} \sinh \sigma \:,
\label{eq:infinite_GKP_string}
\end{eqnarray}
where now $\sigma \in (- \infty, \infty)$ and the plot shows a straight line passing through the centre of $AdS_3$ and extending up to the boundary, rigidly rotating (see Fig. \ref{fig:plot_GKP_infinite}). In particular, when $\sigma$ varies over its full range, the above set of equations only describes one such line, so that we need two copies of this simplified solution, glued together at the endpoints on the boundary, in order to obtain a closed string, recovering the 1-folded GKP solution.

The asymptotics of the vacuum near its endpoints are given by
\begin{eqnarray}
 Z_1 & \simeq & \frac{1}{2} e^{\pm \sigma} e^{i \tau} \:, \qquad \qquad \textrm{as } \sigma \to \pm \infty \nn\\
 Z_2 & \simeq & \pm \frac{1}{2} e^{\pm \sigma} e^{i \tau} \:, \qquad \qquad \textrm{as } \sigma \to \pm \infty \:,
\label{eq:vac_asymptotics_Z_small_spikes}
\end{eqnarray}
or equivalently
\begin{eqnarray}
 t & \to & \tau \:, \qquad \qquad \textrm{as } \sigma \to \pm \infty \nn\\
 \rho & \to & \pm \sigma \:, \qquad \qquad \textrm{as } \sigma \to \pm \infty \nn\\
 \phi & \to & \begin{cases}
               \tau \:, \quad \textrm{for } \sigma \to + \infty \\
               \tau + \pi \:, \quad \textrm{for } \sigma \to - \infty
              \end{cases} \:.
\label{eq:vac_asymptotics_global_small_spikes}
\end{eqnarray}
Since the excitations, or ``small'' spikes, are localised solitonic objects propagating along the GKP string, we expect them not to significantly modify the asymptotic behaviour of the vacuum. In fact, the two solutions we will discuss below reproduce the vacuum asymptotics.

The energy and angular momentum of this solution are infinite. Hence we will regulate the problem by considering instead a closed string of finite length with folds at radial distance $\rho= \Lambda \gg 1$. Up to subleading corrections, this corresponds to the same solution but with the range of the worldsheet coordinate restricted as $-\Lambda\leq \sigma \leq +\Lambda$. We then find
\begin{eqnarray}
 \Delta & = &  \frac{\sqrt{\lambda}}{\pi} \int_{-\Lambda}^{\Lambda} d\sigma \cosh^2 \sigma = \frac{\sqrt{\lambda}}{4\pi} 
  \left( e^{2\Lambda} + 4 \Lambda - e^{-2\Lambda} \right) \nonumber \\
 S & = & \frac{\sqrt{\lambda}}{\pi} \int_{-\Lambda}^{\Lambda} d\sigma \sinh^2 \sigma = \frac{\sqrt{\lambda}}{4\pi}
  \left( e^{2\Lambda} - 4 \Lambda - e^{-2\Lambda} \right) \:.
\label{Delta_S_GKP_vacuum}
\end{eqnarray}
Subtracting the regulated string energy and angular momentum we obtain the standard formula, 
\begin{eqnarray}
E_{0} = \Delta - S & = & \frac{\sqrt{\lambda}}{\pi} 2\Lambda \nn \\
                   & = & \frac{\sqrt{\lambda}}{2 \pi}\left[ 2 \log \left( \frac{\pi S}{\sqrt{\lambda}} \right) + 4 \log 2 +
                         O \left( \frac{\log S}{S} \right) \right] \:,
\label{eq:E-S_for_vacuum}
\end{eqnarray}
which we define as the reference vacuum energy, so that any excitation will raise $\Delta - S$ to a higher value $E > E_0$. Note that the spectrum \eqref{Delta_S_GKP_vacuum} is slightly different from the $K=2$ case of the large $S$ limit of the spectrum \eqref{eq:E_S_omega_to_1_behaviour} of the $N$-folded GKP string. The missing term is moduli-independent and is probably due to the particular choice of regulator we have made.

Finally, the infinite GKP string \eqref{eq:infinite_GKP_string} corresponds to the vacuum solution \eqref{eq:def_sinhG-vacuum} of the sinh-Gordon equation. As anticipated in section \ref{sec:NGKP}, this is due to the fact that we are observing the solution at $S = \infty$, with the two ``large'' spikes touching the boundary of $AdS_3$, and hence the corresponding two static solitons have disappeared from the worldsheet, sucked into the region $\sigma = \pm \infty$. This will also be the case with the solutions describing ``small'' spikes propagating along the GKP string: only the ``small'' spikes will be visible in the sinh-Gordon picture. Nonetheless, as far as the spectrum is concerned, the behaviour will be appropriate for a solution with 2 ``large'' spikes plus some ``small'' spikes.

\subsection{The two-spike solution}
\label{sec:2s}

In \cite{Jevicki:2007aa} several explicit string solutions in $AdS_3$ were constructed, corresponding to solutions of the sinh-Gordon equation describing respectively a single \mbox{(anti-)}soliton $(\bar{s})s$, two solitonic objects of any kind scattering off each other and also the so-called sinh-Gordon breather, which is a bound state of one soliton and one anti-soliton. As we saw in section \ref{sec:review_sinhG_soliton_solutions}, each soliton is identified with a divergence $\alpha \to - \infty$ in the sinh-Gordon field, while for each anti-soliton we have $\alpha \to + \infty$. According to the definition \eqref{eq:def_alpha_Pohlmeyer}, this corresponds to
\begin{eqnarray}
 \partial_+ X_\mu \partial_- X^\mu & = & - e^\alpha \to 0 \:, \qquad \qquad \textrm{soliton} \nn\\
 \partial_+ X_\mu \partial_- X^\mu & = & - e^\alpha \to \infty \:, \qquad \qquad \textrm{anti-soliton} \:.
\label{eq:target_space_singularities_sol_antisol}
\end{eqnarray}
Therefore, it is reasonable to expect that the introduction of an anti-soliton onto the GKP string should lead to the appearance of a new singularity, which would modify the leading large $S$ behaviour \eqref{eq:E-S_for_vacuum} of the spectrum. The excitations we are interested in should instead contribute a finite $O(\sqrt{\lambda})$ amount to the vacuum energy $E_0$. A good candidate for such an excitation is therefore the sinh-Gordon soliton.
\paragraph{}
The 2-soliton soliton scattering solution of \cite{Jevicki:2007aa} is given by
\begin{eqnarray}
 Z_1^{ss} & = & e^{i\tau} \frac{v \mathrm{ch} T \mathrm{ch} \sigma + \mathrm{ch} X \mathrm{ch} \sigma - \sqrt{1-v^2} \mathrm{sh} X \mathrm{sh}
  \sigma - i \sqrt{1-v^2} \mathrm{sh} T \mathrm{ch} \sigma} {\mathrm{ch} T + v \mathrm{ch} X} \nonumber \\
 Z_2^{ss} & = & e^{i\tau} \frac{v \mathrm{ch} T \mathrm{sh} \sigma + \mathrm{ch} X \mathrm{sh} \sigma - \sqrt{1-v^2} \mathrm{sh} X \mathrm{ch}
  \sigma - i \sqrt{1-v^2} \mathrm{sh} T \mathrm{sh} \sigma}{\mathrm{ch} T + v \mathrm{ch} X} \:, \nn\\
  & &
\label{eq:2s_solution}
\end{eqnarray}
where $X = 2 \gamma \sigma$, $T = 2 \gamma v \tau$, $\gamma = 1/\sqrt{1-v^2}$, we have used the abbreviated notation $\mathrm{ch} X \equiv \cosh X$ and $\mathrm{sh} X \equiv \sinh X$, and $0<v<1$ is a parameter related to the centre of mass velocities of the two solitons, which are given by $+v$ and $-v$. The sinh-Gordon solution corresponding to this string through Pohlmeyer reduction is \eqref{eq:def_sinhG_2s2asa_scattering} (first line, upper sign choice), from which we deduce that the worldsheet positions of the two solitons are given by the solutions $\sigma_1 (\tau)$ and $\sigma_2(\tau)$ of the equation $\cosh T = v \cosh X$. As in the vacuum case, the two static solitons located at the endpoints $\sigma = \pm \infty$ are invisible in the Pohlmeyer-reduced picture.

The fact that $v>0$ in the parametrisation of this solution is of particular importance, since it makes the denominators of $Z_1$ and $Z_2$ always greater than zero for finite $\sigma$. The other two scattering solutions ($s \bar{s}$ and $\bar{s} \bar{s}$) and the breather solution discussed in \cite{Jevicki:2007aa} do not have this property and therefore their radial coordinate $\rho$ becomes infinite for finite values of $\sigma$, corresponding to the positions of the anti-solitons. This extra singularity adds a further infinite contribution to the quantity $\Delta - S$ and hence anti-solitons do not seem to be equivalent to excitations of finite energy over the GKP vacuum.

\begin{figure}%
\includegraphics[width=0.83\columnwidth]{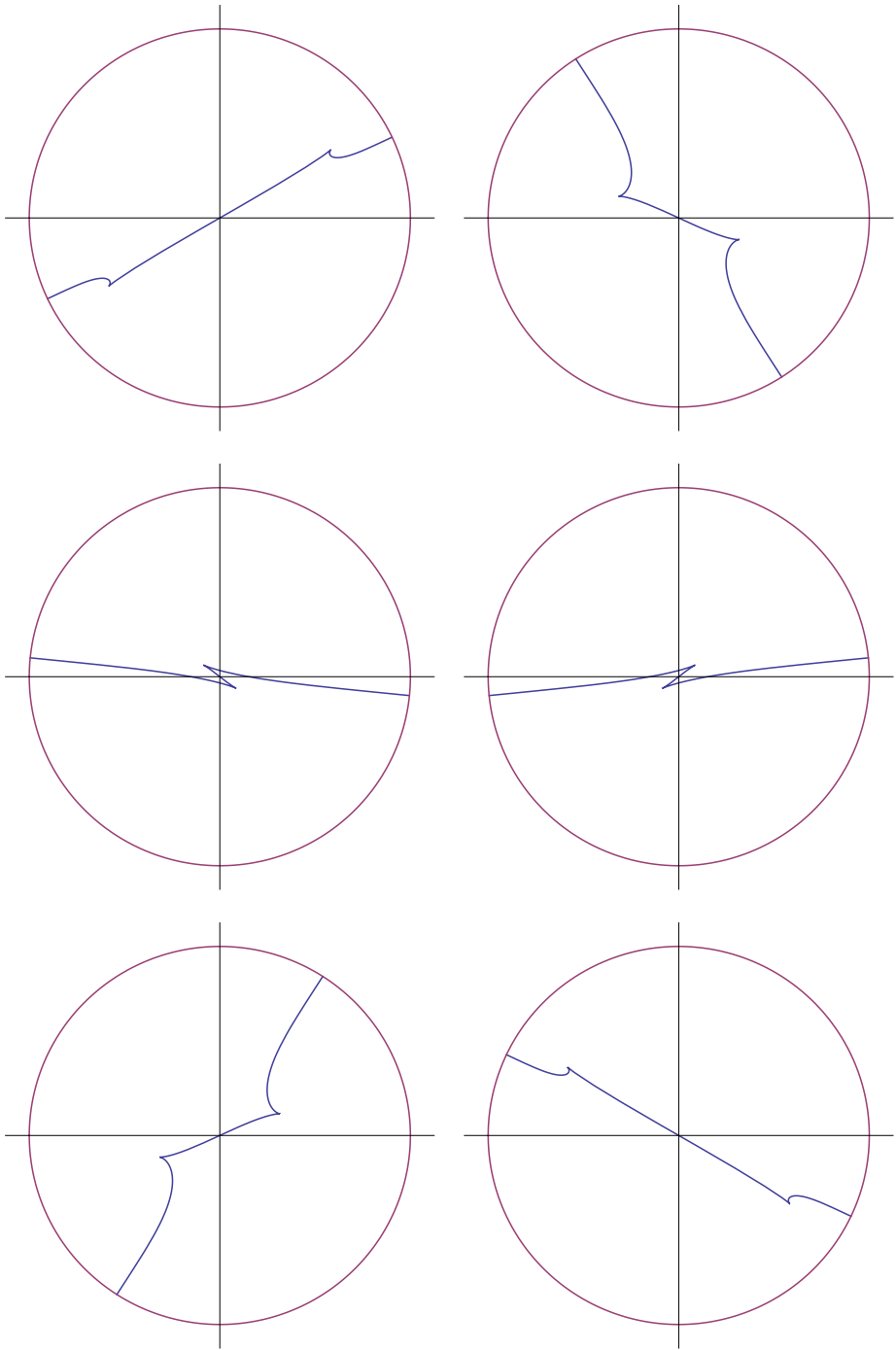}%
\caption{The 2-soliton solution at $t=-2.7,-1,-0.1,0.1,1,2.7$ (from
  left to right and top to bottom). At $t=0$ the string coincides 
with the horizontal axis.}%
\label{fig:Plot_2s_for_6_values_of_t}%
\end{figure}

Several plots of the $ss$ solution at constant global time $t$ are shown in Fig. \ref{fig:Plot_2s_for_6_values_of_t}. This solution has two small spikes located at the positions of the two solitons, which start at the endpoints of the string, then approach each other until they scatter at the origin and then move away towards the endpoints. Interestingly, the same solution plotted at constant worldsheet time $\tau$ has no cusps. As remarked at the beginning of section \ref{sec:JJ}, the real plot of the solution is the one at constant global time and thus this string has spikes. The difference in the plots is due to the fact that $t = {\rm const.}$ is not equivalent to $\tau = {\rm const.}$ in conformal gauge.

The asymptotic behaviour of the solution at the endpoints is given by
\begin{eqnarray}
 Z_{1} & \simeq & \frac{1}{2}e^{\pm \sigma} e^{i\tau} \left( \frac{1-\sqrt{1-v^{2}}}{v} \right)  \nn \\
 Z_{2} & \simeq & \pm \frac{1}{2}e^{\pm \sigma} e^{i\tau}\left( \frac{1-\sqrt{1-v^{2}}}{v} \right)
\label{eq:2s_asymptotics_Z}
\end{eqnarray}
yielding
\begin{eqnarray}
 t & \to & \tau \:, \qquad \qquad \textrm{as } \sigma \to \pm \infty \nn\\
 \rho & \to & \pm \sigma + \alpha \:, \qquad \qquad \textrm{as } \sigma \to \pm \infty \nn\\
 \phi & \to & \begin{cases}
               \tau \:, \quad \textrm{for } \sigma \to + \infty \\
               \tau + \pi \:, \quad \textrm{for } \sigma \to - \infty
              \end{cases} \:.
\label{eq:2s_asymptotics_global}
\end{eqnarray}
This reproduces the vacuum asymptotics \eqref{eq:vac_asymptotics_Z_small_spikes} and \eqref{eq:vac_asymptotics_global_small_spikes}, up to a shift in the radial coordinate, $\alpha = \log [(1- \sqrt{1-v^2})/v]$, which will be important in the following.

We also notice that, since $t = \tau$ at both endpoints, it is possible to glue this solution to a straight infinite GKP string in order to obtain a closed string, thus obtaining a solution which is closed both at constant global time and at constant worldsheet time. Therefore, no timelike segment is required in the contour involved in the definition of the conserved charges for this solution.

The total quantity $\Delta - S$ for such a closed string is given by one half the contribution of the infinite 1-folded GKP string (i.e. one half of \eqref{eq:E-S_for_vacuum}), plus the two-soliton contribution. The latter is also infinite and hence we regulate it in our usual gauge-invariant way, i.e. by imposing $\rho = \Lambda \gg 1$. The asymptotics \eqref{eq:2s_asymptotics_global} then imply we must restrict the spatial worldsheet coordinate to the range $-\tilde{\Lambda}\leq \sigma \leq +\tilde{\Lambda}$ with $\tilde{\Lambda}=\Lambda-\delta\Lambda$, where
\begin{equation}
 \delta \Lambda = \log\left( \frac{1-\sqrt{1-v^{2}}}{v}\right) = \frac{1}{2} \log \left( \frac{1-\sqrt{1-v^{2}}}{1+\sqrt{1-v^{2}}} \right) \:.
\label{eq:def_delta_Lambda_2s}
\end{equation}
Note that the second equality is legitimised by the fact that $v>0$ and that the shift $\delta \Lambda$ vanishes for $v=1$ where the two-soliton solution simply reduces to the vacuum solution \eqref{eq:infinite_GKP_string}. On the other hand, the $\delta \Lambda$ diverges as $v \to 0$, representing an infinite additional contribution to the value of the radial coordinate $\rho$ at fixed $\sigma$, and therefore indicating a change of the asymptotic behaviour of the string solution in this limit. As discussed in section \ref{sec:FGA_interpretation}, the ``small'' spike should become a ``large'' spike in this limit, which is also confirmed by the fact that, for all the explicit solutions we have studied, ``large'' spikes are always associated with static solitons.

We now compute the spectrum $E = \Delta - S$, obtaining
\begin{eqnarray}
 E & = & \frac{1}{2}E_{0} + \frac{\sqrt{\lambda}}{2\pi} \int_{-\tilde{\Lambda}}^{\tilde{\Lambda}} 
        d \sigma \frac{1 - 3 v^2 + \cosh 2T + v^2 \cosh 2X}{2(\cosh T + v \cosh X)^2} \nonumber \\
   & = & \frac{1}{2}E_{0} + \frac{\sqrt{\lambda}}{2\pi} \left[ \sigma - \frac{v \sqrt{1-v^2} \sinh X}{\cosh T + v \cosh X}
          \right]_{-\tilde{\Lambda}}^{\tilde{\Lambda}} \nonumber \\
   & = & \frac{\sqrt{\lambda}}{2\pi} \left[ 2\Lambda+2\tilde{\Lambda} - 2 \sqrt{1-v^2} + O(e^{-2\gamma\Lambda}) \right] \nn \\
   & = & E_{0} + 2E_{\rm sol}(v) + O(e^{-2\gamma\Lambda}) \:,
\label{Delta-S-2s}
\end{eqnarray}
where
\begin{eqnarray}
 E_{\rm sol}(v) & = & \frac{\sqrt{\lambda}}{2\pi} \left[ - \delta \Lambda - \sqrt{1-v^{2}} \right] \nn \\
                & = & \frac{\sqrt{\lambda}}{2\pi} \left[ \frac{1}{2} \log \left( \frac{1 + \sqrt{1-v^{2}}}{1 - \sqrt{1-v^{2}}} \right)
                       - \sqrt{1-v^{2}} \right]                
\label{eq:E_sol(v)_small_spikes}
\end{eqnarray}
is naturally interpreted as the energy of a single soliton of velocity\footnote{Although $0<v<1$ for the two-soliton solution, the result \eqref{eq:E_sol(v)_small_spikes} equally applies to the soliton moving with velocity $v$ and to the soliton moving with velocity $-v$, and certainly the expression is even under $v \to v$. We may therefore state that $E_{\rm sol} (v)$ is the soliton energy independently of the sign of $v$.} $-1\leq v \leq +1$. Note that $E_{\rm sol}$ diverges as the soliton velocity $v$ goes to zero, reflecting the divergent contribution $\delta \Lambda$ to the length of the string mentioned above, and that consistently $E_{\rm sol} (\pm 1) = 0$, so that the excitation disappears in this limit.

Moreover, we notice that the energy and angular momentum of the two-soliton open string have the following leading behaviour:
\begin{equation}
 \Delta_{2s} \simeq S_{2s} = \frac{\sqrt{\lambda}}{8 \pi} e^{2 \Lambda} \:,
\label{eq:Delta_S_2s_lead_order}
\end{equation}
so that, once we take the additional straight GKP string into account, we may rewrite \eqref{Delta-S-2s} as
\begin{equation}
 \Delta - S = \frac{\sqrt{\lambda}}{2 \pi} \left[ 2 \log S + 2 \log \frac{\pi}{\sqrt{\lambda}} + 4 \log 2 \right] + 2 E_{\rm sol} (v) + \ldots \:,
\label{eq:2s_finite-gap_spectrum}
\end{equation}
where the dots indicate terms which vanish as $S \to \infty$. Thus there is agreement with \eqref{eq:Delta-S_FGA_general_K_M}, again modulo a discrepancy in the moduli-independent constant with respect to the results obtained for the ``large'' spikes, which could be due to the regulating procedure (the reference result is still equation \eqref{eq:E_S_omega_to_1_behaviour} with $K=2$ and with two additional solitonic excitations).
\paragraph{}
Having established the existence of excitations of finite energy we now want to determine their dispersion relation. In particular, as the cusps move along the string with velocity $v$, as measured in the spacelike worldsheet coordinate $\sigma$, we want to identify the conserved momentum $P_{\rm sol}(v)$ which is canonically conjugate to the position of the soliton in these coordinates. Here we will follow the same line of reasoning used for the case of Giant Magnons in\footnote{In particular, see discussion around eqns (2.16-2.19) of this reference.} \cite{Hofman:2006xt}.

Consider a configuration with $M$ cusps located at the positions $\sigma = \sigma_{l} (\tau)$ in the worldsheet coordinate $\sigma$ introduced above, moving with velocities $v_{l} = d\sigma_{l} / d \tau$ for $l = 1, \ldots, M$. The total energy of the configuration is
\begin{equation}
 E = E_{0} + \sum_{l=1}^{M} E_{\rm sol} (v_{l})  \:.
\label{eq:total_energy_small_spikes_Hamiltonian_formalism}
\end{equation}
The energy $E = \Delta - S$ is canonically conjugate to the global coordinate $\tilde{t} = (t + \phi)/2$ and we can define a canonical momentum $P_{l}$ for each soliton via Hamilton's equation,
\begin{equation}
 \frac{d{\sigma}_{l}}{d\tilde{t}} = \frac{\partial E}{\partial P_{l}} \:.
\label{eq:Hamilton_eq}
\end{equation}
An important subtlety is that the global time $\tilde{t}$ appearing in the above equation is not equal to the worldsheet time in the string solutions considered above. However, they are equal in the vacuum solution \eqref{eq:infinite_GKP_string} and, as each sinh-Gordon soliton is localised, we have $d \tilde{t} / d \tau \to 1$ exponentially fast away from the centre of each cusp. Thus the differential $\tilde{v}_{l} = d \sigma_{l} / d \tilde{t}$ appearing in \eqref{eq:Hamilton_eq} will be equal to the worldsheet velocity $v_{l}$ up to exponentially small corrections for almost all times\footnote{This only fails to be true during a finite time interval of duration of order $1/v$ when the soliton crosses the origin. This effect will produce a subleading correction to the semiclassical spectrum discussed below.}. Making the replacement $\tilde{v}\rightarrow v$ in Hamilton's equation \eqref{eq:Hamilton_eq} for a single soliton moving at constant velocity $v$ we get
\begin{equation}
 v = \frac{dE_{\rm sol}}{dP_{\rm sol}} = \frac{dE_{\rm sol}}{dv} \left( \frac{dP_{\rm sol}}{dv} \right)^{-1}
\label{eq:relation_v_E_sol_P_sol}
\end{equation}
or equivalently, 
\begin{equation}
 \frac{dP_{\rm sol}}{dv} = \frac{1}{v} \frac{dE_{\rm sol}}{dv} = - \frac{\sqrt{\lambda}}{2 \pi} \frac{\sqrt{1-v^2}}{v^{2}} \:,
\label{eq:dP_sol_dv}
\end{equation}
where we used \eqref{eq:E_sol(v)_small_spikes} in the second equality. As the soliton solutions considered above revert to the vacuum for $|v|=1$, we integrate \eqref{eq:dP_sol_dv} with boundary condition $P_{\rm sol}(\pm 1)=0$ to get,
\begin{equation}
 P_{\rm sol}(v) = \frac{\sqrt{\lambda}}{2\pi}\left[\frac{\sqrt{1-v^{2}}}{v} - {\rm Tan}^{-1}\left(\frac{\sqrt{1-v^{2}}}{v}\right)\right] \:.
\label{eq:P_sol(v)_small_spikes}
\end{equation}
Equations \eqref{eq:E_sol(v)_small_spikes} and \eqref{eq:dP_sol_dv} constitute the dispersion relation of the soliton. Notice that the conserved momentum $P_{\rm sol}(v)$ is an odd function of the velocity $v$ by construction. Thus the total momentum of the two soliton solution considered above is zero. More generally we might expect that an $M$-soliton closed string solution should obey a level-matching condition of the form, 
\begin{equation}
 P = \sum_{l=1}^{M} P_{\rm sol}(v_{l}) = 0 \:.
\label{eq:P_tot_=0_small_spikes}
\end{equation}
The situation is however complicated by the fact that the folds at the end of the string themselves correspond to solitons with zero velocity which therefore yield two infinite contributions of opposite sign to the total momentum which can cancel up to a finite remainder. This consideration presumably accounts for the existence of the one-soliton excitation of the folded string discussed in the next section. This is exactly what happens in the finite-gap picture, where the filling fractions associated with the first surface, $\tilde{\Sigma}_1$, decouple from those associated with the second, $\tilde{\Sigma}_2$, resulting in a level-matching constraint which only applies to $\tilde{\Sigma}_1$ at leading order and not to $\tilde{\Sigma}_2$\footnote{The mechanism is in fact the same in both cases: the filling fractions/momenta of the ``large'' spikes, multiplied by the corresponding mode numbers, cancel at leading order, but they may leave arbitrary subleading contributions, which may then cancel a non-vanishing subleading contribution from the filling fractions/momenta of the ``small'' spikes.}.
\paragraph{}
In order to semiclassically quantise the dispersion relation, we first observe that the semiclassical wavefunction for a soliton of velocity $v$ takes the form
\begin{equation}
 \Psi(\sigma) = \exp \left( i P_{\rm sol} (v) \sigma \right) \:.
\label{eq:semiclassical_soliton_wavefn}
\end{equation}
The quantization condition for the soliton velocity comes from imposing the invariance of the wavefunction under a shift $\sigma \to \sigma + L$, where $L$ is the total length of the string, which in this case amounts to $L = 2 \Lambda + 2 \tilde{\Lambda}$.

For our purpose, the leading order behaviour $L \simeq 2 \log(S) + O (S^{0})$ is sufficient to find the leading-order quantization condition\footnote{In general, there are additional corrections coming from the two-body scattering of solitons leading to a quantisation condition of Bethe Ansatz type. However, we will restrict our attention to the case where the quantized momentum $P_{\rm sol}$ remains of order one as $S\rightarrow \infty$ and the scattering phase is subleading. A similar correction from the inequality of global and worldsheet time near each soliton arises at the same order as the scattering phase.}
\begin{equation}
 P_{\rm sol} (v) \cdot 2 \log S \in 2 \pi \mathbb{Z} \:.
\label{eq:semiclassical_quantisation_condition_small_spikes}
\end{equation}
\paragraph{}
Most importantly, the dispersion relation \eqref{eq:small_spike_disprel} coincides with the dispersion relation \eqref{eq:E(v)}, \eqref{eq:P(v)} of the excitations discovered through the finite-gap analysis and associated with the second Riemann surface $\tilde{\Sigma}_2$ after the factorisation. Moreover, the semiclassical quantisation conditions \eqref{eq:semiclassical_quantisation_condition_small_spikes} and \eqref{eq:semiclassical_quant_condition_P(c)} are also identical. As we already remarked in section \ref{sec:FGA_interpretation}, the quantised dispersion relation also coincides with the one for the ``small'' holes of gauge theory in the large-spin limit and at large momentum $P_{\rm sol} \gg 1$ (we were only able to verify this in the special case of two ``large'' holes, which in fact correspond to the two ``large'' spikes of these strings).

\subsection{The one-spike solution}
\label{sec:1s}

As the consitituent solitons of the two soliton scattering solution are well seperated at very early and late times, it is intuitively clear that string solutions corresponding to individual solitons with vacuum asymptotics must also exist. Indeed such solutions were also presented in \cite{Jevicki:2007aa} (see equations (4.25, 4.26)), but were found to have infinite energy. They also have different asymptotics to the vacuum configuration studied above\footnote{In particular, the asymptotic values of the angular coordinate $\phi$ are the same at both ends of the string while they differ by $\pi$ in the vacuum solution}. In fact this pathological behaviour arises because the solution (4.25, 4.26) of \cite{Jevicki:2007aa} corresponds to a single soliton with zero velocity and is related to the divergence of $E_{\rm sol}(v)$ as $v \to 0$ found above.

A solution corresponding to a single soliton with non-zero velocity $v>0$ can be obtained by space and time translation of the two-soliton scattering solution so that one cusp is located near the origin and the other is sent to infinity. In particular, we choose to work at $\tau \to + \infty$, so that the two solitons are widely separated, and then solve the equation $\cosh T = v \cosh X$ at leading order, focusing on the right-moving soliton, so that $X \to + \infty$ as well: $X \simeq T - \log v$. If we use this relation in order to approximate the two-soliton solution \eqref{eq:2s_solution}, we obtain
\begin{eqnarray}
 Z_1 & = & e^{i\tau} \frac{e^\Sigma (\cosh\sigma - \sqrt{1-v^2} \sinh \sigma) + e^{-\Sigma} (v - i \sqrt{1-v^2}) \cosh \sigma}
           {v e^\Sigma + e^{-\Sigma}} \nonumber \\
 Z_2 & = & e^{i\tau} \frac{e^\Sigma (\sinh\sigma - \sqrt{1-v^2} \cosh \sigma) + e^{-\Sigma} (v - i \sqrt{1-v^2}) \sinh \sigma}
           {v e^\Sigma + e^{-\Sigma}} \:, \nn\\
\label{eq:1s_solution_Rmover_tau+inf}
\end{eqnarray}
where $\Sigma = (X-T)/2 = \gamma (\sigma - v\tau)$ and $X = 2 \gamma \sigma$, $T =  2 v \gamma \tau$ as before. This is in fact another exact solution of the equations of motion and Virasoro constraints. Similar solutions may be obtained by approximating for $\tau \to - \infty$ or focusing on the left-moving soliton.

\begin{figure}%
\includegraphics[width=0.83\columnwidth]{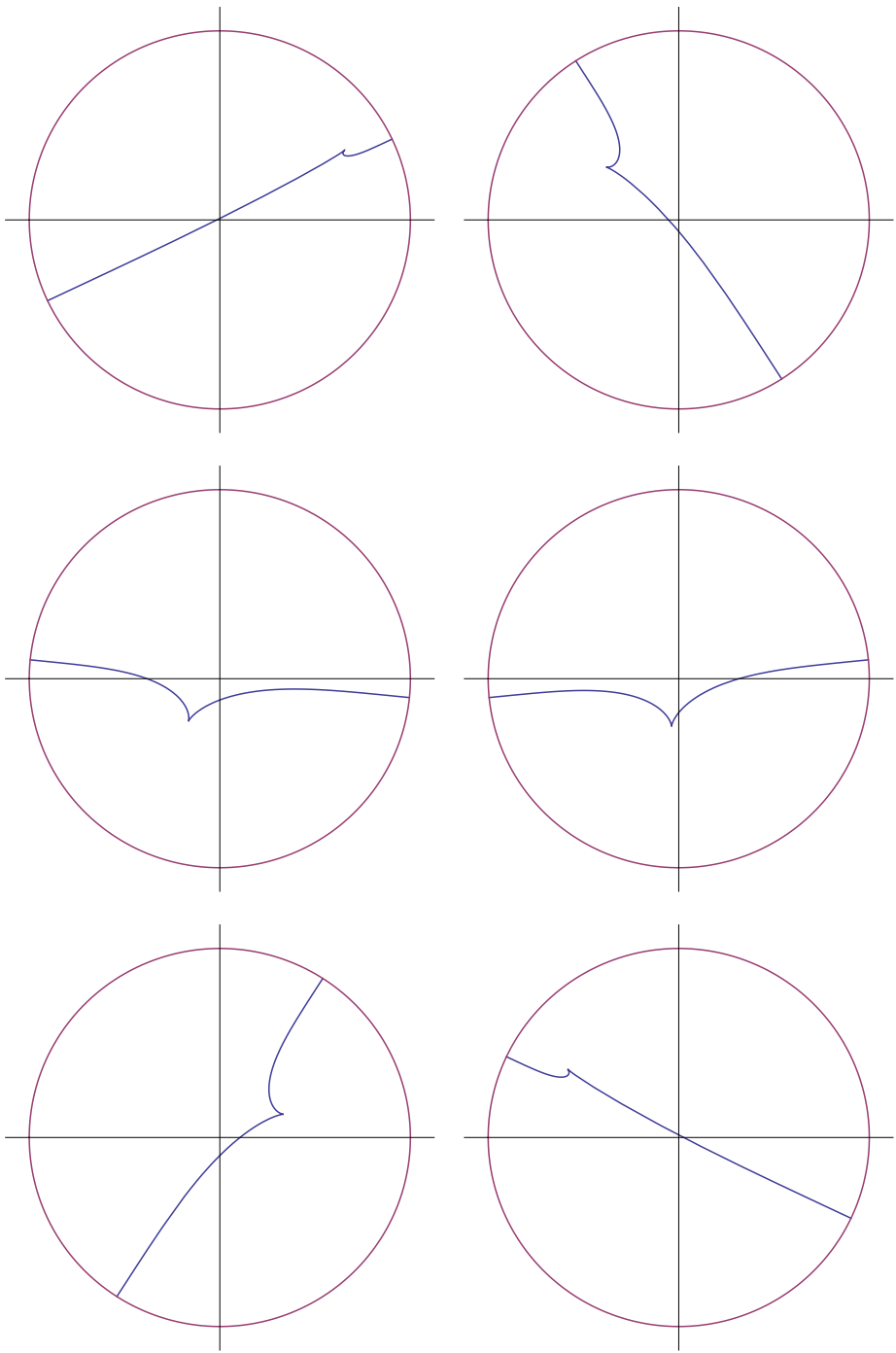}%
\caption{The 1-soliton solution at $t=-3,-1,0,1,3$ (from left to right and top to bottom).}%
\label{fig:Plot_1s_for_5_values_of_t}%
\end{figure}

One may easily check that the corresponding sinh-Gordon angle $\alpha$ coincides with that of the one-soliton solution \eqref{eq:def_sinhG_1s1a} (upper sign) with velocity $v$, up to a constant shift in $\tau$. Again the two static solitons at the endpoints are absent from the sinh-Gordon field due to the fact that we are working at infinite angular momentum $S$.
\paragraph{}
The plot at constant $t$ (see Fig. \ref{fig:Plot_1s_for_5_values_of_t}) shows a single ``small'' spike moving along the infinite GKP vacuum, starting at one endpoint as $\tau \to - \infty$ and ending at the other at $\tau \to + \infty$. As in the two-soliton case, there are no cusps in the constant-$\tau$ plot.

\paragraph{}
The solution (\ref{eq:1s_solution_Rmover_tau+inf}) has asymptotics 
\begin{eqnarray}
 Z_{1} & \simeq & \frac{1}{2}e^{\sigma} e^{i\tau}\left( \frac{1-\sqrt{1-v^{2}}}{v} \right)  \nn \\ 
 Z_{2} & \simeq & \frac{1}{2}e^{\sigma} e^{i\tau}\left( \frac{1-\sqrt{1-v^{2}}}{v} \right) \:,
\label{eq:Z_asymptotics_1s_sigma_+inf}
\end{eqnarray}
as $\sigma\rightarrow\infty$, and
\begin{eqnarray}
 Z_{1} & \simeq & \frac{1}{2}e^{-\sigma} e^{i\tau} \left(v-i\sqrt{1-v^{2}} \right) \nn \\ 
 Z_{2} & \simeq & -\frac{1}{2}e^{-\sigma} e^{i\tau} \left(v-i\sqrt{1-v^{2}} \right) \:,
\label{eq:Z_asymptotics_1s_sigma_-inf}
\end{eqnarray}
as $\sigma \rightarrow -\infty$. Or, in terms of the global coordinates, 
\begin{eqnarray}
 t(\tau,\sigma) & \to & \begin{cases}
                         \tau \:, \quad \textrm{for } \sigma \to + \infty \\
                         \tau - \beta \:, \quad \textrm{for } \sigma \to - \infty
                        \end{cases} \nonumber \\
 \phi(\tau,\sigma) & \to & \begin{cases}
                            \tau \:, \quad \textrm{for } \sigma \to + \infty \\
                            \tau - \beta + \pi \:, \quad \textrm{for } \sigma \to - \infty
                           \end{cases} \nonumber \\ 
 \rho(\tau,\sigma) & \to & \begin{cases}
                            \sigma + \alpha \:, \quad \textrm{for } \sigma \to + \infty \\
                         - \sigma \:, \quad \textrm{for } \sigma \to - \infty
                        \end{cases} \:,
\label{eq:global_asymptotics_1s_solution}
\end{eqnarray}
where $\beta = {\rm Tan}^{-1}(\sqrt{1-v^2}/v)$ and $\alpha=\log[(1-\sqrt{1-v^2})/v]$.

This implies that
\begin{equation}
 \phi \to \begin{cases}
           t \:, \quad \textrm{for } \sigma \to + \infty \\
           t + \pi \:, \quad \textrm{for } \sigma \to - \infty
          \end{cases} \:,
\label{eq:phi_at_endpoints_at_constant_t_1s_solution}
\end{equation}
so that the one-soliton solution reproduces the vacuum asymptotics.
\paragraph{}
As before, we construct a closed string by gluing a straight infinite GKP string at the endpoints of the one-soliton solution. However, the presence of the extra term $\beta$ in \eqref{eq:global_asymptotics_1s_solution} makes the string open at constant $\tau$. Therefore, we need to add a timelike segment to the contour when calculating the conserved charges. In order to regulate the contribution of the latter to the spectrum, we impose the usual condition $\rho = \Lambda \gg 1$. Thus we must restrict the range of the worldsheet coordinate $\sigma$ according to $-\Lambda \leq \sigma \leq \tilde{\Lambda}$, where $\tilde{\Lambda} = \Lambda - \alpha$. Therefore, we obtain the following regulated expression for the spectrum:
\begin{eqnarray}
 \Delta - S & = & \frac{1}{2}E_{0} + \frac{\sqrt{\lambda}}{2\pi} \int_{-\Lambda}^{\tilde{\Lambda}} 
                   d\sigma \left[ 1 - 2 v e^T \frac{e^X}{(e^T + v e^X)^2} \right] + O(e^{-2\gamma\Lambda}) \nonumber \\
            & = & \frac{1}{2}E_{0} + \frac{\sqrt{\lambda}}{2\pi}
                   \left[ \sigma + \frac{e^T}{\gamma(e^T + v e^X)} \right]_{-\Lambda}^{\tilde{\Lambda}} + O(e^{-2\gamma\Lambda}) \nonumber \\
            & = & \frac{\sqrt{\lambda}}{2\pi} \left[ 3 \Lambda + \tilde{\Lambda} - \sqrt{1-v^2} + O(e^{-2\gamma\Lambda}) \right]
                   + O(e^{-2\gamma\Lambda}) \nn \\ 
            & = &  E_{0} + E_{\rm sol}(v) + O(e^{-2\gamma\Lambda}) \:,
\label{eq:evaluating_E-S_analytically_for_1s_solution}
\end{eqnarray}
where the $O(e^{-2\gamma\Lambda})$ term outside the brackets represents the vanishing contribution from the timelike segment, which has no effect on the spectrum in the large $S$ limit.

Thus, we recover again the soliton energy \eqref{eq:E_sol(v)_small_spikes}. From this point onwards, we can repeat all the steps we followed in the previous section and reproduce the soliton momentum \eqref{eq:P_sol(v)_small_spikes} and the associated semiclassical quantisation condition \eqref{eq:semiclassical_quantisation_condition_small_spikes} (where now $L = 3 \Lambda + \tilde{\Lambda}$ which makes no difference at leading order with respect to the two-soliton case). Note that, strictly speaking, from \eqref{eq:1s_solution_Rmover_tau+inf} we may only deduce these results for a right-moving soliton, with velocity $v>0$. However, the analysis could straightforwardly be extended to a left-mover by studying the corresponding solution, which is also obtained from \eqref{eq:2s_solution} through the procedure described above.

Finally, we may also compute the energy and angular momentum of the one-soliton open string at leading order:
\begin{equation}
 \Delta_{1s} \simeq S_{1s} = \frac{\sqrt{\lambda}}{8 \pi} e^{2 \Lambda} \:,
\label{eq:Delta_S_1s_lead_order}
\end{equation}
which predictably coincides with the two-soliton result \eqref{eq:Delta_S_2s_lead_order}. Therefore, we reproduce again the finite-gap spectrum \eqref{eq:Delta-S_FGA_general_K_M}:
\begin{equation}
  \Delta - S = \frac{\sqrt{\lambda}}{2 \pi} \left[ 2 \log S + 2 \log \frac{\pi}{\sqrt{\lambda}} + 4 \log 2 \right] + E_{\rm sol} (v) + \ldots \:,
\label{eq:1s_finite-gap_spectrum}
\end{equation}
this time with only a single additional excitation and up to the usual discrepancy in the moduli-independent constant.

\section[A more general solution at infinite $S$]{Constructing a more general solution at infinite angular momentum}
\label{sec:glue_arcs}

We will now discuss another gluing procedure which allows to construct strings with both ``large'' and ``small'' cusps, such as the one described in Fig. \ref{fig:large_small_spikes}, although subject to certain restrictions.

A very helpful observation, due to Kruczenski and Tseytlin \cite{Kruczenski:2008bs}, is that the arcs of the symmetric spiky string at $\omega = 1$ \eqref{eq:JJ_single_arc_solution_omega=1} satisfy the following equation:
\begin{equation}
 X_0 X_1 + X_2 X_3 - \frac{1}{2} \cos \sigma_0 ( X_0^2 + X_1^2 + X_2^2 + X_3^2) = 0 \:,
\label{eq:KT_eqn}
\end{equation}
where $\sigma_0$ is related to the lowest radial value $\rho_0$ along the arc by $\cot \sigma_0 = \sinh 2 \rho_0$. The open infinite GKP string also satisfies this equation, but only if we introduce shifts in its $t$ and $\phi$ coordinates,
\begin{eqnarray}
 Z_1 & = & X_0 + i X_1 = \cosh \sigma \: e^{i (\tau + t_0)} \nonumber\\
 Z_2 & = & X_2 + i X_3 = \sinh \sigma \: e^{i (\phi + \phi_0)} \:,
\label{eq:infinite_GKP_shifted_t_phi}
\end{eqnarray}
such that $t_0 - \phi_0 = \pi /2 + k \pi$, $k \in \mathbb{Z}$, in which case we have $\sigma_0 = \pi/2$ (which is consistent with the fact that the lowest radial position along the GKP string is $\rho_0 = 0$), so that the second term on the left-hand side in \eqref{eq:KT_eqn} vanishes.

The key point is the fact that, for any solution satisfying
\begin{equation}
 X'_0 X'_1 + X'_2 X'_3 = 0 \:,
\label{eq:GKP_KT_equation}
\end{equation}
we may perform a right $SU(1,1)$-rotation (also known as an $AdS_3$ boost)
\begin{equation}
 g'(\tau, \sigma) = g(\tau, \sigma) e^{\rho_0 s^2} = g(\tau, \sigma) \begin{pmatrix}
                             \cosh \rho_0 & \sinh \rho_0 \\
                             \sinh \rho_0 & \cosh \rho_0
                            \end{pmatrix}
\label{eq:def_right_rotation}
\end{equation}
thus obtaining a new solution
\begin{eqnarray}
 X'_0 & = & X_0 \cosh \rho_0 + X_2 \sinh \rho_0 \nonumber\\
 X'_1 & = & X_1 \cosh \rho_0 + X_3 \sinh \rho_0 \nonumber\\
 X'_2 & = & X_2 \cosh \rho_0 + X_0 \sinh \rho_0 \nonumber\\
 X'_3 & = & X_3 \cosh \rho_0 + X_1 \sinh \rho_0
\label{eq:KT_boost}
\end{eqnarray}
such that its coordinates $X'_\mu$ satisfy \eqref{eq:KT_eqn} with $\cot \sigma_0 = \sinh 2 \rho_0$ now a function of the rotation parameter.

Therefore, if we start from the shifted GKP string \eqref{eq:infinite_GKP_shifted_t_phi} and rotate it with parameter $\rho_0$, we should be able to ``bend'' it into an arc of the type \eqref{eq:JJ_single_arc_solution_omega=1}, with angular separation
\begin{equation}
 \Delta \theta = 2 {\rm Arctan} \frac{1}{\sinh 2 \rho_0} \in (0, \pi] \:.
\label{eq:deltatheta_JJ_omega=1}
\end{equation}
One may check explicitly that this is the case, where the Kruczenski arc obtained through this procedure is given exactly by \eqref{eq:JJ_single_arc_solution_omega=1}, up to a shift in $t$ and $\phi$. Since this solution plays the role of the vacuum state to which we can compare strings with solitons, we also list here the values of its conserved charges:
\begin{equation}
 \Delta \simeq S \simeq \frac{\sqrt{\lambda}}{8 \pi} e^{2 \Lambda} \:,
\label{eq:Delta_S_arc_vac}
\end{equation}
and
\begin{eqnarray}
 \Delta - S & = & \frac{\sqrt{\lambda}}{2\pi} \left( 2 \Lambda - \log w_0 \right) \nn\\
            & = & \frac{\sqrt{\lambda}}{2\pi} \left( \log S + \log \frac{\pi}{\sqrt{\lambda}} + 3 \log 2 - \log w_0 \right) + \ldots
                   \:,
\label{eq:spectrum_arc_vac}
\end{eqnarray}
where $w_0 = \cosh 2 \rho_0$ and the regulator $\Lambda \gg 1$ has been introduced, as usual, by imposing $\rho = \Lambda$ and then solving for $\sigma$. This is, as we might have expected, the same spectrum we would obtain from one half of the closed straight GKP string \eqref{eq:E-S_for_vacuum}, with an additional term due to the fact that the string is ``bent''. Therefore, this result matches \eqref{eq:E-S_omega_to_1_behaviour} with $K=1$ up to the usual moduli-independent constant. The reason why we have to set $K=1$ instead of $K=2$ is that the string we are considering is open: a spike would normally consist of two joining lines, whereas in this case we only have one line for each cusp, and hence the cusps only count as half.

Of course the idea is to glue several of these solutions together in order to recover a proper closed string, but it is interesting to observe that the spectrum already behaves as expected at the level of the individual arcs.

Note that the boost \eqref{eq:KT_boost} leaves the combination $\partial_a X_\mu \partial^a X^\mu$ invariant, and hence the boosted solution has the same sinh-Gordon field as the starting solution. In fact, the Kruczenski arc obtained by boosting the straight GKP string is associated with the sinh-Gordon vacuum, while the one-spike and two-spike arcs which we will obtain shortly correspond to the one-soliton and two-solitons solution respectively, exactly as their straight string equivalents.

\paragraph{}
We can then apply the above technique to the one-soliton and two-soliton solutions discussed in sections \ref{sec:1s} and \ref{sec:2s}. In the first case, we obtain a new solution
\begin{eqnarray}
 Z_1 & = & \frac{e^{i \tau}}{e^T + v e^X} \left\{ i {\rm ch} \rho_0 \left[ - {\rm ch} \sigma \left( e^X + e^T (v - i \sqrt{1-v^2}) \right)
            + e^X \sqrt{1-v^2} {\rm sh} \sigma \right] \right. \nonumber\\
     &   & \left. - {\rm sh} \rho_0 \left[ e^X \sqrt{1-v^2} {\rm ch} \sigma - {\rm sh} \sigma \left( e^X + e^T ( v - i \sqrt{1-v^2}) \right)
            \right] \right\} \nonumber\\
 Z_2 & = & \frac{e^{i \tau}}{e^T + v e^X} \left\{ i {\rm sh} \rho_0 \left[ - {\rm ch} \sigma \left( e^X + e^T (v - i \sqrt{1-v^2}) \right)
            + e^X \sqrt{1-v^2} {\rm sh} \sigma \right] \right. \nonumber\\
     &   & \left. + {\rm ch} \rho_0 \left[ e^X \sqrt{1-v^2} {\rm ch} \sigma - {\rm sh} \sigma \left( e^X + e^T ( v - i \sqrt{1-v^2}) \right)
            \right] \right\}
\label{eq:AdS_arc_1s}
\end{eqnarray}
where we have set $t_0 = \phi_0 + \pi/2$, and, after eliminating $t_0$ in favour of $\phi_0$ and noticing that the latter simply introduced a common shift in $t$ and $\phi$, we have imposed $\phi_0 = 0$.

\begin{figure}%
\includegraphics[width=0.83\columnwidth]{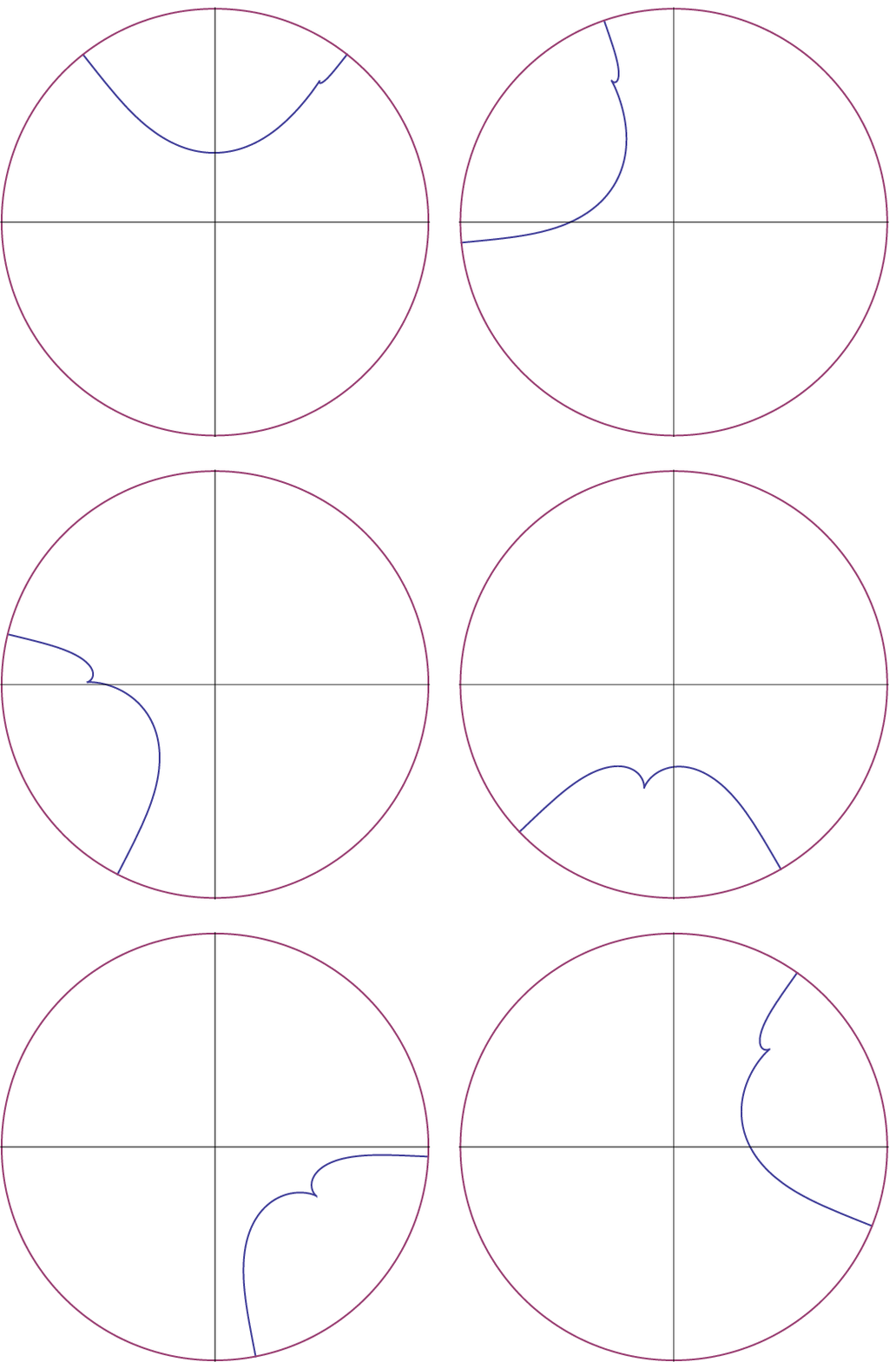}%
\caption{The 1-soliton arc of string with $v=0.5$ and $\rho_0 = 0.53$ at $t = \pi/2 + k$, for $k=0, \ldots,5$ (from left to right and top to bottom).}%
\label{fig:Plot_arc_1s}%
\end{figure}

The plots of the solution at constant $t$ (see Fig. \ref{fig:Plot_arc_1s}) show a ``small'' spike propagating along a Kruczenski arc touching the boundary.

The asymptotic behaviour at the endpoints is given by
\begin{eqnarray}
 t(\tau,\sigma) & \to & \begin{cases}
                         \tau - \delta \:, \quad \textrm{for } \sigma \to + \infty \\
                         \tau + \delta - \beta + \pi \:, \quad \textrm{for } \sigma \to - \infty
                        \end{cases} \nonumber \\
 \phi(\tau,\sigma) & \to & \begin{cases}
                            \tau - \alpha\:, \quad \textrm{for } \sigma \to + \infty \\
                            \tau + \alpha - \beta + \pi \:, \quad \textrm{for } \sigma \to - \infty
                           \end{cases} \nonumber \\ 
 \rho(\tau,\sigma) & \to & \begin{cases}
                            \sigma + C + \frac{1}{2} \log w_0 \:, \quad \textrm{for } \sigma \to + \infty \\
                            - \sigma + \frac{1}{2} \log w_0 \:, \quad \textrm{for } \sigma \to - \infty
                           \end{cases} \:,
\label{eq:global_asymptotics_1s_arc}
\end{eqnarray}
where $\delta = \mathrm{Tan}^{-1} \coth \rho_0$, $\alpha = \mathrm{Tan}^{-1} \tanh \rho_0$, $\beta = \mathrm{Tan}^{-1} (\sqrt{1-v^2}/v)$ and $C = \log [(1- \sqrt{1-v^2})/v]$. It is then easy to verify that the angular separation at constant $t$ between the endpoints is given by \eqref{eq:deltatheta_JJ_omega=1} as a function of $\rho_0$.

\begin{figure}%
\includegraphics[width=0.83\columnwidth]{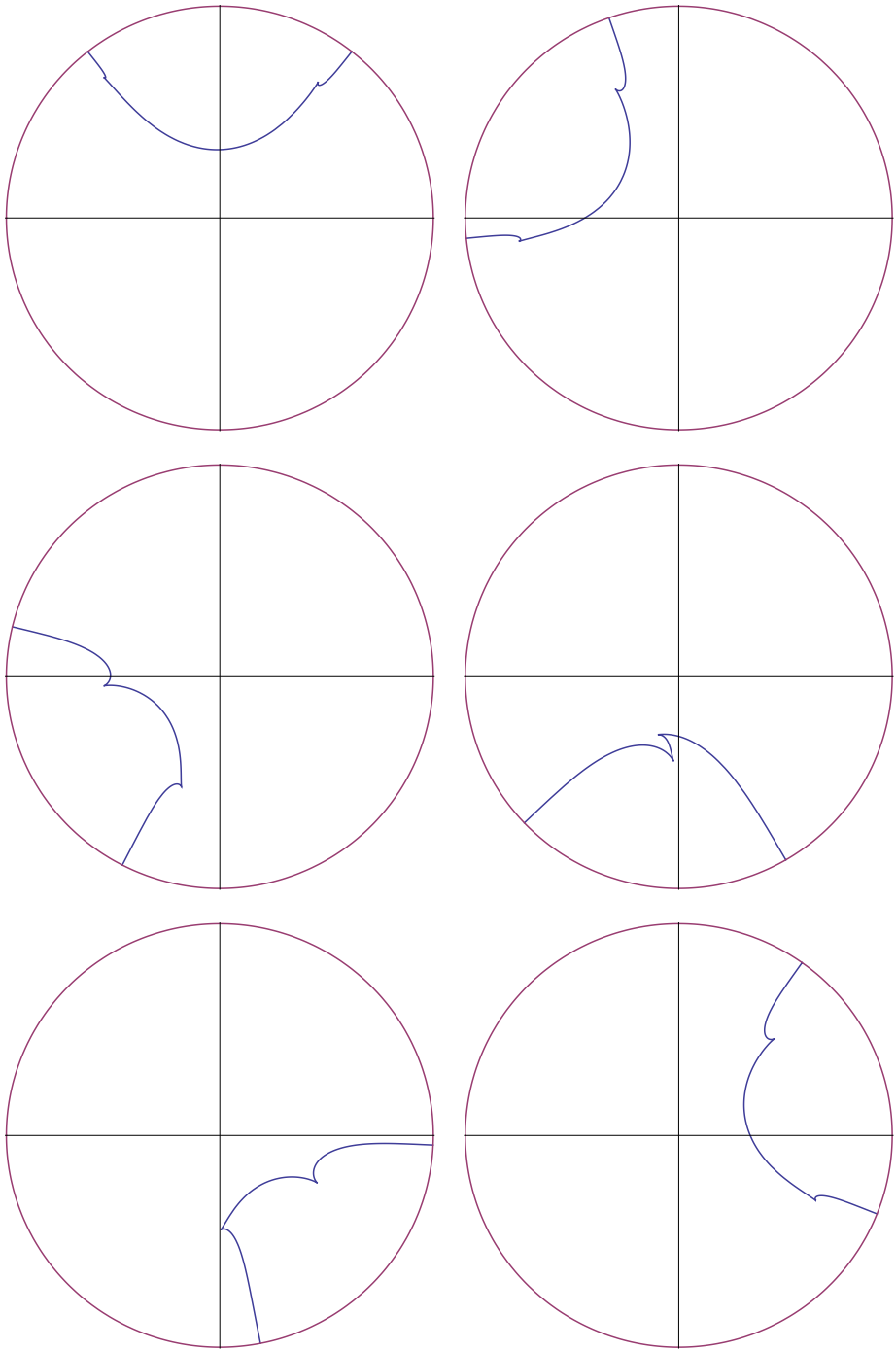}%
\caption{The 2-soliton arc of string with $v=0.5$ and $\rho_0 = 0.53$ at $t = \pi/2 + k$, for $k=0, \ldots,5$ (from left to right and top to bottom).}%
\label{fig:Plot_arc_2s}%
\end{figure}

As always, we calculate the leading behaviour of the conserved charges. If we were working with an open string, we would use an open contour on the worldsheet in order to define its conserved charges. It would then be important, in order to define such a contour, whether we had in mind the string at constant $\tau$ or the string at constant $t$, since they define different contours. As always, we are thinking about the string at constant $t$, and thus both $\tau$ and $\sigma$ have to vary along the appropriate contour. However, an equivalent contour is given by at first letting $\sigma$ vary over its full range, while keeping $\tau$ constant, and then fixing $\sigma$ at the last value reached and letting $\tau$ vary so that $t$ at the final endpoint equals $t$ at the initial endpoint:
\begin{eqnarray}
  \gamma & = & \gamma_1 \cup \gamma_2 \nn\\
  \gamma_1 & : & \tau = \tau_0 \:, \quad \sigma \in [-\infty, + \infty] \nn\\
  \gamma_2 & : & \tau = \tau_0 \in [\tau_0, \tau_1] \:, \quad \sigma = + \infty \:,
\label{eq:charge_contour_for_1s_2s_arc}
\end{eqnarray}
with $t(\tau_0, \sigma_0) = t (\tau_1, \sigma_1)$. 
It turns out that the timelike piece yields contributions which are negligible in the large angular momentum limit. We find:
\begin{equation}
 \Delta \simeq S \simeq \frac{\sqrt{\lambda}}{8 \pi} e^{2 \Lambda} \:,
\label{eq:Delta_S_arc_1s}
\end{equation}
as in the vacuum case, and
\begin{equation}
 \Delta - S \simeq \frac{\sqrt{\lambda}}{2 \pi} \left[ \log S + \log \frac{\pi}{\sqrt{\lambda}} + 3 \log 2 - \log w_0 \right] + E_{\rm sol} (v)
  + \ldots \:,
\label{eq:spectrum_arc_1s}
\end{equation}
where $\rho = \Lambda \gg 1$. Therefore, we have the appearance of a solitonic excitation over the vacuum \eqref{eq:spectrum_arc_vac}. This result generalises \eqref{eq:E-S_omega_to_1_behaviour} to the one-soliton case.
\paragraph{}
If we repeat the calculation starting from the two-soliton solution \eqref{eq:2s_solution}, we obtain
\begin{eqnarray}
 Z_1 & = & \frac{e^{i \tau}}{{\rm ch} T + v {\rm ch} X} \left\{ {\rm ch} \rho_0 \left[ - {\rm ch} \sigma \left( \sqrt{1-v^2} {\rm sh} T +
            i v {\rm ch} T + i {\rm ch} X \right) \right. \right. \nonumber\\
     &   & \left. + i \sqrt{1-v^2} {\rm sh} \sigma {\rm sh} X \right] + {\rm sh} \rho_0 \left[ - \sqrt{1-v^2} {\rm ch} \sigma {\rm sh} X \right.
            \nonumber\\
     &   & \left. \left. + {\rm sh} \sigma \left( - i \sqrt{1-v^2} {\rm sh} T + v {\rm ch} T + v {\rm ch} X \right) \right] \right\} \nonumber\\
 Z_1 & = & \frac{e^{i \tau}}{{\rm ch} T + v {\rm ch} X} \left\{ {\rm sh} \rho_0 \left[ - {\rm ch} \sigma \left( \sqrt{1-v^2} {\rm sh} T +
            i v {\rm ch} T + i {\rm ch} X \right) \right. \right. \nonumber\\
     &   & \left. + i \sqrt{1-v^2} {\rm sh} \sigma {\rm sh} X \right] + {\rm ch} \rho_0 \left[ - \sqrt{1-v^2} {\rm ch} \sigma {\rm sh} X \right.
            \nonumber\\
     &   & \left. \left. + {\rm sh} \sigma \left( - i \sqrt{1-v^2} {\rm sh} T + v {\rm ch} T + v {\rm ch} X \right) \right] \right\} \:,
\label{eq:AdS_arc_2s}
\end{eqnarray}
where we have again set $t_0 = \phi_0 + \pi/2$ and then $\phi_0 = 0$. Some plots of the solution at constant $t$ are given in Fig. \ref{fig:Plot_arc_2s}. This time we have a Kruczenski arc with two ``small'' spikes involved in the same type of scattering process which we saw in the original two-soliton solution.

The asymptotics for this solution are given by:
\begin{eqnarray}
 t(\tau,\sigma) & \to & \begin{cases}
                         \tau - \delta \:, \quad \textrm{for } \sigma \to + \infty \\
                         \tau + \delta + \pi \:, \quad \textrm{for } \sigma \to - \infty
                        \end{cases} \nonumber \\
 \phi(\tau,\sigma) & \to & \begin{cases}
                            \tau - \alpha\:, \quad \textrm{for } \sigma \to + \infty \\
                            \tau + \alpha + \pi \:, \quad \textrm{for } \sigma \to - \infty
                           \end{cases} \nonumber \\ 
 \rho(\tau,\sigma) & \to & \begin{cases}
                            \sigma + C + \frac{1}{2} \log w_0 \:, \quad \textrm{for } \sigma \to + \infty \\
                            - \sigma + C + \frac{1}{2} \log w_0 \:, \quad \textrm{for } \sigma \to - \infty
                           \end{cases} \:,
\label{eq:global_asymptotics_2s_arc}
\end{eqnarray}
where $\delta = \mathrm{Tan}^{-1} \coth \rho_0$, $\alpha = \mathrm{Tan}^{-1} \tanh \rho_0$ and $C = \log [(1- \sqrt{1-v^2})/v]$ as before. Again, the angular separation between the endpoints depends on $\rho_0$ and is given by \eqref{eq:deltatheta_JJ_omega=1}.

The usual considerations apply: a timelike path must be added to the contour in order to make sure that $t$ is identical at its endpoints. The contribution originating from this path is however negligible for large $S$, and hence we have:
\begin{equation}
 \Delta \simeq S \simeq \frac{\sqrt{\lambda}}{8 \pi} e^{2 \Lambda} \:,
\label{eq:Delta_S_arc_2s}
\end{equation}
and
\begin{equation}
 \Delta - S \simeq \frac{\sqrt{\lambda}}{2 \pi} \left[ \log S + \log \frac{\pi}{\sqrt{\lambda}} + 3 \log 2 - \log w_0 \right] + 2 E_{\rm sol} (v)
  + \ldots \:,
\label{eq:spectrum_arc_2s}
\end{equation}
where $\rho = \Lambda \gg 1$, so that there are now two solitons. Again, the result extends \eqref{eq:E-S_omega_to_1_behaviour} to the case of two excitations.
\paragraph{}
We are now ready to glue together several arcs with either one, two or no solitons and with arbitrary angular separations $\Delta \theta_j$ (corresponding to arbitrary parameters $\rho_0^{(j)}$), thus obtaining a generalisation of the $S \to \infty$ limit of the patched solution discussed in section \ref{sec:patching_K}, which allows the presence of ``small'' spikes moving along the arcs connecting the ``large'' spikes. 
We may glue $K-M$ arcs, where $M$ is the total number of solitons present, by setting
\begin{equation}
 \sum_{j=1}^{K-M} \Delta \theta_j = 2 n \pi \:.
\label{eq:arc_total_angular_separation_condition}
\end{equation}
We add a timelike segment at the end of the contour used for computing the conserved charges in order to make it closed. The corresponding contribution is however negligible at large $S$ as usual. Therefore the spectrum of such a glued solution is given by:
\begin{equation}
 \Delta \simeq S \simeq (K-M) \frac{\sqrt{\lambda}}{8 \pi} e^{2 \Lambda}
\label{eq:Delta_S_glued_arcs}
\end{equation}
and
\begin{multline}
 \Delta - S \simeq \frac{\sqrt{\lambda}}{2 \pi} \left[ (K-M) \log S + (K-M) \log \frac{8 \pi}{(K-M) \sqrt{\lambda}}
  - \sum_{j=1}^{K-M} \log w_0^{(j)} \right] \\
  + \sum_{k=1}^M E_{\rm sol} (v_k) + \ldots \:,
\label{eq:spectrum_glued_arcs}
\end{multline}
where $w_0^{(j)} = \cosh 2 \rho_0^{(j)} = 1/ \sin ( \Delta \theta_j /2)$ and $v_k$ is the worldsheet velocity of the $k$-th soliton\footnote{Note that, if we construct the patched solution in this way, whenever we have two solitons on a given arc, we are forced to give them equal and opposite worldsheet velocities.}.
\paragraph{}
Finally, concerning the semiclassical quantisation of the ``small'' spikes appearing on these arcs, we can follow the same steps as in section \ref{sec:2s}. In particular, as we observed above, the sinh-Gordon fields associated with these solutions are the usual one- and two-soliton solutions, so that the worldsheet position $\sigma_l (v_l)$ of a soliton is unchanged with respected to the previous case. It is also still true that $d \tilde{t}/ d \tau \to 1$ exponentially away from the cusps, as we can see by looking directly at the solutions \eqref{eq:AdS_arc_1s}, \eqref{eq:AdS_arc_2s} or from the asymptotics \eqref{eq:global_asymptotics_1s_arc}, \eqref{eq:global_asymptotics_2s_arc}. Hence, we obtain
\begin{equation}
 v = \frac{dE_{\rm sol}}{dP_{\rm sol}}
\label{eq:relation_v_E_sol_P_sol_arcs}
\end{equation}
as before, and from this point onwards the calculation proceeds in the same way. Therefore, there is no difference between these ``small'' spikes and those propagating along the straight GKP vacuum, even at the level of the semiclassical quantisation condition.

\chapter{Spiky strings in $AdS_3$-pp-wave}
\label{sec:pp-wave}

In this chapter, based on as yet unpublished work in collaboration with N. Dorey, we will see how the open GKP string and the Kruczenski arcs with one or two ``small'' spikes may be mapped onto $AdS_3$-pp-wave space. The resulting solutions are arcs of string with endpoints on the boundary and drooping towards the interior. A possible source of interest in this kind of string solutions is the fact that the endpoints describe light-like geodesics and hence, according to the AdS/CFT correspondence, they are dual to infinitely energetic gluons. The interconnecting arc of string is then identified with the chromomagnetic flux tube linking the gluons, representing the strong interaction between them. In this framework, a featureless arc connecting the gluons would be equivalent to the fundamental state of the two-gluon system, while arcs containing spikes would represent excited states. While the smooth arc of string was already known (see \cite{Ishizeki:2008tx} and references therein), the addition of the new one- and two-spike solutions may be helpful to studies concerning high-energy gluons.

\section{String theory in $AdS_3$-pp-wave}

$AdS_3$-pp-wave space is a region close to the boundary of $AdS_3$, corresponding to $\rho \sim \infty$ and $\phi \sim t$. Starting from the $AdS_3$ line element in global coordinates \eqref{eq:AdS3_line_element_global_coords}, one may obtain the $AdS_3$-pp-wave line element
\begin{equation}
 ds^2 = \frac{1}{z^2} (2dx_+ dx_- - \mu^2 z^2 dx_+^2 + dz^2)
\label{eq:AdS3-ppwave_line_element}
\end{equation}
through the coordinate transformation
\begin{eqnarray}
 z & = & \frac{2 \sqrt{2} e^{\rho_1}}{e^\rho} \nonumber\\
 x_\pm & = & e^{\rho_1} e^{\mp \theta} (\phi \pm t) \:,
\label{eq:def_coord_transf_AdS-AdSppwave}
\end{eqnarray}
where $\theta$ and $\rho_1$ are two parameters going to $+ \infty$ in such a way that
\begin{eqnarray}
 \frac{e^\theta}{e^{\rho_1}} = 2 \mu \:,
\label{eq:relation_theta_rho1}
\end{eqnarray}
while $\mu > 0$ is instead a free parameter. Note that $z \geq 0$ (where the boundary is located at $z=0$ and the interior of $AdS_3$ is at $z \to + \infty$), while there are no constraints on $x_+$ and $x_-$.

The bosonic string action in conformal gauge associated with this metric is given by
\begin{equation}
I = - \frac{\sqrt{\lambda}}{4\pi}\int d \sigma d \tau \frac{1}{z^2} \left[ 2 \partial_a x_+ \partial^a x_-
  - \mu^2 z^2 \partial_a x_+ \partial^a x_+ + \partial_a z \partial^a z \right] \:,
\label{eq:AdS3-pp-wave_string_action}
\end{equation}
from which we derive the equations of motion
\begin{eqnarray}
 && \partial_a \partial^a x_+ - \frac{2}{z} \partial_a z \partial^a x_+ = 0 \nn\\
 && \partial_a \partial^a x_- - \frac{2}{z} \partial_a z \partial^a x_- - 2 \mu^2 z \partial_a z \partial^a x_+ = 0 \nn\\
 && \partial_a \partial^a z - \frac{1}{z} \partial_a z \partial^a z + \frac{2}{z} \partial_a x_+ \partial^a x_- = 0
\label{eq:AdS3ppwave_EOM}
\end{eqnarray}
and Virasoro constraints
\begin{eqnarray}
 - \mu^2 (\partial_\pm x_+)^2 + \frac{2}{z^2} \partial_\pm x_+ \partial_\pm x_- + \frac{1}{z^2} (\partial_\pm z)^2 = 0 \:.
\label{eq:AdS3ppwave_Virasoro}
\end{eqnarray}
The metric \eqref{eq:AdS3-ppwave_line_element} is invariant under translations of $x_+$ and $x_-$. The associated conserved charges, named $P_+$ and $P_-$ respectively, are given by:
\begin{eqnarray}
 P_+ & = & \frac{\sqrt{\lambda}}{2 \pi} \int d \sigma \left( \mu^2 \dot{x}_+ - \frac{\dot{x}_-}{z^2} \right) \nn\\
 P_- & = & \frac{\sqrt{\lambda}}{2 \pi} \int d \sigma \frac{\dot{x}_+}{z^2} \:,
\label{eq:P+-_AdS3ppw}
\end{eqnarray}
where $\dot{x}_\pm \equiv \partial_\tau x_\pm$.

\section{The pp-wave GKP string}

We may transport the GKP string to the $AdS_3$-pp-wave region through the following procedure.

First of all, we apply the boost \eqref{eq:def_right_rotation}, \eqref{eq:KT_boost} to the open GKP string of finite length, which is given by the solution \eqref{eq:GKP_rho_of_sigmatilde} for\footnote{From now on, we will rename the unrescaled coordinates $(\tilde{\tau}, \tilde{\sigma})$ as $(\tau, \sigma)$, since we are not going to perform any rescaling in the following.} $\tilde{\sigma} \in [\tilde{L}, 3 \tilde{L}]$. Then we take the boost parameter $\rho_0$ to infinity and use the coordinate transformation \eqref{eq:def_coord_transf_AdS-AdSppwave}, approximating at leading order. We obtain the following solution of the $AdS_3$-pp-wave equations of motion and Virasoro constraints:
\begin{eqnarray}
 z & = & \frac{2 \sqrt{2} \: b \: \mathrm{dn} \! \left( \omega \sigma \left| \frac{1}{\omega} \right. \right)}
          {\sqrt{1 + \frac{1}{\omega^2} \mathrm{sn}^2 \! \left( \omega \sigma \left| \frac{1}{\omega} \right. \right) + \frac{2}{\omega} 
           \cos (\tau - \omega \tau) \mathrm{sn} \! \left( \omega \sigma \left| \frac{1}{\omega} \right. \right)}} \nonumber\\
 x_+ & = & \frac{1}{\mu} \tan^{-1} \left[ \frac{\omega \sin \tau + \mathrm{sn} \! \left( \omega \sigma \left| \frac{1}{\omega} \right. \right)
            \sin \omega \tau}{\omega \cos \tau + \mathrm{sn} \! \left( \omega \sigma \left| \frac{1}{\omega} \right. \right)
            \cos \omega \tau} \right] \nonumber\\
 x_- & = & - 8 \mu b^2 \omega \frac{\mathrm{sn} \! \left( \omega \sigma \left| \frac{1}{\omega} \right. \right) \sin (\tau - \omega \tau)}
          {\omega^2 + \mathrm{sn}^2 \! \left( \omega \sigma \left| \frac{1}{\omega} \right. \right) + 2 \omega \cos (\tau - \omega \tau)
           \mathrm{sn} \! \left( \omega \sigma \left| \frac{1}{\omega} \right. \right)} \:,
\label{eq:GKP_AdS3-pp-wave}
\end{eqnarray}
where $b>0$ is an additional free parameter which we have the freedom of introducing when we identify the boost parameter $\rho_0$ with the coordinate transformation parameter $\rho_1$ as $\rho_0 = \rho_1 - \log b$.

\begin{figure}%
\psfrag{z}{\footnotesize{$z$}}
\psfrag{xm}{\footnotesize{$x_-$}}
\includegraphics[width=\columnwidth]{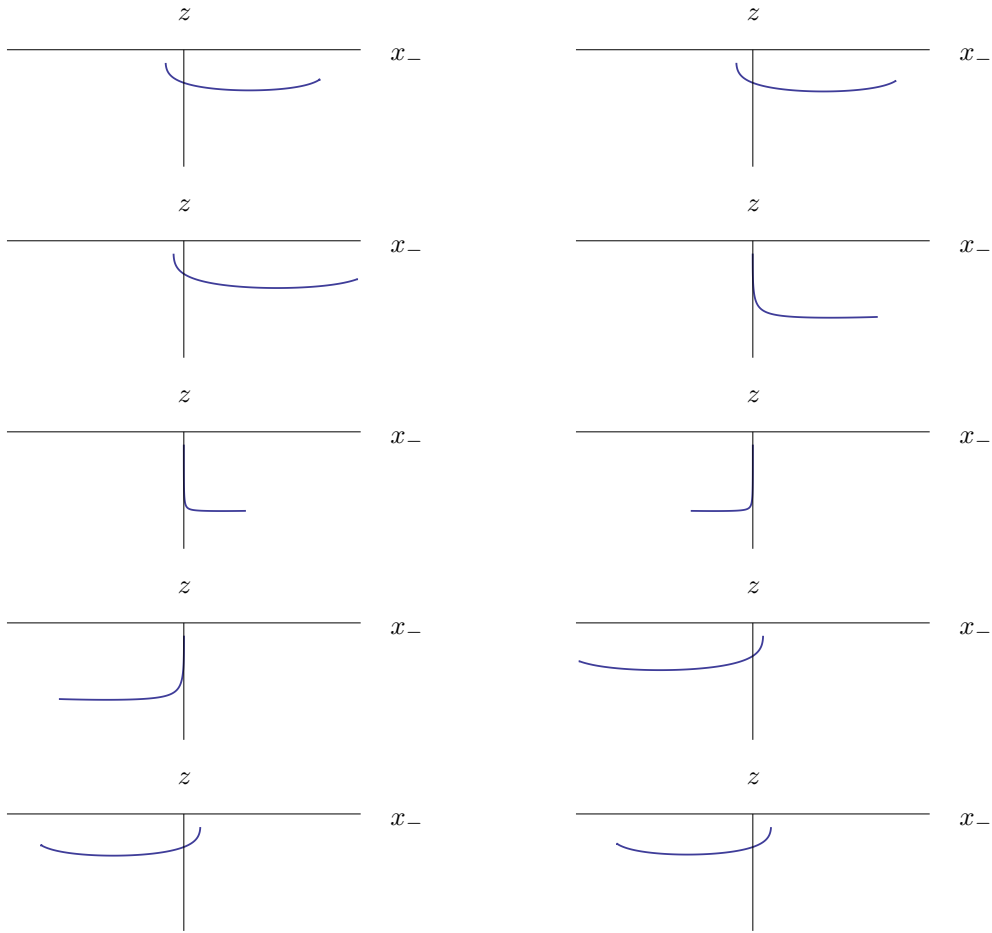}%
\caption{The GKP open string in $AdS_3$-pp-wave, for $\omega = \sqrt{2}$, $\mu = 1.6$, $\tan ( \mu x_+ ) = -5,-3,-1,-0.01,-0.001,$ $0.001,0.01,1,3,5$, from left to right and top to bottom. The boundary is located at the value $z=0$ along the vertical axis.}%
\label{fig:ppw_GKP}%
\end{figure}

We see that the solution is periodic in $\sigma$ with period $4L \equiv 4 \mathbb{K} (1/\omega)$: the full open string may be obtained by imposing the restriction $\sigma \in [L, 3L]$. Various plots of the solution at constant $x_+$, which we choose as our time variable, are shown in Fig. \ref{fig:ppw_GKP}. The string is still rotating, but it is now changing its shape during the motion. In particular, at $\tan(\mu x_+) = 0$ it becomes a straight line segment overlapping with the $z$-axis. As long as we keep $\omega > 1$, its endpoints stay away from the boundary $z=0$ and the interior of $AdS_3$ $z \to \infty$.

As usual, one should add a timelike path to the contour used for computing the conserved charges\footnote{The charge densities along this path are given by \eqref{eq:P+-_AdS3ppw}, with $\partial_\tau$ replaced by $\partial_\sigma$.}, but it turns out that the corresponding contribution vanishes even at $\omega > 1$. Therefore, we obtain:
\begin{eqnarray}
 P_+ & = & \mu \frac{\sqrt{\lambda}}{2 \pi} \int_L^{3L} d \sigma \frac{1 - \frac{1}{\omega} \mathrm{sn}^2 \! \left( \omega
           \sigma \left| \frac{1}{\omega} \right. \right)}{\mathrm{dn}^2 \! \left( \omega \sigma \left| \frac{1}{\omega} \right. \right)}
            \nonumber\\
     & = & \frac{\mu \sqrt{\lambda}}{\pi} \left[ \mathbb{K} \left( \frac{1}{\omega} \right)
            - \frac{\omega}{\omega + 1} \mathbb{E} \left( \frac{1}{\omega} \right)  \right] \nonumber\\
 P_- & = & \frac{1}{8 \mu b^2} \frac{\sqrt{\lambda}}{2 \pi} \int_L^{3L} d \sigma \frac{1 + \frac{1}{\omega} \mathrm{sn}^2 \! \left( \omega
            \sigma \left| \frac{1}{\omega} \right. \right)}{\mathrm{dn}^2 \! \left( \omega \sigma \left| \frac{1}{\omega} \right. \right)}
             \nonumber \\
     & = & \frac{\sqrt{\lambda}}{8 \pi \mu b^2} \left[ \frac{\omega}{\omega - 1} \mathbb{E} \left( \frac{1}{\omega} \right) - \mathbb{K} 
            \left( \frac{1}{\omega} \right) \right] \:.
\label{eq:P+-_ppw_GKP}
\end{eqnarray}
If we compare this result to the original GKP charges $\Delta$ and $S$ \eqref{eq:E_S_Nfolded_GKP} for $K=1$, we see that they are related in the following way:
\begin{eqnarray}
 P_+ & = & \mu (\Delta - S) \nonumber\\
 P_- & = & \frac{1}{8 \mu b^2} \frac{\Delta + S}{\cosh 2 \rho_0} \:.
\label{eq:relation_Delta+-S_P+-_ppw_GKP}
\end{eqnarray}
The first relationship was also observed in the case of the pp-wave limit of an $AdS_3 \times S^1$ string solution with a shape similar to the one of the Kruczenski spiky string \cite{Ishizeki:2008tx} (the main difference being that it has no spikes, i.e. that the protruding ends of the string are rounded).

\begin{figure}%
\psfrag{z}{\footnotesize{$z$}}
\psfrag{xm}{\footnotesize{$x_-$}}
\includegraphics[width=\columnwidth]{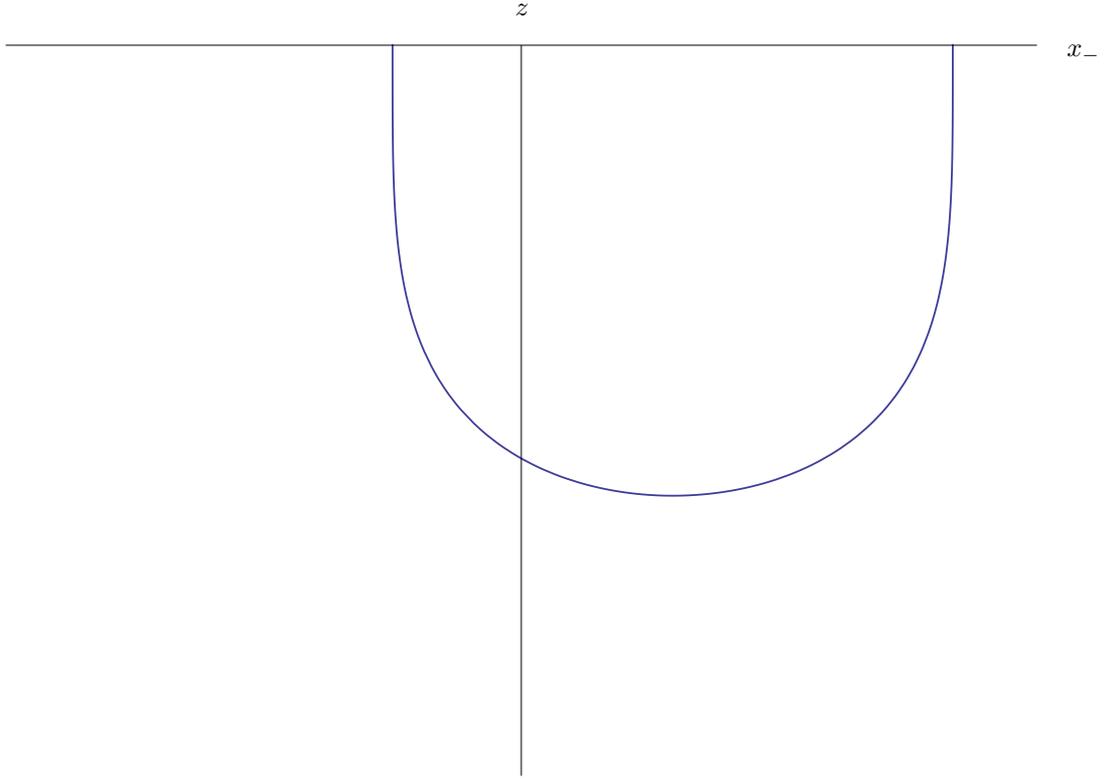}%
\caption{The pp-wave GKP arc of string for $\mu = 0.4$, $\tau_0 = 1$ (and $\omega = 1$).}%
\label{fig:ppw_GKP_omega1}%
\end{figure}

If we take the usual $\omega \to 1$ limit of the conserved charges, with $\omega = 1 + \eta$, we find:
\begin{eqnarray}
 P_+ & \simeq & \frac{\mu \sqrt{\lambda}}{2 \pi} \log \eta \nonumber\\
 P_- & \simeq & \frac{\sqrt{\lambda}}{8 \pi \mu b^2} \frac{1}{\eta} \nonumber\\
 P_+ & \simeq & \frac{\sqrt{\lambda}}{2 \pi} \mu \log P_- \:,
\label{eq:P_+-expansion_as_omegato1}
\end{eqnarray}
where these results, considering the relationship with $\Delta$ and $S$ of the GKP string in $AdS_3$, are typical of all the explicit solutions we have been studying. The third equation was also found in the case of \cite{Ishizeki:2008tx}.
\paragraph{}
We are now going to construct an exact solution at $\omega = 1$ representing an arc drooping from the boundary. The idea is to introduce a shift in the worldsheet time, $\tau = \tau' - \tau_0 / \eta$, and then to approximate \eqref{eq:GKP_AdS3-pp-wave} as $\eta \to 0$ in a specific way\footnote{Note that by taking the ``normal'' $\eta \to 0$ limit one is left with a solution which has one endpoint at $z=0$ and the other endpoint at $z = \infty$, i.e. a straight vertical line.}:
\begin{equation}
 \eta_n = \frac{\tau_0}{2 n \pi} \to 0 \qquad \textrm{as } n \to + \infty \:, \qquad n \in \mathbb{Z} \:.
\label{eq:def_eta_n_for_regularising_omegato1_limit}
\end{equation}
The result is
\begin{eqnarray}
 z & = & \frac{2 \sqrt{2} \: b}{(\cosh \sigma) \sqrt{1 + 2 \cos \tau_0 \tanh \sigma + \tanh^2 \sigma}} \nonumber\\
 x_+ & = & \frac{1}{\mu} \tan^{-1} \left[ \frac{\sin \tau + \tanh \sigma \sin (\tau - \tau_0)}{\cos \tau + \tanh \sigma \cos (\tau - \tau_0)}
            \right] \nonumber\\
 x_- & = & - 8 \mu b^2 \frac{\sin \tau_0 \tanh \sigma}{1 + 2 \cos \tau_0 \tanh \sigma + \tanh^2 \sigma} \:,
\label{eq:omega=1_ppwaveGKP}
\end{eqnarray}
which satisfies the $AdS_3$-pp-wave equations of motion and Virasoro constraints. We notice that running along the full open string now requires to let $- \infty < \sigma < + \infty$. Furthermore, $z$ and $x_-$ are now independent of $\tau$, which implies that the string is now static and its plots at constant $t$ or constant $\tau$ coincide. One such plot is given in Fig. \ref{fig:ppw_GKP_omega1}.

Strictly speaking, one would still need to add a timelike path when computing $P_+$ and $P_-$, but it turns out that it does not contribute in the first case and that the contribution is subleading in the second case. We obtain
\begin{eqnarray}
 P_+ & \simeq & \frac{\mu \sqrt{\lambda}}{2\pi} \left[ 2 \log \Lambda + 2 \log (4 b) \right] \nonumber\\
 P_- & \simeq & \frac{\sqrt{\lambda}}{2 \pi \mu} \Lambda^2 \nonumber\\
 P_+ & \simeq & \mu \frac{\sqrt{\lambda}}{2 \pi} \left[ \log P_- + \log \left( \frac{2 \pi \mu}{\sqrt{\lambda}} \right) + 2 \log (4 b) \right]\:,
\label{eq:P_pm_for_ppwaveGKP_omega=1}
\end{eqnarray}
where the regulator $\Lambda \gg 1$ has been introduced by demanding that $z = 1/\Lambda$.

\section{The one-soliton and two-soliton solutions in $AdS_3$-pp-wave}

We now apply the same procedure to the one-spike Kruczenski arc \eqref{eq:AdS_arc_1s}. Since that solution has already been boosted with boost parameter $\rho_0$, we only need to take $\rho_0 \to \infty$ to obtain a solution to the equations of motion and Virasoro constraints in $AdS_3$-pp-wave:
\begin{eqnarray}
 z & = & 2 \sqrt{2} b (e^T + v e^X) \left[ e^{2X} \left( (2-v^2) {\rm ch} 2 \sigma - 2 \sqrt{1-v^2} {\rm sh} 2 \sigma \right) \right. \nonumber\\
   &   & \left. + 2 e^{T+X} \left( 1 - v^2 + v {\rm ch} 2 \sigma - v \sqrt{1-v^2} {\rm sh} 2 \sigma \right) + e^{2T} {\rm ch} 2 \sigma
          \right]^{-\frac{1}{2}} \nonumber\\
  x_+ & = & \frac{1}{\mu} {\rm Tan}^{-1} \left\{ \left[ \left( (e^X + v e^T) \cos \tau + (e^T + e^X) \sqrt{1-v^2} \sin \tau \right) {\rm ch}
             \sigma \right. \right. \nonumber\\
      &   &  \left. + \left( ( e^T - e^X ) \sqrt{1-v^2} \cos \tau - (e^X + v e^T) \sin \tau \right) {\rm sh} \sigma \right] \nonumber\\
      &   &  \times \left[ \left( ( e^T + e^X ) \sqrt{1-v^2} \cos \tau - (e^X + v e^T) \sin \tau \right) {\rm ch} \sigma \right. \nonumber\\
      &   &  \left. \left. - \left( (e^X + v e^T) \cos \tau + (e^T - e^X) \sqrt{1-v^2} \sin \tau \right) {\rm sh} \sigma \right]^{-1} \right\}
              \nonumber\\
  x_- & = & - 2 \mu b^2 \left[ e^{2X} \left( 4 \sqrt{1-v^2} {\rm ch} 2 \sigma - 2 (2-v^2) {\rm sh} 2 \sigma \right) \right. \nonumber\\
      &   & \left. + 4 e^{T+X} \left( v \sqrt{1-v^2} {\rm ch} 2 \sigma - v {\rm sh} 2 \sigma \right) - 2 e^{2T} {\rm sh} 2 \sigma \right]
             \nonumber\\
      &   & \times \left[ e^{2X} \left( (2-v^2) {\rm ch} 2 \sigma - 2 \sqrt{1-v^2} {\rm sh} 2 \sigma \right) \right. \nonumber\\
      &   & \left. + 2 e^{T+X} \left( 1 - v^2 + v {\rm ch} 2 \sigma - v \sqrt{1-v^2} {\rm sh} 2 \sigma \right) + e^{2T} {\rm ch} 2 \sigma
             \right]^{-1} \:, \nonumber\\
\label{eq:ppw_1s_arc}
\end{eqnarray}
where $X = 2 \gamma \sigma$, $T = 2 \gamma \tau$, $\gamma = 1/ \sqrt{1-v^2}$, ${\rm ch} \equiv \cosh$, ${\rm sh} \equiv \sinh$ and we have identified $\rho_0 = \rho_1 - \log b$ as in the GKP case, thereby introducing a new parameter $b>0$.

\begin{figure}%
\includegraphics[width=\columnwidth]{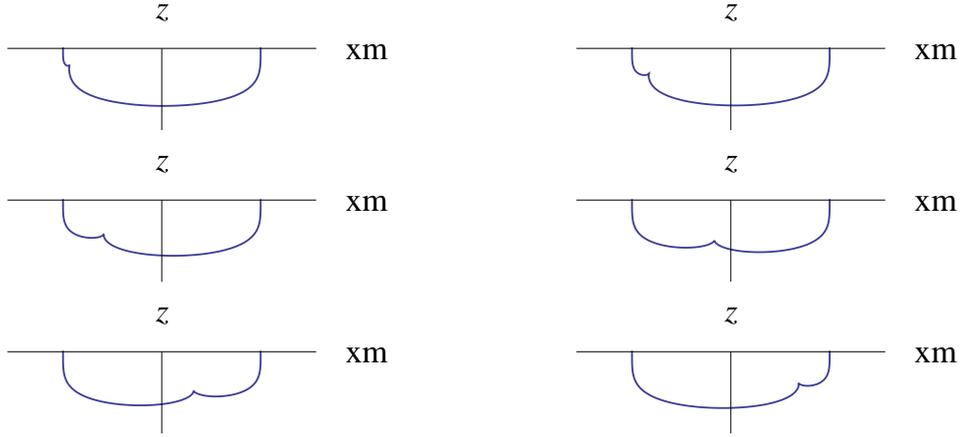}%
\caption{The two-soliton arc of string in $AdS_3$-pp-wave with $v=0.5$, $\mu = 0.6$ and $b = 2$ at $t = -\pi/2 + k$, for $k=0, \ldots,5$ (from left to right and top to bottom).}%
\label{fig:ppwarc1s}%
\end{figure}

Again, the natural range of $\sigma$ covering the whole string has expanded to $- \infty < \sigma < + \infty$. Differently from the GKP string, there is still some time-dependence in $z$ and $x_-$, but, as the plots at constant $x_+$ in Fig. \ref{fig:ppwarc1s} show, it is only related to the motion of the soliton, while the endpoints are static, separated by a distance
\begin{equation}
 \Delta x_- = 8 \mu b^2 \:.
\label{eq:Delta_x-_1s}
\end{equation}

It is also possible to check, by applying the definition \eqref{eq:def_alpha_Pohlmeyer} and replacing $X_\mu$ with the $AdS_3$-pp-wave coordinates $(x_+,x_-,z)$, that the field $\alpha (\tau, \sigma)$ is still given by the sinh-Gordon one-soliton solution on the infinite line \eqref{eq:def_sinhG_1s1a} (upper sign). This could have been expected since, as we already remarked earlier, the expression $\partial_a X_\mu \partial^a X^\mu$ is invariant under the $SU(1,1)$-rotation we are using.

In order to compute the conserved charges, we notice that, once again, the timelike segment does not contribute at leading order, and we introduce the regulator $\Lambda \gg 1$ such that $z = 1/ \Lambda$, finding:
\begin{eqnarray}
 P_+ & \simeq & \frac{\mu \sqrt{\lambda}}{2\pi} \left[ 2 \log \Lambda + 2 \log (4 b) \right] + \mu E_{\rm sol}(v) \nonumber\\
 P_- & \simeq & \frac{\sqrt{\lambda}}{2 \pi \mu} \Lambda^2 \nonumber\\
 P_+ & \simeq & \mu \frac{\sqrt{\lambda}}{2 \pi} \left[ \log P_- + \log \left( \frac{2 \pi \mu}{\sqrt{\lambda}} \right) + 2 \log (4 b) \right]
  + \mu E_{\rm sol}(v) \:,
\label{eq:P_pm_for_ppwave_1s}
\end{eqnarray}
so that the spectrum reproduces the GKP vacuum result \eqref{eq:P_pm_for_ppwaveGKP_omega=1}, with an additional solitonic excitation.

We may also obtain the full dispersion relation for these excitations by following the usual approach. Firstly, the worldsheet position $\sigma_{\rm sol} (\tau)$ of the soliton is still determined by the one-soliton solution of the sinh-Gordon equation. Secondly, the energy of the ``small'' spike is given by $\mu E_{\rm sol}(v)$ and is therefore almost identical to the expression we had in the previous cases. Finally, the Hamiltonian for the excitations is now given by $P_+$, which is conjugate to the coordinate $x_+$. The latter has the following asymptotic behaviour: $x_+ \to \tau / \mu + {\rm const.}$ as $\sigma \to \pm \infty$ for fixed $\tau$. Therefore we have
\begin{equation}
 \frac{d \sigma_{\rm sol}}{d x_+} = \frac{d \sigma_{\rm sol}}{d \tau} \frac{\partial \tau}{\partial x_+}
  = v \left( \frac{\partial x_+}{\partial \tau} \right)^{-1} \simeq \mu v
\label{eq:approx_in_Ham's_eq_pp-wave}
\end{equation}
up to subleading corrections. Thus, Hamilton's equation becomes
\begin{equation}
 \mu \frac{d E_{\rm sol}}{d P_{\rm sol}} = \frac{d \sigma_S}{d x_+} \simeq \mu v \:,
\label{eq:Hamilton_eq_1s_arc}
\end{equation}
where now the two extra factors of $\mu$ cancel out. At this point, the equation is identical to \eqref{eq:relation_v_E_sol_P_sol} and we also impose the same boundary condition $P_{\rm sol} (1) = 0$, due to the fact that the excitation still disappears for $v = 1$ and the one-soliton solution reduces to the GKP vacuum\footnote{In particular, it reduces to the special case $\tau_0 = - \pi/2 + 2n \pi$, for $n \in \mathbb{Z}$, of \eqref{eq:omega=1_ppwaveGKP}.}. Hence, the final result for $P_{\rm sol}$ is the same.

The dispersion relation is thus given by
\begin{eqnarray}
 \tilde{E}_{\rm sol}(v) & = & \mu \frac{\sqrt{\lambda}}{2\pi} \left[ \frac{1}{2} \log \left( \frac{1 + \sqrt{1-v^{2}}}{1 - \sqrt{1-v^{2}}}
                               \right) - \sqrt{1-v^{2}} \right] \nn\\
 \tilde{P}_{\rm sol}(v) & = & \frac{\sqrt{\lambda}}{2\pi}\left[\frac{\sqrt{1-v^{2}}}{v}
                               - {\rm Tan}^{-1}\left(\frac{\sqrt{1-v^{2}}}{v}\right)\right] \:,
\label{eq:ppw_soliton_disprel}
\end{eqnarray}
which agrees with the Giant Hole dispersion relation \eqref{eq:small_spike_disprel}, up to the extra factor of $\mu$ in the energy.
\paragraph{}
We also transport to $AdS_3$-pp-wave the two-soliton solution \eqref{eq:AdS_arc_2s}, through the usual procedure. The resulting expression is
\begin{eqnarray}
 z & = & 2 \sqrt{2} \: b \: ({\rm ch} T + v {\rm ch} X) \Big[({\rm sh} T + {\rm sh} X)^2 (1-v^2) {\rm ch}^2 \sigma \nonumber\\
   &   &  + ({\rm sh} T - {\rm sh} X)^2 (1-v^2) {\rm sh}^2 \sigma + ({\rm ch} X + v {\rm ch} T)^2 {\rm ch} 2 \sigma \nonumber\\
   &   &   - 2 \sqrt{1-v^2} \: {\rm sh} 2 \sigma \: {\rm sh} X ({\rm ch} X + v {\rm ch} T) \Big]^{- \frac{1}{2}} \nonumber
\end{eqnarray}
\begin{eqnarray}
 x_+ & = & \frac{1}{\mu} {\rm Tan}^{-1} \Big\{ \Big[ \cos \tau \Big( - v {\rm ch} T \: {\rm ch} \sigma + \sqrt{1-v^2} ({\rm sh} X - {\rm sh} T)
            {\rm sh} \sigma - {\rm ch} X \: {\rm ch} \sigma \Big) \nonumber\\
     &   & + \sin \tau \Big( v {\rm ch} T \: {\rm sh} \sigma + {\rm ch} X \: {\rm sh}
              \sigma - \sqrt{1-v^2} ( {\rm sh} T + {\rm sh} X ) {\rm ch} \sigma \Big) \Big] \nonumber\\
     &   & \Big[ \cos \tau \Big( v {\rm ch} T \: {\rm sh} \sigma + {\rm ch} X \: {\rm sh} \sigma - \sqrt{1-v^2} ({\rm sh} T + {\rm sh} X)
            {\rm ch} \sigma \nonumber\\
     &   & + \sin \tau \Big( v {\rm ch} T \: {\rm ch} \sigma + \sqrt{1-v^2} ({\rm sh} T - {\rm sh} X) {\rm sh} \sigma
                + {\rm ch} X \: {\rm ch} \sigma \Big) \Big]^{-1} \Big\} \nonumber\\
 x_- & = & 4 \mu b^2 \Big\{ -2 {\rm sh} X ({\rm ch} X + v {\rm ch} T) \sqrt{1-v^2} \: {\rm ch} 2 \sigma \nonumber\\
     &   &  + \Big[ ({\rm ch} X + v {\rm ch} T)^2 + ({\rm sh}^2 T + {\rm sh}^2 X) (1-v^2) \Big] {\rm sh} 2 \sigma \Big\} \nonumber\\
     &   & \times \Big[({\rm sh} T + {\rm sh} X)^2 (1-v^2) {\rm ch}^2 \sigma + ({\rm sh} T - {\rm sh} X)^2 (1-v^2) {\rm sh}^2 \sigma \nonumber\\
     &   & + ({\rm ch} X + v {\rm ch} T)^2 {\rm ch} 2 \sigma - 2 \sqrt{1-v^2} \: {\rm sh} 2 \sigma \: {\rm sh} X ({\rm ch} X + v {\rm ch} T) 
            \Big]^{-1} \:,
\label{eq:ppw_2s_arc}
\end{eqnarray}
with the customary identifications $X = 2 \gamma \sigma$, $T = 2 \gamma \tau$, $\gamma = 1/ \sqrt{1-v^2}$, ${\rm ch} \equiv \cosh$, ${\rm sh} \equiv \sinh$ and $\rho_0 = \rho_1 - \log b$.

\begin{figure}%
\includegraphics[width=\columnwidth]{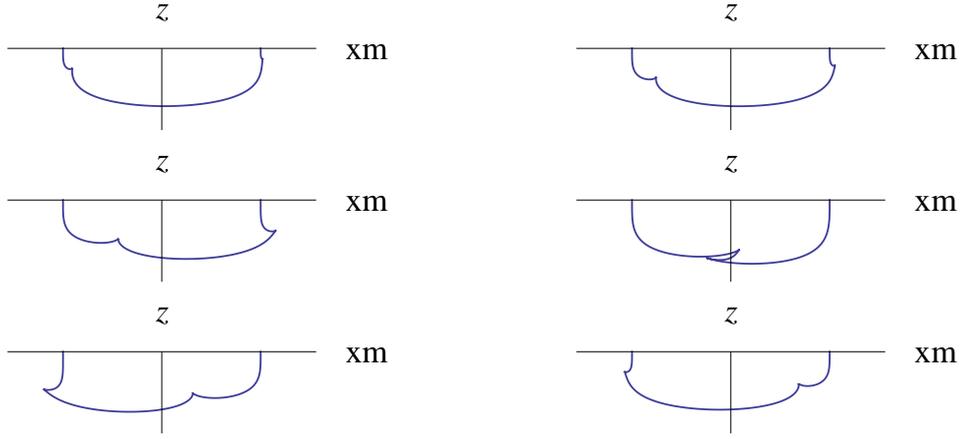}%
\caption{The two-soliton arc of string in $AdS_3$-pp-wave with $v=0.5$, $\mu = 0.6$ and $b = 2$ at $t = 3\pi/2 + k$, for $k=0, \ldots,5$ (from left to right and top to bottom).}%
\label{fig:ppwarc2s}%
\end{figure}

The features of this solution are essentially identical to those of \eqref{eq:ppw_1s_arc}, apart from the fact that is displays two solitons propagating along the drooping arc. In particular, the endpoints are still stationary with the separation \eqref{eq:Delta_x-_1s}. Some plots at constant $x_+$ are given in Fig. \ref{fig:ppwarc2s}. The expressions for the conserved charges have proved to be unwieldy and we have been unable to compute an expression for the spectrum of $P_+$, but the analysis of the one-solution solution should already provide enough evidence in support of the existence of the solitonic excitations and of their properties.

\chapter{Conclusions}

\paragraph{The finite-gap analysis.} The integrability of string theory on $AdS_5 \times S^5$ manifests itself in the existence of a Lax formulation for the equations of motion and Virasoro constraints, which allows to describe the semiclassical conserved charges of a very general class of string solutions in terms of algebraic curves, known as spectral curves. In chapter \ref{ch:FGA}, we have considered the most general family of spectral curves of finite genus associated with strings moving in $AdS_3 \times S^1$ and we have studied their large angular momentum limit, $S \to \infty$. We have seen that, in this limit, the generic $K$-gap curve of genus $K-1$ factorises into two separate Riemann surfaces, $\tilde{\Sigma}_1$ and $\tilde{\Sigma}_2$. The features of the resulting spectrum are controlled by a parameter $M = 0, 1, \ldots, K-2$.

The first surface has genus $K-M-2$ and determines the leading behaviour of the spectrum, $\Delta - S \simeq (\sqrt{\lambda}/2 \pi) (K-M) \log S$, while also introducing an $O(S^0)$ contribution. According to our proposed interpretation \cite{Dorey:2008zy,Dorey:2010id}, the associated string solution should develop $K-M$ ``large'' spikes or cusps, which approach the boundary of $AdS_3$ as $S$ becomes large.

The second surface always has genus $0$ and it contains $M$ simple poles, each of which generates a separate $O(S^0)$ contribution to the spectrum. We have conjectured that each pole introduces a ``small'' spike propagating along the background of ``large'' spikes in the corresponding string solution. Such cusps do not become infinitely long in the large $S$ limit and, due to the integrability of string theory, are expected to be solitonic objects undergoing factorised scattering.

Very importantly, the semiclassical finite-gap spectrum in the large angular momentum limit reproduces the semiclassical spectrum of the integrable spin chain whose Hamiltonian represents the one-loop dilatation operator of SYM theory in the $\mathfrak{sl} (2)$ sector \cite{Belitsky:2006en}, in the large conformal spin limit. In particular, the gauge theory spectrum is also characterised by a spectral curve, which in fact coincides with $\tilde{\Sigma}_1$. An important characteristic of this spin chain is the fact that it may be parametrised in terms of ``small'' and ``large'' holes, which are respectively dual to ``small'' and ``large'' spikes (although the duality only becomes manifest when the hole momentum is large in the case of ``small'' holes).

Due to the generality of the finite-gap construction, these results should apply to any string living in $AdS_3 \times S^1$, in the limit of large angular momentum $S$.

\paragraph{Explicit solutions in $AdS_3$.} The main result outlined in chapter \ref{sec:explicit_solutions} is a series of detailed tests concerning the finite-gap spectrum and its interpretation in terms of ``large'' and ``small'' spikes, carried out on several families of explicit string solutions living in $AdS_3$. In particular, we have seen that the string equivalents of the semiclassical highly excited spin vectors of gauge theory may be extracted from the ``large'' spikes and succesfully used in order to reconstruct the Riemann surface $\tilde{\Sigma}_1$, which coincides with the gauge theory spectral curve. This result also confirms the emergence of a discrete set of degrees of freedom on the string side of the AdS/CFT correspondence in the large $S$ limit.


Specifically, the tests on ``large'' spikes have involved the simpler $N$-folded GKP string, the more general symmetric Kruczenski spiky string, parametrised in terms of a single modulus, and finally a very general family of approximate string solutions, associated with a $(K-1)$-dimensional moduli space. The tests on ``small'' spikes focused instead on strings carrying one or two excitations above the GKP vacuum.

The properties of the ``small'' spikes have induced us to formulate three conjectures. The first is that Giant Holes (or ``small'' spikes) are the string duals of the gauge theory ``small'' holes, extrapolated at strong coupling. Hence, there exists an object which reduces to ``small'' holes at weak coupling and to ``small'' spikes at strong coupling. The second conjecture is then that the prefactor appearing in the large-momentum dispersion relation for such an object is given by the cusp anomalous dimension $\Gamma (\lambda)$. Both these conjectures were motivated by the structure of the ``small'' spike dispersion relation at large momentum. The third conjecture, motivated by the low-momentum behaviour of the dispersion relation, is that ``small'' spikes are continuously connected to the small quadratic fluctuations of the GKP string, where the former are essentially the large-momentum version of the latter. As we previously mentioned, these three conjectures were recently verified in \cite{Basso:2010in}, where the excitations of the GKP string were studied at any value of the coupling $\lambda$.

Moreover, we have also introduced a patched solution displaying both ``large'' and ``small'' spikes at infinite angular momentum $S$, which is consistent with all the previous predictions. Ideally, we would want to be able to construct such a general solution at finite $S$, and then to study the $S \to \infty$ limit, which would allow us to reconstruct the spectral curve.
\paragraph{}
An important property worth mentioning is the fact that, at least in every solution studied here, all the spikes, whether ``large'' or ``small'', correspond to sinh-Gordon solitons. In particular, ``large'' spikes correspond to static solitons, while ``small'' spikes correspond to solitons with non-vanishing velocity. The corresponding solutions of the sinh-Gordon equation are periodic in the case of a closed string such as the Kruczenski symmetric spiky string and are instead non-periodic in the case of an open string with endpoints on the boundary, such as the one-spike and two-spike solutions. This correspondence between spikes in $AdS_3$ strings and sinh-Gordon solitons was proved in \cite{Jevicki:2007aa,Jevicki:2008mm} and later extended in \cite{Jevicki:2009uz}, where the most general solution with $N$ spikes propagating along the infinite GKP string was obtained. Extending such a result to the periodic case, including both static solitons and solitons with non-zero velocities would probably lead to a solution of the type shown in Fig. \ref{fig:large_small_spikes}, exhibiting both ``large'' and ``small'' spikes at large but finite $S$.

Various properties, such as their solitonic nature and their $O(\sqrt{\lambda},S^0)$ contribution to the spectrum, indicate the ``small'' spikes as the $AdS_3$ version of the Giant Magnons living in $\mathbb{R} \times S^2$, which, together with the duality with the holes of the gauge theory spin chain, motivates the proposal of the name ``Giant Holes'' for describing these objects.

\paragraph{$AdS_3$-pp-wave.} Finally, we have also shown how some of the previously studied explicit string solutions may be transported to the $AdS_3$-pp-wave region of $AdS_3$ space, thereby constructing solutions to the corresponding equations of motion and Virasoro constraints. In particular, we have constructed the well-known (see \cite{Ishizeki:2008tx} and references therein) GKP arc with endpoints on the boundary and drooping towards the interior of $AdS_3$ and two of its excited states, respectively carrying one and two ``small'' spikes propagating along the vacuum background. Such solutions may be helpful to studies concerning the strong interaction between highly energetic gluons, since they describe excited states of the two-gluon system, where the endpoints of the string represent the two particles, while the arc of string joining them corresponds to the chromomagnetic flux tube.

\appendix

\chapter{Calculations supplementing the finite-gap analysis}

\section{The matching condition}
\label{sec:matching_condition}

In this section we sketch how to deal with the only period condition which involves both surfaces $\tilde{\Sigma}_1$ and $\tilde{\Sigma}_2$ at the same time:
\begin{equation}
\oint_{\hat{\mathcal{A}}_\frac{K-M}{2}^+} dp = 0 \:.
\label{eq:repeat_match_cond}
\end{equation}
For our initial analysis, we only need the final expression for $d \tilde{p}_1$ \eqref{eq:explicit_dp1tilde_Mn0}. We start by splitting the integral into two separate contributions:
\begin{equation}
 \oint_{\hat{\mathcal{A}}_\frac{K-M}{2}^+} dp = 2 I_1 + 2 I_2 = 0
\label{eq:splitting_integral_for_matching_condition_Mn0}
\end{equation}
with:
\begin{eqnarray}
 I_1 & \equiv & - \frac{1}{2} \int_{b^-}^{b^+} \frac{dx}{y} \sum_{l=M}^{K-2} C_l x^l \nonumber\\
 I_2 & \equiv & - \frac{1}{2} \int_{(a^{(1)}_+)^+}^{(a^{(1)}_+)^-} \frac{dx}{y} \left[ \sum_{l=0}^{M-1} C_l x^l + \frac{\pi J}{\sqrt{\lambda}}
        \right. \nonumber\\
     &        & \left. \phantom{\sum_{l=M}^{K-2}} \times \left( \frac{y_+}{(x-1)^2} + \frac{y_-}{(x+1)^2} + \frac{y'_+}{x-1} + \frac{y'_-}{x+1}
                 \right) \right]
\label{eq:def_I1_I2_Mn0}
\end{eqnarray}
where we have opened up the contour at the left endpoint for $I_1$ and at the right endpoint for $I_2$, turning the integrals into open chains which start at $b$ and respectively $a^{(1)}_+$ on one side of the corresponding cut and end at $b$ and $a^{(1)}_+$ on the other side (the points $b^\pm$ and $(a^{(1)}_+)^\pm$ lie near $b$ and $a^{(1)}_+$ respectively, at the opposide sides of the cut).

Both integrals can be evaluated on $\tilde{\Sigma}_1$, by introducing the change of variables $x = \rho \tilde{x}$. This will introduce factors of the type $\sqrt{\tilde{x} - \epsilon b^{(j)}_\pm}$ into $y$, which can be treated by making use of the binomial expansion\footnote{Strictly speaking, these expansions only converge for $\tilde{x} > \textrm{max } \{ \epsilon b^{(M)}_+, - \epsilon b^{(M)}_- \}$. One way around the problem is to introduce the series, at first restricting ourselves to the region of convergence. After this, we swap the sum with the integral and only then we remove the regulator by letting $\tilde{x}$ reach the endpoint of integration, $\tilde{x} = \epsilon b$. This is what is done below. Another way is to divide the problematic region of the contour into several segments, each joining two consecutive points of the set $\{ \epsilon b^{(j)}_+, - \epsilon b^{(j)}_- \}$. We can then introduce appropriate converging binomial expansions (either of the form \eqref{eq:binomial_expansion_for_sqrt_in_y_on_S1} or with $k_j^\pm$ and $- 1/2 - k_j^\pm$ interchanged) for each interval. The final result is the same.}:
\begin{equation}
 \frac{1}{\sqrt{\tilde{x} - \epsilon b^{(j)}_\pm}} = \sum_{k_j^\pm = 0}^\infty \binom{- \frac{1}{2}}{k_j^\pm} (- \epsilon b^{(j)}_\pm )^{k_j^\pm}
  \tilde{x}^{- \frac{1}{2} - {k_j^\pm}} \:.
\label{eq:binomial_expansion_for_sqrt_in_y_on_S1}
\end{equation}
One can also similarly expand the terms $(x \pm 1)^{-1}$ and $(x \pm 1)^{-2}$ appearing in $I_2$.

In the case of $I_1$, we have the integral of an infinite sum over $2(M+1)$ indices (one for each square root factor we had to expand), which we can write as:
\begin{eqnarray}
 I_1 & = & - \frac{1}{2} \sum_{k_0^\pm, \ldots, k_M^\pm = 0}^\infty \epsilon^{k_\mathrm{tot}} \left[ \prod_{j=0}^M \binom{- \frac{1}{2}}{k_j^+}
  \binom{- \frac{1}{2}}{k_j^-} (- b^{(j)}_+ )^{k_j^+} (- b^{(j)}_- )^{k_j^-} \right] \nonumber\\
     &   & \quad \times \int_{(\epsilon b)^-}^{(\epsilon b)^+} \frac{d \tilde{x}}{\tilde{y}_1} 
            \sum_{l=M}^{K-2} \tilde{C}_l \tilde{x}^{l -(M+1) - k_\mathrm{tot}}
\label{eq:I1_after_expanding_sqrts_Mn0}
\end{eqnarray}
with $k_\mathrm{tot} = \sum_{j=0}^M (k_j^+ + k_j^-)$ and $b^{(0)}_\pm = \pm b$. The remaining integral can now be calculated straightforwardly. For $k_\mathrm{tot} > 0$, the leading order is easily seen to be $O(1)$ ($\tilde{y}_1 (\tilde{x}) \simeq \tilde{y}_1 (0) = \tilde{Q}$). We indicate the sum of all these constant terms as $\hat{I}_1 (c^{(j)}_\pm)$ (at leading order, all the $b^{(j)}_\pm$ reduce to one of the $c^{(j)}_\pm$, according to \eqref{eq:bps_coalesce}, while $b = 1$, thus this expression only depends on the moduli $c^{(j)}_\pm$).

The remaining $k_\mathrm{tot} = 0$ term can be rewritten as:
\begin{equation}
 - \frac{1}{2} \int_{(\epsilon b)^-}^{(\epsilon b)^+} \frac{d \tilde{x}}{\tilde{y}_1} 
            \sum_{l=M}^{K-2} \tilde{C}_l \tilde{x}^{l -M-1}
 = \frac{1}{2} \int_{(\epsilon b)^-}^{(\epsilon b)^+} d \tilde{p}_1
 = \frac{1}{2} [ 2 \tilde{p}_1 (\epsilon b) + 2 \pi (K-M) ]
\label{eq:writing_ktot=0_term_in_I1_in_terms_of_p1tilde_Mn0}
\end{equation}
where we have used the discontinuity properties of $\tilde{p}_1$ in the last step. \eqref{eq:explicit_dp1tilde_Mn0} then implies:
\begin{equation}
 I_1 = - i (K-M) \log \epsilon + i \log \left( \frac{\tilde{q}_{K-M}}{b^{K-M}} \right) + \pi (K-M) + \hat{I}_1 (c^{(j)}_\pm) + O(\epsilon)
\label{eq:final_form_of_I1_Mn0}
\end{equation}
For the moment, the only feature of $I_1$ we are interested in is the fact that it diverges as $\log \epsilon$ in the limit $\epsilon \to 0$. Due to the A-cycle condition \eqref{eq:splitting_integral_for_matching_condition_Mn0}, $I_2$ must then also diverge in the same limit. However, a similar analysis to the one carried out for $I_1$ shows that $I_2$\footnote{This analysis requires us to open up the contour for $I_2$ at $b$, as we did for $I_1$, and not at $a^{(1)}_+$ as indicated above. This is also a legitimate operation. The above definition will instead be useful when we will analyse $I_2$ on $\tilde{\Sigma}_2$.} is $O(1) + O( \epsilon )$. The solution lies in the fact that some of the $O(1)$ terms are proportional to $1/ \sqrt{1 - b^2}$ (these terms originate from $y'(\pm 1)$). $I_2 \sim i \log \epsilon$ then implies:
\begin{equation}
 \frac{1}{\sqrt{1 - b^2}} \sim i \log \epsilon
\label{eq:asymptotic_behaviour_of_sqrt(1-b^2)_from_matching_condition_Mn0}
\end{equation}
which is the result we referred to in equation \eqref{eq:sqrt(1-b^2)_lead_ord_Mn0}. As we saw in the following discussion, this implies that the branch points on $\tilde{\Sigma}_2$ must coalesce and that $d \tilde{p}_2$ must take the form \eqref{eq:explicit_dp2tilde_no_match_cond}.

We can then proceed to evaluate $I_2$ on $\tilde{\Sigma}_2$, i.e. without changing variables to $\tilde{x}$. The steps are similar to what we did in the other coordinates. In particular, we now have to expand the following factors coming from $y$\footnote{Note that ensuring the convergence of the binomial series again requires particular care. We need $|x \epsilon / \tilde{a}^{(j)}_\pm| < 1$, for $j = 1, \ldots, K-M-1$. If we shrink the contour onto the real axis, we can safely assume $|x| \in (0, \rho \tilde{a}^{(1)}_+]$ over the whole domain of integration. The problem arises near the upper limit: $\tilde{a}^{(1)}_+ / |\tilde{a}^{(j)}_\pm|$ is not necessarily less than $1$ for all $j$. However, $\mathrm{min} \{ |\tilde{a}^{(1)}_-| , \tilde{a}^{(1)}_+ \} / |\tilde{a}^{(j)}_\pm| \leq 1$, $\forall j$, and hence we should actually evaluate the matching condition on $\hat{\mathcal{A}}^-_{(K-M)/2}$ instead of $\hat{\mathcal{A}}^+_{(K-M)/2}$ when $|\tilde{a}^{(1)}_-| < \tilde{a}^{(1)}_+$. Nonetheless, all the calculations would work in exactly the same way and the final version \eqref{eq:match_cond_without_infinite_sums} of the matching condition would be identical (as we may guess from the fact that it doesn't depend on $\tilde{a}^{(1)}_+$). Alternatively, we could apply the same reasoning as we discussed for \eqref{eq:binomial_expansion_for_sqrt_in_y_on_S1}.}:
\begin{equation}
 \frac{1}{\sqrt{x \epsilon - \tilde{a}^{(j)}_\pm}} = \sum_{k_j^\pm = 0}^\infty \binom{- \frac{1}{2}}{k_j^\pm}
  (- \tilde{a}^{(j)}_\pm )^{- \frac{1}{2} - k_j^\pm} (x \epsilon)^{k_j^\pm} \:.
\label{eq:binomial_expansion_for_sqrt_in_y_S2}
\end{equation}
Again the integral turns into an infinite sum of integrals. This time, however, the $k_\mathrm{tot} > 0$ part is $O( \epsilon \log \epsilon)$ and hence it does not contribute. We are left with the $k_\mathrm{tot} = 0$ term only:
\begin{eqnarray}
 I_2 & \simeq & - \frac{1}{2} \int_{(\rho \tilde{a}^{(1)}_+)^+}^{(\rho \tilde{a}^{(1)}_+)^-} \frac{dx}{\tilde{y}_2} \left[ 
            \frac{1}{\tilde{Q}} \sum_{l=0}^{M-1} \tilde{C}_l x^l \right. \nonumber\\
     &   & \left. \phantom{\sum_{l=0}^{M-1}} + \frac{\pi J}{\sqrt{\lambda}} \left( \frac{\tilde{y}_2 (1)}{(x-1)^2}
            + \frac{\tilde{y}_2 (-1)}{(x+1)^2} + \frac{\tilde{y}'_2 (1)}{x-1} + \frac{\tilde{y}'_2 (-1)}{x+1} \right) \right] \:.
\label{eq:I2_after_eliminating_sqrts}
\end{eqnarray}
We now observe that the differential appearing in the above integral is \emph{not} the part of $dp$ which contributes to $d \tilde{p}_2$ as $\rho \to \infty$. In particular, it is missing the $l = M$ term in the first sum. We will call the limit as $\rho \to \infty$ of this ``incomplete'' differential $d \hat{p}_2$.

In order to determine $d \hat{p}_2$, one may go through the same steps which led us to $d \tilde{p}_2$, most of which remain identical. The only difference is that now:
\begin{equation}
 \lim_{x \to \infty} h(x) = 0
\label{eq:limit_as_x_to_inf_of_h(x)_for_dp2hat}
\end{equation}
which has the effect of killing the only term in \eqref{eq:explicit_dp2tilde_no_match_cond} that is not regular at infinity:
\begin{equation}
 d \hat{p}_2 = dw_0 + \sum_{j=1}^\frac{M}{2} (dw_j^+ + dw_j^-) \:.
\label{eq:explicit_dp2hat}
\end{equation}
Hence, we may write:
\begin{equation}
 I_2 = \frac{1}{2} \int_{(\rho \tilde{a}^{(1)}_+)^+}^{(\rho \tilde{a}^{(1)}_+)^-} [ d \hat{p}_2 (x) + O(\epsilon \log \epsilon) ]
  \qquad \textrm{as } \epsilon = 1/ \rho \to 0
\label{eq:I_2_in_terms_of_dp2hat}
\end{equation}
where now $O( \epsilon \log \epsilon)$, as a function of $x$, is manifestly integrable on the whole domain of integration, even as $\epsilon \to 0$, due to the fact that both the starting differential appearing in \eqref{eq:I2_after_eliminating_sqrts} and $d \hat{p}_2$ are integrable. This means that its primitive does not diverge as $x \to \infty$ and hence this contribution still vanishes even after integration. Therefore we may neglect it when computing $I_2$ up to $O(1)$:
\begin{eqnarray}
 I_2 & = & \frac{1}{2} \int_{(\rho \tilde{a}^{(1)}_+)^+}^{(\rho \tilde{a}^{(1)}_+)^-} d \hat{p}_2 (x) + \ldots \nonumber\\
     & = & \frac{2 \pi}{\sqrt{\lambda}} \frac{J}{\sqrt{1-b^2}} + \frac{1}{2i} \sum_{j=1}^\frac{M}{2} 
            \left[ T( c^{(j)}_+ ) + T( c^{(j)}_- ) \right] + \frac{\hat{L}_2}{2} + \ldots
\label{eq:I2_up_to_O(1)}
\end{eqnarray}
where $T(c)$ was defined in \eqref{eq:def_T(c)}, the dots denote corrections which vanish in the limit $\rho \to \infty$ and $\hat{L}_2$ is the constant part of the discontinuity of $\hat{p}_2$ across its cut in the region $x > c^{(M/2)}_+$ ($\hat{p}_2 (x + i \epsilon) + \hat{p}_2 (x - i \epsilon) = \hat{L}_2$ for $x \in \mathbb{R}$, $x > c^{(M/2)}_+$).

Thus, the matching condition, up to $O(1)$ in $\epsilon$, yields:
\begin{multline}
 \frac{2 \pi}{\sqrt{\lambda}} \frac{J}{\sqrt{1-b^2}} = i (K-M) \log \epsilon - i \log \left( \frac{\tilde{q}_{K-M}}{b^{K-M}} \right) 
  - \pi (K-M) - \hat{I}_1 (c^{(j)}_\pm) \\
 - \frac{1}{2i} \sum_{j=1}^\frac{M}{2} \left[ T( c^{(j)}_+ ) + T( c^{(j)}_- ) \right] - \frac{\hat{L}_2}{2} \:.
\label{eq:matching_condition_up_to_O(1)_S1}
\end{multline}
Before substituting back into $d \tilde{p}_2$, we will impose the matching condition through a slightly different procedure, which will yield an alternative version of the above expression. By comparing the two versions, we will then see that the expression simplifies.

The idea is now to evaluate \eqref{eq:repeat_match_cond} entirely on $\tilde{\Sigma}_2$, without splitting the integral into two contributions. The first step is to open up the contour at $a^{(1)}_+$:
\begin{equation}
 \int^{(a^{(1)}_+)^+}_{(a^{(1)}_+)^-} dp = 0
\label{eq:opening_up_contour_match_cond_S2}
\end{equation}
where, as before, the points ${(a^{(1)}_+)^\pm}$ lie near $x = a^{(1)}_+$ at the opposite sides of the cut.

Again, the integrand develops factors of the type \eqref{eq:binomial_expansion_for_sqrt_in_y_S2} and we may turn the integral into an infinite sum of integrals by using the same binomial expansion. The $k_\mathrm{tot} > 0$ term is $O(1)$, as well as the part of the $k_\mathrm{tot} = 0$ contribution which depends on the $\tilde{C}_l$, for $l = M+1, \ldots, K-2$. We indicate the total of these two contributions as $2 \hat{I}_2 (\tilde{q}_j)$ ($\hat{I}_2$ depends on the $\tilde{a}^{(j)}_\pm$ and on the $\tilde{C}_l$, for $l = M, \ldots, K-2$; in the $S \to \infty$ limit, both sets of parameters only depend on the $\tilde{q}_j$ through \eqref{eq:explicit_y1tilde} and \eqref{eq:Ctilde(qtilde)_Mn0}).

The remaining integrand reduces to $d \tilde{p}_2$ as $\rho \to \infty$, so that \eqref{eq:opening_up_contour_match_cond_S2} may be written as:
\begin{equation}
 2 \hat{I}_2 (\tilde{q}_j) + \int^{(\rho \tilde{a}^{(1)}_+)^+}_{(\rho \tilde{a}^{(1)}_+)^-} d \tilde{p}_2 = 0
\label{eq:match_cond_with_dp2tilde_S2}
\end{equation}
and we may then use \eqref{eq:explicit_dp2tilde_no_match_cond} to recast the matching condition on $\tilde{\Sigma}_2$ into the following form:
\begin{equation}
 \frac{2 \pi}{\sqrt{\lambda}} \frac{J}{\sqrt{1-b^2}} = - \hat{I}_2 (\tilde{q}_j) + \frac{K-M}{i} \log (2 \rho \tilde{a}^{(1)}_+)
  - \frac{1}{2i} \sum_{j=1}^\frac{M}{2} \left[ T(c^{(j)}_+) + T(c^{(j)}_-) \right] - \frac{\tilde{L}_2}{2}
\label{eq:match_cond_on_S2}
\end{equation}
where the moduli-independent constant $\tilde{L}_2$ is obtained from the discontinuity of $\tilde{p}_2$ at its cut: $\tilde{p}_2 (x+i \epsilon) + \tilde{p}_2 (x - i \epsilon) = \tilde{L}_2$, for $x \in \mathbb{R}$, $x > c^{(M/2)}_+$.

By equating the RHS of \eqref{eq:matching_condition_up_to_O(1)_S1} and \eqref{eq:match_cond_on_S2}, we find:
\begin{multline}
 \hat{I}_2 (\tilde{q}_j) + i (K-M) \log (2 \tilde{a}^{(1)}_+) - i \log ( \tilde{q}_{K-M} ) = \\ \hat{I}_1 (c^{(j)}_\pm) - i (K-M) \log b + \pi (K-M) + \frac{\hat{L}_2}{2} - \frac{\tilde{L}_2}{2}
\label{eq:comparing_2_versions_of_match_cond}
\end{multline}
At this point, we observe that the $K-1$ independent degrees of freedom, which parametrise this class of finite-gap solutions after all the period conditions have been implemented, are given by $\tilde{q}_j$, $j = 2, \ldots, K-M$, and $c^{(k)}_\pm$, $k = 1, \ldots, M/2$. The quasi-momentum $\tilde{p}_1$ on $\tilde{\Sigma}_1$ is completely determined by the $\tilde{q}_j$, while $\tilde{p}_2$ on $\tilde{\Sigma}_2$ is completely determined by the $c^{(k)}_\pm$. The matching condition does not impose any extra constraint on these $K-1$ moduli; instead, it determines the behaviour of $b$ as $\rho \to \infty$, i.e. $b = 1 + \ldots$, where the dots represent vanishing corrections (in other words, it determines one of the parameters of the curve as a function of the moduli, so that in the end the only free parameters left are the moduli themselves).

If we look at equation \eqref{eq:comparing_2_versions_of_match_cond} in the limit $\rho \to \infty$, and we neglect the $\log b$ term, which vanishes at $O(1)$, we easily see that the LHS only depends on the $\tilde{q}_j$, while the RHS only depends on the $c^{(k)}_\pm$\footnote{All of this strictly holds only in the $S \to \infty$ limit.}. As we have just explained, this equation cannot be used in order to eliminate one of the moduli in terms of the others, and hence it can only be satisfied if both sides are equal to a moduli-independent constant, which we call $R'$.

This, together with any of the expressions \eqref{eq:matching_condition_up_to_O(1)_S1} and \eqref{eq:match_cond_on_S2}, yields:
\begin{equation}
 \frac{2 \pi}{\sqrt{\lambda}} \frac{J}{\sqrt{1-b^2}} = - i (K-M) \log \rho - i \log (\tilde{q}_{K-M}) - \frac{1}{2i} \sum_{j=1}^\frac{M}{2} \left[ T(c^{(j)}_+) + T(c^{(j)}_-) \right] - R' - \frac{\tilde{L}_2}{2}
\label{eq:match_cond_final_R'}
\end{equation}
which is easily seen to reduce to \eqref{eq:match_cond_without_infinite_sums} under the identification $R = R' + \tilde{L}_2 / 2$.

\section{Subleading behaviour of the branch points on $\tilde{\Sigma}_2$}
\label{sec:behav_branch_pts_Sigma2tilde}

In this section, we will use the explicit form of $d \tilde{p}_2$, which was derived assuming that $b^{(2j)}_\pm - c^{(j)}_\pm = O (\epsilon^\alpha) = b^{(2j-1)}_\pm - c^{(j)}_\pm$, for $j = 1, \ldots, M/2$ and for some $\alpha > 0$, in order to check that this is actually the limiting behaviour of the branch points as $\rho \to \infty$.

For this purpose, we are now going to have a closer look at the $M$ A-cycle conditions associated with $\hat{\mathcal{A}}_I^\pm$ for $I = (K-M)/2 + 1, \ldots, K/2$. Since these contours lie on $\tilde{\Sigma}_2$, the integrals are easier to deal with if we work with the unrescaled spectral parameter $x$, so that $dp \to d \tilde{p}_2$. We can then write:
\begin{equation}
 0 = \oint_{\hat{\mathcal{A}}_{\frac{K}{2}-j+1}^\pm} dp = \int^{(b^{(2j-1)}_\pm)^+}_{(b^{(2j-1)}_\pm)^-} dp 
\label{eq:A-periods_on_Sigma2tilde_written_as_one-endpoint_integrals_Mn0}
\end{equation}
for $j = 1, \ldots, M/2$, where again by $(b^{(2j-1)}_\pm)^+$ and $(b^{(2j-1)}_\pm)^-$ we mean points infinitesimally close to the branch point $b^{(2j-1)}_\pm$ and on opposite sides of the corresponding cut. We will only be interested in the leading order part of this equation as $\rho \to \infty$, which will be dominated by the diverging contributions from $dw_0$ and $dw_j^\pm$ (the latter is due to the fact that the A-cycle becomes pinched at the pole $c^{(j)}_\pm$ in the limit considered). We may therefore replace $dp$ with $d \tilde{p}_2$; we also introduce $\eta^{(j)}_\pm \gtrsim 0$ as a measure of how fast $b^{(2j-1)}_\pm$ moves towards $c^{(j)}_\pm$: $b^{(2j-1)}_\pm = c^{(j)}_\pm \mp \eta^{(j)}_\pm$. \eqref{eq:A-periods_on_Sigma2tilde_written_as_one-endpoint_integrals_Mn0} then becomes
\begin{equation}
 \int^{(c^{(j)}_\pm \mp \eta^{(j)}_\pm)^+}_{(c^{(j)}_\pm \mp \eta^{(j)}_\pm)^-} d \tilde{p}_2 = 0 \:.
\label{eq:A-periods_on_Sigma2tilde_approx_to_lead_order_Mn0}
\end{equation}
At leading order, we have
\begin{eqnarray}
 \int^{(c^{(j)}_\pm \mp \eta^{(j)}_\pm)^+}_{(c^{(j)}_\pm \mp \eta^{(j)}_\pm)^-} dw_0 & \simeq & \frac{2 \pi}{\sqrt{\lambda}} \frac{J}{\sqrt{1-b^2}} 
  \frac{c^{(j)}_\pm}{\sqrt{(c^{(j)}_\pm)^2 - 1}} \nonumber\\
 \int^{(c^{(j)}_\pm \mp \eta^{(j)}_\pm)^+}_{(c^{(j)}_\pm \mp \eta^{(j)}_\pm)^-} dw_j^\pm & \simeq & \frac{1}{i} \log \eta_j^\pm
\label{eq:diverging_terms_in_A-periods_Sigma2tilde_Mn0}
\end{eqnarray}
and from \eqref{eq:A-periods_on_Sigma2tilde_approx_to_lead_order_Mn0} we then get
\begin{eqnarray}
 \log \eta^{(j)}_\pm & = & - \frac{2 \pi i}{\sqrt{\lambda}} \frac{J}{\sqrt{1-b^2}} \frac{c^{(j)}_\pm}{\sqrt{(c^{(j)}_\pm)^2 - 1}} \nonumber\\
 								 & \simeq & \frac{c^{(j)}_\pm}{\sqrt{(c^{(j)}_\pm)^2 - 1}} (K-M) \log \epsilon
\label{eq:lead_behaviour_of_b_j-branch_points_j_odd_Mn0}
\end{eqnarray}
where in the last step we made use of the matching condition \eqref{eq:match_cond_without_infinite_sums} and the branch of the square root should always be chosen so that the RHS is negative, since $\log \eta^{(j)}_\pm \to - \infty$ (note that $c^{(j)}_- < 0$ while $c^{(j)}_+ > 0$).

We also notice that:
\begin{equation}
 \eta^{(j)}_\pm \sim \epsilon^{\alpha^{(j)}_\pm} \qquad \textrm{as } \epsilon \to 0
\label{eq:lead_vanishing_order_in_b_j-branch_points_j-odd}
\end{equation}
where $\alpha^{(j)}_\pm = (K-M) c^{(j)}_\pm / \sqrt{(c^{(j)}_\pm)^2 - 1} > 0$ so that our previous assumption that corrections to $b^{(j)}_\pm \simeq c^{(j)}_\pm$ are $O(\epsilon^\alpha)$ for some positive $\alpha$ is consistently verified for all $b^{(j)}_\pm$ with odd $j$.

In order to extend the verification to the remaining branch points, we can consider the original A-cycles $\mathcal{A}_{K/2-j+1}^\pm$, for $j=2, \ldots, M/2$. All these contours completely lie on $\tilde{\Sigma}_2$ and are pinched at two different points, $x = c^{(j)}_\pm$ and $x = c^{(j-1)}_\pm$, as $\rho \to \infty$. Accordingly, the corresponding A-periods will reduce to twice the following open chains:
\begin{equation}
 0 = \oint_{\mathcal{A}_{K/2-j+1}^\pm} dp \simeq 2 \int_{b^{(2j-1)}_\pm}^{b^{(2j-2)}_\pm} d \tilde{p}_2
\label{eq:original_A-cycles_on_Sigma2tilde_for_b^j_j_even}
\end{equation}
where we are going to introduce another small regulator: $b^{(2j-2)}_\pm = c^{(j-1)}_\pm \pm \epsilon^{(j-1)}_\pm$, for $j = 2, \ldots, M/2$. These integrals will receive, at leading order, diverging contributions from $dw_0$, $dw_j^\pm$ and $dw_{j-1}^\pm$:
\begin{multline}
 \int_{b^{(2j-1)}_\pm}^{b^{(2j-2)}_\pm} d \tilde{p}_2 = \int_{c^{(j)}_\pm \mp \eta^{(j)}_\pm}^{c^{(j-1)}_\pm \pm \epsilon^{(j-1)}_\pm} 
  d \tilde{p}_2 \\
 \simeq i (K-M) \left[ \frac{c^{(j-1)}_\pm}{\sqrt{(c^{(j-1)}_\pm)^2-1}} - \frac{c^{(j)}_\pm}{\sqrt{(c^{(j)}_\pm)^2-1}} \right] \log \epsilon
  + \frac{1}{i} \log \epsilon^{(j-1)}_\pm - \frac{1}{i} \log \eta^{(j)}_\pm
\label{eq:A-cycle_condition_determines_epsilon_j}
\end{multline}
If we now set this equal to $0$ and substitute in the previous result from \eqref{eq:lead_behaviour_of_b_j-branch_points_j_odd_Mn0}, we obtain:
\begin{equation}
 \log \epsilon^{(j-1)}_\pm \simeq (K-M) \frac{c^{(j-1)}_\pm}{\sqrt{(c^{(j-1)}_\pm)^2-1}} \log \epsilon
\label{eq:lead_behaviour_of_b_j-branch_points_j_even_2_missing_Mn0}
\end{equation}
for $j = 2, \ldots, M/2$, which then implies:
\begin{equation}
 b^{(k)}_\pm - c^{(k)}_\pm \sim \epsilon^\alpha \qquad \textrm{as } \epsilon \to 0
\label{eq:lead_vanishing_order_in_b_j-branch_points_j-even}
\end{equation}
for some $\alpha>0$ and $k = 2, 4, \ldots, M-2$.

We are left with $b^{(M)}_\pm$, whose behaviour can be analysed by studying $\mathcal{A}_{(K-M)/2}^\pm$. These A-cycles lie partly on $\tilde{\Sigma}_1$ and partly on $\tilde{\Sigma}_2$, and thus the corresponding period conditions must be treated similarly to the matching condition. As usual, we turn the integral into twice an open chain:
\begin{equation}
 0 = \oint_{\mathcal{A}_{(K-M)/2}^\pm} dp = 2 \int_{b^{(M)}_\pm}^{a^{(1)}_\pm} dp \:.
\label{eq:turning_A_(K-M)/2^pm_into_open_chain_for_behaviour_of_b^M_pm}
\end{equation}
We then split $dp$ into the two terms which go to $d \tilde{p}_1$ and $d \hat{p}_2$ as $\rho \to \infty$; as we are only interested in the leading order, we can directly substitute the limit of the differential in the integrand:
\begin{equation}
 0 = 2 \int_{\epsilon b^{(M)}_\pm}^{\tilde{a}^{(1)}_\pm} d \tilde{p}_1 + 2 \int_{b^{(M)}_\pm}^{\rho \tilde{a}^{(1)}_\pm} d \hat{p}_2 \:.
\label{eq:splitting_A_(K-M)/2^pm_into_dp1tilde_and_dp2hat}
\end{equation}
The first integral can be evaluated as we did in the case of $I_1$, defined in \eqref{eq:def_I1_I2_Mn0}:
\begin{equation}
 \int_{\epsilon b^{(M)}_\pm}^{\tilde{a}^{(1)}_\pm} d \tilde{p}_1 \simeq \tilde{p}_1 ( \epsilon b^{(M)}_\pm ) \simeq - i (K-M) \log \epsilon
\label{eq:lead_order_of_dp1tilde_part_of_matching_A-cycles}
\end{equation}
and, as we may have expected, the leading order is exactly the same.

The second integral is similar to the previous A-cycles we considered: it receives a diverging contribution from $dw_0$ at both endpoints and a diverging contribution from $dw_{M/2}^\pm$ at $x = b^{(M)}_\pm = c^{(M/2)}_\pm \pm \epsilon^{(M/2)}_\pm$:
\begin{equation}
 \int_{c^{(M/2)}_\pm \pm \epsilon^{(M/2)}_\pm}^{\rho \tilde{a}^{(1)}_\pm} d \hat{p}_2 \simeq i (K-M) \left[ 1 - 
  \frac{c^{(M/2)}_\pm}{\sqrt{(c^{(M/2)}_\pm)^2 - 1}} \right] \log \epsilon - \frac{1}{i} \log \frac{1}{\epsilon^{(M/2)}_\pm} \:.
\label{eq:lead_order_of_dp2hat_part_of_matching_A-cycles}
\end{equation}
\eqref{eq:turning_A_(K-M)/2^pm_into_open_chain_for_behaviour_of_b^M_pm} then implies:
\begin{equation}
 \log \epsilon^{(M/2)}_\pm \simeq (K-M) \frac{c^{(M/2)}_\pm}{\sqrt{(c^{(M/2)}_\pm)^2 - 1}} \log \epsilon
\label{eq:lead_behaviour_of_b_M_pm-branch_points_Mn0}
\end{equation}
which allows us to extend \eqref{eq:lead_vanishing_order_in_b_j-branch_points_j-even} to $k = M$.


\chapter[Calculations concerning explicit solutions in $AdS_3$]{Calculations supplementing the analysis of explicit solutions in $AdS_3$}

\section{Conventions for elliptic integrals and functions}
\label{sec:elliptic_integrals}

The incomplete elliptic integrals of the first, second and third kind are respectively defined as
\begin{eqnarray}
 F (z, k) & = & \int_0^z \frac{dt}{\sqrt{1 - k^2 \sin^2 t}} \nn\\
 E (z, k) & = & \int_0^z \sqrt{1 - k^2 \sin^2 t} \: dt \nn\\
 \Pi (n, z, k) & = & \int_0^z \frac{dt}{(1 - n \sin^2 t)\sqrt{1 - k^2 \sin^2 t}}
\label{eq:def_incomplete_elliptic_integrals}
\end{eqnarray}
and the corresponding complete integrals are given by
\begin{eqnarray}
 \mathbb{K} (k) & = & F \left( \frac{\pi}{2} , k \right) \nn\\
 \mathbb{E} (k) & = & E \left( \frac{\pi}{2} , k \right) \nn\\
 \Pi (n, k) & = & \Pi \left( n, \frac{\pi}{2}, k \right) \:,
\label{eq:def_complete_elliptic_integrals}
\end{eqnarray}
where the elliptic modulus $k$ satisfies $0<k<1$.

We also define:
\begin{eqnarray}
 \mathbb{K}' (k) & = & \mathbb{K} (k') \nn\\
 \mathbb{E}' (k) & = & \mathbb{E} (k') \:,
\label{eq:def_E'_K'}
\end{eqnarray}
with $k'^2 = 1-k^2$.

The Jacobi amplitude function is the inverse function of the elliptic integral of the first kind
\begin{equation}
 z = \mathrm{am} \: (w|k) \quad \Leftrightarrow \quad w = F(z|k)
\label{eq:def_Jacobi_amplitute}
\end{equation}
and it is related to the elliptic sine and cosine functions by
\begin{eqnarray}
 \mathrm{sn} \: (z|k) & = & \sin ( \mathrm{am} \: (z|k) ) \nn\\
 \mathrm{cn} \: (z|k) & = & \cos ( \mathrm{am} \: (z|k) ) \:.
\label{eq:def_sn_cn}
\end{eqnarray}
The derivative of the amplitude function with respect to its first argument defines the $\mathrm{dn}$ function:
\begin{equation}
 \mathrm{dn} \: (z|k) = \frac{\partial \, \mathrm{am} \: (z|k)}{\partial z} \:.
\label{eq:def_dn}
\end{equation}
Starting from this fundamental triplet, we can define the remaining Jacobi elliptic functions:
\begin{equation}
 \mathrm{ns} \: (z|k) = \frac{1}{\mathrm{sn} \: (z|k)} \qquad \mathrm{nc} \: (z|k) = \frac{1}{\mathrm{cn} \: (z|k)} \qquad
  \mathrm{nd} \: (z|k) = \frac{1}{\mathrm{dn} \: (z|k)}
\label{eq:def_ns_nc_nd}
\end{equation}
and
\begin{eqnarray}
 \mathrm{cd} \: (z|k) = \frac{\mathrm{cn} \: (z|k)}{\mathrm{dn} \: (z|k)} & &
  \mathrm{dc} \: (z|k) = \frac{\mathrm{dn} \: (z|k)}{\mathrm{cn} \: (z|k)} \nn\\
 \mathrm{cs} \: (z|k) = \frac{\mathrm{cn} \: (z|k)}{\mathrm{sn} \: (z|k)} & &
  \mathrm{sc} \: (z|k) = \frac{\mathrm{sn} \: (z|k)}{\mathrm{cn} \: (z|k)} \nn\\
 \mathrm{ds} \: (z|k) = \frac{\mathrm{dn} \: (z|k)}{\mathrm{sn} \: (z|k)} & &
  \mathrm{sd} \: (z|k) = \frac{\mathrm{sn} \: (z|k)}{\mathrm{dn} \: (z|k)} \:.
\label{eq:def_cd_dc_cs_sc_ds_sd}
\end{eqnarray}

\paragraph{Properties.} The incomplete elliptic integrals all satisfy the following pseudo-periodicity properties:
\begin{eqnarray}
 F (z + j \pi, k) & = & 2j \mathbb{K} (k) + F(z,k) \nn\\
 E (z + j \pi, k) & = & 2j \mathbb{E} (k) + E(z,k) \nn\\
 \Pi (n, j \pi + z, k) & = & 2j \Pi (n,k) + \Pi(n,z,k) \:.
\label{eq:pseudo-periodicities_of_elliptic_integrals}
\end{eqnarray}
The amplitude function is pseudo-periodic with respect to $z$ with pseudo-period $2 \mathbb{K} (k)$, and it is periodic with respect to $z$ with period $2 i \mathbb{K}' (k)$
\begin{equation}
 \mathrm{am} \: (z + 2j \mathbb{K} (k) + 2 \, i \, l \mathbb{K}' (k) |k) = \mathrm{am} \: (z|k) + j \pi \:, \qquad \forall j,l \in \mathbb{Z} \:,
\label{eq:pseudo_periodicity_of_am}
\end{equation}
while the elliptic functions \eqref{eq:def_sn_cn} and \eqref{eq:def_dn} are all doubly periodic with respect to $z$, although not all of them have the same periods:
\begin{eqnarray}
 \mathrm{sn} \: (z + 2j \mathbb{K} (k) + 2 \, i \, l \mathbb{K}' (k) |k) & = & (-1)^j \mathrm{sn} \: (z|k) \:, \qquad \forall j,l \in \mathbb{Z}
  \nn\\
 \mathrm{cn} \: (z + 2j \mathbb{K} (k) + 2 \, i \, l \mathbb{K}' (k) |k) & = & (-1)^{j+l} \mathrm{cn} \: (z|k) \:, \qquad \forall j,l
  \in \mathbb{Z} \nn\\
 \mathrm{dn} \: (z + 2j \mathbb{K} (k) + 2 \, i \, l \mathbb{K}' (k) |k) & = & (-1)^l \mathrm{cn} \: (z|k) \:, \qquad \forall j,l
  \in \mathbb{Z} \:.
\label{eq:period+half-period_sn_cn_dn}
\end{eqnarray}
We also list the quarter-period transformation rules:
\begin{eqnarray}
 \mathrm{sn} \: (z + \mathbb{K} (k) |k) = \mathrm{cd} \: (z|k) & &
  \mathrm{sn} \: (z + i \mathbb{K}' (k) |k) = \frac{1}{k} \mathrm{ns} \: (z|k) \nn\\
 \mathrm{cn} \: (z + \mathbb{K} (k) |k) = - k' \mathrm{sd} \: (z|k) & &
  \mathrm{cn} \: (z + i \mathbb{K}' (k) |k) = - \frac{i}{k} \mathrm{ds} \: (z|k) \nn\\
 \mathrm{dn} \: (z + \mathbb{K} (k) |k) = k' \mathrm{nd} \: (z|k) & &
  \mathrm{dn} \: (z + i \mathbb{K}' (k) |k) = -i \mathrm{cs} \: (z|k) \:.
\label{eq:quarter-period_sn_cn_dn}
\end{eqnarray}
The corresponding identities for all the remaining elliptic functions may be deduced from these. For Taylor expansions and other properties, see for instance \cite{Byrd::1971}.

\section{Gauge transformation for the Kruczenski solution}
\label{sec:app_gauge_transf_for_K}

The Kruczenski spiky string is described by the following ansatz: $t = \bar{\tau}$, $\rho = \rho(\bar{\sigma})$, $\phi = \omega \bar{\tau} + \bar{\sigma}$, which guarantees that all the equations of motion from the Nambu-Goto action are satisfied if $\rho(\bar{\sigma})$ solves the following:
\begin{equation}
 \rho' = \pm \frac{1}{2} \frac{\sinh 2\rho}{\sinh 2\rho_0} \frac{\sqrt{\sinh^2 2\rho - \sinh^2 2\rho_0}}{\sqrt{\cosh^2 \rho - \omega^2 \sinh^2
  \rho}} \:,
\label{eq:Nambu-Goto_eom_Kruc}
\end{equation}
where $\rho_0$ is an integration constant. The requirement of reality placed upon $\rho$ forces $\rho_0 \leq \rho \leq \rho_1$, where $\coth \rho_1 = \omega$. From now on, we will refer to the function which solves equation \eqref{eq:Nambu-Goto_eom_Kruc} as $\hat{\rho}(\bar{\sigma})$. It is possible to integrate \eqref{eq:Nambu-Goto_eom_Kruc} to get the inverse function $\bar{\sigma}(\hat{\rho})$:
\begin{equation}
 \bar{\sigma} = \pm \frac{\sinh 2\rho_0}{\sqrt{2}\sqrt{w_0 + w_1} \sinh\rho_1} \left\{ \Pi \left( \frac{w_1-w_0}{w_1-1}, \beta, p \right)
  - \Pi \left( \frac{w_1-w_0}{w_1+1}, \beta, p \right) \right\} \:,
\label{eq:sigma_of_rho_Kruc_Nambu-Goto}
\end{equation}
where
\begin{equation}
 p \equiv \sqrt{\frac{w_1-w_0}{w_1+w_0}} \:, \qquad \sin\beta \equiv \sqrt{\frac{w_1-w(\hat{\rho})}{w_1-w_0}}
\label{eq:p_and_beta_in_Kruc_Nambu-Goto}
\end{equation}
($\beta \in [0,\pi/2]$) and we define $w(x) \equiv \cosh (2x)$, $w_0 \equiv \cosh 2\rho_0$ and $w_1 \equiv \cosh 2\rho_1$.

We can construct a spiky string from this object by taking \eqref{eq:sigma_of_rho_Kruc_Nambu-Goto} with the plus sign and then replacing $\beta$ with the new coordinate $\bar{\sigma}'$:
\begin{equation}
 \bar{\sigma} = \frac{\sinh 2\rho_0}{\sqrt{2}\sqrt{w_0 + w_1} \sinh\rho_1} \left\{ \Pi \left(\frac{w_1-w_0}{w_1-1}, \sigma', p \right)
  - \Pi \left( \frac{w_1-w_0}{w_1+1}, \bar{\sigma}', p \right) \right\} \:.
\label{eq:def_sigmaprime_K}
\end{equation}
While \eqref{eq:def_sigmaprime_K} implies that $\bar{\sigma}$ is an increasing function of $\bar{\sigma}'$ (with $\bar{\sigma}(\bar{\sigma}'=0)=0$), \eqref{eq:p_and_beta_in_Kruc_Nambu-Goto} allows us to express $\hat{\rho}$ as a function of $\bar{\sigma}'$:
\begin{equation}
 \sinh^2 \hat{\rho} = \sinh^2 \rho_1 \cos^2 \bar{\sigma}' + \sinh^2 \rho_0 \sin^2 \bar{\sigma}' \:.
\label{eq:rhohat_of_sigmaprime_K}
\end{equation}
From \eqref{eq:def_sigmaprime_K}, we see that, for each increase of $\pi/2$ in $\bar{\sigma}'$, $\bar{\sigma}$ and consequently $\phi$ increase by:
\begin{equation}
 \Delta\phi = \frac{\sinh 2\rho_0}{\sqrt{2}\sqrt{w_0 + w_1} \sinh\rho_1} \left\{ \Pi \left( \frac{w_1-w_0}{w_1-1},p \right) - \Pi \left(
  \frac{w_1-w_0}{w_1+1},p \right) \right\} \:.
\label{eq:def_deltaphi_K}
\end{equation}
Thus, for the string to be closed at fixed $t=\bar{\tau}$, we allow $\bar{\sigma}'$ to vary in $[0,K \pi]$, $K \in \mathbb{N}$ (i.e. $\bar{\sigma} \in [0,2 K \Delta\phi]$), and then demand that the corresponding total increase in $\phi$ be an integer multiple of $2\pi$: $2 K \Delta\phi = 2n\pi$. As expected, \eqref{eq:def_deltaphi_K} matches \eqref{eq:deltaphi_for_JJ}.

We are now ready to discuss the worldsheet coordinate transformation which maps this solution onto the corresponding conformal gauge version \eqref{eq:rho_of_sigmatilde_JJ}, \eqref{eq:f_and_g_of_sigmatilde_JJ}. In order to find it, we just need to impose the equality of the global coordinates $(t,\rho,\phi)$ specified by the two different versions of the ansatz, which leads to the following set of relations:
\begin{equation}
 \tilde{\tau} + f(\tilde{\sigma}) = \bar{\tau} \:, \qquad g(\tilde{\sigma}) - \omega f(\tilde{\sigma}) = \bar{\sigma} \:, \qquad
   \rho(\tilde{\sigma}) = \hat{\rho}(\bar{\sigma}) \:.
\label{eq:equality_of_global_coords_in_two_versions_of_ansatz_K_JJ}
\end{equation}
These are actually three conditions on two unknown functions $\bar{\tau}(\tilde{\tau},\tilde{\sigma})$, $\bar{\sigma}(\tilde{\tau},\tilde{\sigma})$, and we easily see that they give two potentially conflicting expressions for $\bar{\sigma}(\tilde{\tau},\tilde{\sigma})$. For the transformation to exist, these must coincide:
\begin{equation}
 g(\tilde{\sigma}) - \omega f(\tilde{\sigma}) = \hat{\rho}^{-1}(\rho(\tilde{\sigma})) \:.
\label{eq:compatibility_for_coord_transform_K_JJ}
\end{equation}
We already have the inverse of $\hat{\rho}$ from \eqref{eq:sigma_of_rho_Kruc_Nambu-Goto}. We can then compute $w(\rho(\tilde{\sigma})) = \cosh 2 \rho(\tilde{\sigma})$ from \eqref{eq:rho_of_sigmatilde_JJ} and then use it to find:
\begin{equation}
 \sin^2 \bar{\sigma}' = \frac{w_1 - w(\rho(\tilde{\sigma}))}{w_1 - w_0} = \mathrm{sn}^2(v|k) \:.
\label{eq:sin_beta_checking_compatibility_of_coord_transform_K_JJ}
\end{equation}
Remembering that $\bar{\sigma}' \in [0,K \pi]$, it is natural to identify $\bar{\sigma}' = \mathrm{am}(v|k)$. Therefore, by substituting this into \eqref{eq:def_sigmaprime_K}, we get:
\begin{equation}
 \hat{\rho}^{-1} (\rho(\tilde{\sigma})) = \frac{\sinh 2\rho_0}{\sqrt{2} \sinh \rho_1 \sqrt{w_1+w_0}} \left\{ \Pi (n_-, \mathrm{am} (v|k), k) -
  \Pi (n_+, \mathrm{am} (v|k), k) \right\} \:.
\label{eq:sigma_from_rhohatinverse_and_rho}
\end{equation}
It is now only a matter of simple algebra to show that this expression matches $g(\tilde{\sigma}) - \omega f(\tilde{\sigma})$, i.e. that the last two conditions in \eqref{eq:equality_of_global_coords_in_two_versions_of_ansatz_K_JJ} are equivalent, and thus that the coordinate transformation exists. Its explicit form is the following:
\begin{eqnarray}
 \bar{\tau} & = & \tilde{\tau} + \frac{\sqrt{2}\omega\sinh 2\rho_0 \sinh \rho_1}{(w_1+1) \sqrt{w_0+w_1}} \Pi (n_+, \mathrm{am} (v|k), k)
  \nonumber \\
 \bar{\sigma} & = & \frac{\sinh 2\rho_0 }{\sqrt{2} \sqrt{w_0+w_1} \sinh \rho_1} \left\{ \Pi (n_-, \mathrm{am} (v|k), k)
  - \Pi (n_+, \mathrm{am} (v|k),k) \right\} \nn \\
\label{eq:coord_transf_between_NG_and_conf_gauge_K}
\end{eqnarray}

It is also possible to determine the worldsheet metric $h_{ab}$ from Kruczenski's parametrization and then show that \eqref{eq:coord_transf_between_NG_and_conf_gauge_K} brings it to the 2-dimensional Minkowski metric, up to a conformal transformation. We recall that the Nambu-Goto action is obtained from the general $\sigma$-model action by substituting the equations of motion for $h_{ab}$ into it:
\begin{equation}
 \partial_\mu X_a \partial^\mu X_b = \frac{1}{2} h_{ab} h^{cd} \partial_\mu X_c \partial^\mu X_d \:.
\label{eq:eom_for_worldsheet_metric_h}
\end{equation}
We can then invert these equations to find $h_{ab}$ as a function of $\partial_\mu X_a \partial^\mu X_b$, up to an overall rescaling factor (the combination $h_{ab} h^{cd}$ is clearly conformally invariant):
\begin{equation}
 h_{ab} = a(\tau,\sigma) \begin{pmatrix}
                          \dot{X}^2           & \dot{X}^\mu X'_\mu \\
                          \dot{X}^\mu X'_\mu  & X'^2
                         \end{pmatrix} \:.
\label{eq:worldsheet_metric_h_in_a_generic_gauge}
\end{equation}
Now, if a worldsheet coordinate transformation $(\bar{\tau},\bar{\sigma}) \to (\tilde{\tau},\tilde{\sigma})$ brings this metric to conformal gauge, i.e. if it makes it diagonal and traceless (the overall scaling factor can then be eliminated by a conformal transformation), then it must satisfy the following set of conditions:
\begin{eqnarray}
 0 & = & h_{00} \left[ \left( \frac{\partial \bar{\tau}}{\partial\tilde{\tau}} \right)^2
  + \left( \frac{\partial \bar{\tau}}{\partial\tilde{\sigma}} \right)^2 \right]
   + 2h_{01} \left[ \left( \frac{\partial \bar{\tau}}{\partial\tilde{\tau}} \right) \left( \frac{\partial \bar{\sigma}}{\partial\tilde{\tau}}
    \right) + \left( \frac{\partial \bar{\tau}}{\partial\tilde{\sigma}} \right) \left( \frac{\partial \bar{\sigma}}{\partial\tilde{\sigma}}
     \right) \right] 
    \nonumber \\
   & & + h_{11} \left[ \left( \frac{\partial \bar{\sigma}}{\partial\tilde{\tau}} \right)^2 + \left( \frac{\partial
    \bar{\sigma}}{\partial\tilde{\sigma}} \right)^2 \right] \nonumber \\
 0 & = & h_{00} \left( \frac{\partial \bar{\tau}}{\partial\tilde{\tau}} \right) \left( \frac{\partial \bar{\tau}}{\partial\tilde{\sigma}}
  \right) + h_{01} \left[ \left( \frac{\partial \bar{\tau}}{\partial\tilde{\tau}} \right)
   \left( \frac{\partial \bar{\sigma}}{\partial\tilde{\sigma}} \right) + \left( \frac{\partial \bar{\sigma}}{\partial\tilde{\tau}} \right)
    \left( \frac{\partial\bar{\tau}}{\partial\tilde{\sigma}} \right) \right] \nonumber \\
   &   & + h_{11} \left( \frac{\partial \bar{\sigma}}{\partial\tilde{\tau}} \right) \left( \frac{\partial \bar{\sigma}}{\partial\tilde{\sigma}}
    \right)
\label{eq:conditions_on_coord_transf_NG_to_conf_gauge} \nonumber
\end{eqnarray}
All the required derivatives can be obtained from the first two equations \eqref{eq:equality_of_global_coords_in_two_versions_of_ansatz_K_JJ}, and then it is just a matter of algebra to check that, as expected, these equations are verified.


\section[Computing $\hat{q}^{(0)}_2$]{Computing $\hat{q}^{(0)}_2$ for the N-folded GKP and the Kruczenski solutions}
\label{sec:app_computing_qtilde_2_for_N-folded_GKP_and_K}

The easiest way of computing $\hat{q}^{(0)}_2$ in both cases is by using equation \eqref{eq:qktilde_from_spin_chain_patched_K}, which yields all conserved charges for the patched Kruczenski solution. This solution is discussed in detail in section \ref{sec:patching_K}; here we simply recall that it allows arbitrary angular separations $0 < \Delta\theta_j < \pi$ between each pair of consecutive cusps. As we previously observed in section \ref{sec:patching_K}, all results concerning the spectral curve of this generalised solution reduce to those obtained for the Kruczenski spiky string if we set $\Delta\theta_j = 2 n \pi/K$, $\forall j$, where $n$ is a natural number counting how many times the Kruczenski string winds around the centre of $AdS_3$. Furthermore, we saw at the end of section \ref{sec:JJ} that, by setting $n = K/2$, these results in turn reduce to those associated with the N-folded GKP case, where $2 N = K$. Therefore, we can compute the conserved charge $\hat{q}^{(0)}_2$ by specialising the general expression \eqref{eq:qktilde_from_spin_chain_patched_K} to the desired simpler case.

We start by evaluating it for $k=2$:
\begin{eqnarray}
 \hat{q}^{(0)}_2 & = & \frac{4}{K^2} \sum_{1 \leq j_1 < j_2 \leq K} \sin \left( \frac{\theta_{j_2} - \theta_{j_1}}{2} \right)
  \sin \left( \frac{\theta_{j_1} - \theta_{j_2}}{2} \right) \nonumber \\
 & = & - \frac{4}{K^2} \sum_{1 \leq j_1 < j_2 \leq K} \sin^2 \left( \frac{1}{2} \sum_{l = j_1 + 1}^{j_2} \Delta\theta_l \right) \:,
\label{eq:computing_qtilde_2_for_K_1}
\end{eqnarray}
where we have used $\theta_m\equiv \sum_{j=1}^m \Delta\theta_j$.

We now specialise to the Kruczenski case, by setting $\Delta\theta_j = 2 n \pi/K$, $\forall j$, which implies
\begin{equation}
 \sum_{l = j_1 + 1}^{j_2} \Delta\theta_l = \frac{2 n \pi}{K} (j_2 - j_1) \:.
\label{eq:simplification_inside_qtilde_2_in_K_case}
\end{equation}
By substituting this into \eqref{eq:computing_qtilde_2_for_K_1} and introducing the new index $m = j_2 - j_1$, which replaces $j_2$, we obtain:
\begin{eqnarray}
 \hat{q}^{(0)}_2 & = & - \frac{4}{K^2} \sum_{j_1=1}^{K-1} \sum_{m=1}^{K-j_1} \sin^2 \left( \frac{n \pi}{K} m \right)
  = - \frac{2}{K^2} \sum_{j_1=1}^{K-1} \sum_{m=1}^{K-j_1} \left[ 1 - \cos \left( \frac{2 n \pi}{K} m \right) \right] \nonumber \\
 & = & - \frac{2}{K^2} \sum_{j_1=1}^{K-1} \left[ K - j_1 - \sum_{m=1}^{K-j_1} \cos \left( m \frac{2 n \pi}{K} \right) \right] \:.
\label{eq:computing_qtilde_2_for_K_2}
\end{eqnarray}
We now use the general result for the Dirichlet kernel,
\begin{equation}
 1 + 2 \sum_{k=1}^n \cos (k x) = \frac{\sin \left[ \left( n + \frac{1}{2} \right) x \right]}{\sin \left( \frac{x}{2} \right)} \:,
\label{eq:general_result_sum_of_cosines}
\end{equation}
to calculate the last remaining sum over $m$:
\begin{eqnarray}
 \hat{q}^{(0)}_2 & = & - \frac{2}{K^2} \sum_{j_1=1}^{K-1} \left\{ K - j_1 - \frac{1}{2} \left[ \frac{\sin \left[ \left( K - j_1 + \frac{1}{2}
  \right) \frac{2 n \pi}{K} \right]}{\sin \left( \frac{n \pi}{K} \right)} - 1 \right] \right\} \nonumber \\
 & = & - \frac{2}{K^2} \sum_{j_1=1}^{K-1} \left\{ K - j_1 + \frac{1}{2} - \frac{1}{2} \frac{\sin \left(  \frac{n \pi}{K}
  - j_1 \frac{2n \pi}{K} \right)}{\sin \left( \frac{n \pi}{K} \right)} \right\} \nonumber \\
 & = & - \frac{2}{K^2} \sum_{j_1=1}^{K-1} \left\{ K - j_1 + \frac{1}{2} - \frac{1}{2} \left[ \cos \left( j_1 \frac{2 n \pi}{K}
  \right) - \cot \left( \frac{n \pi}{K} \right) \sin \left( j_1 \frac{2 n \pi}{K} \right) \right] \right\} \nonumber \\
 & = & \frac{1}{K^2} \left[ - 2 \left( K + \frac{1}{2} \right) (K-1) + (K-1) K + \sum_{j_1=1}^{K-1} \cos \left( j_1 
  \frac{2 n \pi}{K} \right) \right. \nonumber \\
 & & - \left. \cot \left( \frac{n \pi}{K} \right) \sum_{j_1=1}^{K-1} \sin \left( j_1 \frac{2 n \pi}{K} \right) \right] \:.
\label{eq:computing_qtilde_2_for_K_3}
\end{eqnarray}
At this point, we can evaluate the sums over $j_1$ by using \eqref{eq:general_result_sum_of_cosines} and the analogous result
\begin{equation}
 \sum_{k=1}^n \sin (k x) = \frac{\sin x + \sin (nx) - \sin[(n+1) x]}{2(1 - \cos x)} \:,
\label{eq:general_result_sum_of_sines}
\end{equation}
thus obtaining
\begin{eqnarray}
 \hat{q}^{(0)}_2 & = & - 1 + \frac{1}{K^2} + \frac{1}{2 K^2} \left\{ \frac{\sin \left[ \left( K - \frac{1}{2} \right) \frac{2 n \pi}{K}
  \right]}{\sin \left( \frac{n \pi}{K} \right)} - 1 \right\} \nonumber \\
 & & - \cot \left( \frac{n \pi}{K} \right) \frac{\sin \left( \frac{2 n \pi}{K} \right) + \sin \left[ (K-1) \frac{2 n
  \pi}{K} \right] - \sin (2 n \pi) }{2K^2 \left[ 1 - \cos \left( \frac{2 n \pi}{K} \right) \right]} \nonumber\\
 & = & - 1 \:.
\label{eq:computing_qtilde_2_for_K_4}
\end{eqnarray}
Since the result is independent of $n$, it also holds for the N-folded GKP case.

One may wonder whether a similar relation exists for the patched Kruczenski solution, but the answer is negative: as a counter-example, we study the case $K=3$, with $\Delta\theta_1 = 5\pi/6$, $\Delta\theta_2 = 2\pi/3$, $\Delta\theta_3 = \pi/2$ and consequently $n = 1$. It is easy to check from \eqref{eq:computing_qtilde_2_for_K_1} that:
\begin{equation}
 \hat{q}^{(0)}_2 = - \frac{4}{9} \left[ \sin^2 \left( \frac{\Delta\theta_2}{2} \right) + \sin^2 \left( \frac{\Delta\theta_2 +
  \Delta\theta_3}{2} \right) + \sin^2 \left( \frac{\Delta\theta_3}{2} \right) \right] = - \frac{7 + \sqrt{3}}{9} \:.
\label{eq:counter-example_for_patched_K}
\end{equation}
The only property that continues to hold for the patched solution is the fact that $\hat{q}^{(0)}_2 < 0$, as we can easily see from \eqref{eq:computing_qtilde_2_for_K_1}.

\bibliographystyle{utphys}
\bibliography{Bibliography}

\end{document}